%% file: main.tex
\documentclass[
3p,
singlecolumn,
times,
11pt,
amssymb,
preprintnumbers,
secnumarabic,
nofootinbib,
superscriptaddress
]{elsarticle}

\pdfoutput=1

\usepackage{graphicx}
\usepackage{enumitem}
\usepackage{latexsym}
\usepackage{amsfonts}
\usepackage{amssymb}
\usepackage{color}

\usepackage{amsmath}
\usepackage{slashed}
\usepackage{dcolumn}
\usepackage{verbatim}
\usepackage{float}
\usepackage{multirow}
\usepackage{xspace}
\usepackage[normalem]{ulem}
\usepackage[pdfauthor={Nirmal Raj}]{hyperref}

\definecolor{greeny}{rgb}{0.3,0.7,0.3}

\definecolor{hypershade}{rgb}{0.8,0.3,0.3}
\hypersetup{
  pdfauthor={Nirmal Raj},
  pdftitle={Zwicky's babies},
  pdfsubject={Zwicky's babies},
  colorlinks=true,
  citecolor=magenta,
  urlcolor=blue,
  linkcolor=hypershade
}

\input{universalnewcommands.tex}

\def\RNS{R_{\rm NS}}
\def\MNS{M_{\rm NS}}
\def\mdm{m_{\chi}}

\def\mneff{\bar m_n}
\def\sigmageom{\sigma_0}

\def \tcool {t_{\rm cool}}

\def \Tcoolz {\tilde{T}_{\rm cool}}

\allowdisplaybreaks

\begin{document}

\title{Dark matter in compact stars}

\author{Joseph Bramante}
\ead{joseph.bramante@queensu.ca}
\address{Department of Physics, Engineering Physics, and Astronomy, Queen's University, Kingston, Ontario, K7N 3N6, Canada}
\address{The Arthur B. McDonald Canadian Astroparticle Physics Research Institute, Kingston, Ontario, K7L 3N6, Canada}
\address{Perimeter Institute for Theoretical Physics, Waterloo, Ontario, N2L 2Y5, Canada}

\author{Nirmal Raj}
\ead{nraj@iisc.ac.in}
\address{Centre for High Energy Physics, Indian Institute of Science,
C.V. Raman Avenue, Bengaluru 560012, India}

\date{\today}

\begin{abstract}
White dwarfs and neutron stars are far-reaching and multi-faceted laboratories in the hunt for dark matter.
We review detection prospects of wave-like, particulate, macroscopic and black hole dark matter that make use of several exceptional properties of compact stars, such as ultra-high densities, deep fermion degeneracies, low temperatures, nucleon superfluidity, strong magnetic fields, high rotational regularity, and significant gravitational wave emissivity.
Foundational topics first made explicit in this document include the effect of the  ``propellor phase" on neutron star baryonic accretion, and the contribution of Auger and Cooper pair breaking effects to neutron star heating by dark matter capture.
\end{abstract}

\maketitle

\tableofcontents

\section{Introduction}  \vspace{0.2cm}
\label{sec:intro}

Dark matter is one of the foremost scientific mysteries of our times~\cite{DMreview:Cirelli:2024ssz}.
Given how little is known about its microphysical properties, its possible identities seem limitless.
This is famously encapsulated in the 90+ orders of magnitude that dark matter (DM) masses could span, from $10^{-24}$~eV, set by the maximum possible Compton wavelength containable within a dwarf galaxy, 
to $10^8 M_\odot \simeq 10^{74}$~eV, the mass of DM in a small galaxy.
Over this range of masses DM may be described as a wave/field, a particle, a macroscopic object, or galactic substructure -- including black holes and topological defects.
A promising strategy to confront such remarkable diversity in possibility is to exploit physical systems with remarkable diversity in characteristics.

Compact stars -- white dwarfs (WDs) and neutron stars (NSs) typically formed as relics of nuclear-powered stars -- afford such an environment.
Since their quantum properties were first described in the 1920s$-$30s by 
Fowler~\cite{WDhistory:Fowler:1926}, 
Anderson~\cite{WDhistory:Anderson:1929},
Stoner~\cite{WDhistory:Stoner:1930}, 
Chandrasekhar~\cite{WDhistory:Chandra:1931}, 
Zwicky and Baade~\cite{NShistory:BaadeZwicky:1934zex} (and possibly Landau~\cite{NShistroy:LandauInterpreted2013}), our understanding of compact stars has been enriched at the intersection of several branches of physics: astrophysics, general relativity, particle physics, nuclear physics, statistical physics, thermodynamics, and plasma physics.
It is little wonder that they feature in numerous tests of fundamental physics~\cite{WDfundamental:Isern:2022vdx, NSfundamental:Nattila:2022evn}, and it should come as no surprise that they are also ideal laboratories to search for dark matter.
Indeed, DM hunters would do well to take advantage of their striking properties: they have very high densities, with accompanying steep gravitational potentials, sometimes deeply degenerate constituent fermions, often very low temperatures, the presence of nucleon superfluidity, ultra-strong magnetic fields, extreme regularity in rotation rivaling the precision of atomic clocks, and powerful gravitational radiation emitted during binary mergers, to name a few.   

The use of stars to look for evidence of DM dates to proposals that weakly interacting particle DM might alter nuclear reaction rates in the Sun~\cite{Press:1985ug,Gould:1987ir}. 
Shortly after, it was realized that NSs were useful for seeking out certain models of DM that could form black holes in their interior~\cite{Goldman:1989nd}.
One immediate difference between a search for DM in compact stars and a terrestrial detector is that, since DM is accelerated in the deep gravitational well of a compact star, its interactions with stellar constituent particles occur at semi-relativistic velocities: $\Oc(10^{-2}-10^{-1})~c$ for a WD and $\Oc(0.5)~c$ for a NS.
This high DM speed provides enhanced sensitivity to theoretical models with velocity-suppressed rates for scattering on Standard Model (SM) states, since in the Milky Way's Galactic halo (and by extension in terrestrial detectors) the velocity of DM particles is only $\Oc(10^{-3}) c$. 
In particular, the environs of a NS are greatly suited to testing the origin of DM, since the kinetic energy of DM at speeds $\sim 0.7c$ are similar to that during cosmological production, particularly for ``freeze-out" processes~\cite{Bertone:2004pz}.
See Ref.~\cite{Leane:2024bvh} for a study of the optimality of compact stars vs other celestial bodies for discovering particle DM.

This review is organized as follows.
In Section~\ref{sec:physicscompact} we provide an overview of the properties of NSs and WDs, emphasizing aspects that will be important for dark matter searches.
In Section~\ref{sec:WDvDM}, we describe WD searches for dark matter, treating dark matter annihilation and heating of WDs, conversion of WDs to black holes, ignition of Type Ia supernovae, and effects of dark matter on WD equations of state.
In Section~\ref{sec:NSvDM}, we describe NS searches for dark matter, including dark matter heating NSs kinetically and via annihilations, models of dark matter that convert NSs to black holes, exotic compact stars that constitute dark matter, NSs admixed with dark matter,  models of dark matter that lead to internal heating of NSs, signals of dark matter in NS-related gravitational waves and pulsar timing, and the utility of NSs in discovering axion-like and primordial black hole dark matter.
In Section~\ref{sec:concs}, we briefly discuss future research directions for dark matter in compact stars.

\section{The physics of compact objects} \vspace{0.2cm}
\label{sec:physicscompact}

A detailed account of the physical characteristics of WDs and NSs is beyond the scope of this review, and for these we refer the reader to Refs.~\cite{Saumon:2022gtu} and~\cite{LattPrakReview}. 
Here we outline key properties of these stars that make them useful dark matter detectors.

{\bf White dwarfs} are stellar remnants formed from main sequence stars that undergo a red giant phase not hot enough to fuse carbon.
Depending on its mass, a WD will be composed of some proportion of helium, carbon, oxygen, neon and magnesium, which make up the bulk of the mass.
A sea of electrons co-habiting with nuclei provide, as we will see, the Fermi degeneracy pressure that supports the WD against gravitational collapse.

Super-giant progenitors of mass around 10--25 $M_\odot$ that undergo core-collapse leave behind short-lived ``proto-NSs" through which neutrinos diffuse out carrying away 99\% of the star's binding energy, following which {\bf neutron stars} are born.
They are composed mainly of Fermi degenerate neutrons formed by electron-proton capture, $e^- + p \ra n + \nu_e$, at extreme densities and temperatures.
Due to beta chemical equilibrium, NSs are also thought to contain populations of protons, electrons, and muons; it is in fact the filled Fermi seas of these fermionic fields that keep neutrons from decaying to protons, electrons, and muons inside NSs. 
The supernova collapse is generically expected to be hydrodynamically asymmetric, resulting in a natal ``kick" to the NS at 450-1000 km/s speeds in a random direction~\cite{kick:Lyne:1994az,kick:Scheck:2003rw,kick:Scheck:2006rw,kick:Ng:2007aw,kick:Nordhaus:2010ub}; a 1\% fractional anisotropy in the momenta of escaping neutrinos could be another source of the asymmetric kick~\cite{kicknu:Kusenko:1996sr,kicknu:Kusenko:1998bk,kicknu:Barkovich:2002wh}. 
Pulsar kicks are reviewed in Ref.~\cite{ReviewKicks:LambiasePoddar:2024cjy}.

\subsection{Fermi gas model and maximum masses} \vspace{0.2cm}

Compact stars, especially WDs, are prevented from collapsing under their own gravity by Fermi degeneracy pressure.
In a low-temperature Fermi gas of Fermi momentum $p_F$, the number of fermions (of spin degeneracy $g_s =2$) filling up a real volume $V$ and Fermi sphere volume $V_F = 4\pi p_F^3/3$, is $N_f = g_s V V_F/(2\pi)^3$, from which we obtain:
\beq
p_F = (3\pi^2 n)^{1/3}~,
\label{eq:fermimomentum}
\eeq
where $n$ is the fermion number density. 
The total energy of the Fermi gas given the energy of a state $e(p)$ is
\beq
E = 4\pi g_s V \int_0^{p_F} dp p^2 e(p)~,
\label{eq:EtotFermigas}
\eeq
and for an energy density $\varepsilon = E/V$ the pressure is obtained as
\beq
P = - \bigg( \frac{\partial E}{\partial V} \bigg)_{N_f} = n^2 \frac{{\rm d}}{{\rm d}n} \bigg(\frac{\varepsilon}{n}\bigg)~.
\label{eq:pressuregen}
\eeq

Setting $e(p) = m_f + p^2/(2m_f)$ in the non-relativistic limit and $e(p) = p$ in the relativistic limit, and using Eqs.~\eqref{eq:fermimomentum},\eqref{eq:EtotFermigas} and \eqref{eq:pressuregen}, we get the Fermi degeneracy pressure of a species as
\beq
P = \begin{cases}
    [(3\pi^2)^{2/3}/5 m_f] \ n^{5/3}~,~~p_F \ll m_f~, \\ 
    [(3\pi^2)^{1/3}/4] \ n^{4/3}~~~~~,~~p_F \gg m_f~.
\end{cases}
\label{eq:pressureFermi}
\eeq
The net pressure of the compact star is the sum of the contributions of constituent species. 
In WDs, the electrons are unbound -- the electron-nucleus Coulomb energy $\simeq Z e^2 (n_e/Z)^{1/3}$ is $\Oc(10^{-2})$ times the typical kinetic energy, $(3\pi^2 n_e)^{1/3}$ -- and thus form their own Fermi gas system.
It may be seen from Eq.~\eqref{eq:EtotFermigas} that due to their lightness and abundance, it is the {\em electrons} that contribute the greatest to the pressure of WDs.
In contrast, in neutron stars the constituent neutrons contribute the most to both the stellar mass and pressure.

The total energy of the star in the non-relativistic limit is given by 
\bea 
\nn E^{\rm non-rel}_{\rm tot} &\simeq& ({\rm total \ kinetic}) - ({\rm gravitational \ binding}) \\
\nn &=&  \frac{3}{5} N_{\rm f} \frac{p_F^2}{2m_{\rm f}} - \frac{3}{5}\frac{G M_\star^2}{R_\star} \\
   &=&  \bigg(\frac{27\sqrt{3}\pi}{40\sqrt{10}}\bigg)^{2/3} \frac{1}{m_f R_\star^2} \bigg(\frac{Z M_\star}{A m_N} \bigg)^{5/3} - \frac{3}{5}\frac{G M_\star^2}{R_\star}~,
   \label{eq:Etotnonrel}
\eea
and in the relativistic limit,
\bea 
\nn E^{\rm rel}_{\rm tot} &=&  \frac{3}{4} N_{\rm f} p_F - \frac{3}{5}\frac{G M_\star^2}{R_\star} \\
   &=&  \bigg(\frac{243\pi}{256}\bigg)^{1/3} \frac{1}{R_\star} \bigg(\frac{Z M_\star}{A m_N} \bigg)^{4/3} - \frac{3}{5}\frac{G M_\star^2}{R_\star}~.
   \label{eq:Etotrel}
\eea

Eq.~\eqref{eq:Etotnonrel} shows that, in the ground state of the compact star where the virial theorem (potential energy = -2 $\times$ kinetic energy) applies, we have $R_\star \propto M^{-1/3}_\star$ as a mass-radius relation\footnote{This shows that more compact stars are generally denser, which we will see is relevant to determining the speed of DM passing through and the density of DM collected in compact stars.}. 
This implies that WDs, modeled accurately as a Fermi gas system, become smaller with increasing mass.
Hence the heaviest WDs are the densest, and thus one expects electrons in them to be ultra-relativistic and Eq.~\eqref{eq:Etotrel} to apply.
In Eq.~\eqref{eq:Etotrel} both the potential and kinetic terms fall as $R_\star^{-1}$, however the former grows faster with $M_\star$ than the latter, implying a maximum WD mass above which the star will collapse.
This ``Chandrasekhar limit"~\cite{chandlimit} (see also Sec.~\ref{subsection:thermonucrunaway}) is given by 
\beq
M_{\rm max-WD-rel} = \sqrt{\frac{5\pi}{G^3}} \frac{15}{16} \bigg(\frac{Z}{A m_N}\bigg)^2  \simeq 1.7 \ \bigg(\frac{2Z}{A}\bigg)^2 \ M_\odot~.
\label{eq:MChWDrelapprox}
\eeq
A similar limit may be obtained for NS masses by setting $A \ra 1$, $Z \ra 1$:
\beq
M_{\rm max-NS-rel} = 6.8 \ M_\odot~.
\label{eq:MChNSrelapprox}
\eeq
These estimates are not physically motivated: they assume relativistic fermions constituting the entire volume of the star (true for neither WDs nor NSs) and a non-interacting Fermi gas (not true for NSs). 
Nevertheless they ballpark the true limit to within $\Oc(1)$ factors.
A more precise treatment must account for the stellar structure, which we will discuss below, but first let us make two more estimates of the maximum mass of NSs.

(i) If we assume non-relativistic neutrons, the virial theorem using Eq.~\eqref{eq:Etotrel} gives a mass-radius relationship:
\beq
R_{\rm NS} \simeq 12 \ {\rm km} \ \bigg(\frac{M_\odot}{M_{\rm NS}}\bigg)^{1/3}~.
\eeq
In this picture the NS radius is a decreasing function of its mass; however it cannot become smaller than the Schwarzschild radius corresponding to a maximum mass, $R_{\rm Schw} = 3 \ {\rm km} \ (M/M_\odot) $.
This condition gives  
\beq
M_{\rm max-NS-nonrel} = 2.8 \ M_\odot~.
\label{eq:MChNSrelapprox}
\eeq

(ii) Due to super-nuclear densities in the NS cores, strong interactions cannot be neglected in considerations of NS structure. 
A maximum mass can be obtained in the (unphysical) limit where such interactions solely support the star against gravitational collapse~\cite{Burrows:2014fla}.
Strong interactions become repulsive at inter-nucleon distances roughly shorter than the reduced Compton wavelength of the mediating pion, $m_\pi^{-1}$.
This gives a maximum neutron density $m_N m_\pi^3$, corresponding to a mass-radius relation of $M_{\rm NS} = 4\pi m_N m^3_\pi R^3_{\rm NS}/3$. 
For a surface escape speed $v_{\rm esc}$, we have $R_{\rm NS} = R_{\rm Schw} v_{\rm esc}^{-1} =  3~{\rm km}~(M/M_\odot)~v_{\rm esc}^{-1}$.
Putting these together yields the maximum NS mass as
\beq
M_{\rm max-NS-strong} = \sqrt{\frac{3}{32\pi}} v^{3/2}_{\rm esc} \bigg(\frac{M^6_{\rm Pl}}{m_N m_\pi^3} \bigg)^{1/2} \simeq 2~M_\odot \ \bigg(\frac{v_{\rm esc}}{0.5 c}\bigg)^{3/2}~.
\label{eq:MChNSstrongapprox}
\eeq
As we will see below, this turns out to be an excellent estimate.

As argued in, e.g., Ref.~\cite{ReddySilbar:2003wm}, a more accurate reason for the existence of a maximum mass for NSs is that the sound speed $c_s$ in NS material cannot be arbitrarily large.
In particular, $c^2_s/c^2 = (\partial P/\partial \varepsilon)_{\bar s} \leq 1$ must be satisfied everywhere in the NS, where $\bar{s}$ is the specific entropy.
Physically, increments in the self-gravitating energy density result in increments in equilibrium-restoring pressure, however this trend cannot extend forever due to the sound speed limitation, putting a cap on NS masses.
This is also an important criterion in modelling the equation of state (EoS) of high-density NS matter.

Briefly, we review the argument for a \emph{minimum} NS mass. 
For a given EoS of the NS core fluid, as the central density (hence mass) of the NS is decreased, the gravitational binding energy will decrease, and at some minimum density, the NS will be unstable to small radial perturbations. 
This EoS-dependent minimum NS mass is typically $\sim 0.1 M_\odot$~\cite{ShapiroTeukolsky,Haensel:2002cia}. 
Such an NS would be primarily composed of an $\Oc(100)$ km crust zone, with a percent-level fraction of mass in the central degenerate neutron fluid~\cite{Haensel:2002cia}. 
Be that as it may, a realistic minimum mass of NSs is about $1 M_\odot$, after neutrino pressure and other thermal effects during the formation of a NS in a core collapse supernova are considered\footnote{A compact object of mass $0.77 \substack{+0.20 \\ -0.17} M_\odot$ has been observed in the supernova remnant HESS J1731$-$347~\cite{NSHESS1}, exciting speculations as to its nature and the EoS of nuclear matter~\cite{NSHESS:DiClemente:2022wqp,NSHESS:Sagun:2023rzp}.}~\cite{Suwa:2018uni}.

\subsection{Structure equations and equation of state} \vspace{0.2cm}
\label{subsec:deadstaranatomy}

Detailed reviews of NS structure and the role of EoSs may be found in Refs.~\cite{Lattimer:2004pg,OzelFreireMRNS:2016oaf,Lattimer:2021emm}, while we present here the essentials.
Accurate estimates of compact star macroscopic properties are best obtained by solving the spherically symmetric stellar structure equations:
\bea
\nn \frac{dP}{dr} &=& -\frac{G m \varepsilon}{c^2 r^2} \bigg(1+ \frac{P}{\varepsilon} \bigg) \bigg(1 + \frac{4\pi r^3 P}{m c^2} \bigg)\bigg( 1 - \frac{2Gm}{c^2r} \bigg)^{-1}~, \\
\frac{dm}{dr} &=& 4\pi \frac{\varepsilon}{c^2} r^2~,
\label{eq:TOV}
\eea
Here $m$ is the mass enclosed within a radial distance $r$, and all other quantities are as defined above.
The first equation,  the {\bf Tolman-Oppenheimer-Volkoff (TOV) equation}, describes hydrostatic equilibrium, and the second describes the conservation of mass in the star.
Given an EoS $P(\varepsilon)$ and the boundary conditions $m(0) = 0$, $\varepsilon (0) = \varepsilon_c$ (a ``central density"), the structure equations can be solved to obtain useful quantities: a mass-radius relation (which would capture the maximum allowed mass), radial profiles of pressure, energy or number density, chemical potential, and so on. 

\begin{figure*}[t]
    \centering
    \includegraphics[width=0.47\textwidth]{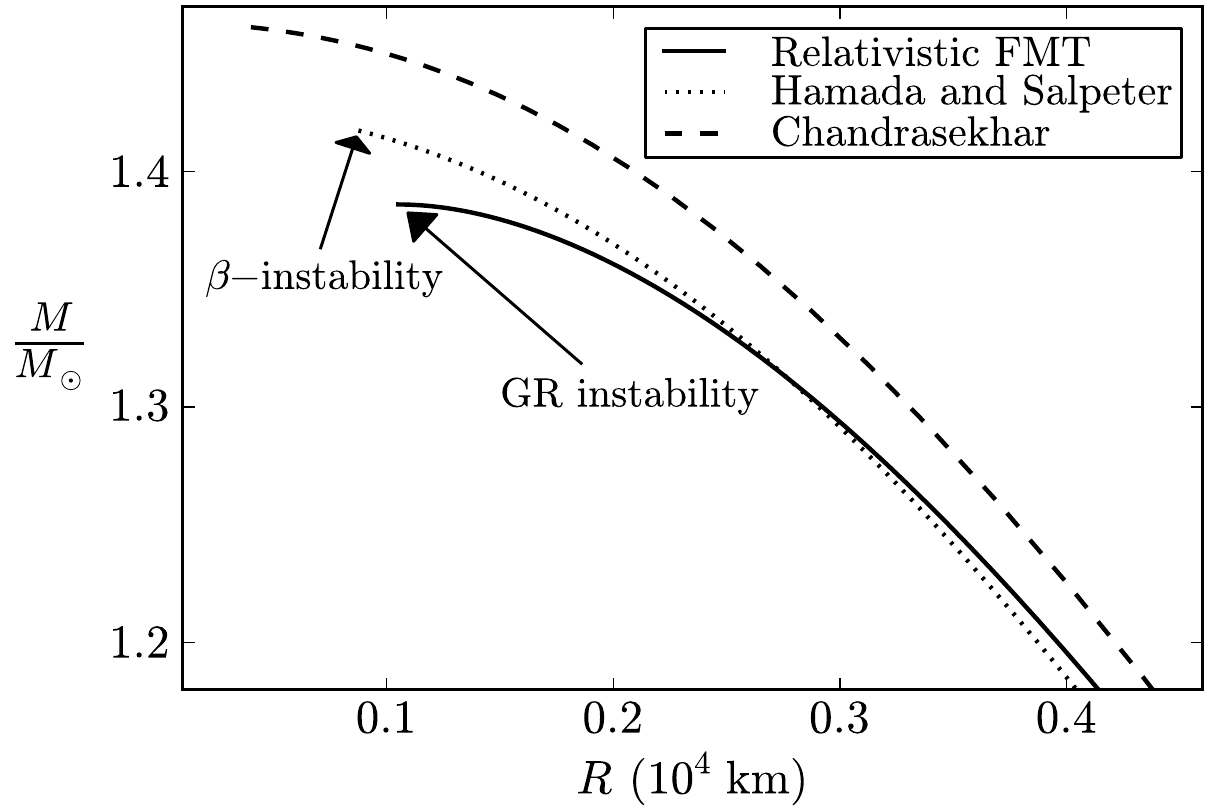}  \includegraphics[width=0.47\textwidth]{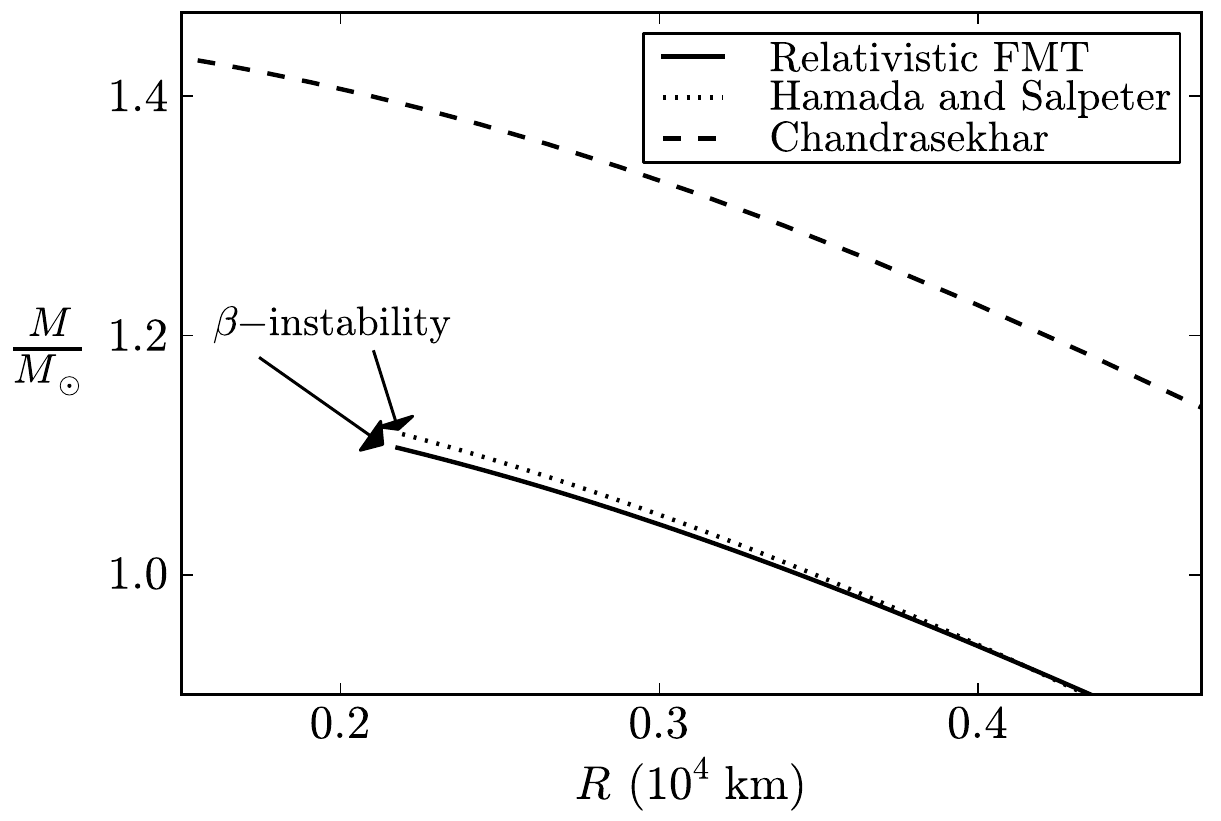} \\
      \includegraphics[width=0.47\textwidth]{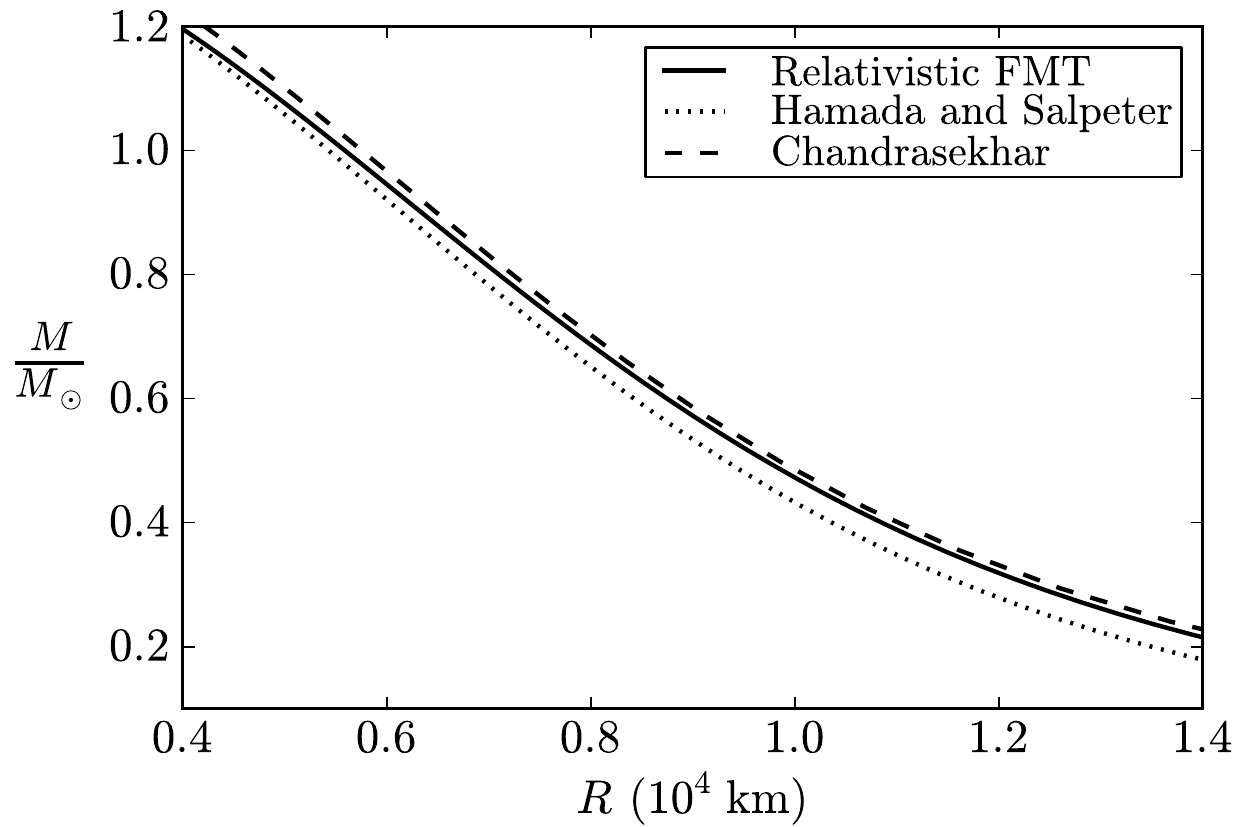}  \includegraphics[width=0.47\textwidth]{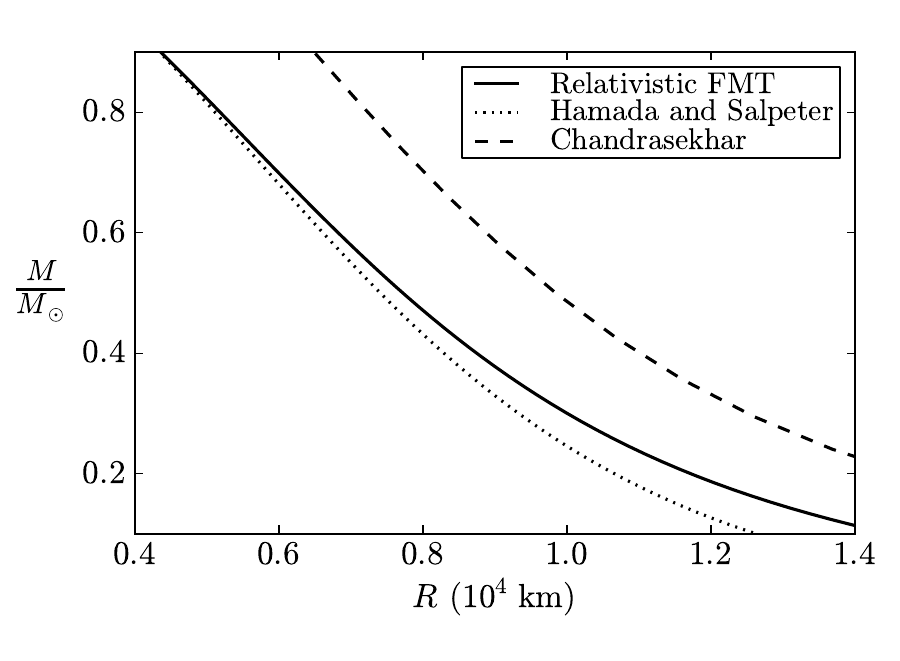} \\
    \caption{{\bf \em Left.} WD mass-radius relations derived from three different equations of state for a $^{12}$C WD. The Chandrasekhar solution treats electrons as a free gas. The Hamada and Salpeter equation of state incorporates Coulomb corrections to the free gas approximation. The Relativistic FMT model additionally accounts for WD relativistic corrections to the Wigner-Seitz cell (Coulomb-corrected) treatment of the equation of state,~\cite{Rotondo:2011zz}. Quantitatively similar curves are obtained for $^4$He and $^{16}$O WDs.
    The point marked ``$\beta$ instability" corresponds to where the central density is high enough that the WD is unstable against electron capture of nuclei resulting in $(Z, A) \to (Z-1, A)$ conversions.
    The point marked ``GR instability" is where the general relativistic corrections shift an otherwise infinite central density to a finite value as the point at which the WD becomes gravitationally unstable.
    {\bf \em Right.} Same as the left panels, but for $^{56}$Fe WDs.
    These plots are taken from Ref.~\cite{Rotondo:2011zz}.}
    \label{fig:WDMvR}
\end{figure*}

\begin{figure}[t]
\centering
\includegraphics[width=0.7\textwidth]{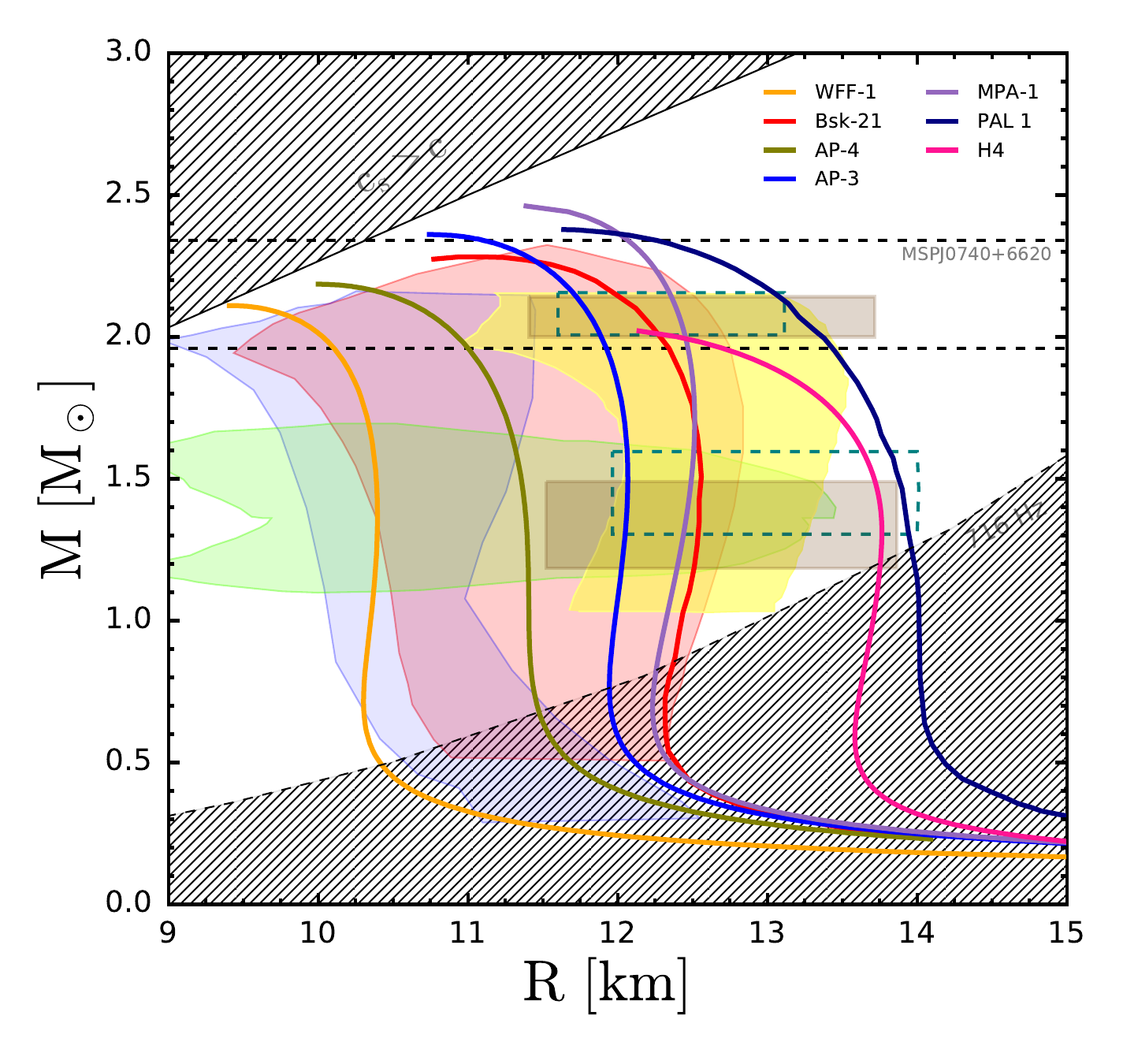}
\caption{NS mass-radius relations for various equations of state for nuclear matter at high densities. 
 The blue shaded region is preferred by pulsar observations~\cite{OzelFreireMRNS:2016oaf}, the yellow region is a fit to the observation of binary NS mergers using a hadronic EoS~\cite{Most:2018hfd}, the green region is the 90\% C.L. preferred region from an EoS-insensitive fit to GW170817~\cite{LIGOScientific:2018cki}, and the red regions are 
    Bayesian fits at 90\% C.L. from a combination of gravitational wave and low-energy nuclear and astrophysical data~\cite{Raithel:2018ncd}.
    The horizontal thick-dashed lines depict the measured mass of the heaviest observed pulsar MSP J0740+6620~\cite{NANOGrav:2019jur}. 
    The line-shaded bottom right region is excluded by centrifugal mass loss, with the limit coming from observations of the fastest-spinning (716 Hz) pulsar~\cite{Hessels:2006ze}.
    The line-shaded top left region is excluded by the condition of causality: for any EoS the sound speed $c_s \leq c$.
    The rectangular regions are simultaneous fits at 68\% C.L. of NS mass and radius by NICER (light brown by Refs.~\cite{riley2019nicer,riley2021nicer} and dashed-green-enclosed by Refs.~\cite{miller2019psr,miller2021radius}).
     This plot is taken from Ref.~\cite{NSvIR:IISc2022}.}
     \label{fig:NSMvR}
\end{figure}

A reliable estimate of WD properties may be gained by assuming a polytropic EoS: $P(\varepsilon) = K \varepsilon^\gamma$.
For WDs, one can set the second and third terms to unity on the right-hand side of the first equation in Eq.~\eqref{eq:TOV}, as the $c$-dependent terms depict general relativistic corrections that are only important for NSs.
It is then straightforward to solve Eq.~\eqref{eq:TOV} for polytropes~\cite{ReddySilbar:2003wm,orderofmag:2015crq}.
In particular, the cases of $\gamma = 5/3$ and $\gamma = 4/3$, applicable respectively to the limit of non-relativistic and relativistic electrons, result in the $M$-$R$ scaling relations we derived from the virial theorem in Eqs.~\eqref{eq:Etotnonrel} and \eqref{eq:Etotrel} with more refined numerical factors.
Notably, for the relativistic case we obtain the Chandrasekhar mass as:
\beq
M_{\rm Ch-WD} \simeq 1.4 M_\odot~.
\label{eq:Mchpolytrope}
\eeq
Realistic EoSs are non-polytropes accounting for Coulomb corrections arising from electron-ion interactions, e.g., the Feynman-Metropolis-Teller EoS~\cite{Rotondo:2011zz}. 
Figure~\ref{fig:WDMvR} shows representative $M$-$R$ relations for WDs of various nuclear compositions, taken from Ref.~\cite{Rotondo:2011zz}.
A simple analytical fit to translate between $\rho$ and WD masses $M_{\rm WD} \in [0.1, 1.35] M_\odot$ is~\cite{Fedderke:2019jur}
\beq
 \label{eq:rhoWDvMWD}
 \bigg(\frac{\rho_{\rm WD}}{1.95\times 10^6 \ {\rm g/cm}^3}\bigg)^{2/3} +1 \approx \bigg[ \sum_{i=0}^6 c_i \bigg(\frac{M_{\rm WD}}{M_\odot}\bigg)^i\bigg]^{-2}~,
\eeq
with $\{c_i\} = \{1.003, -0.309, -1.165, 2.021, -2.060, 1.169, -0.281\}$.

In NSs the EoS of nuclear matter is non-trivial due to the non-perturbative nature of QCD at the densities of the core.
EoSs must account for nucleon-nucleon interactions, far more uncertain than Coulomb interactions, and must fit data on the 
per-nucleon binding energy in symmetric nuclear matter, 
the so-called symmetry energy that accounts for the energy above the $N=Z$ ground state,
the nuclear compressibility, and much else.
For these reasons a wide range of EoSs has been proposed, resulting in multiple predictions for NS configurations.
Figure~\ref{fig:NSMvR} displays $M$-$R$ curves obtained from a few popular EoSs.
The top left is a region where $c_s > c$ and hence causality is violated; the NS mass for various EoSs is seen to reach a maximum close to this region.

\subsection{Spin periods} \vspace{0.2cm}

Celestial bodies have a maximum angular speed: the gravitational force on a mass element on the equator must exceed the centrifugal force on it, giving a minimum spin period
\beq
P_{\rm min} \simeq \sqrt{\frac{3 \pi}{G \rho}} = 10^4~{\rm s} \sqrt{\frac{{\rm g}/{\rm cm}^3}{\rho}}~.
\eeq
Thus for WDs with $\rho = \Oc(10^6)$~g/cm$^3$, $P_{\rm min} \simeq 10$~s, and for NSs with $\rho = \Oc(10^{14})$~g/cm$^3$, $P_{\rm min} \simeq 10^{-3}$~s.
And indeed, the first pulsars historically espied were identified as such by their gradually lengthening sub-second spin periods.
Moreover, no pulsars with spin periods smaller than $\Oc(\rm ms)$ have been observed; those observed near this limit are called millisecond pulsars.
The bottom right region of Fig.~\ref{fig:NSMvR} is excluded by the fastest spinning pulsar observed with rotation frequency 716 Hz~\cite{Hessels:2006ze}, a limit given by ($R$/10 km)$^{3/2} \geq$~1280~Hz~$(M/1.5~M_\odot)^{1/2}$~\cite{Lattimer:2004pg}. 

\begin{figure*}[t] 
\centering
\includegraphics[width=0.7\textwidth]{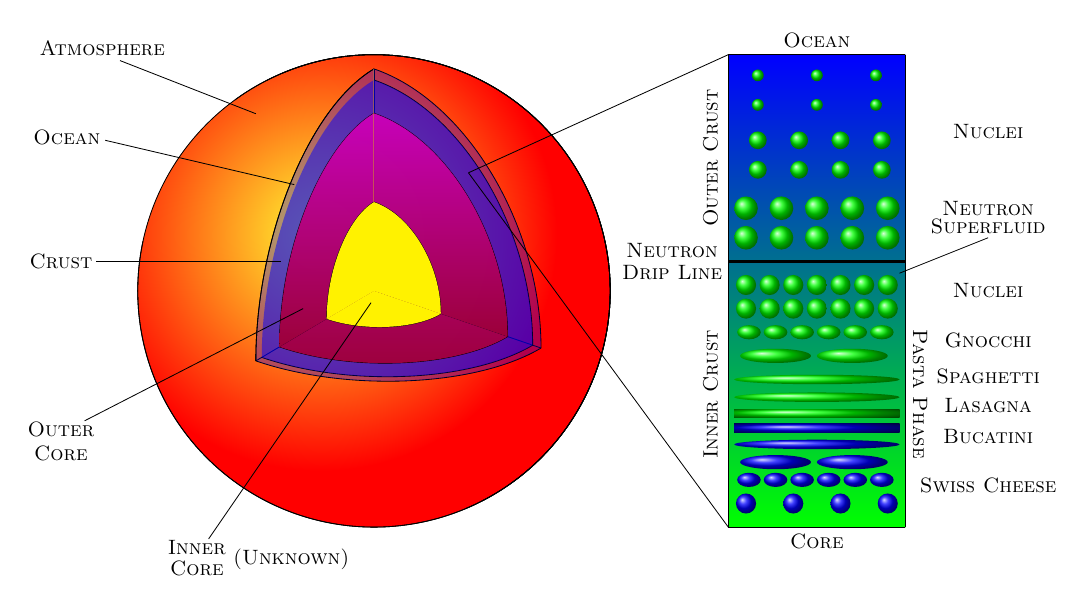} \\
\includegraphics[width=0.47\textwidth]{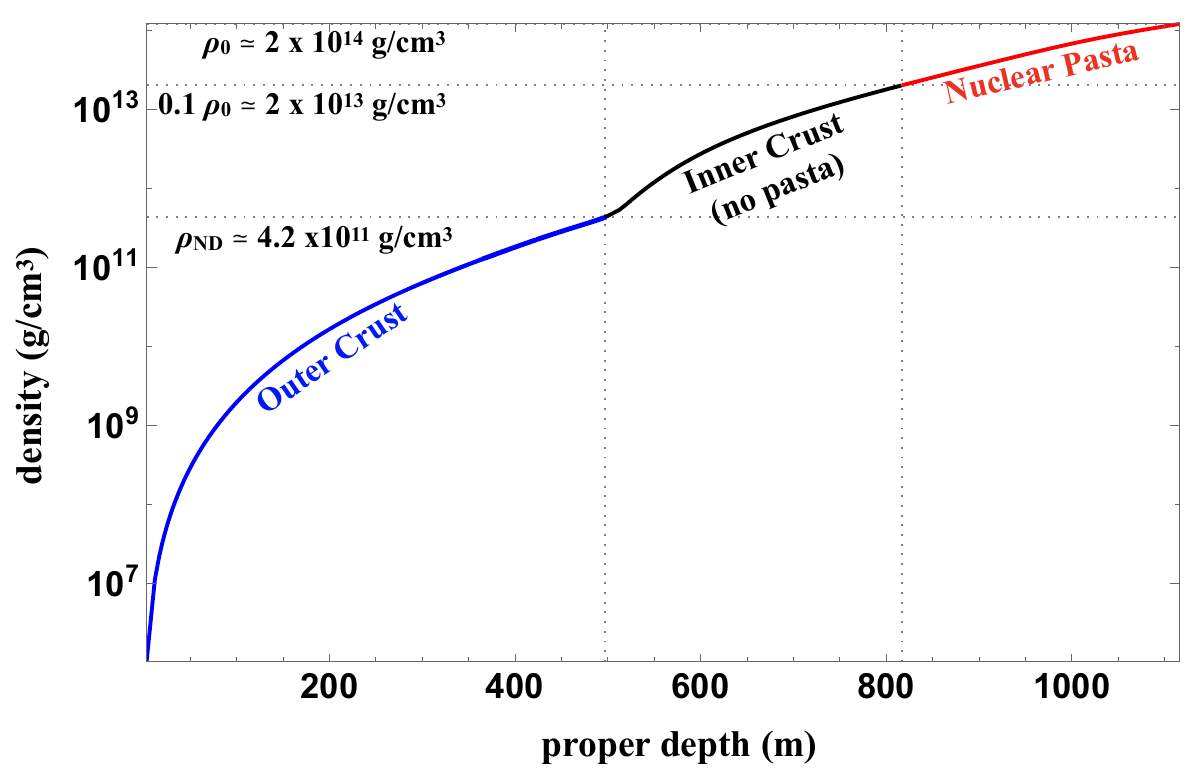} \includegraphics[width=0.47\textwidth]{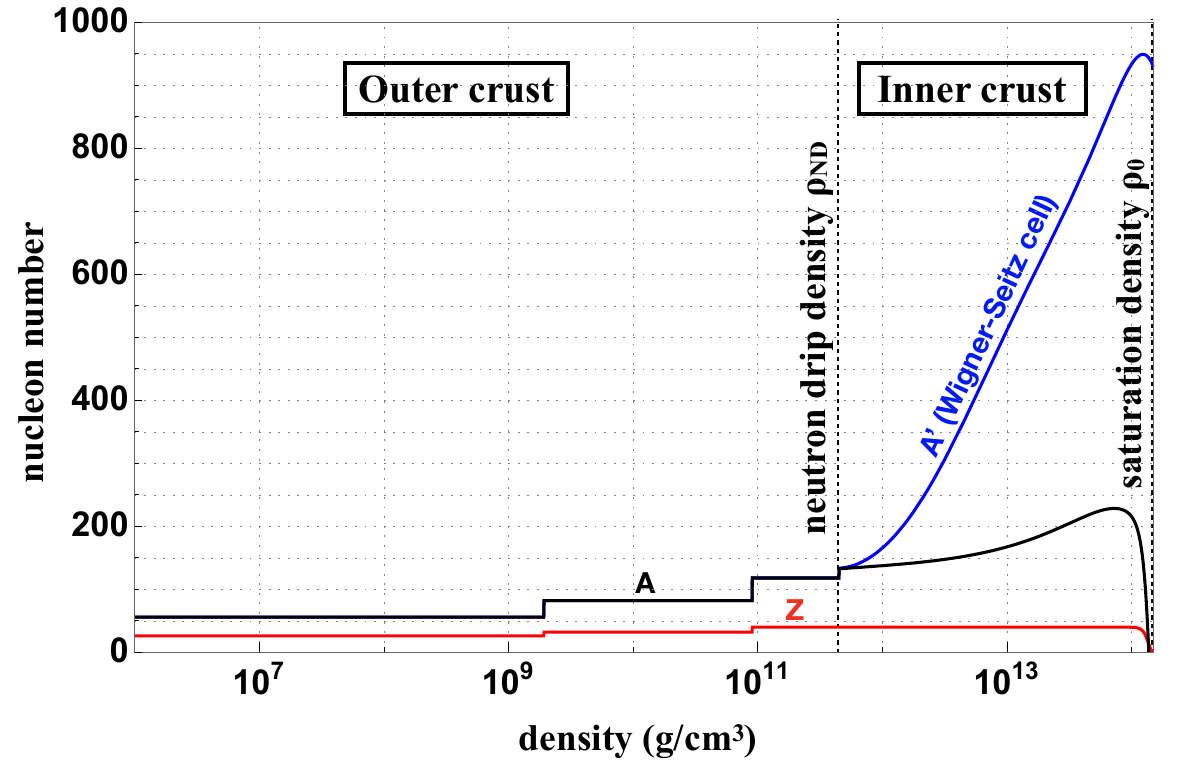} 
\caption{{\bf \em Top.} Schematic of the internal structure of a NS, taken from Ref.~\cite{NSvIR:Pasta}. 
The layers of the crust are shown in the zoom. 
{\bf \em Bottom left.} Density profile of (various layers of) a NS crust. {\em \bf Right.} Nucleon number as a function of NS crust density.
See Sec.~\ref{subsec:NSsubstruct} for further details.
}
\label{fig:NSsubstruct}
\end{figure*}

\subsection{Neutron star substructure} \vspace{0.2cm}
\label{subsec:NSsubstruct}

In the top panel of Figure~\ref{fig:NSsubstruct} we show a schematic of the interior structure of a NS. 
The physics of substructure is obtained by solving Eq.~\eqref{eq:TOV} with the appropriate EoS for each stellar region.
For illustration here we will make use of the Brussels-Montreal ``unified" equation of state (``BSk") accounting for all regions/densities in the NS, expressed in terms of analytic fits~\cite{Pearson:2018tkr}. 
What follows is an overview of NS substructure; interested readers may gain further details from Ref.~\cite{Chamel2008} and the references listed in Ref.~\cite{NSvIR:Pasta}.

The {\em crust}, about $1 \ \rm{km}$ thick, spans over 10 decades in density and consists of several distinct layers corresponding to different phases of nuclear matter.
The bottom left panel of Figure~\ref{fig:NSsubstruct} shows the density of material as a function of the proper depth for the various crustal layers, and the bottom right panel shows nucleon numbers of nuclei as a function of densities spanning the entire crust; both plots were made using the EoS BsK21~\cite{NSvIR:Pasta}.
These plots do not show the {\em atmosphere} (density $ < 10^4$~g/cm$^3$, thickness $\mathcal{O}(\mu$m), composed of hydrogen and lighter elements) and {\em ocean} (density $< 10^{10}$~g/cm$^3$, thickness $\Oc(10)$~m, composed of carbon and heavy metals); these layers affect the star's thermal spectrum, and are influenced by the star's magnetic field. 

The {\em outer crust} (density $10^4-10^{11}$~g/cm$^3$) is composed of nuclei forming a body-centered-cubic Coulomb crystal, interspersed with a degenerate and nearly-free relativistic gas of electrons. 
{\em \` A la} WDs, electron degeneracy contributes dominantly to the pressure, while nuclei contribute dominantly to the mass. 
Deep in the crust, where the electron chemical potential is higher, nuclei become increasingly neutron-rich due to inverse beta decay.
The outer crust terminates when the density and pressure become so high that free neutron states begin to appear.

The transition to the {\em inner crust} is marked by the neutron drip line, marked by density $\rho_{\rm drip} \simeq 4.2 \times{10^{11} \ \rm{g/cm^{3}}}$~\cite{Potekhin:2013qqa}, beyond which a fraction of neutrons becomes unbound from nuclei. 
Up to densities about $0.1$ times the nuclear saturation density $\rho_0 \simeq {2\times{10^{14}} \ \rm{g/cm^{3}}}$, the inner crust comprises of heavy, neutron-rich nuclei (also known as proton clusters) forming a lattice, along with both an electron gas and a dripped-neutron gas. 
Such a system is inaccessible to terrestrial experiments, hence the composition of the inner crust is far more uncertain than the outer crust, and studies of this region are limited to theoretical calculations, {\em e.g.}, 
the Compressible Liquid Drop Model, 
the Thomas-Fermi approximation, and many-body quantum calculations. 
As the NS cools down, the dripped neutrons are expected to form a superfluid phase. 

Further down, the inner crust density approaches the nuclear saturation point, and homogeneous nuclear matter appears~\cite{BAYM1971225,LATTIMER1985646}.
This has led to the prediction of the so-called nuclear ``{\em pasta}" phase at the bottom of the inner crust~\cite{Ravenhall:1983uh,10.1143/PTP.71.320,10.1143/PTP.72.373,Williams:1985prf,Lorenz:1992zz,Oyamatsu:1993zz}.
Intricate competition between nuclear attraction and Coulomb repulsion forms these extended non-spherical phases of nuclear matter; as the density increases, gnocchi, then spaghetti, and then lasagna pasta phases become more prevalent. 
In the deepest layer of the inner crust there are ``inverted pasta phases'' where nuclear density material predominates over sparser, sub-nuclear density voids. 
This includes bucatini (anti-spaghetti) and swiss cheese (anti-gnocchi) phases.
Nuclear pasta is confined to a thin layer, yet they constitute a significant fraction of the crustal mass as they span densities of $0.1 - 1 \ \rho_{0}$. 
They may also impact several properties of the NS such as its thermal and electrical conductivity, and the elasticity and neutrino opacity of the crust.

The inner crust terminates when the density reaches $\rho_{0}$, beyond which nuclei ``melt" into uniform nuclear matter, which form the core of the NS.
The core is further sub-divided into the {\em outer core} (densities 0.5$-$2~$\rho_0$), where the nuclear matter is expected to be comprised of neutrons, protons, and electrons, and the {\em inner core} (densities 2$-$10~$\rho_0$), where exotic states of matter may possibly be present.
These could be meson and hyperon condensates ~\cite{1975ApJ...199..471H,1982ApJ...258..306H,BROWN19761,sfluid:PageReddyReview:2006ud}.
These could also be deconfined quark matter, which is possible when the bag constant is large in the QCD bag model~\cite{sfluid:PageReddyReview:2006ud}, either as $ud$ matter~\cite{Holdom:2017gdc} or $uds$ matter~\cite{Ivanenko1965,Ivanenko1969,Collins:1974ky,Weber:1999qj,Lastowiecki2015,Chatterjee:2015pua}. 
Deconfined quark matter is believed to be in a color superconducting (``CSC") phase~\cite{BARROIS1977390,Frautschi:1978rz,Alford:2007xm}, which could be crystalline~\cite{Anglani2014}.
If strange-quark pairing is suppressed, a two-flavor superconducting (``2SC") phase is formed involving $u$ and $d$ quarks.
If not, the $uds$ matter may exist in in a color-flavor-locked (``CFL") phase~\cite{Alford:1998mk,Alford:2007xm} which may co-exist with confined states~\cite{Glendenning:1995rd,Bedaque:2001je,Kaplan:2001qk}.

\subsection{Thermonuclear explosions} \vspace{0.2cm}
\label{subsection:thermonucrunaway}

Astrophysical situations may arise in which a WD exceeds its Chandrasekhar mass (Eq.~\eqref{eq:MChWDrelapprox}).
For carbon-oxygen WDs, this would lead to ignition of runaway carbon fusion that unbinds the star.
This is how Type Ia supernovae, conventionally used as ``standard candles" in cosmological distance measurements, have been theorized to originate -- via accreting material from a binary companion and going super-Chandrasekhar. 
This picture, however, is disputed by the lack of a specific ``trigger" of the thermonuclear process along with a number of other observational inconsistencies~\cite{Maoz:2011iv}. 
As will be discussed later, other possible Type Ia progenitors include WD mergers and pyconuclear reactions in sub-Chandrasekhar mass WDs.

Yet another setting in which thermonuclear chain reactions create an explosion is in the ocean layer of NS crusts, and in particular the carbon component, which could be ignited by mass accretion from a binary companion. 
For accretion rates $> 10\%$ of the Eddington limit,
the result is ``superbursts", x-ray bursts that spew $\mathcal{O}(10^{35})$ J of energy, lasting for hours, and in some cases recurring about every year~\cite{supburst:Cflashes:Cumming:2001wg,supburst:underZanding:2017ugu,supburst:MINBARcatalog:2020,supburst:catalog2023}. 
This must be distinguished from regular Type-I bursts in NSs, typically ignited by surface accretion, emitting $10^3$ times less energy and lasting $10^3$ times shorter.

Ref.~\cite{TimmesWoosley1992} provides extended discussion on the physics of thermonuclear runaway fusion, while we provide here a brief summary.
Two generic conditions must be satisfied:
(1) a mimimum energy $Q_{\rm dep}$ must be deposited to raise the temperature of a critical mass $M_{\rm crit}$ of density $\rho$ to a critical temperature $T_{\rm crit}$ which can sustain fusion:
\bea
\nn && \textsc{Condition 1}   \\
&& Q_{\rm dep} \geq M_{\rm crit} (\rho, T_{\rm crit}) \bar c_p (\rho, T_{\rm crit}) T_{\rm crit}~.
\label{eq:cond1}
\eea
The temperature prior to heating is here assumed $\ll T_{\rm crit}$, and $\bar c_p \simeq c^{\rm e}_p/2 + c^\gamma_p/4 + c^{\rm ion}_p$ is the average isobaric specific heat capacity, with 
\beq
 c^\ell_p (\rho, T_{\rm crit}) = \frac{a_\ell b_\ell}{u} \bigg(\frac{T_{\rm crit}}{E_{\rm F}}\bigg)^{\alpha_\ell} \bigg[1 - \bigg(\frac{m_e}{E_{\rm F}}\bigg)^2 \bigg]^{\beta_\ell}~. 
\label{eq:cp}
\eeq
Here $u$ is the atomic mass unit, 
$m_e$ the electron mass,  
and for
the \{electronic, radiative, ionic\} contributions, $a_\ell = \{\pi^2,4\pi^4/5,5/2\}$, 
$b_\ell = \{\sum X_i Z_i/A_i, \sum X_i Z_i/A_i,\sum X_i/A_i, \}$ (with $X_i$, $Z_i$, $A_i$ the mass fraction, charge and atomic number of the ion species $i$ respectively),
$\alpha_\ell = \{1, 3, 0\}$, and
$\beta_\ell = \{-1, -3/2, 0\}$ .
The Fermi energy $E_{\rm F} = [m^2_e+(3\pi^2 n_e)^{2/3}]^{1/2}$ with $n_e = \rho b_{\rm e}/u$ (Eq.~\eqref{eq:fermimomentum}).
The trigger energy in Eq.~\eqref{eq:cond1} ranges $\mathcal{O}(10^{17})$~GeV $\ra$ $\mathcal{O}(10^{24})$~GeV for WD central densities corresponding to WD masses ranging 1.4 $M_\odot \ra$ 0.8 $M_\odot$. 

Eq.~\eqref{eq:cond1} is necessary but not sufficient for runaway fusion.
There is a second condition, through which the critical mass $M_{\rm crit} = 4\pi \rho \lambda_{\rm trig}^3/3$ is also defined. 
To wit, the rate of energy gain via nuclear fusion must exceed the rate of energy loss via diffusion over the volume set by the ``trigger length" $\lambda_{\rm trig}$:\\
\bea
\nn && \textsc{Condition 2}  \\
&& \dot Q_{\rm nuc} > \dot Q_{\rm diff}~.
\label{eq:cond2}
\eea
Here we have $\dot Q_{\rm nuc} = M_{\rm crit} \dot S_{\rm nuc}$ and $\dot Q_{\rm diff} \simeq 4\pi k \lambda_{\rm trig} T_{\rm crit}$ for a nuclear energy deposition rate per mass $\dot S_{\rm nuc}$ and thermal conductivity $k$.
Conductive diffusion from relativistic electrons provides the dominant source of diffusion in WDs at the temperatures and densities relevant for igniting thermonuclear fusion; see Ref.~\cite{Bramante:2015cua,Acevedo:2019agu} for analytic expressions for  $\dot Q_{\rm diff}$. 

The estimation of $\dot S_{\rm nuc}$ involves numerical simulations of flame propagation with a nuclear reaction network~\cite{TimmesWoosley1992}.
From this,
\bea
\nn \lambda_{\rm trig} &=& \sqrt{\frac{3 k T_{\rm crit}}{\rho \dot S_{\rm nuc}(\rho, T_{\rm crit})}}~ \\
&=& \begin{cases} 
\lambda_1~(\frac{\rho}{\rho_1})^{-2} \ \ \ \ \ \ \ \ \ \ \ \ \ \ \ \ \ \ \  \ , \rho \leq \rho_1 \\
\lambda_1~(\frac{\rho}{\rho_1})^{\ln(\lambda_2/\lambda_1)/\ln(\rho_2/\rho_1)}~, \rho_1 < \rho \leq \rho_2
\end{cases}
\label{eq:lambdatrig}
\eea
where for WDs  
$\{\lambda_1^{\rm WD},\lambda_2^{\rm WD}\} = \{1.3 \times 10^{-4}~{\rm cm}, 2.5 \times 10^{-5}~{\rm cm} \}$ 
and $\{\rho_1,\rho_2\} = \{2 \times 10^{8}~{\rm g/cm}^3, 10^{10}~{\rm g/cm}^3 \}$.
This analytic form was obtained in Ref.~\cite{Fedderke:2019jur}
by fitting to Figure 6 of Ref.~\cite{TimmesWoosley1992} -- that is restricted to $\rho_1 \leq \rho \leq \rho_2$ -- and  extrapolating to lower densities assuming plausible density-scalings of $k$ and $\dot S_{\rm nuc}$.
The fit is for $T_{\rm crit}$ = 0.5 MeV and assumes equal carbon and oxygen masses in WDs.
In the NS ocean, the mass fraction of carbon is 10\%~\cite{supburst:Cflashes:Cumming:2001wg}, implying $\rho \to 0.1 \rho$ in Eq.~\eqref{eq:lambdatrig} if Eq.~\eqref{eq:lambdatrig} holds for pure carbon burning\footnote{It probably does, for the scalings of Eq.~\eqref{eq:lambdatrig} are seen to be similar to those in Table 3 of Ref.~\cite{TimmesWoosley1992}, for conductive burning.}.
One could also fit a relation among the WD central density, critical temperature and trigger mass~\cite{Bramante:2015cua}:
\beq
   T_{\rm crit} \gsim 
   10^{9.7}~{\rm K}
   \bigg( \frac{\rho}{10^8~{\rm g/cm^3}} \bigg)^{3/140} 
   \bigg( \frac{M_{\rm crit}}{{\rm g}} \bigg)^{3/70}~.
\eeq

\begin{figure*}[t]
\includegraphics[width=0.46\textwidth]{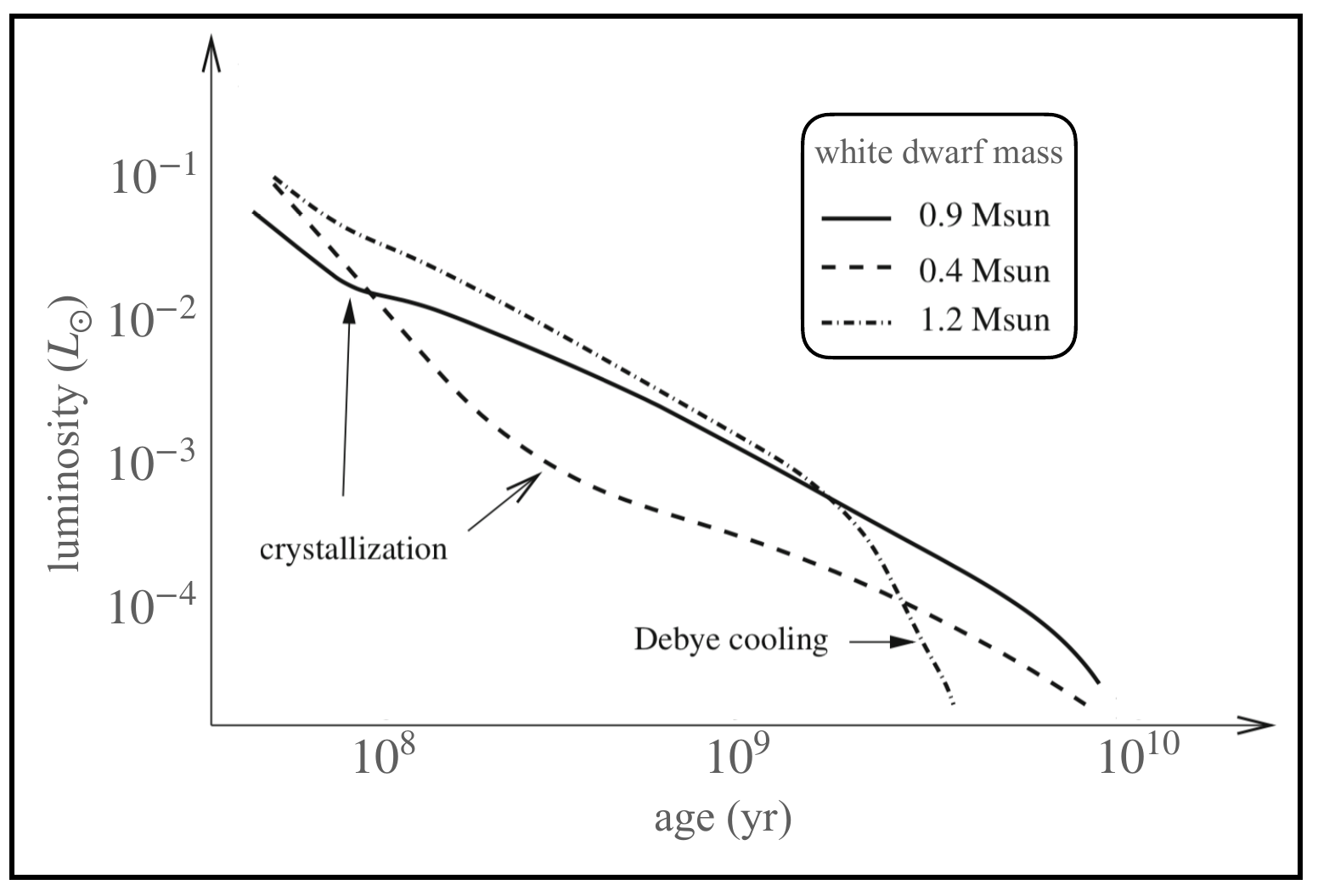} \
\includegraphics[width=0.49\textwidth]{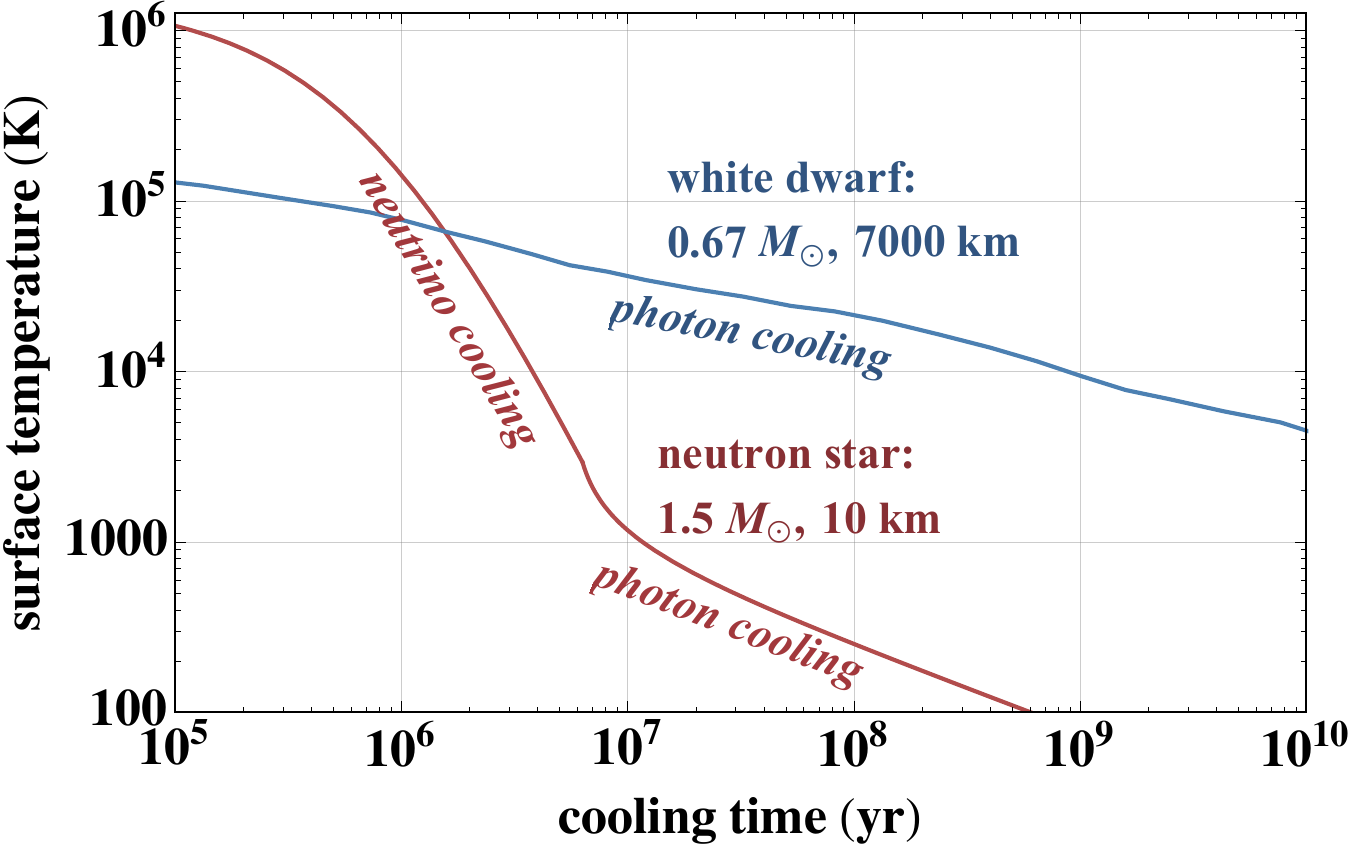}
\caption{Cooling curves. {\bf \em Left.} Luminosity versus time of WDs of various masses, taken from Ref.~\cite{WDcooling:Althaus2010}. 
The onset of crystallization at about $10^8$ yr takes cooling from the regime of thermal ions to the Debye regime. 
{\bf \em Right.} Surface temperature versus time of a benchmark WD and NS. 
Early cooling dominated by emission of neutrinos is distinctly faster than that of photons. 
See Sec.~\ref{subsec:cooling} for further details.}
\label{fig:coolingcurvesNSWD}
\end{figure*}
 
\subsection{Cooling} \vspace{0.2cm}
\label{subsec:cooling}

As no nuclear fuel is burnt in compact stars, they cool down continually from the moment of their birth unless energy is deposited into them by some means, as discussed in Sections~\ref{sec:WDvDM} and \ref{sec:NSvDM}.
Observations of compact star cooling are an important handle on the physics governing their internal dynamics.

\subsubsection{White dwarf cooling.}

WDs initially cool by shedding the thermal energy of constituent ions. 
Given the specific heat per ion $c_v = 3/2$, the total WD energy in thermal ions is 
\begin{equation}
U = \frac{3 T}{2} \bigg( \frac{M_{\rm WD}}{A m_N}\bigg)~.    
\end{equation}
The WD luminosity $L = - dU/dt$, and the cooling curve can be obtained from an independent expression for the luminosity in terms of the WD internal temperature $T_{\rm int}$:
\begin{equation}
  L = 0.2 \ {\rm J/s} \ \bigg(\frac{M_{\rm WD}}{M_\odot}\bigg) \bigg(\frac{T_{\rm int}}{{\rm K}}\bigg)^{7/2}~, 
\end{equation}
derived from photon diffusion in the WD surface layers assuming Kramer's opacity, and combining it with the EoS; see Ref.~\cite{ShapiroTeukolsky} for a detailed treatment.
The cooling timescale is then obtained as
\begin{equation}
\label{eq:wdcoolingtime}
t_{\rm cool} \simeq {\rm Gyr}~\bigg(\frac{M/M_\odot}{L/(10^{-3} L_\odot)}\bigg)^{5/7}~.
\end{equation}
Thus the cooling times are long enough to keep WDs from becoming invisibly faint today, yet short enough to make them fainter than main-sequence stars.
The above relation only holds for WDs with $T_{\rm int} > T_{\rm Debye} \simeq 10^{7}~$K, the typical Debye temperature below which the ions crystallize.
For smaller temperatures, corresponding to $L \lsim 10^{-4} L_\odot$, the specific heat comes from the vibration of the crystal lattice as opposed to thermal motion of the ions. 
Obtaining WD cooling times accounting for this effect involves a non-trivial treatment~\cite{ShapiroTeukolsky} that is beyond our scope.
In Fig.~\ref{fig:coolingcurvesNSWD} left panel we show a luminosity-vs-time cooling curve indicating the point at which crystallization effects become important.
In the right panel we show a temperature-vs-time curve for a benchmark WD of mass 0.67 $M_\odot$ corresponding to a 7000 km radius.

\subsubsection{Neutron star cooling.}

NSs cool by emitting neutrinos (generated in weak processes) and photons; the rate of neutrino cooling rate is initially larger and hence dominates up to a point, before photon cooling takes over. 
In describing the cooling of NSs, where GR effects are significant, it is necessary to distinguish between the temperature in the frame of the NS, $T$, and in the frame of a distant observer, $\tilde T$, related by 
\bea
\nn \tilde T &\equiv& T/(1+z)~,\\
1 + z &=& \frac{1}{\sqrt{1-2G \MNS/\RNS}}~.
\eea
The temperature evolution during passive cooling is given by
\beq
c_{\rm v}(\tilde T) \frac{d\tilde T}{dt} = - L_{\nu}^\infty (\tilde T) - L_\gamma^\infty (\tilde T)~,
\label{eq:dTdtfull}
\eeq
where the neutrino luminosity of the NS as measured by a distant observer of our benchmark NS is zenzizenzizenzic in NS temperature~\cite{heatcapacity:ReddyPageHorowitz:2016weq}:
\beq
L_\nu^\infty (\tilde T) = 1.33 \times 10^{39}~{\rm J/yr}~\bigg(\frac{\tilde T}{10^9~{\rm K}} \bigg)^8~,
\eeq
applicable for slow/modified Urca (``Murca") processes such as $N + n \ra N + p e^- \bar\nu_e$ and $N + p e^- \ra N + n \nu_e$ (with $N = n, p$), the neutrino cooling mechanism as prescribed by the ``minimal cooling" paradigm~\cite{coolingminimal:Page:2004fy}.
In principle there could also be fast/direct Urca (``Durca") processes such as $ n \ra p e^- \bar\nu_e$ and $p e^- \ra  n \nu_e$~\cite{cooling:Yakovlev:2004iq}.
These processes dominate the NS cooling down to $\tilde T = 10^8$~K.
It has also been suggested that cooling via $N\gamma \to N\nu \bar \nu$ induced by QCD anomaly-mediated interactions are comparable to Murca processes in early-stage NS cooling~\cite{Harvey:2007rd,Chakraborty:2023wgl}.
The luminosity of photon blackbody emission from the NS surface is:
\beq
L_\gamma^\infty (\tilde T_s) = 4 \pi (1+z)^2 R_{\rm NS}^2 \tilde T^4_s~.
\label{eq:blackbodylum}
\eeq
The NS heat capacity $c_V$ is given by~\cite{cvanalytic:Ofengeim:2017xxr} 

\bea
\nn c_V (\tilde T) &=& 4.8 \times 10^{26}~{\rm J/K}~\bigg(\frac{\tilde{T}}{10^4~{\rm K}}\bigg) \\
               &=& 2.7 \times 10^{-21}~M_\odot/{\rm K}~\bigg(\frac{\tilde{T}}{10^4~{\rm K}}\bigg)~.
\label{eq:heatcapelec}
\eea

Solving Eq.~\eqref{eq:dTdtfull} requires a relation between the surface ($T_s$) and internal ($T$) temperatures.
Such a relation is highly sensitive to the composition of the NS' outermost envelope, which acts as an insulating layer for temperatures $\gsim \Oc(10^3)$~K, and becomes too thin for insulation at smaller temperatures~\cite{cooling:Yakovlev:2004iq,coolingcatalogue:Potekhin:2020ttj}.
For an iron envelope at high temperatures~\cite{TbTs-Fe-Pethick1983,NSenvelope:Beznogov:2021ijc},
\beq
T_s = 10^6~{\rm K} \bigg[\bigg( \frac{M_{\rm NS}}{1.5~M_\odot}\bigg) \cdot \bigg(\frac{10~{\rm km}}{R_{\rm NS}}\bigg)^2\bigg]^{1/4} \bigg[\frac{T}{9.43\times 10^7~{\rm K}}\bigg]^{0.55}~.
\label{eq:TsTbFe}
\eeq
One can then identify the thin-envelope regime by solving for $T_s = T$ in the above equation, which gives $T_{\rm env} = 3908$~K, below which one can simply set $T_s = T$. 

The solution of Eq.~\eqref{eq:dTdtfull} can now be written down as the time for the NS to cool to a temperature $\Tcoolz$ ($\ll$ the initial temperature)~\cite{coolinganalytic:Ofengeim:2017cum}: 
\begin{equation}
\tcool (\tilde T_9)/{\rm yr} = \begin{cases}
t_{\rm env} = s_1^{-k} q^{-\gamma} \big[\big(1+ (s_1/q)^k \tilde T_9^{2-n} \big)^{-\gamma/k} - 1 \big], \ \Tcoolz > \tilde T_{\rm env}~, \\
t_{\rm env} + (3s_2)^{-1} (\tilde T_9^{-2} - \tilde T_{\rm env}^{-2}), \ \ \ \ \ \ \ \ \ \ \ \ \ \ \ \ \ \ \ \ \ \ \  \Tcoolz \leq \tilde T_{\rm env}~,
\end{cases}
\label{eq:tcoolvTfull}
\end{equation}
where $\tilde T_9 = \Tcoolz/(10^9~{\rm K})$,
$q = 2.75 \times 10^{-2}$,
$s_1 = 8.88 \times 10^{-6}$,
$s_2 = 8.35 \times 10^4$,
$k = (n-2)/(n-\alpha)$ and
$\gamma = (2-\alpha)/(n-\alpha)$ with
$\alpha = 2.2$ and
$n=8$; $\tilde T_{\rm env} \simeq 4000~$K corresponds to the time after which the surface and internal temperatures equalize. 
The right-hand panel of Figure~\ref{fig:coolingcurvesNSWD} shows the NS cooling curve plotted using the above expression, with the distinct regimes of neutrino and photon cooling labelled.

While in standard cooling scenarios NSs are expected to cool down to $\Oc(10^2)$ K over Gyr timescales (as in Fig.~\ref{fig:coolingcurvesNSWD}), temperatures as high as $10^4$ K have been conjectured to persist if some additional astrophysical source of NS heating is present: we discuss such speculative reheating mechanisms in Sec.~\ref{subsubsec:NSheatNOTDM}.
We note that in the cases of magnetic field decay and rotochemical heating, the expectation is that late-stage NSs will still glow at below $10^3$ K.

\subsubsection{Comparison of white dwarf and neutron star late-stage cooling}

From the discussion above, and from the right-hand panel of Fig.~\ref{fig:coolingcurvesNSWD}, we see that the temperature of an NS expected at late stages ($t_{\rm cool} \gtrsim$ Gyr), about $10^2$ K, is much smaller than the late-stage temperature of a WD, about $10^{3.3-4}$ K. 
It is natural to wonder why, as one may expect both to have a similar late-stage temperature
since both are low-temperature degenerate stars.
The crucial difference is that the late-stage WD heat capacity is determined by vibrational modes of the nuclear ionic Coulomb lattice forming its interior, while the NS heat capacity is that of a degenerate Fermi gas. 

In the WD case, the heat capacity per ion is~\cite{ShapiroTeukolsky}
\begin{equation}
    c_v^{\rm WD} 
    \simeq \int_0^{\kappa_{\rm max}} \frac{\kappa^2 d\kappa}{2\pi^2 n} \sum_{\lambda = 1}^3  \frac{e^{\omega_{\lambda}(\kappa)/T}(\omega_{\lambda}(\kappa)/T)^2}{(e^{\omega_{\lambda}(\kappa)/T}-1)^2}~,
\end{equation}
where $\omega_\lambda$ is the vibrational energy, $\kappa $ is the wavenumber of the normal modes of the Coulomb lattice, and $\lambda$ labels transverse and longitudinal modes. 
Using the linear approximation $\omega_\lambda = \kappa c_s$, where $c_s$ is the sound speed in the lattice, and changing variables in the integral to $\kappa \ra y T/c_s$, it can be immediately seen that $ c_v^{\rm WD} \propto T^3$.

In the case of the NS, in the limit where the temperature $T$ drops below the Fermi energy $E_F$, only a fraction  $T/E_F$ of the fermions close to the Fermi surface will be excited and raise the bulk temperature.
Hence the energy per fermion $\propto T (T/E_F)$, and it follows that the heat capacity $c_v^{\rm NS} \propto T$.

 From Eq.~\eqref{eq:dTdtfull}, setting the photon luminosity $\propto T^4$, we obtain $t_{\rm cool} \propto \log T$ for WDs and $t_{\rm cool} \propto 1/\sqrt{T}$ for NSs.
Thus, WDs indeed cool much slower than NSs at later stages. 

\subsection{Nucleon superfluidity} \vspace{0.2cm}
\label{subsec:sfluid}

It has been recognized since 1959 that nucleons in a NS could be in a superfluid state~\cite{sfluid:Migdal1959}, a hypothesis supported by observational fits to cooling curves~\cite{coolingcatalogue:Potekhin:2020ttj}.
Neutron superfluidity and proton superconductivity arise due to their Cooper pairing with a 0.1 MeV energy gap, corresponding to a critical temperature $T_c \simeq 10^{10}$~K~\cite{sfluid:Yakovlev:1999sk,sfluid:PageReddyReview:2006ud,sfluid:Haskell:2017lkl}.
Pairing occurs mainly close to the Fermi surface, hence superfluidity does not influence the EoS of NS matter (therefore bearing no consequence on NS mass-radius relations), but does play a major role in setting the NS' heat capacity and neutrino emissivity.
This is because these quantities are sensitive to particle-hole excitations close to the Fermi surface: the energy gap exponentially suppresses the nucleon contribution to the heat capacity for NS temperatures $\ll T_c$.

Neutrons in the NS inner crust are expected to form Cooper pairs in the singlet $^1S_0$ state, and in the core in a triplet $^3P_2$ state: at higher densities, singlet pairing becomes repulsive~\cite{sfluid:Haskell:2017lkl}.
The less dense protons are expected to pair in the singlet state in the NS core.
A quark core in the NS could give rise to ``color superconductivity" with $ud$, $ds$, $su$ Cooper pairs carrying color charges~\cite{sfluid:colorAlfordRajagopalWilczek:1997zt}.
Nucleon pairing models play a central role in the possibility of rotochemical heating, as discussed in Sec.~\ref{subsubsec:NSheatNOTDM}.
The presence of superfluidity in NSs also gives rise to macroscopic vortices and flux tubes, the former of which may play a role in late-stage reheating of NSs (Sec.\ref{subsubsec:NSheatNOTDM}).

One evidence for nucleon superfluidity in NSs is the role it is thought to play in the observations of ``glitches"~\cite{glitch:Anderson:1975zze,glitch:Haskell:2015jra,glitchreview:Zhou:2022cyp,glitch:Antonelli:2023vpd} in about 200 pulsars~\cite{glitch:Manchester2017}, which are sudden increases in rotational frequency followed by recovery over days or months to the pre-glitch frequency.  

\begin{figure*}
    \centering
        \includegraphics[width=0.53\textwidth]{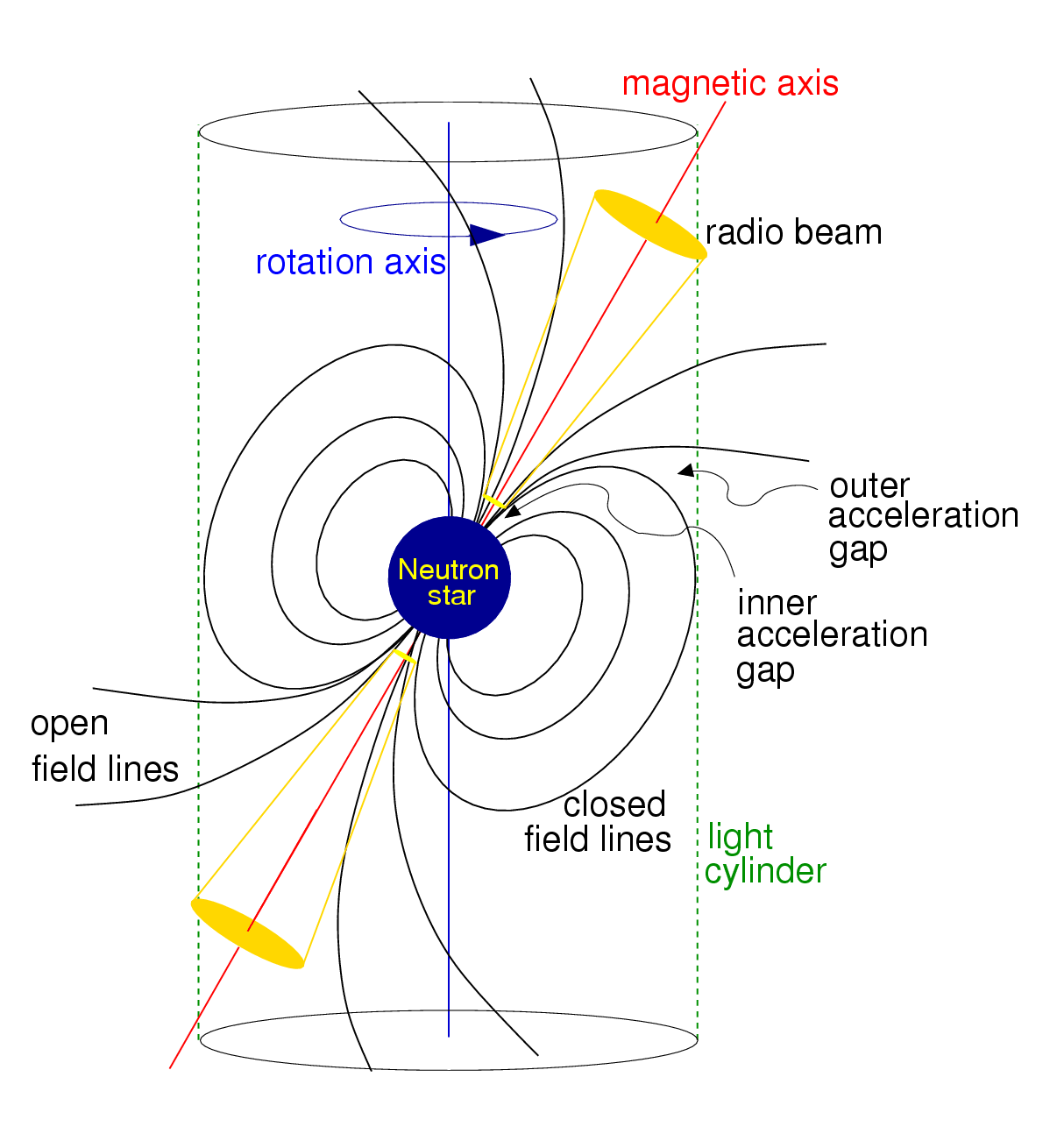} 
         \includegraphics[width=0.41\textwidth]{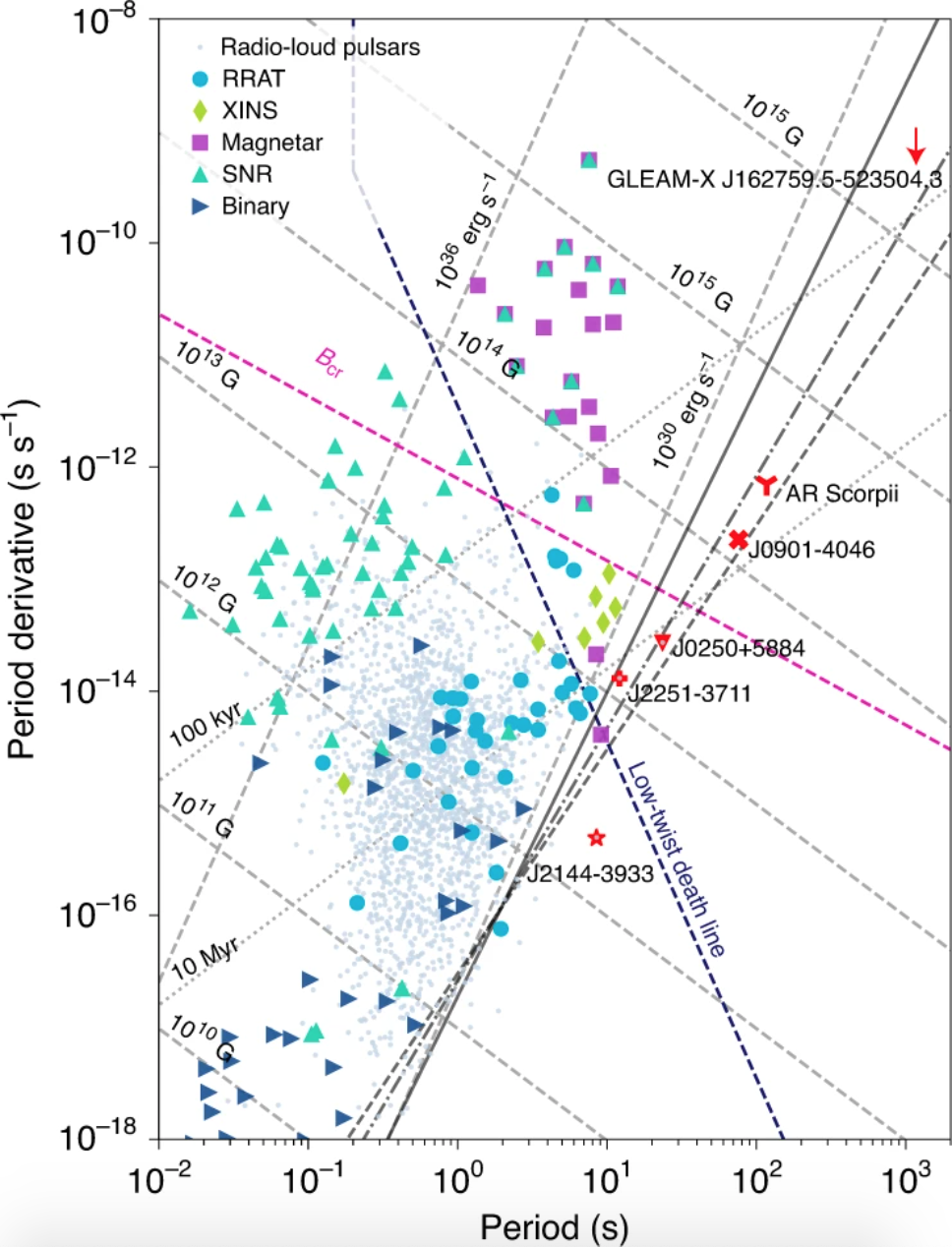} 
    \caption{{\bf \em Left.} The ``light cylinder" around a NS within which the co-rotating magnetosphere is confined~\cite{NSBfieldspindown}. Acceleration of charges in this region are thought to produce electromagnetic beams that are detected terrestrially as regular pulses, making young NSs ``pulsars". {\bf \em Right.} $P$-$\dot P$ diagram taken from Ref.~\cite{PPdot:Caleb:2022xyo}, illustrating the evolution of pulsars. For a description of the types of pulsars displayed here, see Ref.~\cite{NSBfieldspindown}. See Sec.~\ref{subsec:Bspindown} for further details.}
    \label{fig:pulsation}
\end{figure*}

\subsection{Neutron star magnetic field and spin-down} \vspace{0.2cm}
\label{subsec:Bspindown}

When a progenitor star turns into an NS, its surface area shrinks by a factor of about $10^{10}$.
As a result, thanks to conservation of magnetic flux (Gauss' law for magnetism, $B \times R_{\rm NS}^2$ = constant) the stellar magnetic fields increase by this factor, and thanks to conservation of angular momentum the rotational speed also rises by this factor. 
Flux conservation also implies that the total energy in the NS due to the magnetic field decreases with the NS size: 
\beq
E^{\rm NS}_B = \frac{B^2}{8\pi} \cdot \frac{4\pi R_{\rm NS}^3}{3} = \frac{\rm const.}{R_{\rm NS}}~,
\eeq
hence the presence of $B$ fields tends to enlarge the NS. 
However $E^{\rm NS}_B$ is bounded by the gravitational binding energy of the NS, giving the condition
\beq
B \leq \sqrt{\frac{18}{5}\frac{GM^2_{\rm NS}}{R^4_{\rm NS}}} \simeq 10^{18}~{\rm Gauss}~\bigg(\frac{\MNS}{M_\odot}\bigg)\bigg(\frac{10~{\rm km}}{\RNS}\bigg)^2~.
\eeq
A stricter upper limit can be obtained from considerations of hydromagnetic stability~\cite{ReiseneggerBfield2009}.
Measurements from pulsar spin-down (discussed here) find that millisecond pulsars typically have $B$ field strengths of about $10^{8}$~Gauss, classical pulsars about $10^{12}$~Gauss, and magnetars about $10^{15}$~Gauss.
NSs have a ``magnetosphere", a region of plasma surrounding the NS and co-rotating with it due to their coupling through the $B$ field.
One can see that this region is finite by simply locating the equatorial radius at which its tangential speed $= c$ for a spin period $P$:
\beq
R_\perp^{\rm LC} = \frac{c P}{2\pi} = 48~{\rm km} \bigg(\frac{P}{\rm ms}\bigg)~.
\eeq
This region defines the ``light cylinder" shown in Figure~\ref{fig:pulsation}, left panel.
The electrodynamics of this region is key to understanding the pulsar nature of most NSs observed. 
In a region just above the polar surface where a vacuum gap in the plasma is thought to exist, electrodynamic force-free criteria do not apply, and non-zero residual electric fields are allowed.
These accelerate charged particles, leading to curvature radiation sourcing energetic photons which in turn interact with the magnetic field to produce electron-positron pairs, which in turn emit more photons, setting up a cascade~\cite{PulsarB:1975RudermanSutherland,PulsarB:reviewMichel:1982fj}.
This results in the emission of electromagnetic beams at radio frequencies from near the magnetic poles of the NS. 
This beam, as we will soon see, is ultimately powered by the rotational energy of the NS. 
The lighthouse-like sweep of the beam, detected as regular pulses on Earth, serves to reveal NSs as pulsars\footnote{At the time of writing, ``white dwarf pulsars" have also been discovered~\cite{WDpulsar:Rea:2023stb}.}. 
This is how NSs were historically discovered by Bell and Hewish, and continues to be the primary method for finding NSs in the sky~\cite{ATNF:2004bp}.

The NS spin varies over the lifetime of the NS due to a number of factors, chief among which is magnetic dipole radiation extracting rotational kinetic energy, an effect known as pulsar spin-down.  
The radiation power of a rotating magnetic dipole of moment $m$, with a component $m_\perp$ perpendicular to the NS spin axis, and angular speed $\omega = 2 \pi/P$, is given by~\cite{NSBfieldspindown}
\beq
\dot E_{\rm rad, B} = \frac{2}{3 c^3} m_{\perp}^2 \omega^4 = \frac{2}{3 c^3} (B_\perp R_{\rm NS})^3 \bigg( \frac{2\pi}{P} \bigg)^4~,
\label{eq:powerdipoleB}
\eeq
where in the second equation we have used the expression for a sphere uniformly magnetized with field strength $B$.
The rotational power of an NS of moment of inertia $I = 2 M_{\rm NS} R^2_{\rm NS}/5$  is given by
\beq
\dot E_{\rm rot} = I \omega \dot\omega = - 4\pi^2 \frac{I \dot P}{P^3}~.
\label{eq:powerrotKE}
\eeq
For sub-kHz frequencies this radiation cannot penetrate the ISM nebula surrounding the NS, and is hence deposited in it; the observed $P$, $\dot P$ and luminosities of supernova remnants such as the Crab Nebula ($P$ = 0.033~sec, $\dot P$ = 1 sec/80,000 yr, luminosity = $10^5 L_\odot$, much higher than that of the Crab Pulsar within) bears out the supposition that $- \dot E_{\rm rot} \simeq \dot E_{\rm rad, B}$~\cite{NSBfieldspindown}. 

NS spin-down provides a remarkably valuable handle on the age of an NS through measurement of just its $P$ and $\dot P$, i.e. without requiring knowledge of its radius, mass and $B$ field.
Assuming the $B$ field remains constant, by equating Eqs.~\eqref{eq:powerdipoleB} and \eqref{eq:powerrotKE} we see that $P\dot P$ is constant over time.
For an initial spin period $P_0$,
\bea
\nn \int_0^\tau dt (P' \dot P') &=&  \int_{P_0}^P dP' P'   \\
\nn \Rightarrow P \dot P \tau &=& \frac{P^2 - P_0^2}{2} \\
\Rightarrow \  \ \ \tau &=& \frac{P}{2\dot P}~,
\label{eq:spindowncharage}
\eea
where the last equality assumed that the initial period $P_0 \ll P$.
This {\em characteristic age} $\tau$ due to spin-down is often an excellent order-of-magnitude estimate of an observed NS's true age.
It slightly overestimates the latter for young NSs as an NS' spin may initially decelerate via gravitational radiation due to an oblate shape.
For instance, for the Crab Pulsar, whose supernova was observed in 1054 A.D., one finds $\tau$ = 1300 years.  
In the case of older pulsars, Eq.~\eqref{eq:spindowncharage} must again be used with special care, specifically when being applied to NSs that are thought to have spun up at some point in their life. 
These could be, {\em e.g.}, millisecond pulsars that are modelled as accreting mass and angular momentum from a binary companion; these have been observed with a characteristic age older their actual age~\cite{Tauris}. 
In particular, there are millisecond pulsars with $\tau > $13.8 Gyr~\cite{ATNF:2004bp}, the measured age of the universe~\cite{Planck:2018vyg}.

We note in passing that for NSs for which precise data on distances and proper motions are available, their {\em kinematic age} may also be estimated by tracing back their trajectories and locating a plausible birth site~\cite{NSkinematicageMag7}.
This technique is possible thanks to the kick velocity imparted to the NS by the asymmetric explosion of the progenitor, as mentioned in the beginning of Sec.~\ref{sec:physicscompact}.

The pulsar braking index $n$ is defined via $\dot \omega \propto \omega^n$. 
With a little elementary calculus, it may be seen that 
\beq
n \equiv \frac{\omega \ddot \omega}{\dot \omega^2} = 2 - \frac{P \ddot P}{\dot P^2}~.
\label{eq:brakingindex}
\eeq
For spin-down induced by magnetic dipole radiation, one finds by equating Eqs.~\eqref{eq:powerdipoleB} and \eqref{eq:powerrotKE} that $n = 3$, although pulsars with braking indices of 1.4$-$3 have been observed, suggesting other spin-down mechanisms~\cite{NSBfieldspindown}.

It is useful to place observed pulsars on a $P$-$\dot P$ diagram such as the one shown in Fig.~\ref{fig:pulsation} right panel. 
Pulsars typically begin life at the north-west region of the diagram, and move south-east along contours of constant $B$ strengths while crossing contours of constant spin-down age. 
Eventually as they age to about 10 Myr the rotational energy is insufficient to provide the potential drop required for Schwinger pair production that generates the pulsar beam, and they cross the ``pulsar death line", sometimes referred to as the ``death valley".
However, the death line is not well-understood, for the exact mechanism by which pulsar beams are created is still unknown and is an active area of research.
This is evident in the $P$-$\dot P$ diagram: quite a few pulsars lie beyond various models of the death line~\cite{PulsarDeathLineAnomaly:PSRJ0250+5854,PulsarDeathLineAnomaly:PSRJ0901-4046,PulsarDeathLineAnomaly:PSRJ2144-3933,PulsarDeathLineAnomaly:PSRJ2251-3711}, with PSR J2144-3933 lying well beyond all the canonical death lines.
We will re-encounter this oddball pulsar, which also happens to be the coldest NS observed, in Sec.~\ref{subsec:DMkinannheat}.

\section{The white dwarf as a dark matter laboratory} \vspace{0.2cm}
\label{sec:WDvDM}

WDs have been used as DM detectors via a number of mechanisms. 
There are four main effects, which we will detail in the rest of the section:
(1) DM can collect and annihilate inside WDs, heating them to above the temperature that would be expected from a standard WD cooling curve such as in Sec.~\ref{subsec:cooling}.
(2) So much non-annihilating DM accumulates in a WD that the DM collapses and forms a black hole deep in the WD interior. This small black hole can grow to accrete the entire WD, thereby converting its host into a solar mass black hole. 
(3) DM encounters with and collection in the WD can cause it to explode. 
(4) WDs' internal structure could be altered if a substantial fraction of its mass were comprised of DM.

In addition, resonant conversion of axion-like particle DM to photons in the corona of a magnetic WD may be observed~\cite{Wang:2021wae}; we relegate discussion of this phenomenon in the context of NSs to Sec.~\ref{subsec:NSvALP}.

\begin{figure*}[t]
    \centering
        \includegraphics[width=0.47\textwidth]{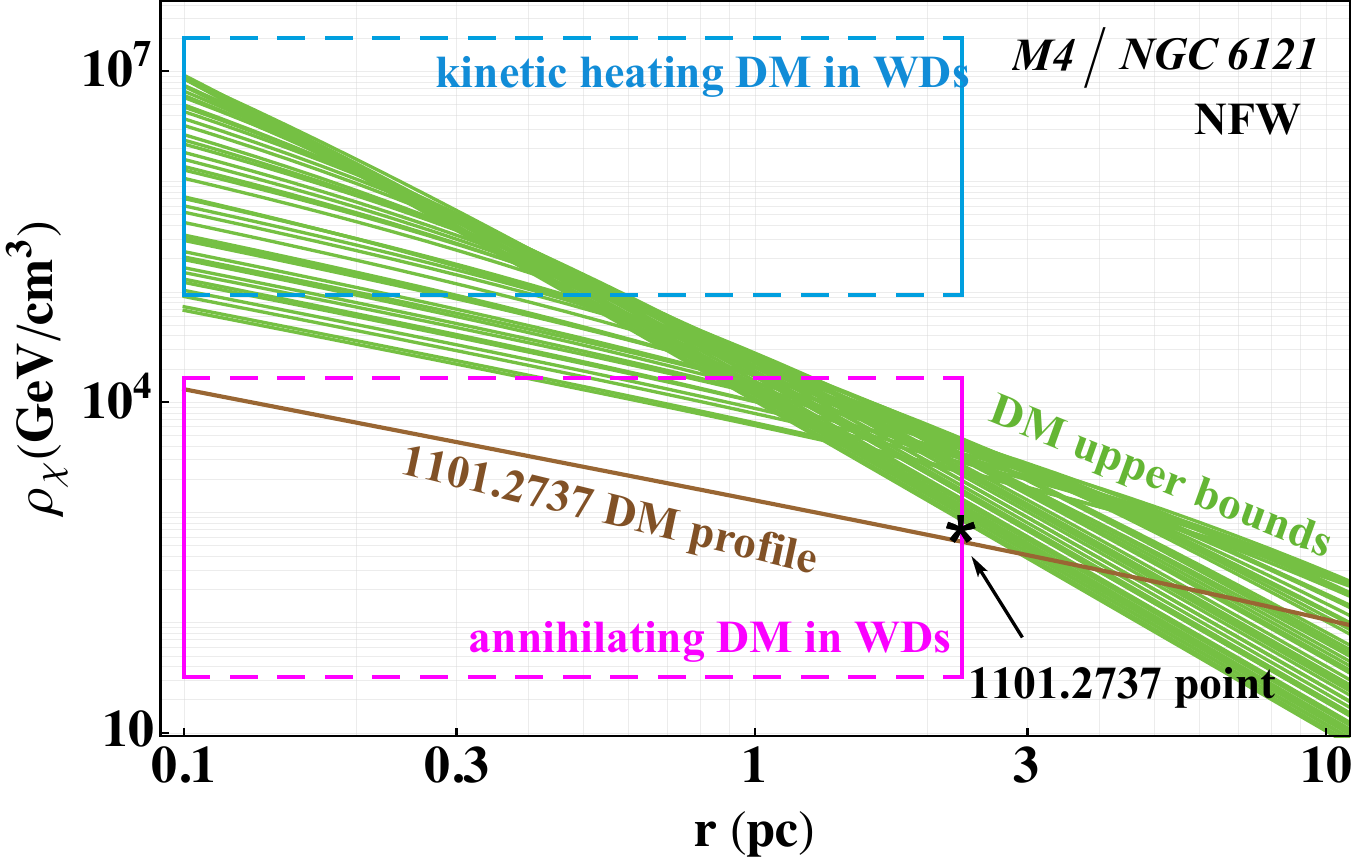} 
         \includegraphics[width=0.47\textwidth]{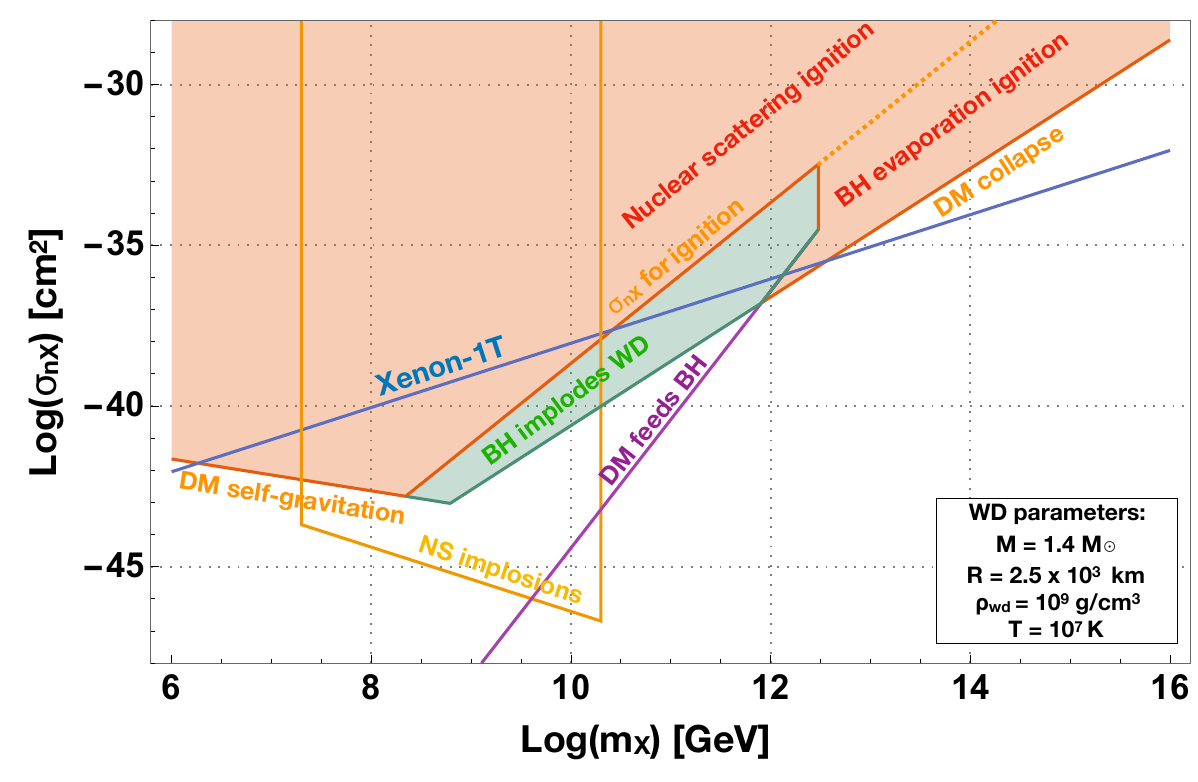} 
    \caption{{\bf \em Left.} Upper bounds from Ref.~\cite{Garani:2023esk} on DM density distributions in the globular cluster M4, compared with an estimate of the DM densities (labelled ``1101.2737") from Ref.~\cite{McCullough:2010ai} using a spherical collapse model. Also shown are the range of DM densities required to match the observed luminosities of WDs in M4 via DM annihilations within the WD as well as kinetic heating by infalling DM; the horizontal range of the rectangles spans the uncertainty in the positions of the WDs. {\bf \em Right.} Bounds on dark matter using an old WD in the Milky Way taken from~\cite{Acevedo:2019agu}.
    See Secs.~\ref{subsec:WDheating} and \ref{subsec:WDtoBH} for further details.}
    \label{fig:DMvWDlimits}.
     \end{figure*}

\subsection{Dark matter annihilation inside and heating white dwarfs} \vspace{0.2cm}
\label{subsec:WDheating}

The possibility that dark matter can accumulate inside and change the internal thermal properties of stars has long been appreciated~\cite{Press:1985ug,Gould:1987ir}. 
A number of works has proposed that old WDs could have their late-time temperature altered through accumulation and annihilation of DM in the interior~\cite{Bertone:2007ae,McCullough:2010ai,Hooper:2010es,Horowitz:2020axx}.
To a good approximation the amount of collisionless DM (for local DM density $\rho_\chi$ and average DM-WD relative speed $v_{\rm rel}$) flowing through a WD with mass $M_{\rm WD} = 1.2~ {\rm M_\odot}$, radius $R_{\rm WD}=4000$ km, and surface escape velocity $v_{\rm esc} =\sqrt{2G M_{\rm WD}/R_{\rm WD}}$ is
\begin{align}
    \dot M &= \rho_\chi v_{\rm rel} \times \pi\left(  \frac{R_{\rm WD} v_{\rm esc}}{v_{\rm rel}} \right)^2  =  10^{-7} ~{\rm \frac{M_\odot}{\rm Gyr}}~ \left(\frac{R_{\rm WD}}{4000~{\rm km}}\right)^2 \left(\frac{M_{\rm WD}}{1.2~{\rm M_\odot}}\right) \left( \frac{\rho_\chi}{0.4 ~{\rm GeV/cm^3}} \right),
    \label{eq:wdcap}
\end{align}
where we have normalized to the mass accumulated over a gigayear to emphasize that the DM mass accumulated inside the WD over the lifetime of the universe is only a tiny fraction of the stellar mass. 
This expression assumes that all DM incident on the WD is captured; for the DM-nucleon or DM-electron cross section dependence of the capture rate, see Refs.~\cite{Bramante:2017xlb,Acevedo:2019agu}. 

The late-time temperature of a benchmark WD described above, assuming it is determined by the capture and annihilation of all DM transiting the WD, is given by~\cite{Garani:2023esk}
\begin{equation}
    T_{\rm WD} \approx 4000~{\rm K}\left( \frac{350~{\rm km/s}}{v_{\rm rel}}\right)^{1/4} \bigg( \frac{\rho_\chi}{10^3~{\rm GeV/cm^3}}\bigg)^{1/4},
    \label{eq:dmwdheatapprox}
\end{equation}
where here we have normalized this expression to a typical $v_{\rm rel}$, but have chosen $\rho_\chi$ more than three orders of magnitude greater than the inferred DM density near most WDs whose temperatures have been determined. 
This is the DM density required for heating WDs above their expected late-time temperature shown in Figure~\ref{fig:coolingcurvesNSWD}. 
In practice, this means that in order to find or exclude DM this way, one would need to find an ancient WD in a region that conclusively has a high DM density. 

Reference~\cite{McCullough:2010ai} studied the heating effect that certain inelastic DM models would have on the late-stage temperature of WDs, and found that for a background DM density of $\rho_\chi \simeq 3 \times 10^4 ~{\rm GeV/cm^3}$, they would be sensitive to inelastic inter-state mass splittings of about $10-10^3$ keV and per-nucleon scattering cross sections $\sigma_{n \chi} \gsim 10^{-41}~{\rm cm^2}$. 
These authors proceeded to investigate whether WDs observed in a very dense self-bound stellar system, the globular cluster NGC 6121, a.k.a. Messier 4 (M4), might reside in a background density of DM large enough to observe heating from DM. 
Assuming that M4 was formed from a subhalo that was then tidally stripped by the Milky Way parent halo, using a spherical collapse model first derived in Ref.~\cite{Bertone:2007ae}, adopting an NFW density profile, and accounting for the slight adiabatic contraction of densities from the baryon potential, they estimated that the DM density was approximately 800 GeV/cm$^3$ at a cluster-centric distance $r = 2.3$~pc, where the farthest WDs were observed in the Hubble Space Telescope.
Following this, a number of authors investigated the implications of DM in globular clusters capturing in celestial bodies, under the assumption of a large (10$^3$-10$^4$ GeV/cm$^3$) DM density~\cite{Kouvaris:2010jy,Cermeno:2018qgu,Dasgupta:2019juq,Panotopoulos_2020,LongLivedMedNS:Leane:2021ihh,Bell:2021fye,Biswas:2022cyh,Ramirez-Quezada:2022uou}.

A recent study~\cite{Garani:2023esk} set empirical limits on the DM densities in M4 using measurements of stellar line-of-sight velocities and performing a spherical Jeans analysis; Figure~\ref{fig:DMvWDlimits} shows these limits on various DM density profiles corresponding to upper bounds on NFW scale parameters.
The density estimate of Ref.~\cite{McCullough:2010ai}, denoted by an asterisk, is safe from these limits. 
Nevertheless, it was argued that the use of globular clusters as copious sources of DM resulting in far-reaching conclusions about its microscopic properties is problematic for several reasons.
\begin{enumerate}
    \item The origin story of globular clusters is unclear. 
     While Ref.~\cite{McCullough:2010ai} echoed a popular theory -- corraborated in $N$-body simulations -- that globular clusters originate in DM subhalos that then strip tidally~\cite{SearleZinn:1978ApJ,Peebles:1984ApJ,Diemand:2005MNRAS}, alternative simulations suggest they may form with no aid from dark matter via the collapse of giant molecular clouds~\cite{Kravtsov:2003sm,Claydon_2019,Ashman2001,vandenBergh2001}.
     \item The V-band mass-to-light ratios of globular clusters in solar units is 1--5, which is equivocal about the presence of DM in them, unlike, say, dwarf galaxies (10--100), the Coma Cluster of galaxies (660) or the Milky Way (25) which are known to harbor significant amounts of DM. In fact, a structure defined as a stellar ``cluster" is {\em defined} as a system whose dynamics need not be explained by DM, unlike a ``galaxy"~\cite{clustervgalaxyWillman:2012uj}. 
     Accordingly, studies of more than 20 globular clusters looking for DM in them have either failed to detect DM or come to ambiguous conclusions~\cite{Garani:2023esk}. 
     \item There is no guarantee that any invisible mass favored in globular cluster data is in fact DM, as it may also be from faint stellar remnants~\cite{EvansStrigari:2021bsh}.
     \item The interpretation of the presence or absence of DM in ambiguous datasets is sensitive to priors and parametrizations. 
      Ref.~\cite{Ibata2012} found no evidence for DM when analyzing NGC 2419 by fitting a Michie model for the stellar and a generalized NFW profile for the DM distributions, but found strong evidence for DM when  fitting these quantities with {\em no} analytic form, floating instead 389 free parametes.
\end{enumerate}
One could conclude that, due to these significant uncertainties and the related infeasibility of determining DM density distributions in globulars with current and imminent measurement sensitivities, globular clusters are sytems that are far from robust for making statements about DM interactions.
On that note, there are proposals for finding WDs in dwarf galaxies like Segue I and II~\cite{Krall:2017xij}.

\subsection{Non-annihilating dark matter converting white dwarfs into black holes} \vspace{0.2cm}
\label{subsec:WDtoBH}

If enough non-annihilating DM accumulates in WDs, the DM can collapse, and subsequently form a small black hole that accretes surrounding WD material, eventually consuming the entire WD~\cite{Kouvaris:2010jy,Acevedo:2019agu,Janish:2019nkk}. 
Typically the DM is assumed to be ``asymmetric" since in such models DM typically does not self-annihilate~\cite{Petraki:2013wwa}. If in the process of accumulation and collapse DM self-annihilates efficiently, too much of it may be lost to form a black hole in the WD core. 

The routine by which DM could form a small black hole in the interior of a WD is very similar to the more studied case of DM  forming black holes in NSs\footnote{See also Ref.~\cite{Acevedo:2020gro,Ellis:2021ztw,Diamond:2021scl}, which study black hole formation in other astrophysical bodies like the Earth, Sun, and Population III stars.}, which is detailed in length in Section~\ref{subsec:BHinNS}. 
To avoid repetition, here we will emphasize aspects that are distinct from the case of the NS.
The WD-to-BH conversion process is as follows. 
First, DM accumulates in the WD over time, through scattering on nuclei or electrons in its interior.
Then, the captured DM thermalizes with the WD interior, i.e., after repeated scattering it is expected to localize within a small volume determined by the WD's internal temperature and gravitational potential.

One chief difference here between WDs and NSs is that during thermalization, DM will scatter with a Coulomb lattice of ions in the core of the WD, which is stabilized by relativistic electron degeneracy pressure. 
This effect considerably suppresses DM-nucleus scattering rates at low momentum transfers, the regime that determines the thermalization timescale $t_{\rm th}^{\rm WD}$.
For a carbon WD, this is given by~\cite{Acevedo:2019agu}
\begin{equation}
    t_{\rm th}^{\rm WD} \simeq 20~{\rm yr}~ \left(\frac{10^{-40} ~{\rm cm^2}}{\sigma_{n \chi}} \right) \left(\frac{m_\chi}{10^6~{\rm GeV}} \right)^2 \left(\frac{10^7~{\rm K}}{T_{\rm WD}} \right)^{5/2}~.
    \end{equation}
Thus for $m_\chi > 10^{10}$~GeV, it can take $>$ Gyr for DM to thermalize with the WD interior.
An exception to this mode of thermalization is low-mass helium WDs, which do not crystallize due to small Coulomb binding energy.
Conversion of $\Oc(0.1)~M_\odot$ helium WDs to black holes by dark matter over $10^{14}$~yr timescales can have implications for recurrence cosmology in models where new universes are spawned in the interior of black holes, with the relevant DM parameter space testable in direct detection and astrophysical experiments~\cite{BramanteRajSelfReplic:2024idl}.

Another difference between DM collapsing to form black holes in WDs and NSs is that, during the collapse inside a WD, DM may trigger a star-destroying thermonuclear explosion.
We now turn to this topic.

\begin{figure}
    \centering
         \includegraphics[width=0.8\textwidth]{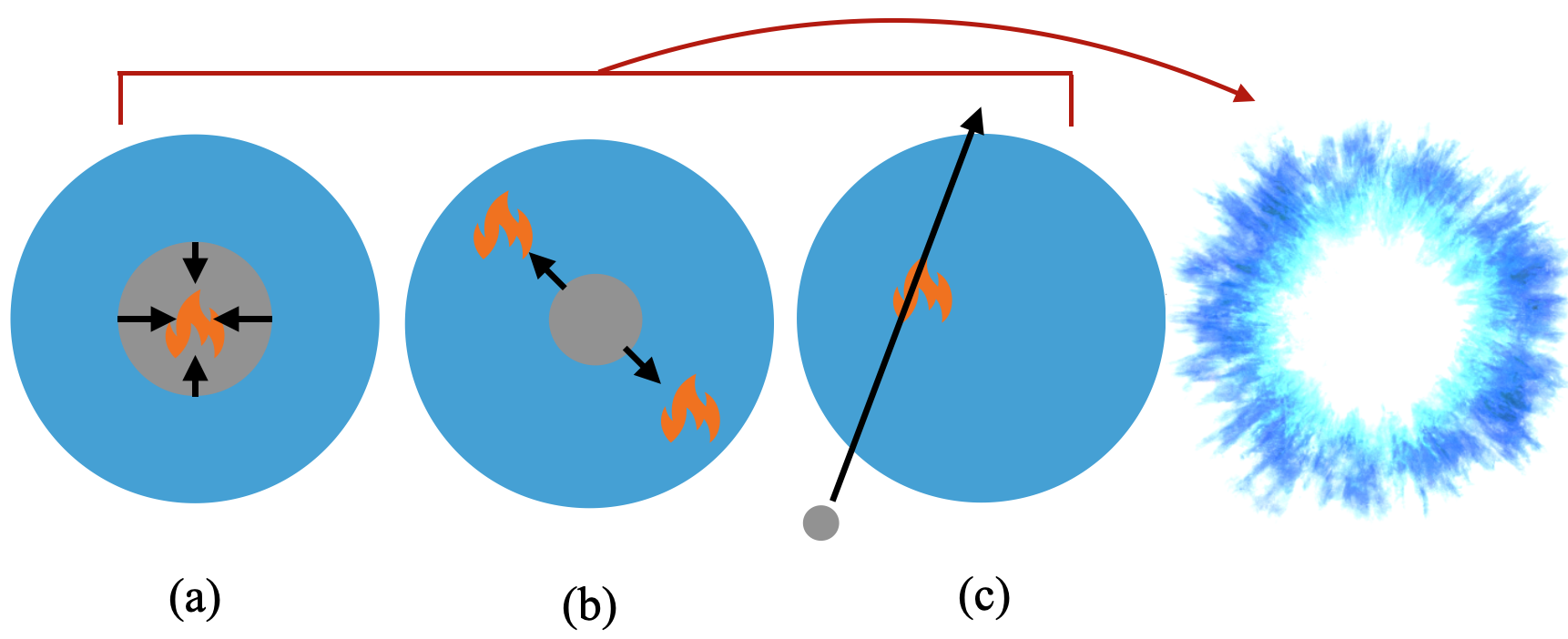} 
    \caption{Illustration of mechanisms by which WDs may be prompted to explode by dark matter. 
    (a) DM acumulates to the point of collapse in the center of the WD, then while collapsing (or after collapsing and forming a black hole) heats the WD to a temperature inducing a thermonuclear chain reaction. 
    (b) The internal potential or mass energy of a large composite DM state is transferred as WD nuclei enter the DM's interior, prompting local heating that initiates the thermonuclear runaway.
    (c) Macroscopic DM transiting the WD transfers explosive kinetic energy via scattering on WD constituents.}
    \label{fig:wdexplo}
\end{figure}

\subsection{White dwarf explosions via dark matter} \vspace{0.2cm}
\label{subsec:DMvWDboom}

Dark matter accumulated inside WDs might trigger a Type Ia-like supernova explosion through the deposition of enough energy to prompt runaway fusion reactions in the carbon/oxygen/neon interior of the WD~\cite{Bramante:2015cua,Graham:2015apa}; see also Ref.~\cite{Leung:2013pra} for an early discussion of DM cores affecting Type Ia supernovae.

More generally, DM triggering WDs into supernovae can proceed in a number of ways:

 \begin{itemize}
 
    \item Attendant to DM converting WDs to black holes, DM can collect into a core region of the WD, collapse, and as a result of the collapse, deposit enough energy to ignite the WD. 
    Ignition can occur either directly through nuclear scattering during the collapse of the DM core~\cite{Bramante:2015cua,Graham:2015apa,Fedderke:2019jur} or through the evaporation of a small black hole that forms out of the collapsed DM core~\cite{Acevedo:2019gre,Janish:2019nkk,Fedderke:2019jur}.
        
   \item DM can have internal properties that result in energy being liberated as WD particles enter the DM state. 
   A simple example of this is captured (and possibly thermalized) DM annihilating and depositing energy in the WD medium with which it is admixed~\cite{Graham:2018efk}. 
   Other interesting possibilities are composite DM with an internal potential for baryons~\cite{Acevedo:2020avd},
   solitonic Q-ball DM that absorbs baryonic charge and dissociates nuclei in the process~\cite{Graham:2018efk}, 
   monopoles that possibly induce nucleon decay in similar fashion (Sec.~\ref{subsubsec:NSheatDManns}),
   and accretion of WD carbon ions onto a black hole formed from collapse of electrically charged DM states~\cite{Fedderke:2019jur}. 

       \item During an initial transit through the WD, DM can deposit kinetic energy gained by falling into the WD's gravitational potential. 
       The DM could be in the form of primordial black holes (PBHs), in which case energy is transferred via dynamical friction~\cite{Graham:2015apa,Montero-Camacho:2019jte}, or particles inducing nuclear scatter recoils~\cite{Graham:2018efk,MACROSidhu2020,WDNSboom:Raj:2023azx}. 
       Tightly bound asteroid-like DM triggering WD explosions via stellar shocks has also been suggested~\cite{DasEllisSchusterZhou:2021drz}.
    \end{itemize}

However the WD is heated, a number of requirements must be met for thermonuclear reactions sparked to sustain themselves and cause the WD to explode. 
These requirements are described in Sec.~\ref{subsec:DMvWDboom}.
We now discuss some subtle aspects of this phenomenon as explored in the literature.

A detailed simulation of PBH energy deposition in a WD, including the effect of turbulent flows in the wake of the passing PBH, found that heavier PBHs were required to ignite WDs~\cite{Montero-Camacho:2019jte} compared to initial estimates~\cite{Graham:2015apa}. 
This study employed a 1D+1D hydrodynamic simulation of the shock front created by a transiting PBH, and found that the development of hydrodynamic instabilities dissipating heat deposited through dynamical friction appeared to occur more rapidly than ignition processes, which were modeled using the same carbon fusion reaction rates used in Ref.~\cite{TimmesWoosley1992}. Instead of a burning reaction, the prompt detonation of WD material by transiting PBHs was further studied in Ref.~\cite{Steigerwald:2021vgi}.
Another study investigated ignition during DM core collapse using a system of differential equations that track the evolution of per-particle energies~\cite{Steigerwald:2022pjo} - this tracked carbon fusion inside the collapse region, and found carbon depleted before the WD ignition temperature in Ref.~\cite{TimmesWoosley1992} was reached.
Future work could build on this result in a number of directions, $e.g.$ by studying C-O burning  and employing the full nuclear reaction network used in Ref.~\cite{TimmesWoosley1992} to obtain WD ignition temperatures.
In addition, future WD ignition estimates should also consider convective flows of heated WD material moving carbon through the collapse region replenishing carbon and oxygen, and whether WD ignition occurs via thermal energy transported out of the collapse region. 
This is especially important, since studies on carbon fusion occurring inside DM bound states found that fusion can be induced in the region surrounding the collapsing region that is the source of heat, either through the evaporation of black holes of size much smaller than the ignition region, or through effluence of thermal energy outside of the transiting DM composite~\cite{Acevedo:2019agu,Janish:2019nkk,Acevedo:2020avd,Fedderke:2019jur}. 
Finally, the ignition of WD supernovae via oxygen burning typically requires a temperature somewhat higher than that of carbon\footnote{We thank Melissa~Diamond for correspondence on this point.}~\cite{TimmesWoosley1992}, and future detailed treatments of WD ignition by collapsing DM should account for this possibility.

In Fig.~\ref{fig:NSvclumps} bottom left panel we show the masses and radii of DM mini-clusters constrained by the observed existence of WDs in our Galaxy, taken from Ref.~\cite{WDNSboom:Raj:2023azx}.
Overlaid here are contours of the minimum DM-nucleus elastic scattering cross sections required to transfer sufficient kinetic energy to the WD trigger volume to induce stellar explosion.
The cross section ceilings here, coming from the overburden of the stellar outer layers, are discrepant with Refs.~\cite{Graham:2018efk,MACROSidhu2020} by several orders of magnitude, possibly due to these studies committing errors in reading off values of relevant quantities from textbooks; see Ref.~\cite{WDNSboom:Raj:2023azx} for an explanation.

A number of phenomena have been linked to the DM-induced ignition of thermonuclear explosions in WDs. 
It has been posited that DM core collapse in WDs might account for a large fraction of observed Type Ia supernovae~\cite{Bramante:2015cua}, as a solution to the Type Ia progenitor problem~\cite{Maoz:2011iv} and consistent with the apparent observation of sub-Chandrasekhar WDs as the origin of most Type Ia supernovae~\cite{Scalzo:2014wxa}. Reference~\cite{Bramante:2015cua} also found that a trend in existing Type Ia data~\cite{Pan:2013cva}, showing that more massive progenitors explode sooner, is consistent with certain DM models that induce WD explosions through DM core collapse, where this would occur sooner for heavier WDs. 
Dark matter-induced supernovae may be partly responsible for galactic stellar feedback and subsequent alterations in star formation rates and the halo density profile~\cite{Acevedo:2023cab}.
The accumulation in certain sub-Chandrasekhar WDs of charged massive particles (CHAMPs) making up DM, which might occur preferentially outside galaxies with magnetic fields that serve to deflect CHAMPs, could be an explanation of the distribution of calcium-rich gap transient WD supernovae~\cite{Fedderke:2019jur} that do explode preferentially on the outskirts of galaxies~\cite{Kasliwal:2011se}. The distribution of Type Ia supernovae in galaxies could be tied to local properties like velocity dispersion, especially in the case of PBH-ignition~\cite{Steigerwald:2021bro}.
Finally, a separate study has investigated whether WD explosions from DM could explain the aforementioned Ca-rich gap transient distribution, through the ignition of WDs in dwarf spheroidal galaxies expected to be located at some distance from larger galactic DM halos~\cite{Smirnov:2022zip}.

\subsection{Dark matter's influence on white dwarf equations of state} \vspace{0.2cm}
\label{subsec:DMvWDEOS}

WD mass-radius relationships can also be observably impacted by DM. 
If a substantially massive core of DM accumulated in the interior of a WD, its stable configurations would be altered through revised TOV equations~\cite{Leung:2013pra,Leung:2019ctw,Zha:2019lxw}.
For a typical circumambient DM density, the amount of collisionless DM required to induce these effects, $10^{-4} M_\odot - 10^{-1}~M_\odot$, well exceeds what could be collected in the WD over the lifetime of the universe; see Eq.~\eqref{eq:wdcap}. However, future studies could investigate whether such a large quantity of DM might be collected through collisional accretion, analogous to the NS treatment in Ref.~\cite{NSvIR:clumps2021} (discussed in Sec.~\ref{subsec:DMsubstruct}). 
Another effect comes through the axion: its existence implies that its non-derivative coupling to nucleons would displace it from its usual minimum in a finite-density medium. 
This results in a reduction of the nucleon mass and alters the TOV equations~\cite{Balkin:2022qer}. 

\section{The neutron star as a dark matter laboratory} 
\label{sec:NSvDM}

\begin{figure*}
    \centering
    \includegraphics[width=0.33\textwidth]{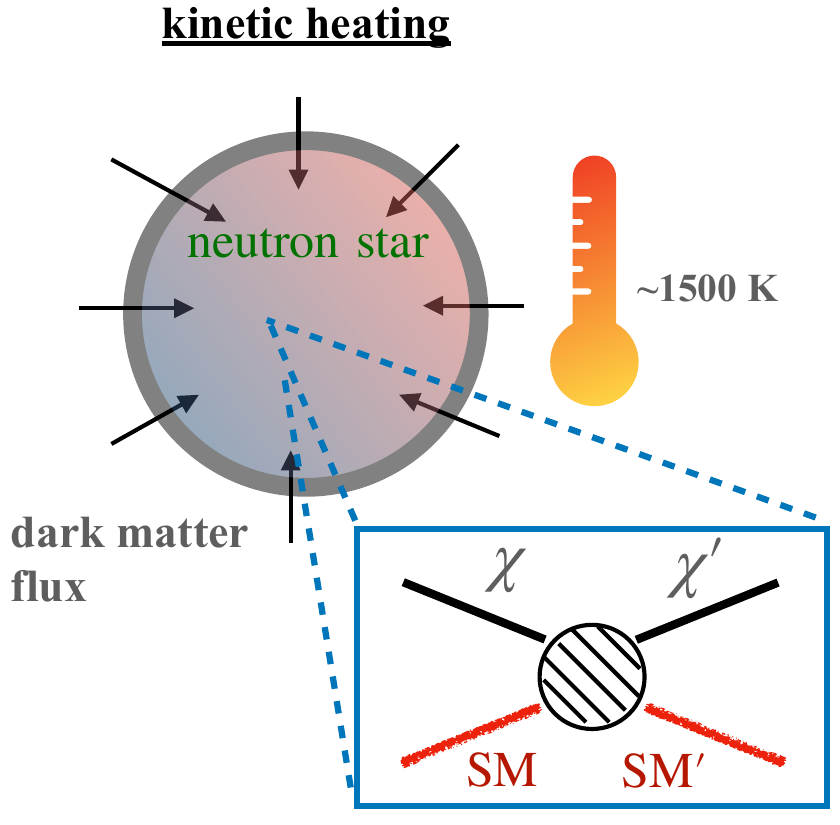}  \ 
     \includegraphics[width=0.28\textwidth]{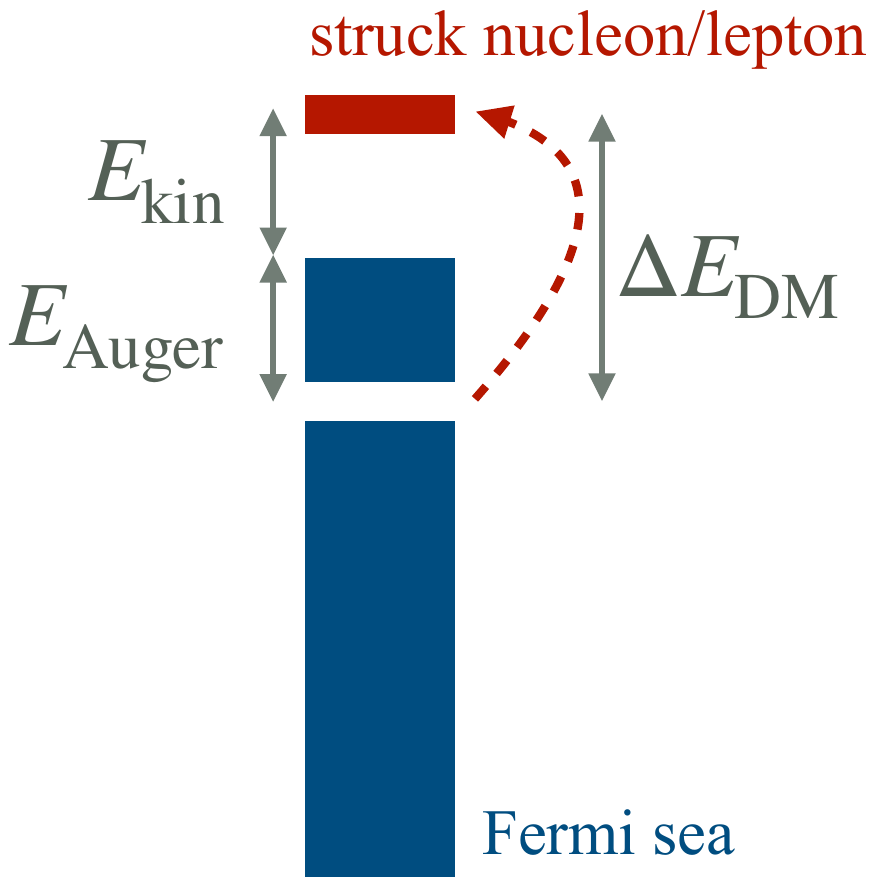}  \ 
         \includegraphics[width=0.28\textwidth]{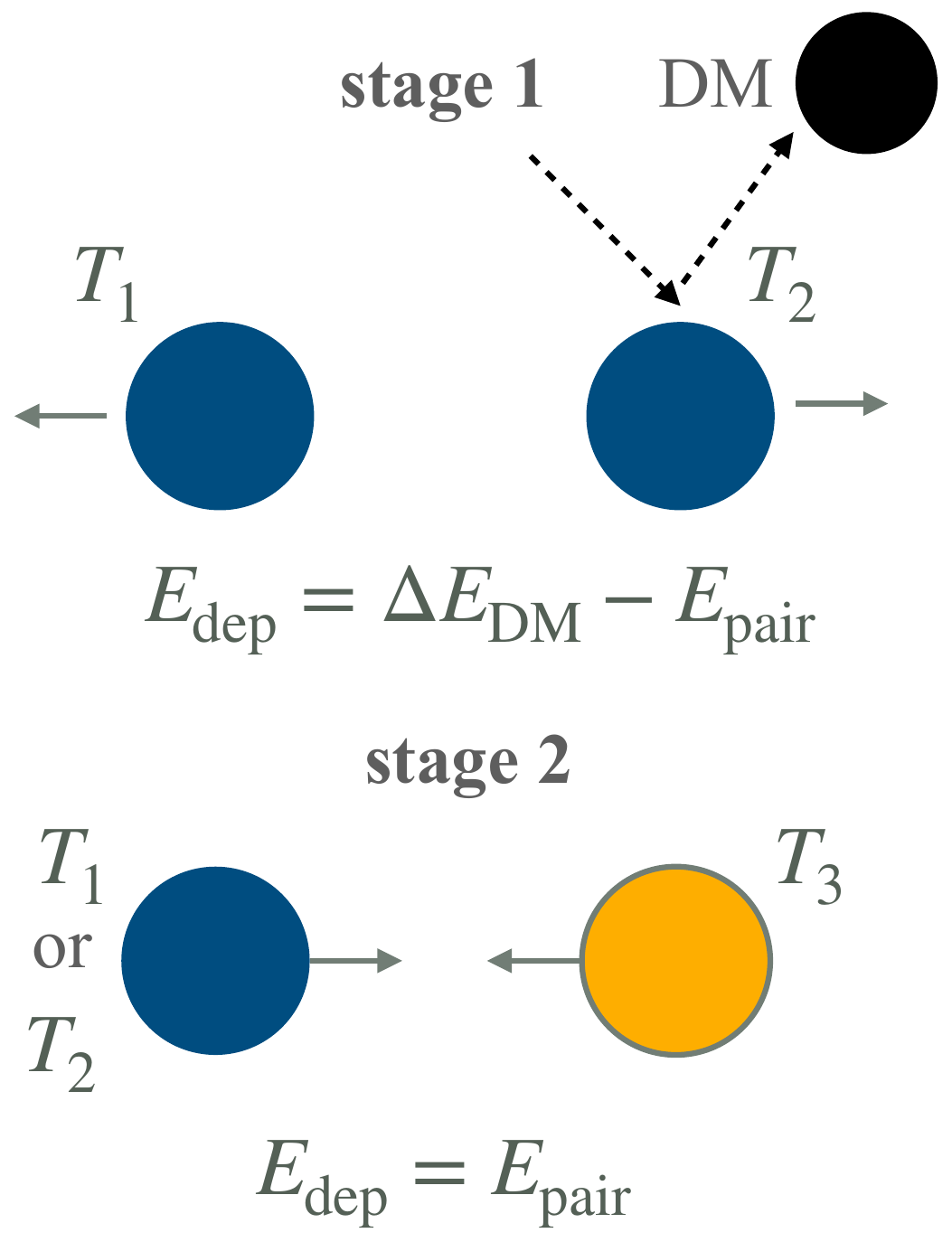} \\
    \includegraphics[width=0.33\textwidth]{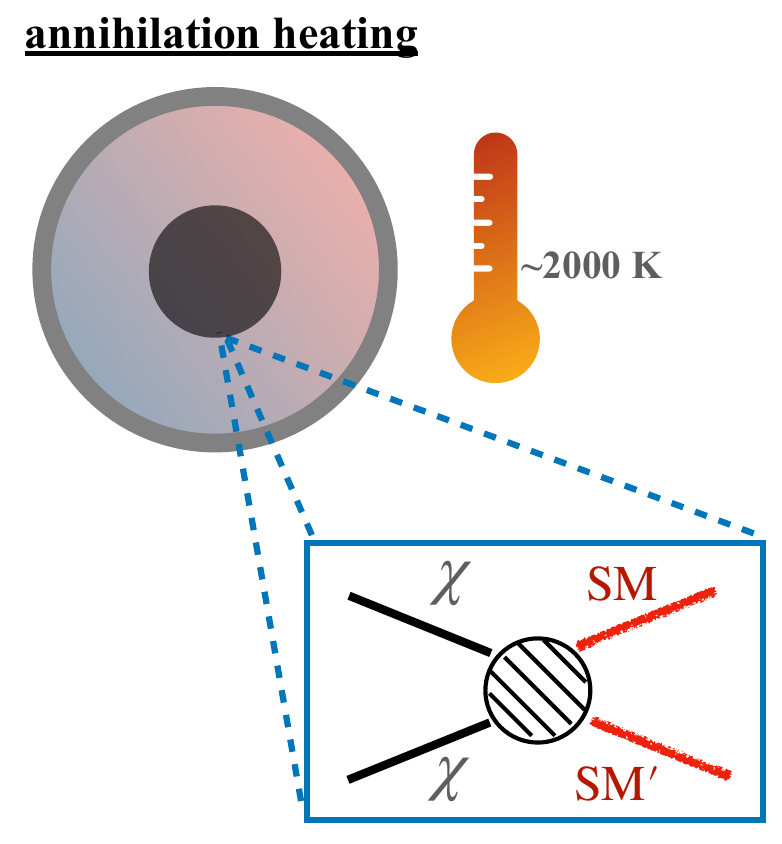}  \
    \includegraphics[width=0.45\textwidth]{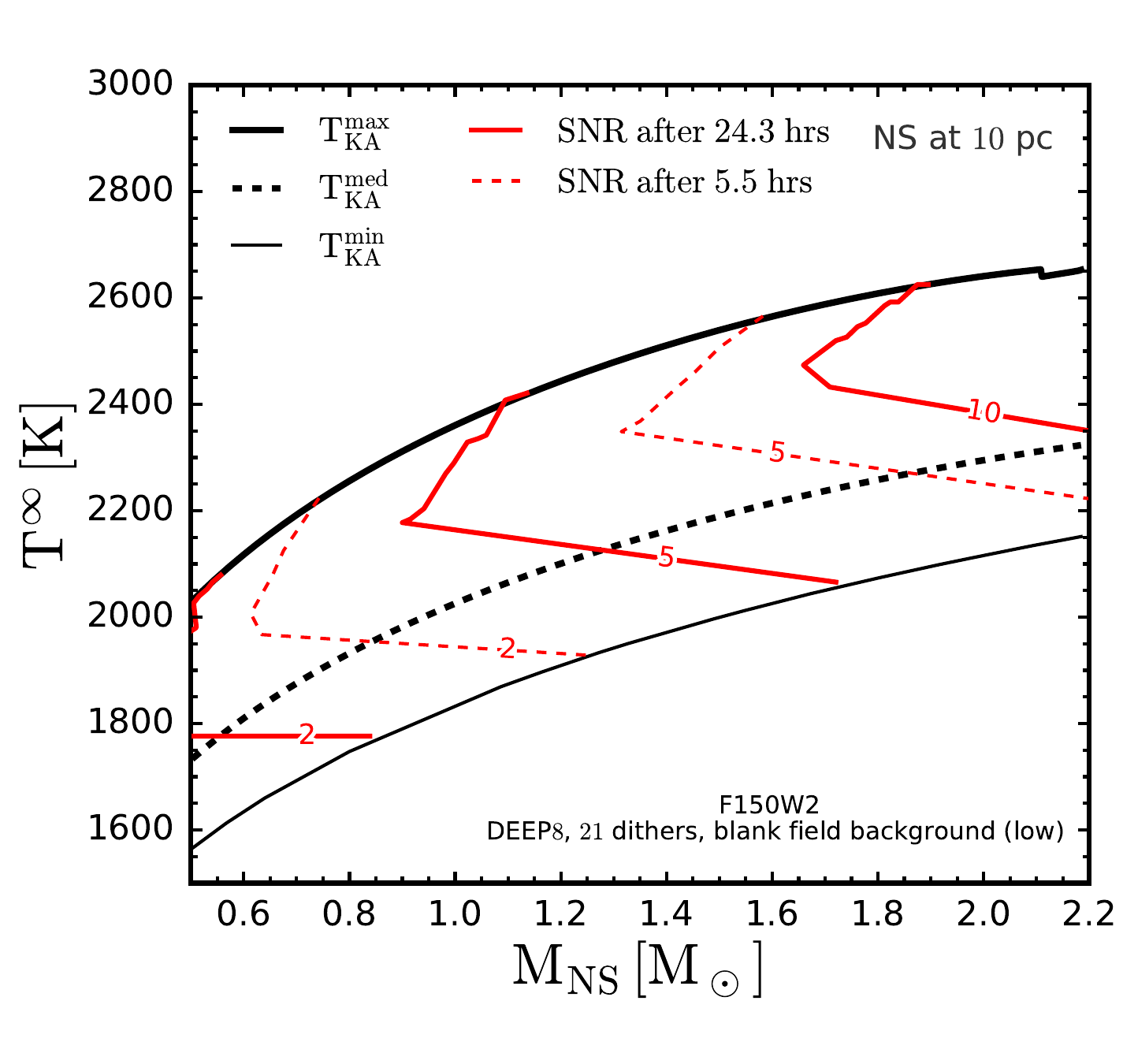}  \ 
    \caption{{\bf \em Top left.} Cartoon showing the dark kinetic heating effect in NSs. Scattering interactions of the infalling dark matter flux contribute to the luminosity of a typical NS at the level of a $1500$~K blackbody temperature.  
    {\bf \em Top middle.} The nucleon Auger effect that contributes to kinetic (and possibly annihilation) heating by dark matter in NSs. The total energy deposited after scattering turns out to be the dark matter energy transfer, although physically it comes as the sum of two contributions: the energy spilled during the rapid filling of the hole left behind by the struck target, and the energy carried by the target in excess of the Fermi energy.
 {\bf \em Top right.} The breaking and re-pairing of Cooper pairs that contributes to kinetic (and possibly annihilation) heating by dark matter in NSs. 
 This phenomenon takes place for dark matter with mass above about 35 MeV; for smaller masses, dark matter capture proceeds through collective excitations in the nucleon superfluid medium.
 {\bf \em Bottom left.} Cartoon showing possible additional heating of NSs via self-annihilations of dark matter possibly collected in a thermalized volume. This highly model-dependent process could heat the NS to blackbody temperatures around 2000~K.
 {\bf \em Bottom right.} As a function of NS mass, NS effective temperatures imparted by dark kinetic+annihilation heating that can be measured at the James Webb Space Telescope at various signal-to-noise ratios, taken from Ref.~\cite{NSvIR:IISc2022}.
 The band denotes variations over NS radii predicted by numerous equations of state as well as NS-DM relative velocities from  estimates by various NS population models.
 See Sec.~\ref{subsubsec:DMcapturekineticheat} for further details.}
    \label{fig:kineticaugerpairbreak}
\end{figure*}

\subsection{Dark matter kinetic and annihilation heating of neutron stars} \vspace{0.2cm}
\label{subsec:DMkinannheat}

\subsubsection{Capture and kinetic heating}
\label{subsubsec:DMcapturekineticheat}
\begin{figure*}
    \centering
         \includegraphics[width=0.47\textwidth]{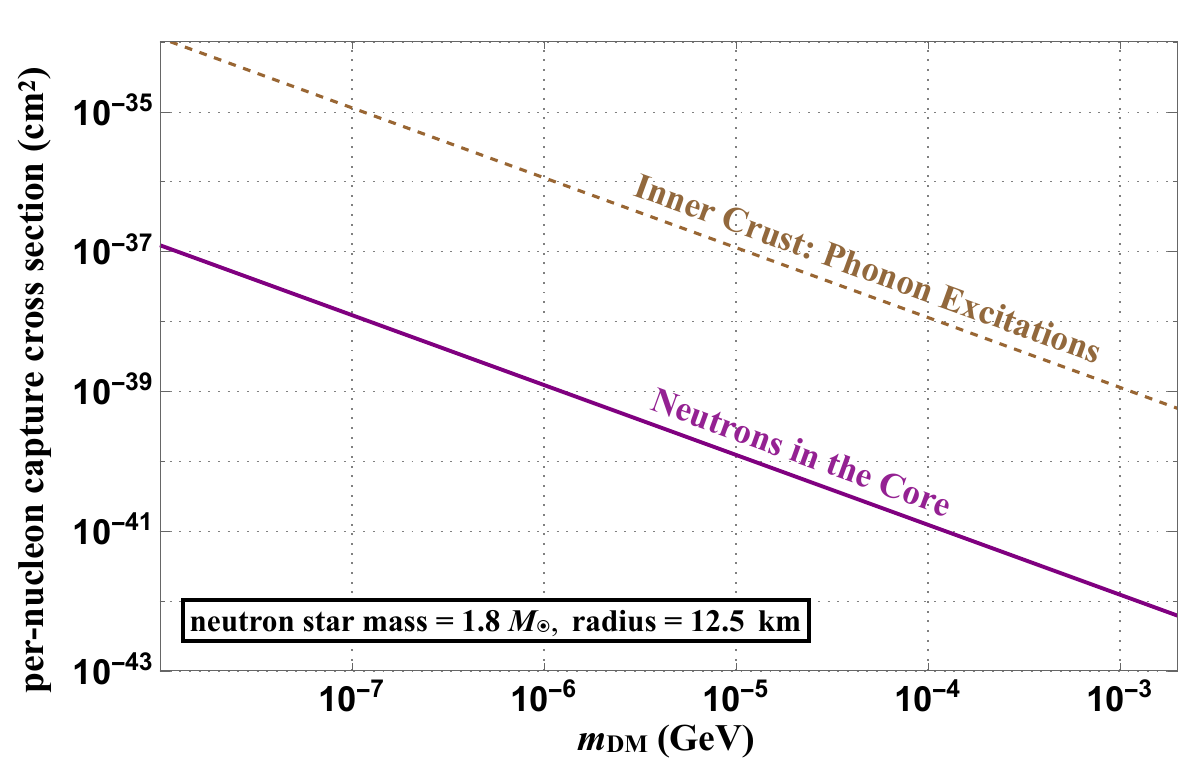}  \ \ \ \ \ \ \ 
         \includegraphics[width=0.47\textwidth]{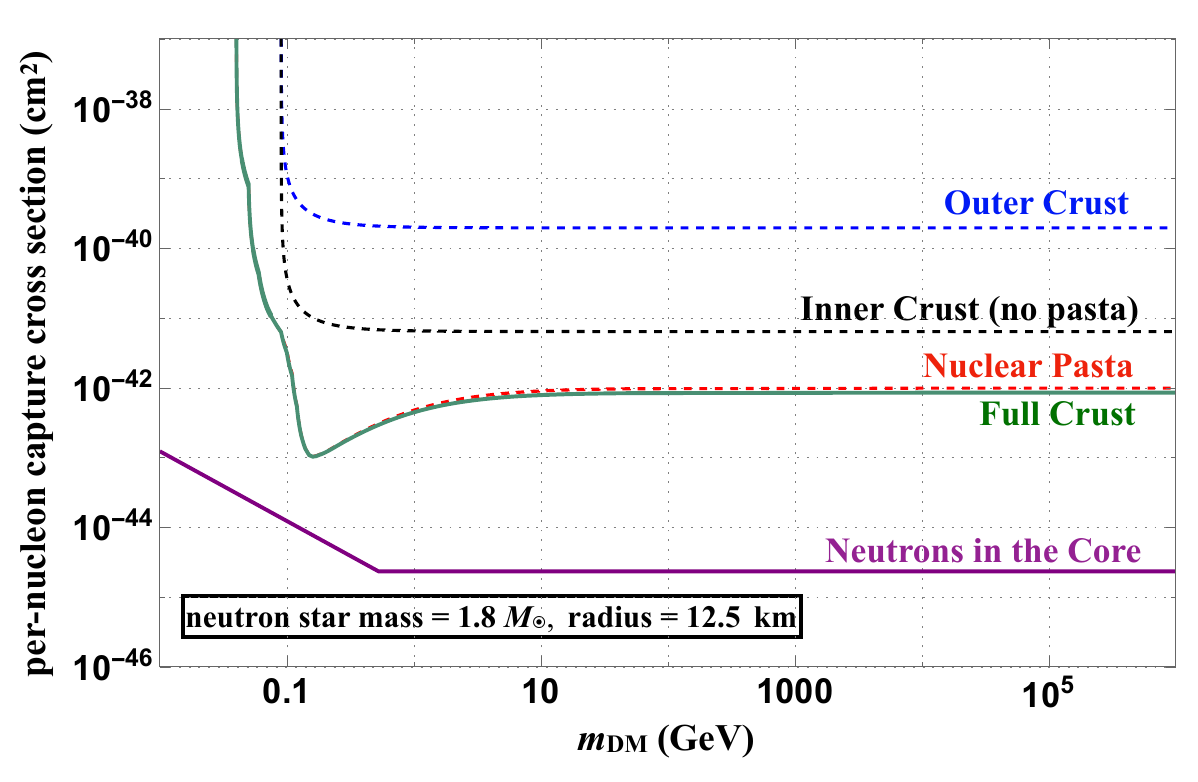} \\
          \includegraphics[width=0.47\textwidth]{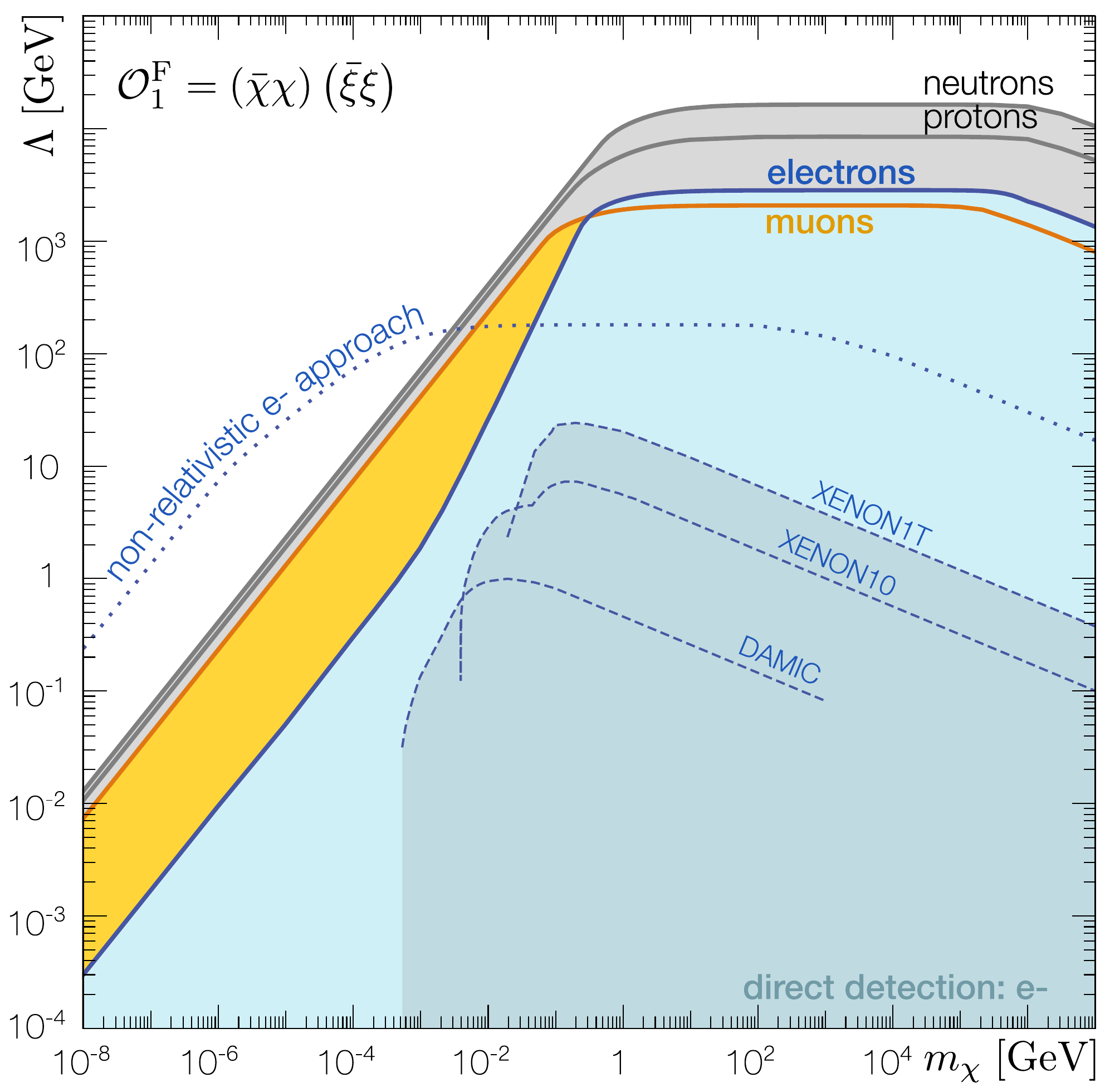}  \ \ \ \ \ \ \ 
         \includegraphics[width=0.47\textwidth]{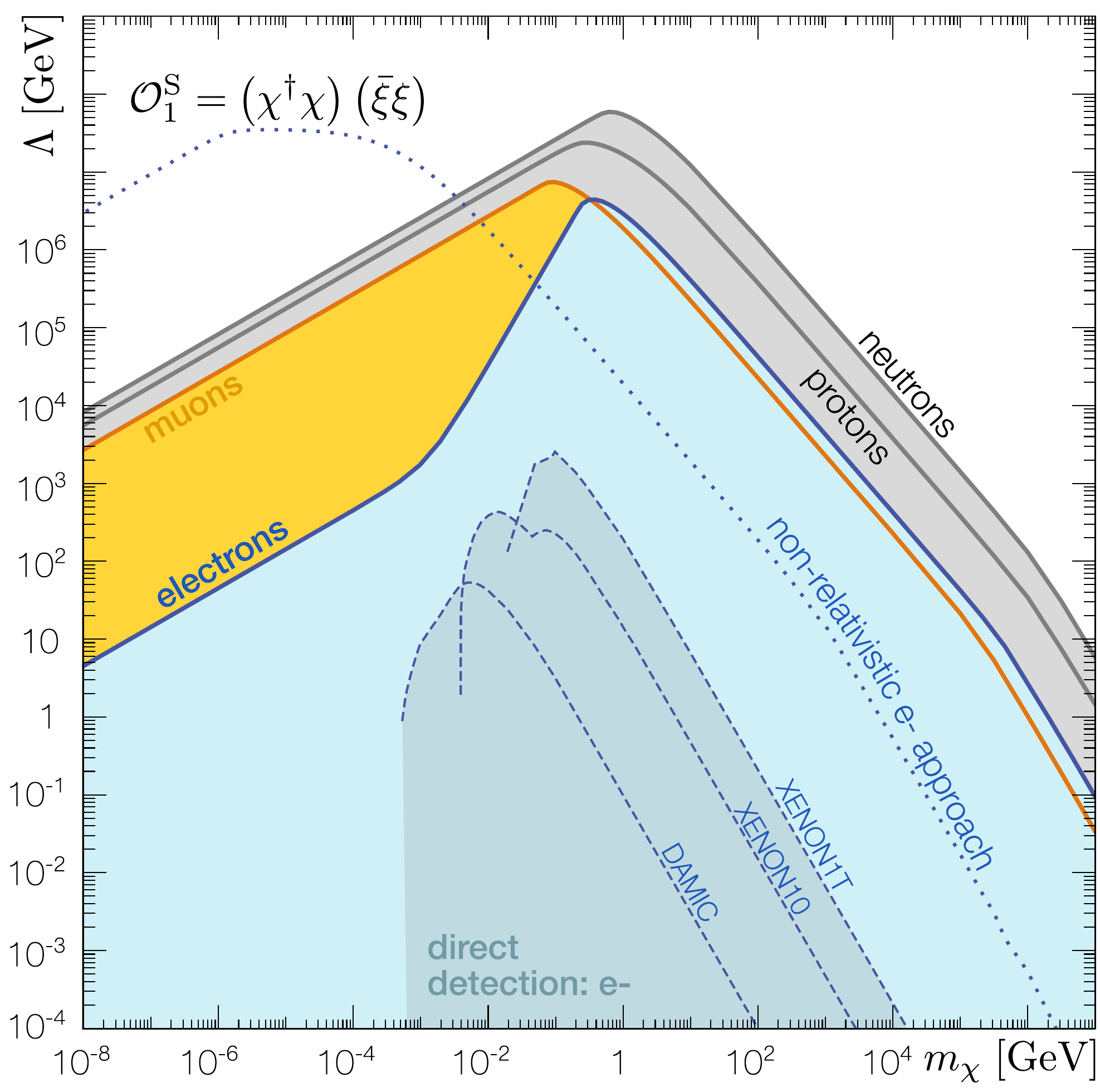} 
    \caption{{\bf \em Top.} Capture cross section sensitivities for light dark matter scattering in a NS crust ({\bf \em left}) (via excitation of superfluid phonons in the inner core) and in the NS core (via Pauli-blocked contact scattering on neutrons, although see Sec.~\ref{subsubsec:DMcapturekineticheat} for a discussion on scattering in the superfluid core), and for heavier dark matter scattering in various layers of the crust and the core ({\bf \em right}). 
  These two plots are taken from Ref.~\cite{NSvIR:Pasta}. 
  See Sec.~\ref{subsubsec:DMcapturekineticheat} for further details.
 {\bf \em Bottom.} Sensitivities to the cutoff of effective CP-even scalar interactions of dark matter with relativistic, degenerate electrons in a NS, for DM that is spin-1/2 ({\bf \em left}) and spin-0 ({\bf \em right}). 
 Also shown are the sensitivities for interactions with muons, protons and neutrons. 
 The electron scattering limits are seen to widely complement terrestrial searches.
 These two plots are taken from Ref.~\cite{NSvIR:Riverside:LeptophilicLong}.
 See Sec.~\ref{subsubsec:DMcapturekineticheat} for further details.}
    \label{fig:NScapcrustcore}
\end{figure*}

\begin{table*}[t]
    \centering
    \begin{tabular}{|c|c|c|c|}
\hline
        effect & change in capture rate & applicability & reference \\
         \hline
         \hline
       \multirow{2}{*}{EoS of star effects} & $\mathcal{O}$(1): BSk20 $\to$ 21 & all $m_\chi$ &~\cite{NSvIR:GaraniGenoliniHambye} \\
         & none: QMC-2 $\to$ BSk24 & all $m_\chi$ &~\cite{NSvIR:anzuiniBell2021improved} \\
        \hline
        mass-radius configuration   & $\mathcal{O}(100)$ as $1 \to 2.2 M_\odot$ & all $m_\chi$ &~\cite{NSvIR:Bell:Improved} \\
        \hline
       nuclear self-energies  &\multirow{2}{*}{30$-$100} & $m_\chi >$~100 MeV, any EoS &\cite{NSvIR:Bell2020improved}  \\
    nucleon structure & & $\mathcal{O}(10^3)$ for 2 $M_\odot$ NSs &~\cite{NSvIR:anzuiniBell2021improved} \\
        \hline
        non-elastic scattering & comparable & $>2 M_\odot$ NSs &~\cite{NSvIR:anzuiniBell2021improved,NSvIR:DIS:Su:2024flx}\\
        \hline
        ``collective" effects & $\mathcal{O}(1-10^3)$ & 2 $M_\odot$ NS, &~\cite{collectiveDeRoccoLasenby:2022rze} \\
        &  &  $m_\chi < 100$~MeV, & \\
         & &  $A'$ mediator & \\
        \hline
        superfluidity: energy gap & maybe $\mathcal{O}$(1) & $m_\chi \lesssim 35$~MeV, &~\cite{NSvIR:Pasta} \\
        & & single phonon excitation &~\cite{NSvIR:clumps2021} \\
        \hline
       NS opacity/ extinction factor & $\mathcal{O}(1)$ & $m_\chi >$~GeV &~\cite{NSvIR:Bell:Improved} \\
        \hline
       \multirow{2}{*}{relativistic kinematics} &  $\sim 4$ & $m_\chi >$~GeV &~\cite{NSvIR:Bell:Improved} \\
       &  $\sim 10$ & $m_\chi <$~GeV &~\cite{NSvIR:Bell:Improved}\\
        \hline
      gravitational focusing  & $< 2$ & all $m_\chi$&~\cite{NSvIR:Bell:Improved} \\
        \hline
    \multirow{2}{*}{light mediator kinematics} &  $\mathcal{O}(1)$ & $m_\phi/\mu_{\rm red} < 10^{-1}$ &  \multirow{2}{*}{\cite{NSvIR:DasguptaGuptaRay:LightMed}} \\
    & voided & $m_\phi/m_\chi < 10^{-4}$ & \\
        \hline
        DM halo velocity distribution  & $< 2$ & all $m_\chi$&~\cite{Bose:2022ola} \\
        \hline
    \end{tabular}
    \caption{Known effects that modify the rates of dark matter capture in NSs. See Sec.~\ref{subsubsec:improvementsNScap} for further description.}
    \label{tab:NSsideuncerts}
\end{table*}

NSs are excellent captors of particle dark matter by virtue of their extreme densities and steep gravitational potentials, and also quite serviceable as thermal detectors thanks to their typically small temperatures. 
While the capture of DM in NSs and its subsequent thermal relaxation was first treated in Ref.~\cite{Goldman:1989nd},
it was only recently realized that this could be a minimal probe of dark matter scattering on Standard Model (SM) states: the transfer of DM kinetic energy to the NS's constituent particles during the infall of DM at semi-relativistic speeds overheats the NS~\cite{NSvIR:Baryakhtar:DKHNS}.
It was also proposed that upcoming infrared telescopes, 
{\em e.g.}, 
the Thirty Meter Telescope (TMT)~\cite{TMT:2015pvw} 
and the Extremely Large Telescope (ELT)~\cite{ELT:neichel2018overview} 
are sensitive to this ``dark kinetic heating" mechanism~\cite{NSvIR:Baryakhtar:DKHNS} for NSs out to about $100$ pc from Earth; a study has also been dedicated to the sensitivity at the recently launched James Webb Space Telescope (JWST)~\cite{JWST:Gardner:2006ky,NSvIR:IISc2022}, which has shown that finding an NS much closer than 100 pc would likely be required.
Thermal observations of nearer pulsars could be made following the discovery of old, isolated NSs in radio telescopes such as FAST~\cite{FAST2011}, CHIME~\cite{CHIME2021} and SKA~\cite{SKA:2004nx}.
Though their $B$ fields and rotational velocities are expected to be low, implying they populate regions near the ``pulsar death line" in $P$-$\dot P$ space beyond which NSs are supposed to stop pulsing, NSs have been observed beyond the death line~\cite{PulsarDeathLineAnomaly:PSRJ2144-3933,PulsarDeathLineAnomaly:PSRJ0250+5854,PulsarDeathLineAnomaly:PSRJ2251-3711,PulsarDeathLineAnomaly:PSRJ0901-4046} calling into question models of NS pulsation (as also discussed in Sec.~\ref{subsec:Bspindown}).
It is estimated that about $10^5$ NSs in the Galaxy lie beyond the death line~\cite{PulsarDeathLineAnomaly:PSRJ2144-3933}.

To illustrate the idea of dark kinetic heating let us consider the following representative NS configuration:
\bea
\nn M_{\rm NS} &=& 1.5 \ M_\odot,~~R_{\rm NS} = 12.85 \ {\rm km} \\ 
\Rightarrow v_{\rm esc} &=& \sqrt{\frac{2 G \MNS}{\RNS}} \simeq 0.59 \ . 
\label{eq:BMNS}
\eea
where $v_{\rm esc}$ is the escape speed at the surface.
This configuration is obtained for a Quark Meson Coupling~(QMC) EoS of matter~\cite{NSvIR:anzuiniBell2021improved}. 

For local DM density $\rho_\chi$ and average DM-NS relative speed $v_{\rm rel}$ (which in the solar vicinity are $0.4$~GeV/cm$^3$ and 350 km/s~\cite{vcircMW}), the DM mass capture rate is given by~\cite{Goldman:1989nd} 
\bea
\nn \dot M = \mdm C_{n\chi} &=& \rho_\chi v_{\rm rel} \times \pi b_{\rm max}^2  \times p_v \times p_\sigma \ ,\\
&=& p_v  p_\sigma~\times~1.76 \times 10^{25}~{\rm GeV/s}~,
\label{eq:masscaprate}
\eea
where $b_{\rm max} = \RNS (1+z) (v_{\rm esc}/v_{\rm rel})$ is the maximum impact parameter of DM intersecting the NS, with $1+z = (1-v_{\rm esc}^2)^{-1/2}$ a blueshift factor magnifying the NS radius to a distant observer, and $p_v$ is the probability that a scattered DM particle loses sufficient energy to be captured.
For instance, this probability $\simeq 1$ for scalar- or vector-mediated scatters, but may be suppressed for pseudoscalar-mediated interactions that favor soft forward scatters~\cite{NSvIR:DasguptaGuptaRay:LightMed}.
Eq.~\eqref{eq:masscaprate} is, of course, the DM capture rate for an isolated NS; an NS in a binary system could capture DM at a rate greater by up to a factor of a few thanks to gravitational assist~\cite{BrayeurTinyakovBinaryCap:2011yw}. 

The probability that incident DM is scattered is given by
$p_\sigma = 1 - e^{-\tau} \simeq \tau = \sigma_{n \chi}/\sigma_{\rm cap}$
where, for optical depth $\tau$, the approximate equality in the first line holds in the optically thin limit.
The ``capture cross section" above which $\tau > 1$ in the NS core is:
\begin{align}
\displaystyle
\sigma_{\rm cap} =
 \begin{cases}
	\sigmageom (\mneff/\mdm)\
	&, \ \ m_{\rm evap} < \mdm < \mneff \ , 
	\\
	\sigmageom \
	&, \ \ \mneff \leq \mdm \leq \text{PeV} \ , 
	\\
	\sigmageom (\mdm/{\rm PeV})\
	&, \ \ \mdm > {\rm PeV} \ ,
 \end{cases}
 \label{eq:sigmathreshold}
\end{align}
where the NS geometric cross section $\sigmageom = \pi (\mneff/\MNS) \RNS^2 \simeq 2.2 \times 10^{-45}\,{\rm cm}^2$.
One understands the dependence on $\mdm$ in Eq.~\eqref{eq:sigmathreshold} by considering the typical neutron recoil energy in the neutron rest frame:
\begin{align}
\Delta E_{\rm DM} &\simeq \frac{\mneff \mdm^2 (1+z)^2 v^2_{\rm esc}}{(\mneff^2+\mdm^2+2(1+z) \mneff\mdm)} \ ,
\label{eq:Erec}
\end{align}

The above expression is a good approximation to describe DM-neutron scattering in the stellar rest frame as well, since the neutrons are typically non-relativistic: their Fermi momenta, varying over a few 100 MeV across the NS, are smaller than their $\sim$GeV mass.  
For $\mdm \!<\!\mneff$, only a fraction $\simeq 3 \Delta p/ p_F$
		of degenerate neutrons close enough to their Fermi surface receive the typical momentum transfer $\Delta p = \sqrt{2\mneff \Delta E_{\rm DM}}$ to scatter to a state above the Fermi momentum $p_F \simeq 0.4~\text{GeV}$.
	This ``Pauli-blocking" effect gives	$\sigma_{\rm cap} \propto \Delta E_{\rm DM}^{-1/2} \propto \mdm^{-1}$.
The so-called evaporation mass,
\beq
m_{\rm evap} \simeq 20~{\rm eV} \ \bigg(\frac{T_{\rm NS}}{10^3~{\rm K}}\bigg)~,
\eeq
is the DM mass below which the thermal energy of the NS would kinetically eject the captured DM from the stellar potential well~\cite{NSvIR:GaraniGenoliniHambye,EvaporationGaraniSergio:2021feo}.
 For $\mneff\!\leq\! \mdm \!\leq\! 10^6~\!\text{GeV}$, a single scatter suffices for capture: $\Delta E_{\rm DM} \simeq \mneff v_{\rm esc}^2 \gamma^2 >$ KE$_{\rm halo}$, the DM halo kinetic energy. 
For $\mdm > \text{PeV}$, multiple scatters are required for capture, so that approximately $\sigma_{\rm cap}\propto {\rm KE}_{\rm halo}/\Delta E_{\rm DM} \propto \mdm$. 
The expression in Eq.~\eqref{eq:masscaprate} can be refined to account for the velocity distribution of DM far from the NS~\cite{Kouvaris:2007ay}.

The heating of the NS comes not only from the recoil of incident DM but from two other secondary effects.
As depicted in Fig.~\ref{fig:kineticaugerpairbreak}, a target neutron (or a lepton) that is upscattered by DM leaves behind a hole in the Fermi sea.
The hole is filled up immediately by a nearby neutron from a higher energy level, which in turn leaves a hole, and so on. 
This process spills over energy in the form of radiation and kinetic energy, and is reminiscent of the Auger effect observed in electron levels in superconductors; we will re-encounter this effect as a means of NS internal heating in Sec.~\ref{subsec:nucleonAuger}.
The net energy deposited in the NS by this effect, $E_{\rm Auger}$, is simply the difference in energy between the Fermi surface and the position of the original hole.
The energy carried by the struck nucleon/lepton in excess of the Fermi energy, $E_{\rm kin}$, is dissipated as kinetic energy above the Fermi surface. 
Thus the total energy deposit $E_{\rm Auger} + E_{\rm kin}$ comes out to be simply the DM recoil energy $\Delta E_{\rm DM}$.
Yet another effect comes from the superfluidity of nucleons (see Sec.~\ref{subsec:sfluid}).
For $\mdm \gsim 35$~MeV, DM participates in elastic scattering by first breaking a nucleon Cooper pair, which is bound with an energy given by the superfluidity energy gap $\sim$ MeV.
The absorbed $\sim$ MeV energy is redeposited into the NS when the free nucleon finds another and pairs up, liberating the gap energy.
For $\mdm \lsim 35$~MeV nucleons in the NS might not scatter elastically as there isn't enough energy transfer to break nucleon Cooper pairs, leaving DM to capture via collective excitations instead~\cite{NSvIR:Pasta,NSvIR:clumps2021}. 
Light DM capture in certain models through collective effects in NSs has been studied~\cite{collectiveDeRoccoLasenby:2022rze}.
The presence of DM self-interactions can enhance the capture rate by orders of magnitude as initially captured DM particles can serve as captors of ambient DM~\cite{NSvIR:SelfIntDM}.

Once captured in the potential well, a DM particle repeatedly scatters on and thermalizes with the NS until its orbit shrinks to within the radius of the star, by which times most of its kinetic energy is transferred.
Under equilibrium, the kinetic power of the infalling dark matter, constituting the NS heating rate, equals the rate at which photons are emitted from the NS surface, constituting the NS cooling rate.
The latter is dominated by such photon emission for NSs older than $\sim$Myr, as we saw in Sec.~\ref{subsec:cooling}. 
The NS luminosity corresponding to a temperature $T$ (in the NS frame) is then  
$  L = z \dot M = 4 \pi \RNS^2 T^4$,
which attains a maximum value $L_{\rm max}$ for unit capture probabilities $p_\sigma$ and $p_v$.
For our representative NS configuration (Eq.~\eqref{eq:BMNS}),
$L_{\rm max} = 7.6 \times 10^{24}~{\rm GeV/s}$, corresponding to a NS temperature seen by a distant observer $\tilde T = T/(1+z)$ of
$\tilde T = 1400~{\rm K}$.
Temperatures in this range are measurable within reasonable integration times at current and imminent infrared telescope missions~\cite{NSvIR:Baryakhtar:DKHNS,NSvIR:Raj:DKHNSOps}, in particular at the recently launched JWST~\cite{NSvIR:IISc2022}, and the forthcoming ELT and TMT.
For instance, the NIRCam instrument at JWST could constrain the surface NS temperature at 1750~K with a signal-to-noise ratio (SNR) of 2 in $27.8~{\rm hr} (d/10 \ {\rm pc})^4$, where $d$ is the distance to the NS~\cite{NSvIR:Baryakhtar:DKHNS}; the IRIS instrument at TMT could do the same in $19.4~{\rm hr} (d/10 \ {\rm pc})^4$.
In the bottom right panel of Fig.~\ref{fig:kineticaugerpairbreak} are displayed the NS effective temperatures constrainable at JWST at various SNRs for integration times of 5.5 hr and 24.3 hr, using the F150W2 filter on NIRCam.
In this plot taken from Ref.~\cite{NSvIR:IISc2022}, the band spans the range of the NS radii (which determines the range of DM capture rates) predicted by various EoSs, and integrates over the NS-DM relative velocities predicted by various NS population models in Ref.~\cite{NSdistribs:Ofek2009,NSdistribs:Sartore2010}.
These sensitivities are for the case of NSs being heated not only by the kinetic energy of infalling DM but also by DM annihilations, which we will discuss in Sec.~\ref{subsubsec:NSheatDManns}.
Searches for DM using NS thermal emissions are best carried out with NSs whose ``standard" temperatures are expected to be below approx. $1000$~K.
Thus one would need NSs older than 10~Myr (Fig.~\ref{fig:coolingcurvesNSWD}), making the determination of their age via spin-down or kinematic considerations (Sec.~\ref{subsec:Bspindown}) crucial.
One would also need them sufficiently isolated to ensure no accretion of material from a binary companion.

For a detailed study on the sensitivities of various filters on the imaging instruments of JWST, ELT, and TMT to NSs reheated to 2000$-$40000 Kelvin in their late stages (corresponding to various astrophysical and non-standard heating mechanisms discussed in this document), see Ref.~\cite{NSHeatObs:Raj:2024kjq}.
It is also possible that through campaigns of parallax measurement we may find that NSs that are inferred to be too distant via dispersion measure-based estimates are actually close enough for their luminosities to be measured at these telescopes over short integration times~\cite{NSHeatObs:Bramante:2024ikc}.

The DM-nucleon scattering cross section may be so large that DM scatters dominantly with the $\sim$km-thick low-density crust of the NS before reaching the $\sim$20 km-diameter denser core.
Moreover, the core may consist of exotic phases of high-density matter such as meson condensates and deconfined $ud$ or $uds$ quark matter, the latter of which may exist in any of the multiple phases discussed in Sec.~\ref{subsec:NSsubstruct}; in such cases, the dynamics governing DM scattering cannot be unambiguously computed, whereas the better understood crust can be treated robustly as a DM captor.
DM scattering with the NS crust leads to surface emission of photons under thermal equilibrium analogous to capture in the NS core discussed above, hence the observational signatures of NS heating are unchanged.
In Figure~\ref{fig:NScapcrustcore} we show the DM capture cross section $\sigma_{\rm cap}$ for every layer of the NS described in Sec.~\ref{subsec:NSsubstruct}, derived in Ref.~\cite{NSvIR:Pasta} for a 1.8~$M_\odot$ mass, 12.5~km radius NS.
For DM masses below about 10 MeV (left panel), DM capture can occur by scattering on superfluid neutrons in the inner crust, and exciting phonons.
The single-phonon emission mode is expected to dominate, which proceeds via a static structure function = $\Delta p/(2m_n c_s)$ that relates the per-nucleon cross section to the phonon-excitation cross section. 
Here $c_s$ is the phonon speed.
Due to the proportionality to the transfer momentum, $\sigma_{\rm cap} \propto \mdm^{-1}$ similar to the Pauli-blocking regime of the NS core discussed above.
The latter sensitivity (applicable to when the core is populated mainly by neutrons) is also shown for comparison in the plot.
For DM masses above about 100 MeV (right panel), DM capture can occur by scattering on individual nucleons locked up in nuclei in the outer crust by transferring energies greater than their $\sim$MeV binding energy.
Scattering on nuclei is generally suppressed: large $\Delta p$ leads to loss of nuclear coherence over multiple nucleons, and small $\Delta p$ leads to loss of coherence over multiple nuclei, described by a lattice structure function.
Deeper down in the inner crust, heavier-than-100-MeV DM capture proceeds by scattering on loosely bound nucleons, and even further down, by scatterig on the pasta phase.
Pasta scattering may either be on individual nucleons at high DM masses or on multiple nucleons at low DM masses as described by response functions accounting for inter-nucleon correlations.
A resonant peak in the response function is seen to enhance the capture sensitivity near $\mdm \simeq$ 100~MeV.
For comparison is also shown the DM capture cross section for scattering in an NS core dominated by neutrons.

Even in the absence of exotic phases, NS cores are expected to contain $\sim$10\% level populations of protons, electrons, and muons thanks to beta chemical equilibrium.
DM may be possibly be leptophilic, such that scattering at tree level is solely on $e^-$ and/or $\mu^-$, or iso-spin violating, such that scattering is dominantly on protons.
NS capture and heating applies to these scenarios, too~\cite{NSvIR:Baryakhtar:DKHNS}.
While the Fermi momenta of protons and muons are smaller than their mass, making them non-relativistic and amenable to the above treatment, that of electrons are 1--2 orders of magnitude greater than $m_e$, warranting relativistic kinematics to treat their DM capture in the stellar rest frame~\cite{Bertoni:2013bsa,NSvIR:GaraniGenoliniHambye,NSvIR:Bell2019:Leptophilic,Garani:2019fpa,NSvIR:Riverside:LeptophilicShort,NSvIR:Bell:ImprovedLepton,NSvIR:GaraniGuptaRaj:Thermalizn,NSvIR:Riverside:LeptophilicLong}.
This also makes the treatment of Pauli-blocking non-trivial~\cite{NSvIR:Riverside:LeptophilicShort,NSvIR:Riverside:LeptophilicLong}.
In particular, the capture probability accounting for Pauli-blocking, relativistic scattering and summing over multiple scatters is~\cite{NSvIR:Riverside:LeptophilicLong} 
\beq
  df = 
  \sum_{N_\text{hit}} 
  \;
  d\sigma_{\rm CM} 
  \, v_{\rm Mol} \, dn_\text{T}  \, \frac{\Delta t}{N_\text{hit}} 
    \,
  \Theta
  \left(
    \Delta E - \frac{E_\text{halo}}{N_\text{hit}}  
  \right)
    \Theta
    \left(
      \frac{E_\text{halo}}{N_\text{hit}-1}   - \Delta E
    \right)
  \Theta
  \left(
    \Delta E + E_p - E_\text{F}
  \right)
  \ ,
  \label{eq:df:dsig:CM:v:dnt:dt:total:step 1}
  \eeq
  where $v_{\rm Mol}$ is the M\" oller velocity that relates the cross section in any frame to that in the center of momentum frame ($d\sigma_{\rm CM}$), $dn_\text{T}$ is the differential volume of the target momentum space normalized to the Fermi volume, $E_{\rm halo}$ is the DM halo kinetic energy, and $\Delta E$ is the energy transfer. 
 We refer the reader to Ref.~\cite{NSvIR:Riverside:LeptophilicLong} for a detailed formalism. 
 In Figure~\ref{fig:NScapcrustcore}'s bottom panels we show the NS capture sensitivity to contact interaction cutoffs versus $\mdm$ for scalar-type operators involving spin-1/2 and spin-0 DM.
 For electron scattering the NS capture reach is seen to be orders of magnitude greater than that of terrestrial direct searches for $\mdm >$ MeV, and indeed completely complements the latter for sub-MeV DM masses.  

NS capture-and-heating can also provide  orders-of-magnitude improvement over Earth-bound searches for DM with scattering that is 
\begin{enumerate}
\item spin-dependent, since scattering directly on fermions instead of nuclei does not lead to the loss of nuclear coherence that limits spin-dependent searches at direct detection~\cite{NSvIR:Raj:DKHNSOps,NSvIR:Bell:Improved,NSvIR:PseudoscaTRIUMF:2022eav},
\item and/or velocity-dependent~\cite{NSvIR:Raj:DKHNSOps,NSvIR:Bell:Improved,NSvIR:PseudoscaTRIUMF:2022eav}, since semi-relativistic DM speeds at the point of capture overcome velocity-suppressed scattering rates,
\item inelastic~\cite{NSvIR:Baryakhtar:DKHNS,NSvIR:Bell2018:Inelastic,NSvIR:InelasticJoglekarYu:2023fjj}, since again the high DM speeds ensure that $\Oc(100)$~MeV mass splittings between the DM and its excited state can be probed, as opposed to $\Oc(100)$~keV at direct detection, and
\item below the so-called neutrino floor at direct searches, coming from irreducible neutrino backgrounds that are irrelevant for NS capture; see Fig.~\ref{fig:NScapcrustcore} top right panel,
\item with heavier-than-PeV DM, where DM capture proceeds mainly through multiple scattering in transit~\cite{NSMultiscat:Bramante:2017xlb,NSMultiscat:Ilie:2020vec,NSMultiscat:Ilie:2024sos,NSvIR:Baryakhtar:DKHNS}.
\end{enumerate} 

\subsubsection{Dark matter self-annihilations, nucleon co-annihilations, and induced nucleon decay}
\label{subsubsec:NSheatDManns}

While the discussion above focused NS heating from the transfer of captured DM kinetic energy, applicable to any particulate dark matter model -- in particular to non-annihilating DM such as asymmetric DM -- certain scenarios may lead to DM annihilation inside the NS that further brightens it~\cite{Kouvaris:2007ay,deLavallaz:2010wp} and thereby facilitate observations~\cite{NSvIR:Baryakhtar:DKHNS,NSvIR:Raj:DKHNSOps,NSvIR:Pasta,NSvIR:IISc2022}, in some cases reducing telescope integration times by a factor of 10.
For instance, JWST/NIRCam could constrain a 2480 K NS, heated by local DM kinetic energy + annihilations, with SNR~2 in 2.5 hr $(d/10 {\rm pc})^4$, and TMR/IRIS could do so in 0.56 hr $(d/10 {\rm pc})^4$~\cite{NSvIR:Baryakhtar:DKHNS}; compare these with kinetic heating-only exposure times in Sec.~\ref{subsubsec:DMcapturekineticheat}.
Fig.~\ref{fig:kineticaugerpairbreak} shows JWST sensitivities in more detail, as discussed in Sec.~\ref{subsubsec:DMcapturekineticheat}.

Self-annihilations of DM into most SM states would result in NS heating, the exception being neutrinos with sub-100 MeV energies as their optical depth in the NS material is too small to be trapped~\cite{Reddy:1997yr}.
In any case, this phenomenon relies intricately on whether or not the DM thermalizes with the NS within its lifetime, since DM may possibly annihilate much more efficiently if it is collected within a small volume in the NS core; this is a highly model-dependent question~\cite{Bertoni:2013bsa,NSvIR:GaraniGuptaRaj:Thermalizn,NSvIR:Pasta,NSvIR:Bell:Thermalization:2023ysh,UnthermalizedHiggsinoAcevedo:2024ttq} as discussed in Sec.~\ref{subsubsec:DMNSthermalizn}. 
To understand this, consider the evolution of the number of DM particles $N_\chi$ within a volume $V$ of the NS self-annihilating with a thermally averaged cross section $\langle \sigma_{\rm ann} v \rangle$, and its solution:
\bea
\nn \frac{dN_\chi}{dt} &=& C_\chi - \frac{\langle \sigma_{\rm ann} v \rangle  N_\chi^2}{V}~,\\
 N_\chi(t) &=& \sqrt{\frac{C_\chi V}{\langle \sigma_{\rm ann} v \rangle}} \tanh \bigg( \frac{t}{\tau_{\rm eq}} \bigg)~, \\
\nn  \tau_{\rm eq} &=&  \sqrt{\frac{V}{C_\chi \langle \sigma_{\rm ann} v \rangle}}~,
\eea
where $C_\chi = C_{n\chi} + C_{\chi \chi}$ is the total DM capture rate via scattering on nucleons (Eq.~\eqref{eq:masscaprate}) and, through self-interactions, on DM already accumulated in the NS, and
$\tau_{\rm eq}$ is the characteristic timescale for equilibrium between capture and annihilation to establish, after which $N_\chi (t)$ achieves a steady state ($dN_{\chi}/dt \ra 0$). 
Thus for $t > \tau_{\rm eq}$, the total annihilation rate equals the capture rate.
When $V$ is the thermal volume (Eq.~\eqref{eq:rtherm}), one can then compute the minimum annihilation cross section required for capture-annihilation equilibrium to occur well within the age of an observed NS, $\tau_{\rm NS}$.
Using a partial-wave expansion $\langle \sigma_{\rm ann} v \rangle = a + b v^2$, the condition may be written for $s$-wave and $p$-wave domination as~\cite{NSvIR:GaraniGuptaRaj:Thermalizn}
\bea
\nn a > 7.4 \times 10^{-54} \ {\rm cm^3/s} \ 
 \bigg(\frac{{\rm Gyr}}{\tau_{\rm NS}}\bigg)^2 
 \bigg(\frac{C_{\rm max}}{C_\chi}\bigg)  \bigg(\frac{\rm GeV}{m_\chi}\frac{T_{\rm NS}}{10^3~{\rm K}} \bigg)^{3/2}~,\\
 b > 2.9 \times 10^{-44} \ {\rm cm^3/s} \
 \bigg(\frac{{\rm Gyr}}{\tau_{\rm NS}}\bigg)^2 
 \bigg(\frac{C_{\rm max}}{C_\chi}\bigg)  \bigg(\frac{\rm GeV}{m_\chi}\frac{T_{\rm NS}}{10^3~{\rm K}} \bigg)^{1/2}~,
\eea
where $C_{\rm max}$ is the maximum capture rate achieved at the saturation cross section.

Interestingly, a thermal Higgsino of 1.1 TeV mass, a largely unconstrained true electroweak WIMP~\cite{Krall:2017xij}), would thermalize with just the NS crust rapidly enough to heat a reasonably old NS through annihilations in equilibrium with the rate of capture~\cite{NSvIR:Pasta}.
We also remark that due to different scalings of the NS luminosity from kinetic or annihilation heating on the NS mass and radius, in principle it must be possible to distinguish between the two heating mechanisms using an ensemble of NSs~\cite{NSvIR:Baryakhtar:DKHNS,NSvIR:IISc2022}.

An interesting way to probe DM self-annihilations in NSs is possible if the primary annihilation products are feebly interacting DM-SM mediators that live long enough to exit the star before decaying to SM states.
One could search for a flux of these states sourced by DM ``focused" in celestial bodies via capture.
For gamma-ray final states, limits have been imposed with Fermi and H.E.S.S. data on DM-nucleon scattering and DM self-annihilation cross sections using brown dwarfs for sub-GeV DM and NSs for TeV mass DM~\cite{LongLivedMedNS:Leane:2021ihh}.
For neutrino final states, limits in the TeV-PeV range come from IceCube, KM3NeT and ANTARES~\cite{LongLivedMedNS:BoseMaity:2021yhz,LongLivedMedNS:NguyenTait:2022zwb}.
Dark matter with inelastic mass splittings that is partially thermalized with the NS (see Sec.~\ref{subsubsec:DMNSthermalizn}) can self-annihilate {\em outside} the star to produce signals in gamma rays and neutrinos~\cite{UnthermalizedHiggsinoAcevedo:2024ttq}.

Dark matter species that carry negative baryon number, arising for instance from ``hylogenesis" models, could annihilate with baryons in a NS post-capture leading to possibly observable heating signals~\cite{NSvIR:hylogenenesis1:2010,NSvIR:hylogenenesis2:2011,NSvIR:coann-nucleons:JinGao2018moh}. 
Such co-annihilations with nucleons are also possible in models of ``dark baryons" that undergo quantum mixing with the neutron~\cite{NSvIR:Marfatia:DarkBaryon}, and baryon-number violating interactions that destruct nucleons in a chain reaction~\cite{NSvIR:baryodestruct:Ema:2024wqr}.
Yet another co-annihilation-like scenario resulting in NS heating is when a component of the DM comes in the form of magnetically charged black holes (MBHs)~\cite{NSvIR:magneticBH:2020}.
Subspecies that come with electroweak-symmetric coronas are expected to be near-extremal in mass, however upon encountering NSs they may become non-extremal: first they may capture in the NS by stopping due to the Fermi-degenerate gas, then they could absorb nucleons that are emitted back as (baryon number-violating) Hawking radiation, overheating NSs.
A smaller deposit of heat could come from mergers of captured MBHs that enhance Hawking radiation, mimicking a self-annihilation process.
Energy depositions from DM annihilations and decays may also possibly nucleate quark matter in the NS interior, resulting in emission of radiation, cosmic rays and gravitational waves~\cite{DMann:strangelet:Silk2010xlt,DMdecayNS:Silk2014dra,DMann:bubblenucleation:Silk2019}.

The production of x-rays and other high-energy effluents emitted from NSs, resulting from monopoles passing through and catalyzing nucleon decay, have been studied~\cite{Kolb:1982si,Dimopoulos:1982cz}.
This provides a strong bound on the abundance of monopole species that induce nucleon decay, which is a well-motivated class of monopoles arising from symmetry-breaking in Grand Unified Theories. 
\\

\subsubsection{Improvements and uncertainties}
\label{subsubsec:improvementsNScap}

The above treatment has been improved by accounting for a number of physical effects in the NS, which in some cases leads to observational uncertainties; these effects are collected in Table~\ref{tab:NSsideuncerts}.
The largest uncertainty in the capture rate, spanning  two orders of magnitude, comes from the unknown mass of the NS candidate that will be observed~\cite{NSvIR:Bell:Improved}, unless some precise mass-radius measurement is performed.
Other effects that may modify the DM capture rate, applicable to different DM and NS mass ranges, are variations in the EoS of NS matter, self-energies from the nuclear potential, nucleon structure that suppresses coherent scattering, nucleon superfluidity, extinction in the NS in the optically thick regime, scattering on relativistic nucleons, gravitational focusing in the NS interior layers, suppression of scattering via the propagator of a mediator of mass smaller than the transfer momentum, and the Galactic velocity distribution of DM.
Table~\ref{tab:NSsideuncerts} lists the appropriate references that treat these effects.
\\

\begin{figure*}
    \centering
    \includegraphics[width=0.47\textwidth]{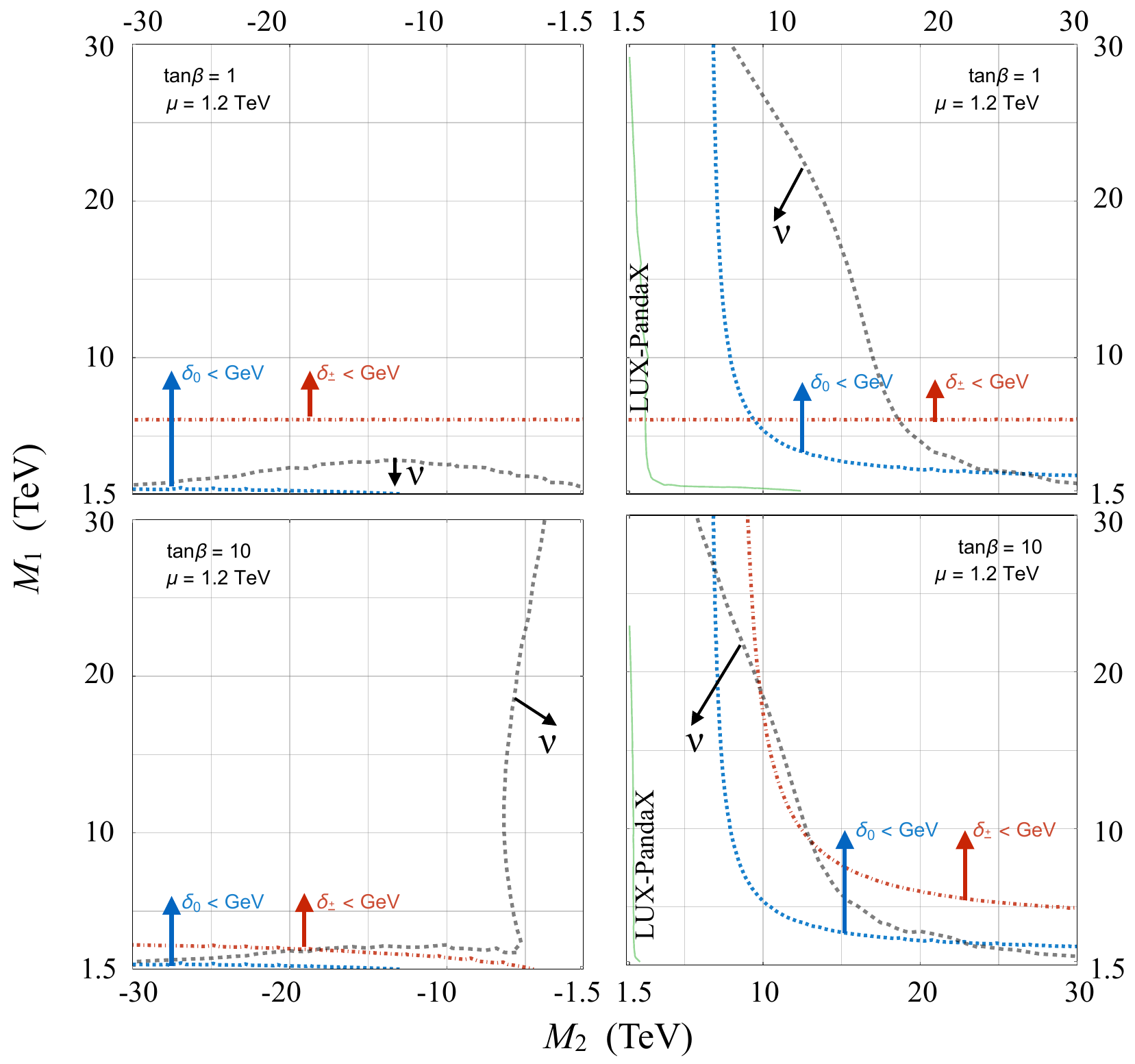} \ \includegraphics[width=0.45\textwidth]{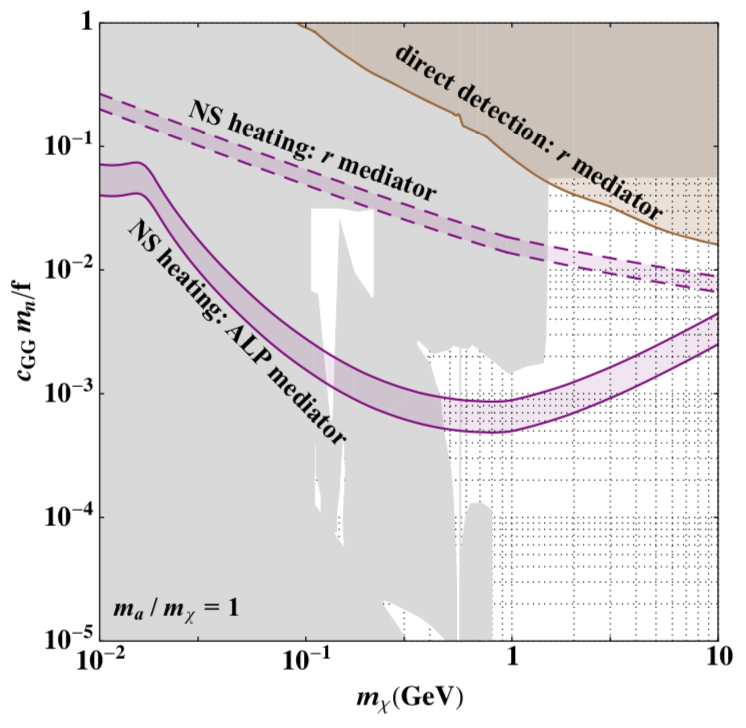}   \\
 \includegraphics[width=0.48\textwidth]{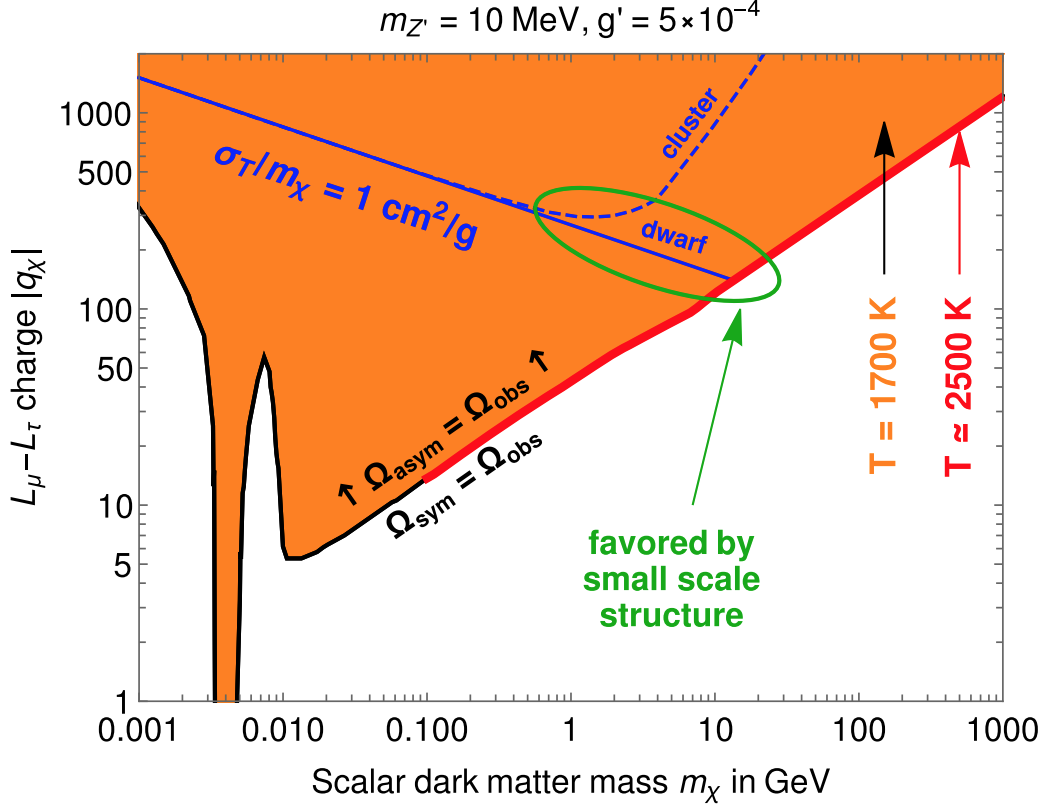} \  \includegraphics[width=0.48\textwidth]{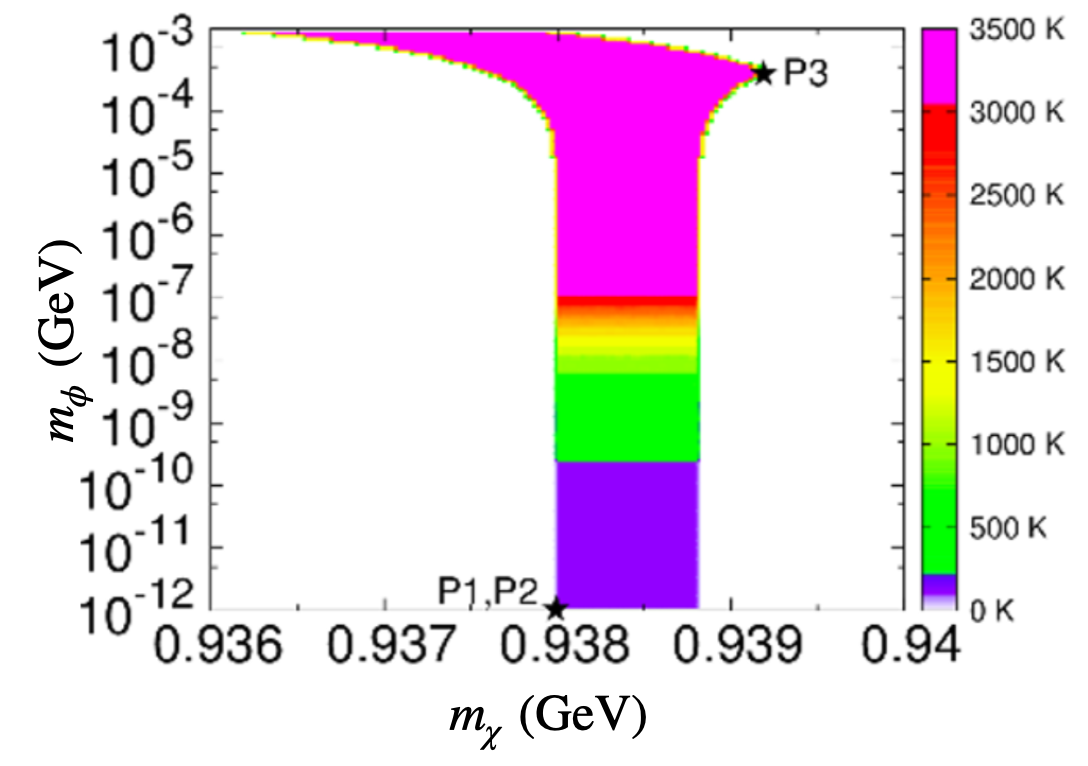}  
    \caption{Sensitivities of self-consistent dark matter models to NS kinetic heating; see Sec.~\ref{subsubsec:NSheatDMmodels}.
    {\bf \em Top left}.~\cite{NSvIR:Baryakhtar:DKHNS} Electroweakino singlet and doublet mass parameters for various $\tan \beta \equiv$ ratio of Higgs VEVs, that may be cornered through inelastic scattering of thermal Higgsino DM in the NS via excitation to charged and neutral states (regions marked by ``$\delta <$~GeV"). 
    {\bf \em Top right.}~\cite{NSvIR:PseudoscaTRIUMF:2022eav} As a function of DM mass, gluonic coupling to an axion-like particle that mediates velocity-dependent scattering interactions. The gray region depicts limits from beam dumps, rare meson decays, and astrophysics. NS capture can also proceed through mediation by a CP-even scalar in the theory, which gives rise to limits from direct detection.
    {\bf \em Bottom left.}~\cite{NSvIR:GaraniHeeck:Muophilic} The orange region can be probed for spin-0 DM scattering on muons in the NS by exchanging a $U(1)_{L_\mu - L_\tau}$ gauge boson. Also shown are constraints from DM self-interactions. 
    {\bf \em Bottom right.}~\cite{NSvIR:Marfatia:DarkBaryon} NS temperatures achieved by capture and heating of the anti-particle of DM carrying baryon number = 1, in a scenario where DM self-interacts repulsively and annihilates to the mediator $\phi$ that then decays to SM states that deposit heat. 
        }
    \label{fig:DMmodelsvNS}
\end{figure*}

\subsubsection{Dark matter models that heat neutron stars through scattering and annihilation}
\label{subsubsec:NSheatDMmodels}

Making use of the general effects discussed above, specific UV-complete and self-consistent DM models have been explored in the context of NS capture and heating.
These include the supersymmetric partner of the Higgs field, the Higgsino, that captures through inelastic scattering to electrically neutral and charged excited states~\cite{NSvIR:Baryakhtar:DKHNS,NSvIR:Pasta}, 
a generalization of this to electroweak multiplets~\cite{NSvIR:HamaguchiEWmultiplet:2022uiq},
a model of DM with a vector force-carrier of a gauged $L_\mu - L_\tau$ interaction~\cite{Garani:2019fpa},
DM in the form of a GeV-scale ``dark baryon" that mixes with the neutron~\cite{NSvIR:Marfatia:DarkBaryon},
simplified models of DM (specifying a single state each for DM and the mediator) with various mediator species~\cite{NSvIR:Queiroz:Spectroscopy,NSvIR:Queiroz:BosonDM,NSvIR:Lin2021:spin1med}, 
DM that arises as a pseudo-Goldstone boson~\cite{NSvIR:zeng2021PNGBDM}, 
models of dark sectors that can explain the muon $g-2$ anomaly~\cite{NsvIR:HamaguchiMug-2:2022wpz},
consistent models of DM interacting with nucleons through a pseudoscalar mediator: axion-like particles and a CP-odd state that arises in a Two-Higgs doublet model~\cite{NSvIR:PseudoscaTRIUMF:2022eav},
and $>$keV mass sterile neutrinos that mix with active neutrinos~\cite{NSvIR:sterilenu:Das:2024thc}.
The sensitivities to parameters of some of these scenarios are shown in Fig.~\ref{fig:DMmodelsvNS}.

\subsubsection{Neutron star reheating mechanisms not involving dark matter}
\label{subsubsec:NSheatNOTDM}

A search for DM reheating NSs in the late stages of their cooling must encompass understanding other astrophysical mechanisms that could possibly do the same.
We discuss below those that feature prominently in the literature.

\begin{enumerate}

\item DM capture in NSs would not encounter a ``neutrino floor" due to very dilute ambient neutrino densities that produce suppressed recoils/absorption on NS constituents, owing to low cross sections and Pauli-blocking.
However, it is natural to ask if there is an ``ISM floor" from accretion of interstellar material.
It turns out that old, isolated NSs that have spin periods $<$ 1000 seconds do not accrete ISM as they are in an {\em ejector phase}~\cite{NSvIR:ISMaccretionphases:Treves:1999ne}: a ``pulsar wind" of ISM outflow powered by the NS' magnetic field, being much denser than the inflowing material attempting accretion, would  pre-empt accretion via kinetic pressure.
Even if the pulsar wind happens to be weak enough for the ISM to overcome it, there is a second barrier to accretion: the magnetosphere co-rotating with the NS will impute centrifugal acceleration to the ISM, spraying away the gas -- the {\em propellor phase}.
For NSs with unusually large spin periods of $> 1000$ seconds, these arguments do not apply, instead infalling ISM would be deflected along the magnetic field lines of the NS and accretion will be confined to a small polar region, which can be distinguished from all-surface thermal emission.
In any case, the ISM density in the local 100 pc is 10$^{-3}$ GeV/cm$^3$~\cite{ISMBubble:Jenkins2009} so that any ISM accretion will be outdone by present-day DM capture near geometric cross sections.

\item  {\em Rotochemical heating}~\cite{Rotochem:FernandezReisenegger:2005cg,Rotochem:Reisenegger:2006ky,Rotochem:PetrovichReis:2009yh,Rotochem:GonzalezJimenezReis:2014iia,Rotochem:GusakovReisenegger2015:CrustNSE,Rotochem:Gusakov2021:Refined} could result from an imbalance in chemical potentials as the NS material is driven out of beta chemical equilibrium by deceleration in the rotation of NSs. 
Reactions that seek to restore chemical equilibrium deposit heat in the NS.
This mechanism could occur for NSs with small (sub-7 ms) pulsar spin periods at birth for certain nucleon pairing models~\cite{NSvIR:Hamaguchi:RotochemicalvDM2019} -- a requirement in tension with studies that find that natal spin periods are likely $\Oc(10-100)$~ms (see the references listed in Ref.~\cite{NSvIR:Hamaguchi:RotochemicalPure2019}).

\item {\em Vortex creep} heating could arise as a consequence of nucleon superfluid vortex lines in the inner crust that may be pinned to the nuclear lattice. 
Differences in rotational speed between the normal and superfluid matter may induce a Magnus force on the vortex lines, pushing them outward, in turn inducing friction that dissipates energy~\cite{NSvIR:Fujiwara:VortexCreepPure,NSvIR:Fujiwara:VortexCreepvDM2023}.

\item Other astrophysical late-time NS reheating mechanisms include~\cite{NsvIR:otherinternalheatings:Reisenegger,NSvIR:otherinternalheatings:2022} {\em magnetic field decay} that dissipates energy into the NS material, and
{\em crust cracking} which arises when the NS crust breaks as the NS relaxes from an oblate to spherical shape, releasing accumulated strain energy.

\end{enumerate}

We note that these mechanisms are speculative, and none have been unequivocally observed.
An exclusion set by non-observation of DM-induced heating at imminent telescopes would also rule out these mechanisms. 
Another notable point is that while the rotational power of NSs goes into dipole radiation, which in turn illuminates the nebula surrounding the pulsar as we saw in Sec.~\ref{subsec:Bspindown}, it very likely does not contribute to the NS thermal luminosity.
This is already apparent in the Crab Nebula example discussed in Sec.~\ref{subsec:Bspindown}, but can also be inferred from x-ray emission bounds on observed pulsars which have thermal luminosities 5-6 orders smaller than the rotational power: see Table 5 of Ref.~\cite{RotationPowervXRayLumi:PrinzWerner2015}.
Further, the diffusion of the $B$ field in the NS is also unlikely to heat the NS; as argued in Ref.~\cite{magnetothermalheating:Pons2008fd}, NSs older than about Myr are cool enough for magnetic diffusion timescales to exceed the NS age, effectively shutting off $B$ field dissipation regardless of the initial strength of the field.

\begin{figure*}
 \includegraphics[width=0.38\textwidth]{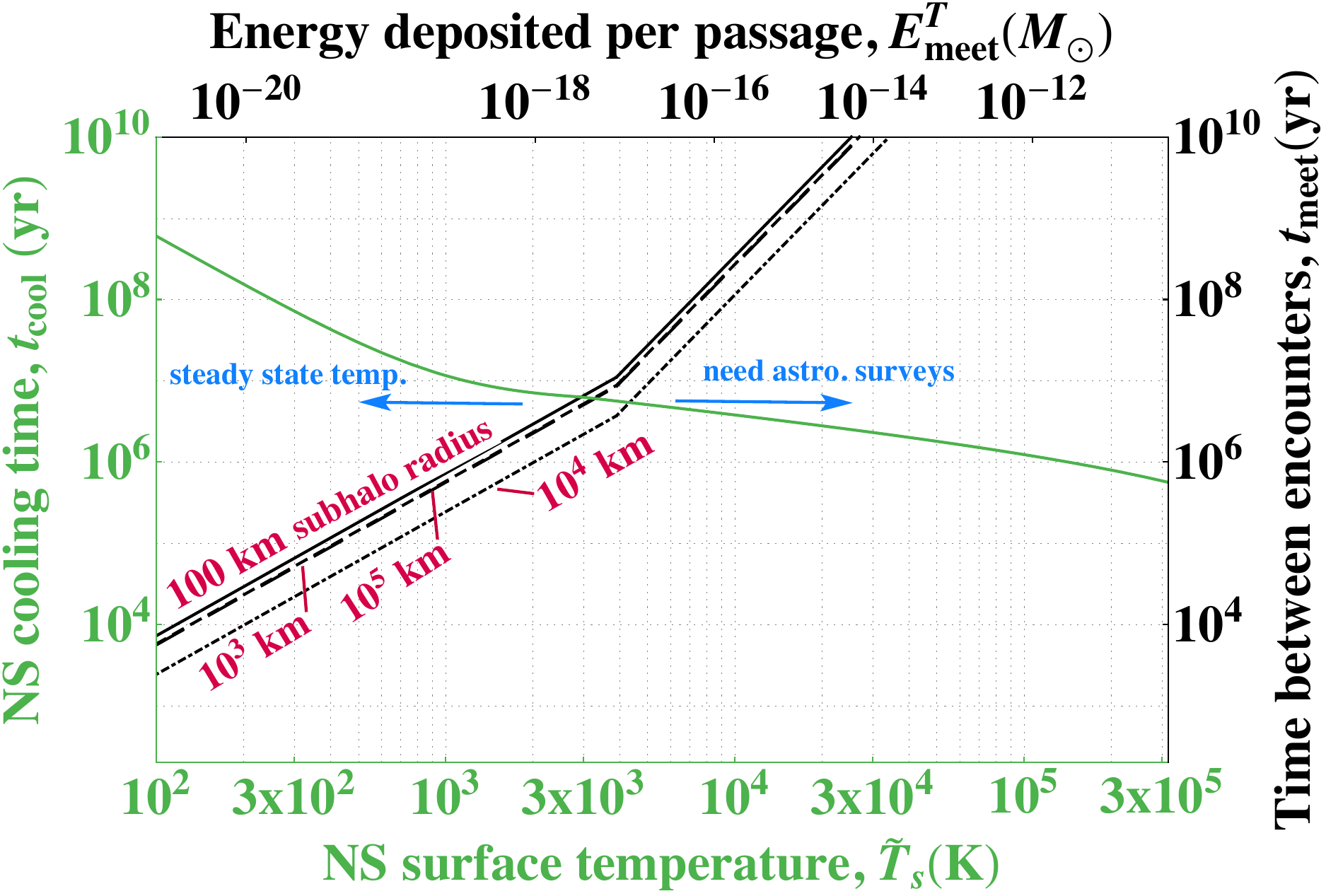} \  \includegraphics[width=0.60\textwidth]{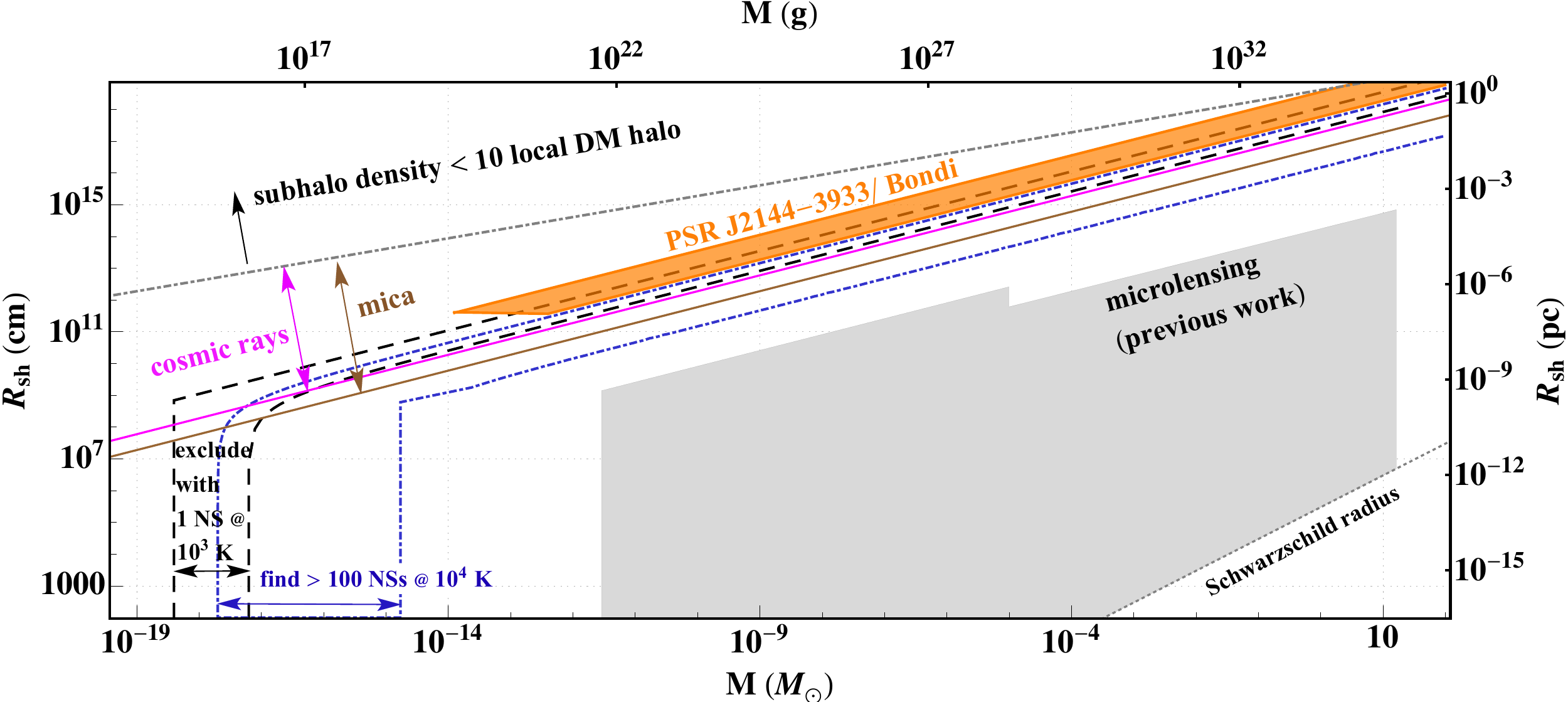}  \\ 
    \includegraphics[width=0.4\textwidth]{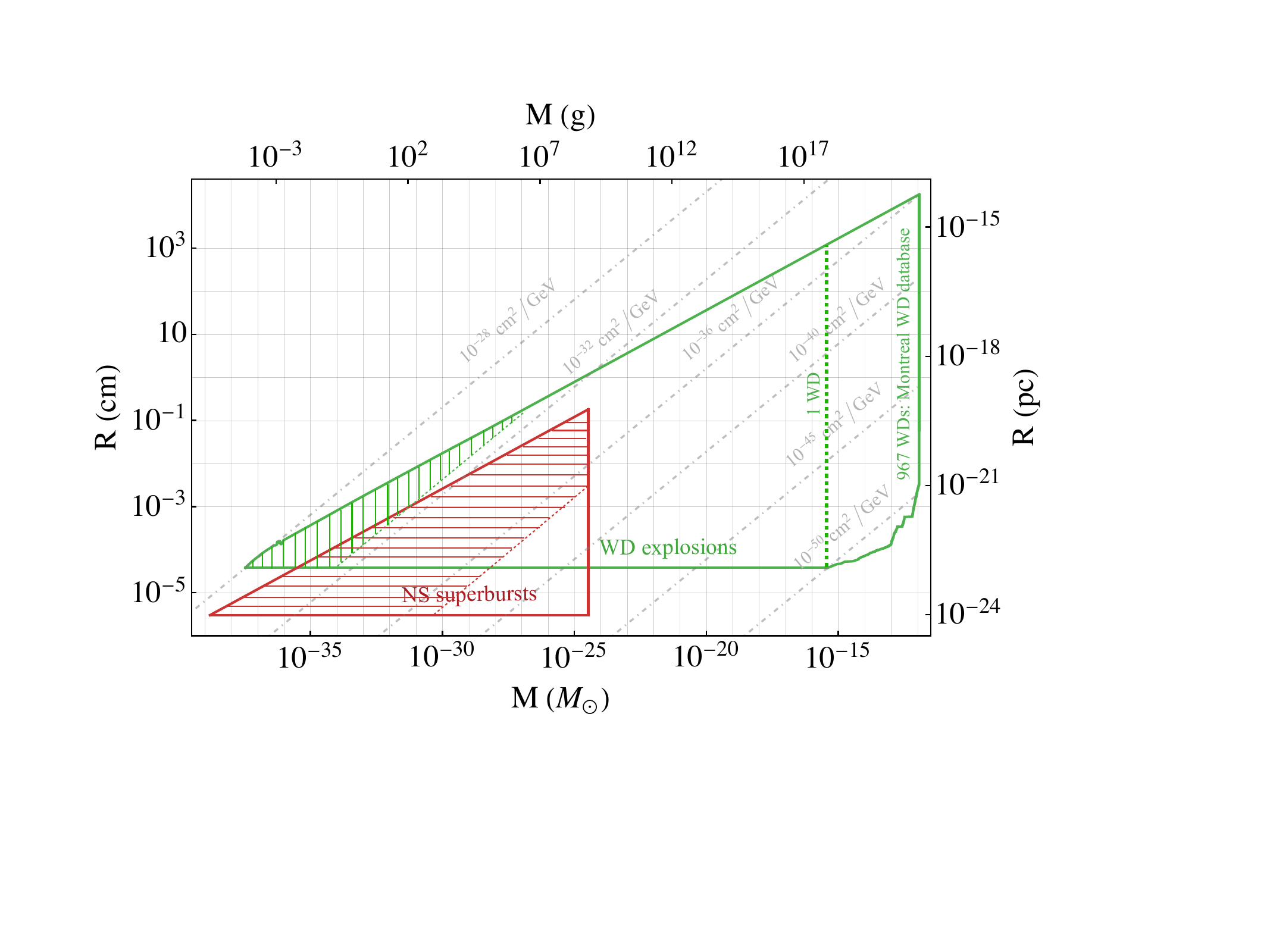} \   \includegraphics[width=0.59\textwidth]{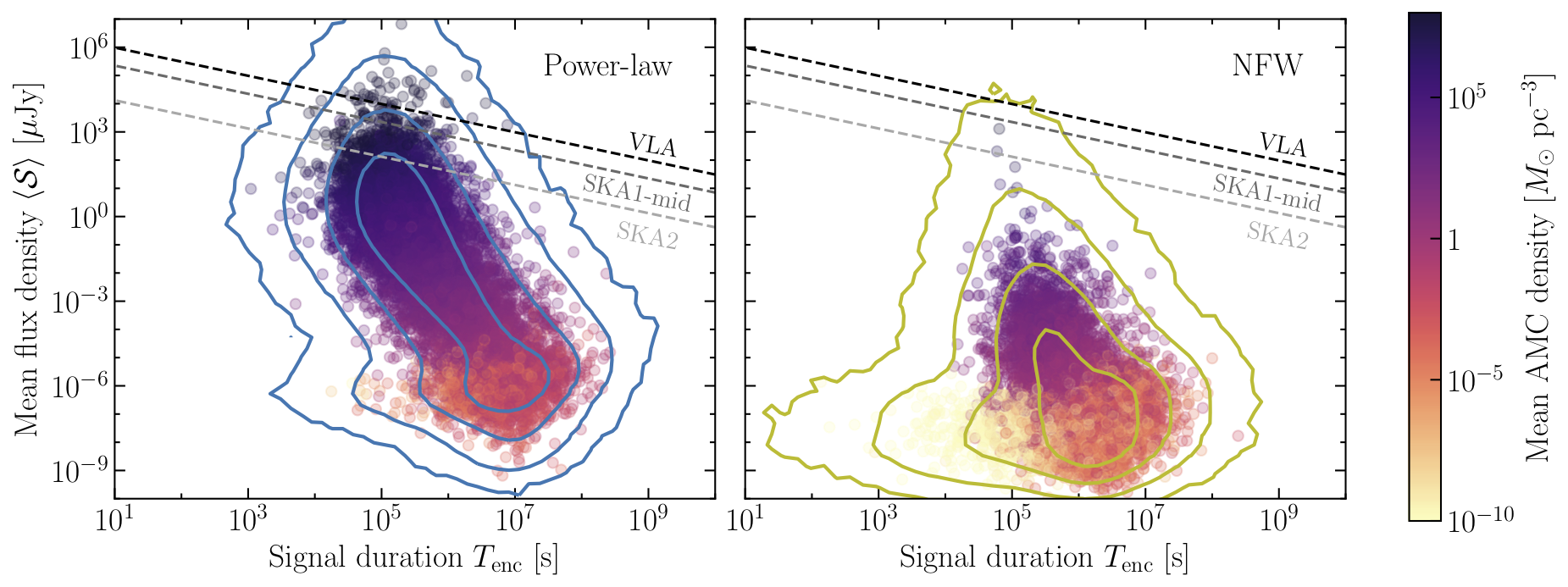}
    \caption{
 {\bf \em Top.} NS cooling timescale versus surface temperature (obtained from Eq.~\eqref{eq:tcoolvTfull}), superimposed on a plot of time between DM clump-NS encounters versus the energy deposited by kinetic heating during the passage of a clump for various NS radii ({\bf \em left}). The ticks on either x-axis and either y-axis correspond one-to-one to each other.
 This plot shows the region in which NSs are expected to glow at a steady temperature, so that observing a single NS is enough to set constraints, and the region where overheated NSs cool down rapidly between clump encounters, so that astronomical surveys are required to observe the fraction of overheated NSs in an ensemble.
 On the {\bf \em right} are future sensitivities of astronomical observations of NSs on DM clump radii and masses, exploiting dark kinetic heating, seen to be complementary to limits from other experiments.
 These limits are valid for DM-nucleon cross sections greater than the values for which the effects in these searches are relevant.
 These two plots are taken from Ref.~\cite{NSvIR:clumps2021}; see Sec.\ref{subsec:DMsubstruct} for further details.
{\bf \em Bottom left} Dark clump masses and radii constrained by compact stellar thermonuclear explosions, occurring for the minimum DM-nucleus cross sections per DM mass overlaid. Line-shaded regions rely on the ``saturated overburden effect" valid only for certain dark matter models; see Ref.~\cite{WDNSboom:Raj:2023azx} and Sec.~\ref{subsec:DMsubstruct}.
{\bf \em Bottom right.} For two different internal density profiles, the mean flux density of transient radio signals at various telescopes from encounters of axion miniclusters as a function of the transit time (= signal burst duration), taken from Ref.~\cite{Edwards:2020afl}.
See Ref.~\ref{subsec:NSvALP} for further details.
    }
    \label{fig:NSvclumps}
\end{figure*}

\subsection{Neutron stars and halo substructure} \vspace{0.2cm}
\label{subsec:DMsubstruct}

Numerous cosmologies predict enhanced small-scale power, for instance via an early matter-dominated era or DM self-interactions assisting primordial density perturbations, resulting in a substantial fraction of DM surviving in substructure termed variously as clumps, subhalos, minihalos and miniclusters~\cite{ErickcekSigurdson,Barenboim:2013gya,FanWatson,drorcodecay,inflatflucs,Buckley:2017ttd,nussinovcluster,Barenboim:2021swl}. 
If DM has scattering interactions with the SM, and if the interacting component resides in clumps, direct searches may have observed no conclusive signal simply because the Earth has yet to encounter a subhalo since their inception.
In this senario, subhalo DM may be observed by its heating of old, nearby NSs: the latter may travel through DM clumps and capture constituent DM particles, giving rise to kinetic and/or annihilation heating.

In the top left panel of Fig.~\ref{fig:NSvclumps}, taken from Ref.~\cite{NSvIR:clumps2021}, is shown the cooling time of NSs as a function of the NS surface temperature in green, and in the same plot is shown the energy deposited by clumps in NSs during encounters, $E_{\rm meet}^T$, as a function of the time between NS-clump encounters for various clump sizes. 
The $E_{\rm meet}^T$ in the top x-axis correspond to the NS temperatures imparted in the bottom x-axis immediately following the encounter.
For encounter times shorter than cooling times, the NS will glow at a steady-state luminosity, whereas for those longer than cooling times, NSs would be expected to glow brightly for short durations following encounters before dimming.
In the latter case, sky surveys of large populations of NSs may be able to pick out the fraction that is still above some temperature to which the telescope is sensitive.
In the top right panel, also taken from Ref.~\cite{NSvIR:clumps2021}, are shown clump mass vs radius regions that may discovered by observing more than 100 NSs above 10$^4$~K in the local kiloparsec, {\em e.g.,} by Roman/WFIRST and Rubin/LSST, and excluded by observing a single NS with temperature $<$ 1000 K, e.g. by JWST, ELT and TMT.
Also shown is a region that is already excluded by the observation of the coldest ($<$30,000 K) known NS PSR J2144$-$3933 by the Hubble Space Telescope (HST)~\cite{coldestNSHST} for clumps made of dissipative or strongly self-interacting DM, which would acrrete onto NSs through the Bondi–Hoyle–Lyttleton mechanism~\cite{BondiHoyle1944,Bondi1952,Begelman1977}.

In addition, in the presence of a long-range fifth force, NS heating by clumps may be enhanced by greater focusing effects, greater DM kinetic energies upon arrival at the NS surface, and seismic oscillations induced by an effective tidal force.
In the bottom right panel of Fig.~\ref{fig:GWPTA}, taken from Ref.~\cite{NANOGrav:2023hvm}, is shown the limit from overheating PSR J2144$-$3933 on the effective NS-clump coupling versus clump mass, for four values of the range of the fifth force arising from a Yukawa potential~\cite{NSvIR:tidalfifthforce:Gresham2022}.
(We do note that the DM need not be in the form of a clump for these limits to apply, but could also be a tightly bound composite.)
The curve labelled ``NS kinetic heating" corresponds to having an additional short-range interaction enhance DM capture.
These limits are complementary to those coming from the Bullet Cluster on DM self-interactions mediated by the light mediator, from weak equivalence principle tests using Galacto-centric motions of celestial bodies on the inter-baryonic force mediated by the same, and from the 15 year dataset of the NANOGrav pulsar timing array (see also Sec.~\ref{subsubsec:PTAs}).

Yet another signature of clumps with nucleon scattering interactions is thermonuclear explosions induced in compact stars, as discussed in Sec.~\ref{subsec:DMvWDboom}.
These could be Type Ia-like supernovae in carbon-oxygen WDs or x-ray superbursts in the carbon ocean layer in NS crusts (Sec.~\ref{subsection:thermonucrunaway}).
Constraints from the observed frequency of NS superbursts (Sec.~\ref{subsection:DMvsupbursts}) and from the existence of WDs (Sec.~\ref{subsec:DMvWDboom}) are shown in the left bottom panel of Fig.~\ref{fig:NSvclumps} in the plane of clump size and mass; the contours overlaid are the minimum reduced nuclear cross sections required to ignite a trigger mass of the stellar material.
This method of constraining clumps could be extended to those with baryonic long-range forces discussed above.
In that case, limits on the effective coupling apply to far smaller values (all the way to unity) than shown in Fig.~\ref{fig:GWPTA} bottom right panel, and to much higher clump masses.
See Ref.~\cite{WDNSboom:Raj:2023azx}.

Clumps encountering NSs can also be made of axions, leading to interesting signatures depicted in the right bottom right panel of Fig.~\ref{fig:NSvclumps}, which we discuss in Sec.~\ref{subsec:NSvALP}.
We also note that the phenomenology of black hole formation inside NSs (Sec.~\ref{subsec:BHinNS}) would be applicable here if NS-clump encounters are frequent enough. 

\subsection{Dark matter inducing superbursts in neutron stars} \vspace{0.2cm}
\label{subsection:DMvsupbursts}

Superbursts in NS carbon oceans, described in Sec.~\ref{subsection:thermonucrunaway}, can be induced by transiting DM if it is sufficiently heavy to deposit the requisite trigger energy.
Ref.~\cite{MACROSidhu2020} set limits on the cross sections and (super-Planckian) masses of macroscopic DM opaque to nuclei by satisfying the runaway criteria (Eqs.~\eqref{eq:cond1} and \eqref{eq:cond2}) and requiring that the time between DM-NS encounters is smaller than the inferred maximum recurrence time of the superburst 4U 1820+30.
Ref.~\cite{WDNSboom:Raj:2023azx} set limits on the masses, radii and interaction strengths of dark clumps (shown in Fig.~\ref{fig:NSvclumps} bottom left panel) and nuggets with long-range baryonic forces, using inferred recurrence times of the six superbursts (out of 16 detected in total) that have been observed to repeat~\cite{supburst:MINBARcatalog:2020,supburst:catalog2023}.

\subsection{Dark matter that implodes neutron stars into black holes} \vspace{0.2cm}
\label{subsec:BHinNS} \vspace{0.2cm}

Dark matter that is captured by an NS, after repeated re-scattering with the NS medium, will settle into a small thermalized region at the center of the NS. 
As more DM is collected, this spherical agglomeration can grow to a large enough mass that it collapses and forms a small black hole, which may (depending on its mass) subsequently accrete the entirety of the NS, transforming it into a solar mass black hole~\cite{Goldman:1989nd}. 
The processes of DM capture in NSs, thermalization, accumulation to the point of collapse, collapse, formation of a black hole, and its possible evaporation via Hawking radiation or growth consuming the NS, have been investigated in Refs.~\cite{Gould:1989gw, Kouvaris:2007ay, Bertone:2007ae, deLavallaz:2010wp, Kouvaris:2010vv, McDermott:2011jp, Kouvaris:2011fi, Kouvaris:2011gb, Bramante:2013hn, Bell:2013xk, Bramante:2014zca, Bramante:2015cua,Bramante:2016mzo,Bramante:2017ulk,Garani:2018kkd,Kouvaris:2018wnh, Kopp:2018jom, Acevedo:2019gre,Janish:2019nkk,East:2019dxt,Tsai:2020hpi,Takhistov:2020vxs,Dasgupta:2020mqg,Acevedo:2020gro,Acevedo:2020avd,Garani:2021gvc,Giffin:2021kgb,Ray:2023auh,Liu:2024qbe,Dutta:2024vzw,Diks:2024cww}. 
In addition, possible astrophysical signatures of DM converting NSs to black holes have been identified in, $e.g.$, Refs.~\cite{Bramante:2014zca,Fuller:2014rza,Bramante:2017ulk,Garani:2018kkd}. 

The kind of DM that is by and large studied in this context is ``asymmetric dark matter", DM primarily made of its {\em particles} as opposed to a symmetric population of {\em particles and anti-particles}. 
This emulates the visible universe, which is primarily matter (electrons, nucleons) and not anti-matter; 
indeed, the asymmetry in DM may be linked to that of the visible sector~\cite{Petraki:2013wwa,Zurek:2013wia}, but this is not necessary for the discussion that follows. 
The primary feature that permits asymmetric DM to convert NSs into black holes is that it is typically\footnote{For the exception, see Ref.~\cite{asymmetricannih:Kumar:2013vba}.} non-annihilating, and so as it collects inside the NS, it is not expected to annihilate to Standard Model states. 
This may be compared with symmetric, annihilating DM discussed in Sec.~\ref{subsubsec:NSheatDManns}. 
Investigation into what fraction of the DM may self-annihilate or co-annihilate with nucleons, while still forming a black hole inside the NS, was undertaken in Refs.~\cite{Bramante:2013hn,Bell:2013xk,Bramante:2013nma}.

Another kind of DM which could convert NSs into black holes is primordial black holes~\cite{PBHreviewGreenKavanagh:2020jor,PBHreviewCarr:2016drx}. 
A PBH captured in an NS can settle inside, accrete NS material, and convert the NS into another black hole~\cite{Capela:2012jz,Capela:2013yf,Pani:2014rca,Kurita:2015vga,Abramowicz:2017zbp,Bramante:2017ulk,Takhistov:2017nmt,Takhistov:2017bpt,PBHvsNSGenoliniTinyakov:2020ejw,Diamond:2021scl,Baumgarte:2021thx,Estes:2022buj,Zenati:2023ckz}; 
this is detailed in Section \ref{subsec:pbhconvertnstobh}. 
For the remainder of this sub-section we will focus on particle DM. 

\begin{figure*}
 \includegraphics[width=0.45\textwidth]{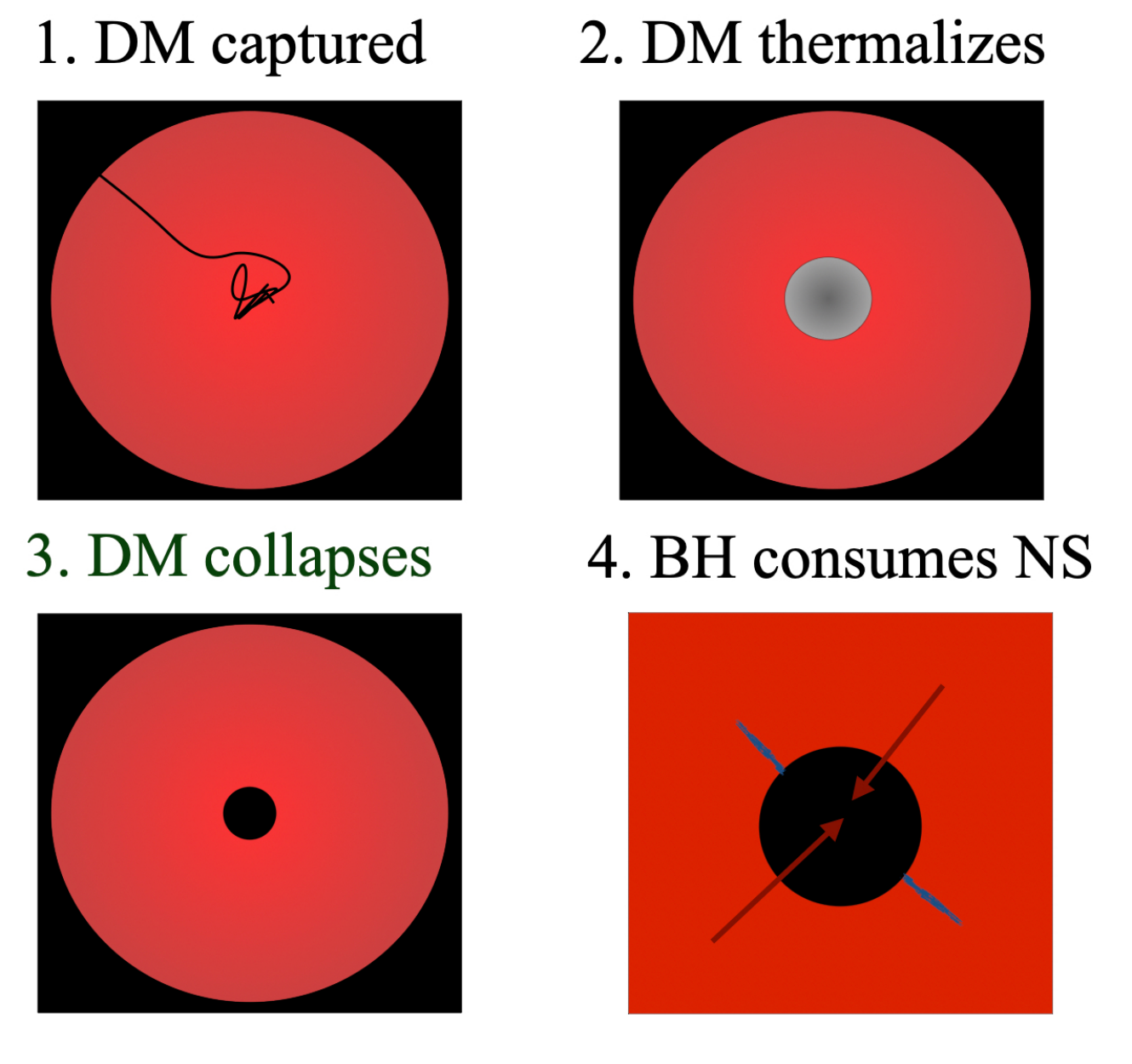} \  \includegraphics[width=0.55\textwidth]{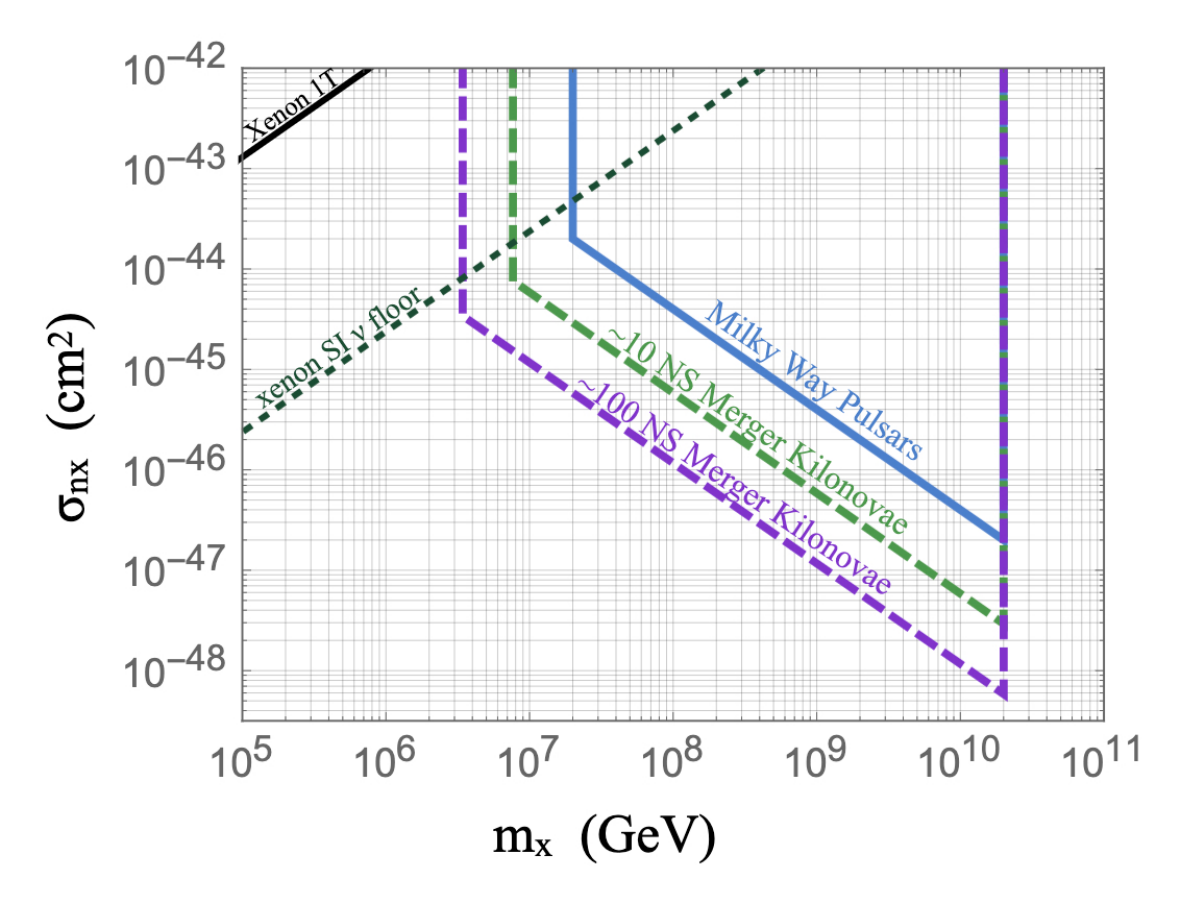}  
    \caption{{\bf \em Left.} Schematic of asymmetric dark matter converting a NS into a black hole. 
    {\bf \em Right.} Dark matter per-nucleon scattering cross section versus mass bounds on heavy fermionic asymmetric dark matter from the observation of old Milky Way pulsars that have not been converted to black holes~\cite{Bramante:2017ulk}, compared with terrestrial direct search limits and their neutrino floor. 
    Also shown are prospects for observating NS mergers with accompanying kilonovae, localized to 1 kpc precision inside Milky Way-like spiral galaxies. A detailed discussion of Milky Way pulsar ages, and in particular PSR J1738+0333, which has a characteristic age confirmed by the age of its WD companion, can be found in Ref.~\cite{Bramante:2015dfa}.}
    \label{fig:NSvBHintro}
\end{figure*}

We now turn to details of the processes leading asymmetric dark matter to convert NSs into black holes.
They proceed as follows: (1) DM is captured in the NS and thermalizes with the NS interior, forming a small ball of DM at the center, (2) the DM ball reaches a critical mass at which it collapses, and through some cooling process continues to collapse until (3) a small black hole forms which, provided accretion of NS material outstrips Hawking radiation, will result in the conversion of the NS to a black hole. Figure~\ref{fig:NSvBHintro} left panel shows a simple schematic of this process. 

\subsubsection{Dark matter thermalization in neutron stars}
\label{subsubsec:DMNSthermalizn}

\begin{figure*}
 \includegraphics[width=0.45\textwidth]{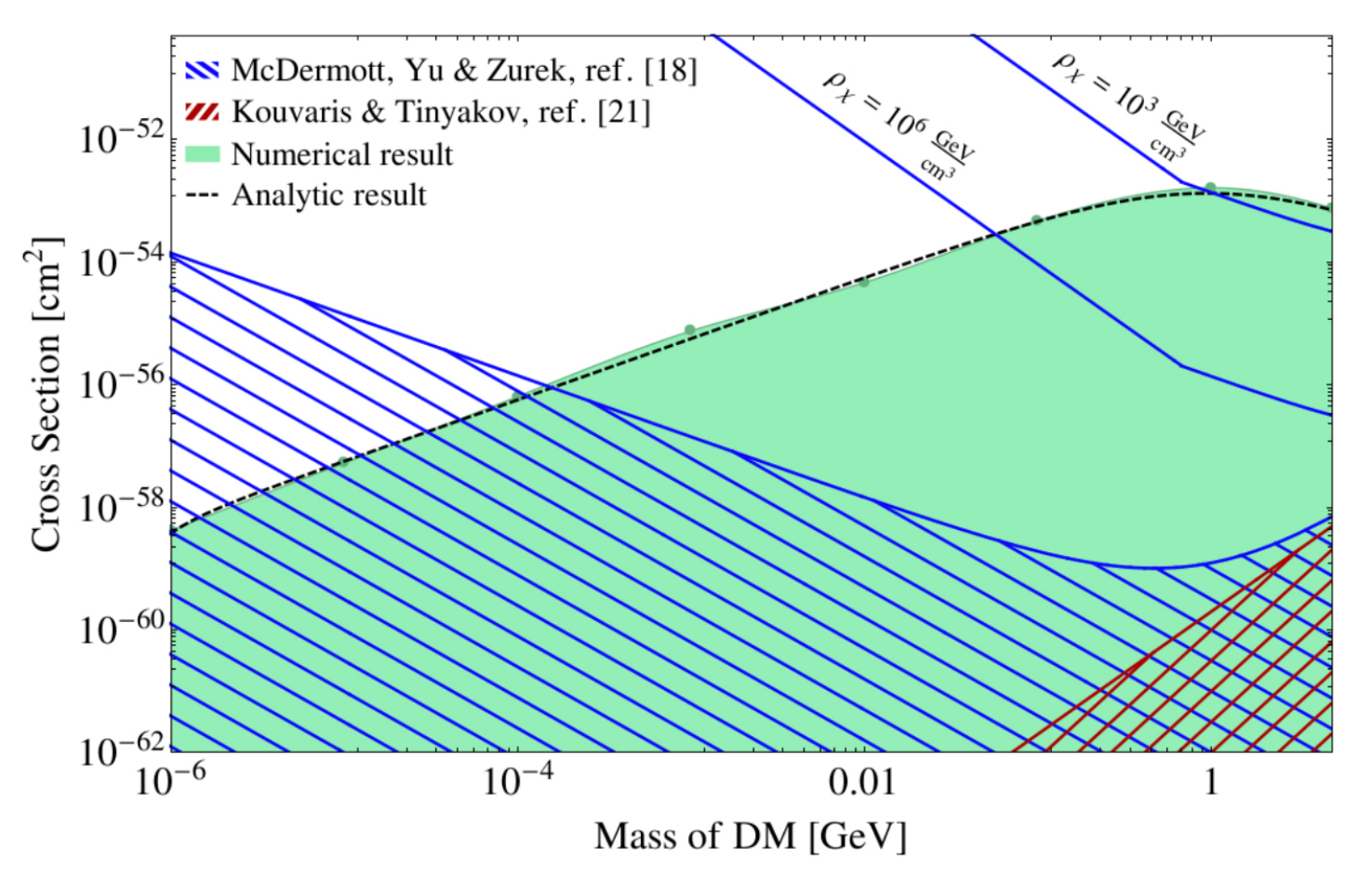} \  
 \includegraphics[width=0.45\textwidth]{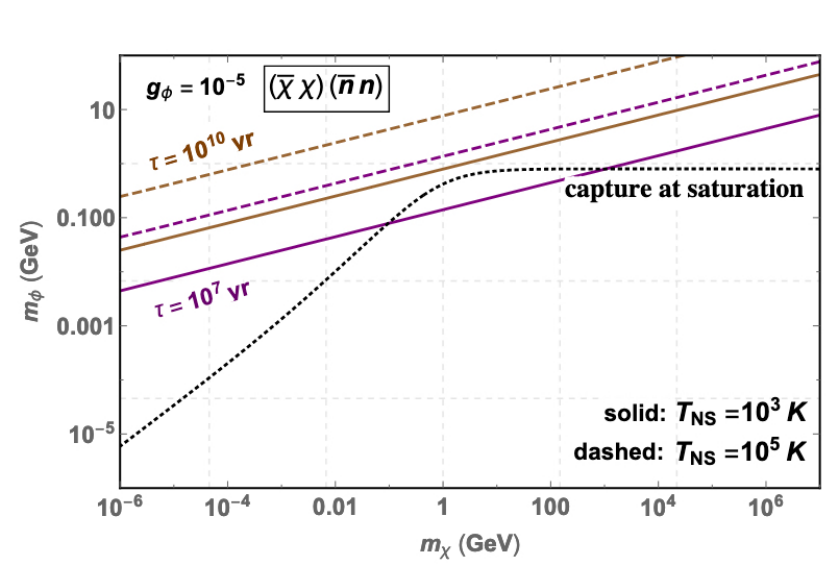}  
    \caption{{\bf \em Left.}~\cite{Bertoni:2013bsa} Cross section for dark matter to thermalize in a neutron star in 10 billion years, assuming a momentum-independent cross section with neutrons.  
    {\bf \em Right.}~\cite{NSvIR:GaraniGuptaRaj:Thermalizn} The case of scattering on neutrons through the scalar current operator indicated in the figure with mediator mass $m_\phi$. Thermalization in 10 Myr and 10 Gyr for different NS temperatures, and a curve indicating parameters that lead to DM capture in the NS through geometric cross sections, are shown.}
    \label{fig:thermtimes}
\end{figure*}

In step (1) above, the size of the thermalized region is determined by the temperature of the NS, which sets the final temperature of the DM particles, and by the central density of the NS, which sets the gravitational potential binding energy. 
A simple application of the virial theorem yields an estimated DM thermal radius of~\cite{Bramante:2013hn}
\begin{equation}
    r_{\rm th} \approx 20 ~{\rm cm} \left(\frac{{\rm GeV}}{m_\chi}\right)^{1/2}\left(\frac{T_{NS}}{10^3~\rm K}\right)^{1/2}\left(\frac{10^{15}~{\rm g/cm^3}}{\rho_{\rm NSc}}\right)^{1/2},
    \label{eq:rtherm}
\end{equation}
where $\rho_{\rm NSc}$ is the NS central density. 
The time it takes for DM to sink to this region depends on a few timescales (see $e.g.$, Ref.~\cite{Acevedo:2020gro}, Section 3 for a review), but usually the longest is the time it takes for DM to scatter with its lowest velocities/temperatures on nucleons, after having mostly settled inside the NS. 
A detailed calculation of this timescale requires modeling the NS core, and so the result will depend on the density, degeneracy, and possibly even new QCD phases in the NS interior. 
For  neutrons treated as a degenerate fluid, we have~\cite{Bertoni:2013bsa}
\begin{equation}
    t_{\rm th} \approx 3000 ~{\rm yr}~\frac{\frac{m_\chi}{m_n}}{\left(1+\frac{m_\chi}{m_n} \right)^2}\left(\frac{2\times 10^{-45}~{\rm cm^2}}{\sigma_{n \chi}}\right)\left(\frac{10^5~{\rm K}}{T_{NS}}\right)^{2},
\end{equation}
where this expression assumes a momentum-independent cross section for spin-1/2 DM scattering on nucleons via a heavy mediator. 
Extensions to spin-0 DM, Lorentz structures of DM-nucleon interactions leading to momentum-dependent cross sections, and light mediators were investigated in Refs.~\cite{NSvIR:GaraniGuptaRaj:Thermalizn} and \cite{NSvIR:Bell:Thermalization:2023ysh}.
In the above expression, the thermalization timescale counter-intuitively {\em decreases} with increasing DM mass above $m_n$: one would naively expect that heavier DM takes {\em longer} to thermalize.
But the effect comes about because $t_{\rm th}$ is set by the inverse of the energy loss rate (in turn depending on the DM-nucleon scattering rate) in the NS degenerate medium with phase space restrictions, and this rate goes as positive powers of the (continually degrading) DM momentum $k_{\rm cold}$.
For DM energies close to the NS temperature, $k_{\rm cold} \simeq \sqrt{3 m_\chi T_{\rm NS}}$, implying energy is lost faster in the last few scatters for heavier DM, i.e., implying quicker thermalization.
In Figure~\ref{fig:thermtimes} we show the per-nucleon cross section or effective field theory coupling necessary for DM to thermalize inside an NS on 10 Gyr year timescales for certain models.

As discussed in Sec.~\ref{subsubsec:NSheatDManns}, depending on the DM annihilation cross section, thermalized DM collected within $r_{\rm th}$ can annihilate efficiently enough to yield interesting signals.

\subsubsection{Collapse of dark matter and formation of small black hole}

In step (2), after enough DM has collected in the thermalized region in the NS, it will reach a critical mass at which it collapses. The exact density of DM required to initiate collapse will depend on its self-interactions, and by extension its EoS and sound speed while contained in the NS. 
Assuming negligible self-interactions, the critical mass required for collapse is 
\begin{equation}
        M_{\rm crit} \approx 2\times 10^{33} ~{\rm GeV} \left(\frac{10^7~{\rm GeV}}{m_\chi}\right)^{3/2}\left(\frac{T_{\rm NS}}{10^3~\rm K}\right)^{3/2}\left(\frac{10^{15}~{\rm g/cm^3}}{\rho_{\rm NSc}}\right)^{1/2}.
\end{equation}
For a detailed review of the conditions for collapse see $e.g.$, Section 4 of~\cite{Acevedo:2020gro}. 
It is generally the case that if DM thermalizes rapidly through scattering with neutrons in the NS interior, then when it reaches the point of collapse it will also rapidly shed the gravitational energy required to form a black hole.
This is because the temperature is higher during collapse and hence the time to shed gravitational energy is typically shorter.
As the shortness of this timeframe is common, this part of the collapse dynamics is not always treated explicitly, but Refs.~\cite{Bramante:2015cua,Acevedo:2020gro,Acevedo:2019gre,Goldman:1989nd} provide more detailed treatment, both in compact stars and other astrophysical bodies. 
The time for the DM sphere to collapse below its Schwarzschild radius will depend on whether it cools via scattering with neutrons or through other radiative process, $e.g.$, emission of a light particle in the dark sector~\cite{Bramante:2015cua}.

An additional consideration is whether enough DM will have collected to exceed the dark sector Chandrasekhar mass (analogous to Eq.~\eqref{eq:MChWDrelapprox}), parametrically of order 
\begin{equation}
    M_{\rm Chand,f} \approx \frac{M_{\rm Pl}^3}{m_\chi^2} \approx M_\odot \left( \frac{{\rm GeV^2}}{m_\chi^2}\right)~,
\end{equation}
while for bosons this is 
\bea
\nn     M_{\rm Chand,b} &\approx& \frac{2 M_{\rm Pl}^2}{m_\chi} \left(1+ \frac{\lambda }{32 \pi} \frac{m_{pl}^2}{m_\chi^2}  \right)^{1/2} \\
 & \ra & \begin{cases}
     2 \frac{M_{\rm Pl}^2}{m_\chi}~, \ \ \lambda \ll 1, \\
     \frac{\lambda}{2\sqrt{2}}  \frac{M_{\rm Pl}^3}{m_\chi^2}~,\lambda > 100 m_\chi/M_{\rm Pl}~,
 \end{cases}
\eea
where $\lambda$ is the boson $\phi$'s repulsive self-interaction coupling arising in the Langrangian $\mathcal{L} \supset - (\lambda/4!) \phi^4$.

Attractive DM self-interactions could alter the amount of asymmetric fermionic DM necessary for collapse to a black hole~\cite{Kouvaris:2011gb,Bramante:2013nma,Bramante:2015dfa}. 
The collapse of light fermionic DM is in principle permitted by the attractive self-interaction mediated by a light scalar, however, a detailed study of the final stage collapse to a black hole has pointed out an important caveat~\cite{Gresham:2018rqo}: for a simple scalar field potential consisting only of a mass term and a coupling to the fermions, the effective mass of the scalar could grow during DM fermion collapse, preventing collapse to a black hole.
Whether bosonic self-interactions let DM form black holes in NSs is a non-trivial question. 
In particular, a large value of $\lambda$ can shift bosonic asymmetric DM bounds from old NSs to higher DM masses~\cite{Kouvaris:2011fi,Bramante:2013hn,Bell:2013xk}. 
 Bosonic asymmetric DM forming black holes inside a NS do so by forming a Bose-Einstein condensate (BEC)~\cite{Kouvaris:2011fi,McDermott:2011jp,Kouvaris:2012dz,Bramante:2013hn,Bell:2013xk,Garani:2018kkd}, from which collapse will proceed for GeV-mass DM. 
 The dynamics of the BEC prior to and following collapse would affect whether a black hole is produced and is an area of investigation~\cite{Kouvaris:2012dz,Bramante:2013hn,Bell:2013xk,Garani:2018kkd,Garani:2022quc}. 

\subsubsection{Growth or evaporation of dark matter-formed black hole in the neutron star}

After a black hole is formed inside the NS, step (3) is to  determine whether it is so small that it will rapidly evaporate away via Hawking radiation, or whether it is so large that through accumulating surrounding baryonic material it will grow to consume the NS in a relatively short timeframe. 
Initial studies of this process estimated whether Bondi accretion by the black hole would proceed faster than Hawking radiation, which is entirely determined by the initial mass of the black hole~\cite{Kouvaris:2011fi,McDermott:2011jp}. 
Later studies incorporated the accumulation of DM particles additionally collected into the NS and onto the black hole, finding that this can substantially influence whether the black hole would grow to consume the NS~\cite{Bramante:2013hn,Bell:2013xk,Bramante:2013nma}. 

Altogether, the requirement that the black hole grows in the NS is given by
\begin{equation}
     \dot{M}^{(\rm NS~accretion)} + \dot{M}^{(\rm DM~accretion)} -\dot{M}^{(\rm Hawking)} > 0,
\end{equation}
where the first term is the NS accretion rate onto the black hole, the second is the DM accretion rate onto the black hole, and the third is the Hawking radiation rate. 
Each of these terms has been individually studied in the context of asymmetric DM which causes NSs to implode: 
\begin{enumerate}
    \item {\em NS accretion}: The simplest treatment of NS accretion onto the black hole assumes Bondi accretion. 
     In practice, angular momentum of the NS fluid around the black hole, for a rapidly spinning NS, can diminish accretion relative to na{\"i}ve Bondi accretion, but the high viscosity of the NS fluid results in infall rates consistent with spherical Bondi accretion despite angular momentum effects~\cite{Kouvaris:2013kra}. 
     Sufficiently small black holes will have a quantum penalty to accumulation of neutrons due to the neutron de Broglie wavelength;  this effect can be pronounced for black hole masses near the edge of growth vs.~evaporation~\cite{Giffin:2021kgb}. 
     The accretion of the NS fluid onto the black hole inside a NS has been studied in a detailed simulation that accounts for hydrodynamic and general relativistic effects~\cite{East:2019dxt}, finding that in the final stages of accretion, the mass of NS fluid ejected from the accretion zone is likely less than about $10^{-4}~M_\odot$. 
    \item {\em DM accretion}: The accretion of DM onto the small black hole inside the NS can substantially affect whether it grows or shrinks, especially when the DM accretion rate onto the NS is maximized~\cite{Bramante:2013hn,Bell:2013xk,Bramante:2013nma,Garani:2018kkd}.
    \item {\em Hawking radiation}: There is a correction to the evaporation rate of black holes inside NSs, coming from the Pauli blocking of Hawking radiation, since the region around the black hole will be inhabited by a degenerate sea of SM fermions~\cite{Autzen:2014tza}. 
    The assumption of Hawking radiation may itself break down after the black hole has lost about half its mass due to quantum backreaction on the black hole itself, which can result in much stronger constraints on the DM parameter space~\cite{Basumatary:2024uwo}.
\end{enumerate}

\subsubsection{Signatures of dark matter that implodes neutron stars} \vspace{0.2cm}

\begin{figure*}
 \includegraphics[width=0.57\textwidth]{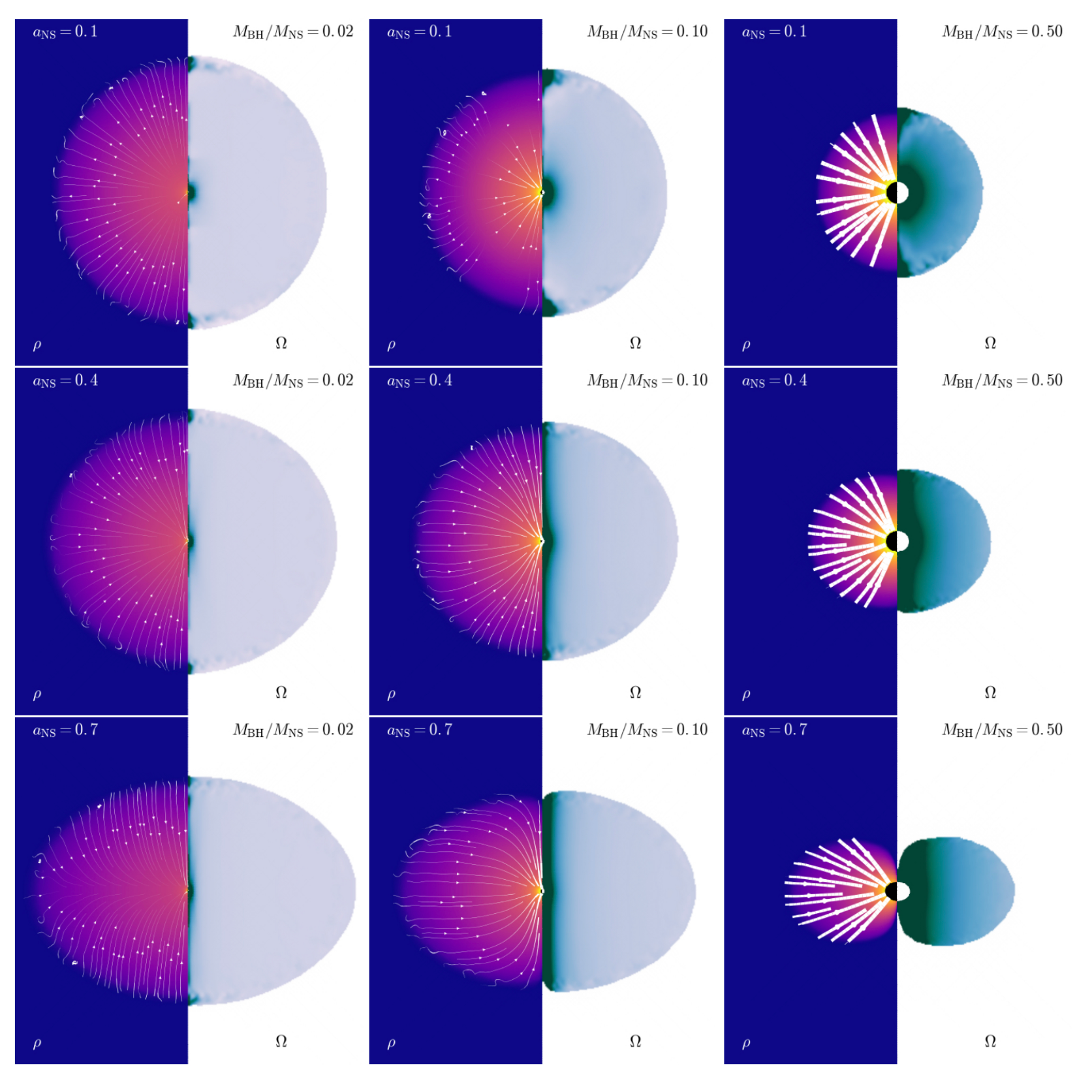} \  ~~
 \includegraphics[width=0.4\textwidth]{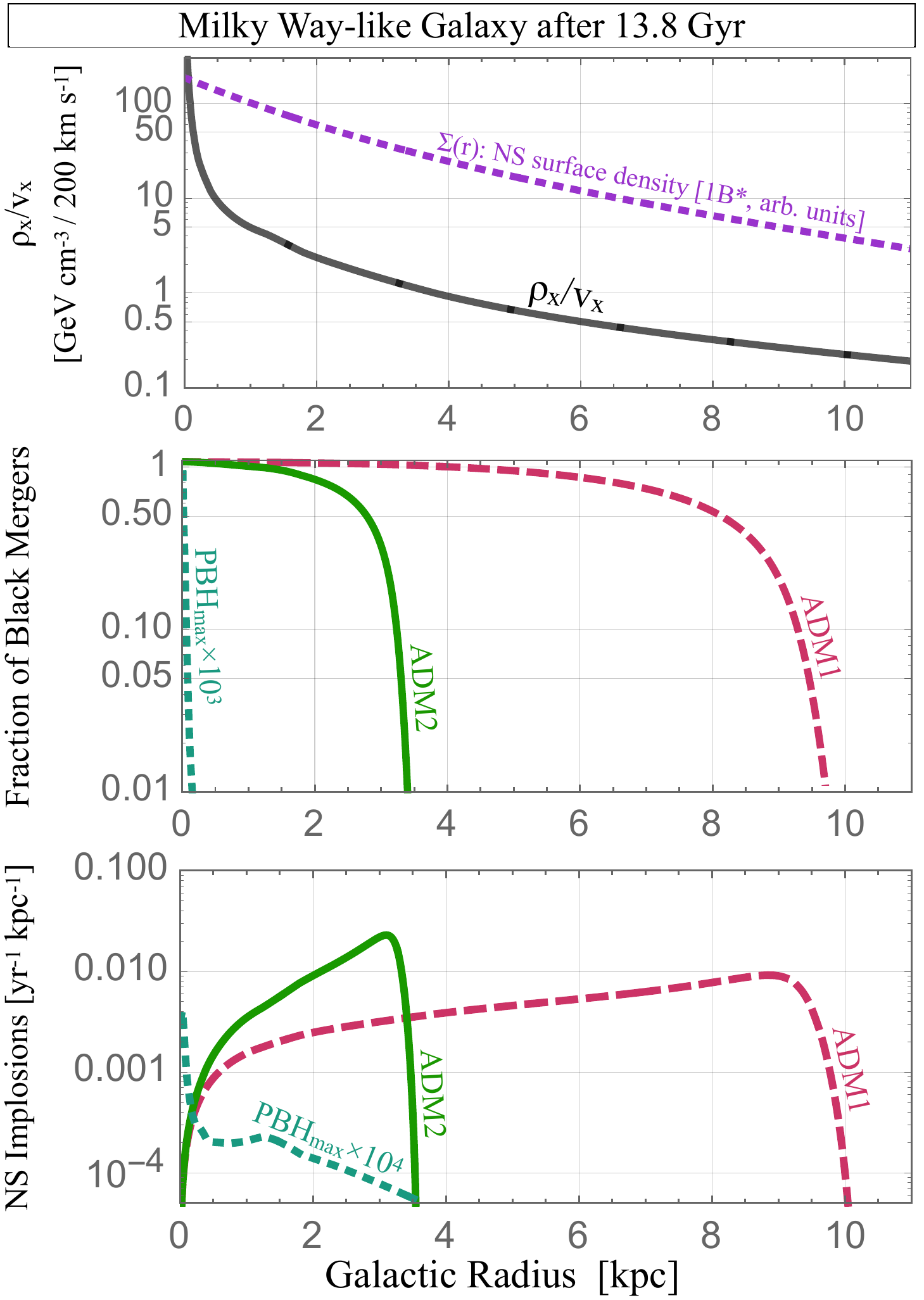}  
    \caption{{\bf \em Left.} From a simulation of a NS accreting onto a black hole at is center~\cite{East:2019dxt}. NS fluid density and velocity vectors are shown in the left half of these figures, with angular momentum density shown on the right half, for a different NS spin parameters $\Omega$. 
    {\bf \em Right.} Number of NSs converted to black holes in a Milky Way-like galaxy, along with the current NS implosion rate for dark matter models that would cause NSs near Earth to implode in 10 Gyr (``ADM1") and 50 Gyr (``ADM2")~\cite{Bramante:2017ulk}.}
    \label{fig:nsimpsig}
\end{figure*}

A number of striking astrophysical signatures arise from DM that converts NSs to black holes. 
Firstly, the oldest known pulsars can be used to set limits on asymmetric DM, since for a given background DM density, the existence of these pulsars limits the accumulation of DM~\cite{Gould:1989gw, Kouvaris:2010vv, McDermott:2011jp, Kouvaris:2011fi, Kouvaris:2011gb, Bramante:2013hn, Bell:2013xk, Bramante:2014zca,Kouvaris:2007ay, deLavallaz:2010wp}. 
However, the ``characteristic age" of pulsars comes with caveats: it is not always a good indicator of the actual age of the pulsar as discussed in Sec.~\ref{subsec:Bspindown}. 
One particular pulsar, PSR J1738+0333, is in a binary system with an old WD, and thus to go with its characteristic age, has an additional age-marker in its WD companion, both of which point to a $\gtrsim 5$ Gyr-old NS. 
Hence this pulsar has been used to set bounds on asymmetric DM~\cite{Bramante:2015dfa,Bramante:2017ulk}.
A recent work~\cite{Liang:2023nvo} integrates over the density of DM that NSs traverse during their orbits around the Milky Way, refining bounds that use characteristic ages of old, nearby pulsars.

A number of prompt and delayed signatures may be sought if DM is converting Gyr-old NSs to black holes in regions where DM is denser than in the outer regions of the Milky Way.  
In particular, the absence of millisecond pulsars in the Galactic Center~\cite{Dexter:2013xga,Suresh:2022vmf} has been linked to models of DM that would convert old NSs to black holes~\cite{Bramante:2014zca,deLavallaz:2010wp}. 
These studies predict a maximum pulsar age that increases with Galacto-centric distance, corresponding to a decrease in the amount of DM accumulating in pulsars. 
It has been shown that collapsing NSs can shed their magnetospheres and emit a radio pulse about a millisecond in duration, and thus the implosion of pulsars via DM is a possible origin of non-repeating fast radio bursts (FRBs)~\cite{Fuller:2014rza}. 
The Galactic localization of FRBs sourced from DM-induced NS implosions could be used to confirm or limit this hypothesis~\cite{Bramante:2017ulk}. 
Based on these estimates, approximately ten FRBs localized in Milky Way-equivalent galaxies would be required to differentiate between an FRB population sourced by DM-induced NS implosions and one that simply matches the NS distribution in these galaxies.

The implosion of NSs into black holes has also been explored as a source of $r$-process elements.
NS fluid ejected during the implosion could undergo $r$-process enrichment, sourcing early $r$-process elements observed in ultra-faint dwarf spheroidal galaxies~\cite{Bramante:2016mzo,Bramante:2017ulk}. 
$r$-process enrichment has also been studied for rapidly spinning NSs and NSs that capture primordial black holes~\cite{Fuller:2017uyd}, along with associated high-energy events~\cite{Takhistov:2017nmt,Takhistov:2017bpt}. 
However, the dynamics of ejecta from NS implosions later investigated in simulations~\cite{East:2019dxt,Baumgarte:2021thx,Richards:2021upu} disfavor large amounts of NS material being ejected in any single NS implosion event. 
In particular, Ref.~\cite{East:2019dxt} found in their simulations that even maximally spinning NSs would eject less than $10^{-4} ~M_\odot$ of their material during implosion. 
Recently, neutrinos produced during DM-induced NS implosions, and the prospects for detecting them as a diffuse background were studied~\cite{NeutrinosNSBHZenatiSilk:2023ckz}.

Figure~\ref{fig:nsimpsig}, from Ref.~\cite{East:2019dxt}, shows a detailed simulation of the final stages of NS accretion onto a black hole in its interior, and the expected rate of these NS implosions in Milky Way-like galaxies for a generalized NS-implosion-DM framework laid out in Ref.~\cite{Bramante:2017ulk}. 
Since the accumulated mass of DM onto NSs will be proportional to DM density and inversely proportional to the DM-NS relative velocity, one can parameterize how quickly particle DM will convert NSs into black holes:
\begin{equation}
    \bigg(\frac{t_{\rm convert}}{\rm Gyr}\bigg) \bigg(\frac{\rho_\chi}{{\rm GeV/cm^3}}\bigg) \bigg(\frac{200~{\rm km/s}}{v_\chi}\bigg) \lesssim 2~,
\end{equation}
where we have quoted the bound from Ref.~\cite{Bramante:2017ulk} using the $\gtrsim 5$ Gyr-old PSR J1738-0333. 

Many additional pathways for discovering DM that implodes NSs arise from the population of solar mass black holes that would be created, which would reside preferentially in the centers of galaxies~\cite{Bramante:2017ulk,Kouvaris:2018wnh}. 
The number of NS merger events with accompanying kilonovae required to increase sensitivity to these DM models was estimated for 10$-$100 kilonovae in Ref.~\cite{Bramante:2017ulk}.
As stated in that study, kilonovae would not occur following an apparent ``neutron star merger" if both solar mass compact objects were black holes converted from NSs. 
The prospects of gravitational wave observatories like LIGO/VIRGO and their successors for finding a population of NSs converted to black holes have been detailed in Refs.~\cite{Bramante:2017ulk,Kouvaris:2018wnh,Takhistov:2020vxs,Tsai:2020hpi,Dasgupta:2020mqg,Steigerwald:2022pjo,Bhattacharya:2023stq,Bhattacharya:2024pmp}. 
The capability of future gravitational wave observatories to differentiate solar mass black hole mergers from solar mass NS mergers, depending on minute variations in the waveform just prior to the merger, was examined in Ref.~\cite{Yang:2017gfb}, which noted that a large population of mergers may be needed to find evidence for a solar mass black hole population.

\subsubsection{Primordial black hole dark matter and neutron stars} \vspace{0.2cm}
\label{subsec:pbhconvertnstobh}

Black holes could form in the infant universe through some mechanism producing sub-horizon overdensities, and for masses above $5 \times 10^{14}$~g these ``primordial black holes" do not evaporate away within the age of the universe~\cite{PBHreviewCarr:2016drx,PBHreviewGreenKavanagh:2020jor} -- thus constituting non-baryonic dark matter.
PBHs are constrained by a number of phenomena they give rise to: evaporation, gravitational (micro)lensing, disruption of dwarf galaxies and wide binaries, gravitational waves, and accretion~\cite{PBHreviewGreenKavanagh:2020jor}. 
Wide-ranging as they are, these limits nevertheless leave open a mass window in which PBHs could make up all the DM: roughly $10^{17}-10^{22}$~g (or $10^{-16}-10^{-11}~M_\odot$).

PBHs encountering NSs may capture through the energy loss mechanism of dynamical friction: the NS constituent particles absorb the momentum of the transiting PBH~\cite{PBHvsNSCapelaTinyakov:2013yf}. 
In this picture there are no collective excitations of the NS medium, as the PBH travels at supersonic speeds after being accelerated by the NS' gravity. 
An alternative treatment of the energy loss is by considering oscillations of the NS medium excited by the passing PBH\footnote{This analysis was reused in Ref.~\cite{NSvIR:tidalfifthforce:Gresham2022} to treat the tidal heating of NSs by composite particulate DM.}, which seemingly extracts more energy from the PBH~\cite{PBHvsNSLoebPani:2014rca}.
However, these approaches were shown to be equivalent by modelling the NS as a semi-infinite incompressible fluid~\cite{PBHvsNSDefillonTinyakov:2014wla}.
All said and done, the captured PBH proceeds to grow via Bondi accretion of the NS material~\cite{PBHvsNSCapelaTinyakov:2013yf,PBHvsNS-Bondi-Richards:2021upu,PBHvsNS-survivaltime-Baumgarte:2021thx} (see Section \ref{subsec:BHinNS} for additional considerations regarding black hole growth in NSs), eventually destroying the NS in a catastrophic event that emits telltale electromagnetic signals and gravitational waves~\cite{Bramante:2017ulk,PBHvsNSGenoliniTinyakov:2020ejw}.
Collisions between PBHs and NSs may also explain fast radio bursts~\cite{Fuller:2014rza,Bramante:2017ulk,PBHvNS-FRB-Kainulainen:2021rbg}.
In the case that black holes are charged, their capture and consumption of the NS is treated differently~\cite{Diamond:2021scl}, where specifically this work employed the capture rate for monopoles and a different accretion rate appropriate for extremal black holes.
Finally, as mentioned in Sec.~\ref{subsec:DMkinannheat}, magnetically charged black holes that may constitute DM could capture in NSs and heat them via Hawking radiation of absorbed nucleons.

\subsection{Neutron stars as sources of gravitational microlensing}

Gravitational microlensing is the transient and typically all-wavelength magnification of a background star due to the gravitational potential of a transiting object, with images usually unresolved~\cite{microlens:NarayanBartelmann}. 
It has excluded massive compact halo objects and primordial black holes in the mass
range from 10$^{-11} - 10$ solar masses as the entire content of dark matter, using stars as sources in the Milky Way and neighboring galaxies~\cite{PBHreviewGreenKavanagh:2020jor}.
With some modifications, the technique can also be
applied to extended non-compact objects that may constitute dark matter~\cite{microlens:erosogle,microlens:subaru}.
Below 10$^{-11}~M_\odot$ the effects of wave optics and the finite extent of the stellar source suppress the magnification and conventional optical microlensing loses sensitivity~\cite{Nakamura:1999:waveoptix,Matsunaga2006FiniteSourceWaveOptix,SugiyamaWaveptics:2019dgt}.

Ref.~\cite{microlens:xray:Bai:2018bej} recognized that, to probe dark matter structures lighter than 10$^{-11}$ solar mass, accreting x-ray NSs in the
Magellanic Clouds are an appropriate source for microlensing.
These furnish a steady flux above which transient magnification of microlensing can be discerned, and are located far enough from Earth to provide appreciable optical depth of intervening dark matter structures.
This idea was applied in Ref.~\cite{microlens:xray:Tamta:2024pow} to possible searches in the NICER x-ray telescope~\cite{NICERDesign2016}, whose primary mission is to precisely measure NS masses and radii.
It was found that with 60 days of exposure on the x-ray pulsar SMC-X1 the PBH mass window may be reached.
Successor missions such as Strobe-X~\cite{STROBE-XScienceWorkingGroup:2019cyd} with larger effective areas may further probe this region.
Another technique for probing the PBH mass window is ``parallax microlensing" or ``picolensing" that uses differences in magnification as seen simultaneously by two observing instruments separated by a finite distance~\cite{Parallaxmicrolens1995,Parallaxmicrolens1998}.
The source here is energetic gamma ray bursts (GRBs) thought to arise from mergers of NSs or core-collapse supernovae.
Designed to observe more than 1000 GRBs, the future {\em Daksha} system~\cite{dakshabhalerao2022science} may perform this search with one satellite placed near the Earth and Moon each~\cite{Gawade:2023gmt}.
Such a search, however, may be impeded by uncertainties on the size of the GRB source, and for robust conclusions an inter-satellite distance larger than the Earth-Moon distance may be needed~\cite{Fedderke:2024wpy}.
Observing more than 100 GRBs may also probe the PBH mass window via ``femtolensing", the measurement of interference fringes in the energy spectrum~\cite{femtolensing:Katz:2018zrn}.

\subsection{Neutron stars admixed with dark sectors} \vspace{0.2cm}
\label{subsec:admixedNS}

NSs that contain an appreciable fraction of exotic particle species could give rise to interesting observational signatures such as modified NS mass-radius relations and distinct gravitational wave signals.
But to obtain such ``admixed" NSs one cannot rely on their capture of ambient DM.
One can see this immediately from Eq.~\eqref{eq:masscaprate}: for an NS of age 10 Gyr $\simeq 3 \times 10^{17}$~s in the solar vicinity, the total DM mass accumulated over its lifetime is only about $5 \times 10^{-15}~M_\odot$.
Thus other mechanisms must be in play, like a conversion of a large fraction of the NS's interior to dark sector particles. We detail some such exotic DM accumulation mechanisms below.

\begin{figure*}
         \includegraphics[width=0.47\textwidth]{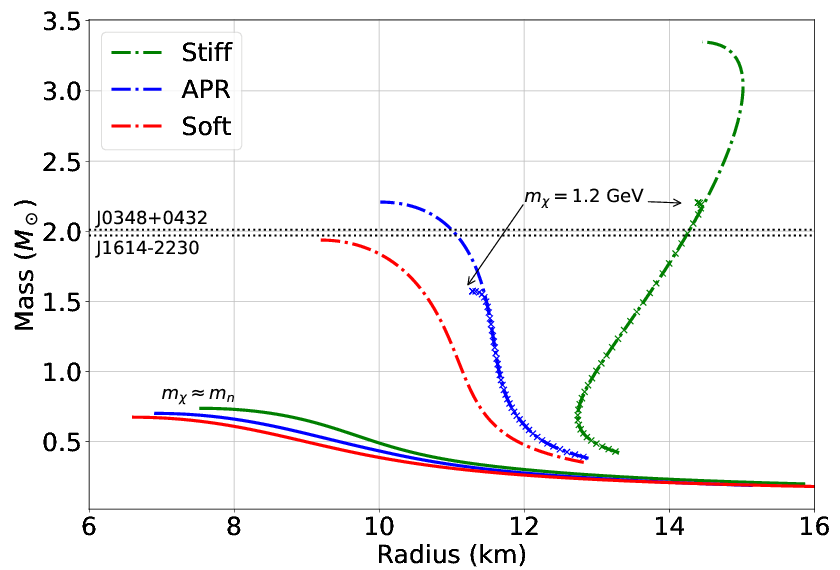}  \ \includegraphics[width=0.47\textwidth]{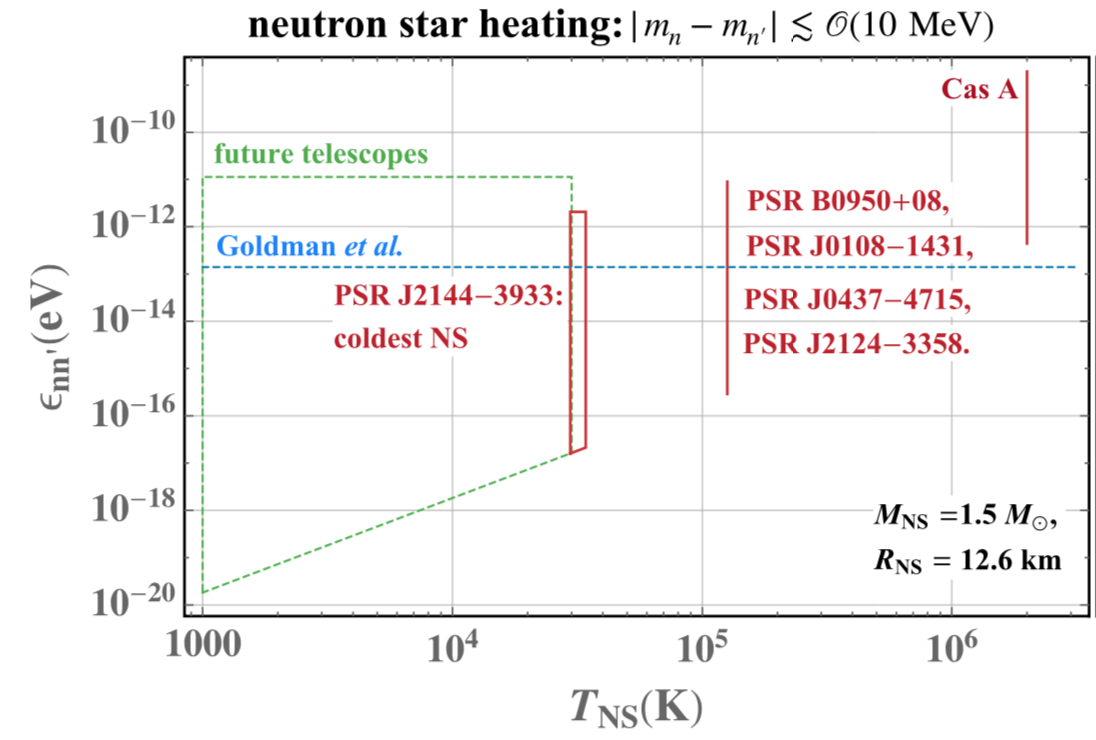}
    \caption{{\bf \em Left.} Mass-radius relations of NSs admixed with a large population of dark baryons, taken from Ref.~\cite{McKeenNelsonReddyZhouNS}. 
    The presence of dark baryons without self-interactions softens the EoS, resulting in a maximum NS mass smaller than those observed above 2~$M_\odot$ (dashed horizontal lines). 
    [Since the appearance of Ref.~\cite{McKeenNelsonReddyZhouNS}, an even heavier 2.35~$\pm$~0.17~$M_\odot$ NS has been observed~\cite{NSFastestHeaviest:Romani:2022jhd}.] 
    Analogous constraints could apply to dibaryons/hexaquarks, a QCD bound state that could make up dark matter~\cite{McDermottReddydibaryons:2018ofd}.
    {\bf \em Right.} Limits on the neutron mixing amplitude of dark/hidden/mirror neutrons from the heating of NSs via the ``nucleon Auger effect", taken from Ref.~\cite{NSheat:Mirror:McKeen:2021jbh}. These are shown for NSs of various surface temperatures, with the ceiling for each set by the timescale of neutron-to-dark-neutron conversions being smaller than the NS age.
    While these limits are valid for neutron-dark neutron mass splittings of up to the NS nuclear self-energy $\simeq$ 10--100 MeV, terrestrial limits from ultracold neutron facilities are only valid for mass splittings up to the Zeeman splitting from the Earth's magnetic field, which is 19 decades smaller. 
    Also shown is a weaker bound from Ref.~\cite{Goldman:2019dbq} from NS mass loss due to neutron-to-dark-neutron conversions, measured with binary pulsar timing (Sec.~\ref{subsubsec:binarypulartiming}).
    See Secs.~\ref{subsubsec:admixedEoS} and \ref{subsec:nucleonAuger} for further details.}
    \label{fig:darkbaryoneffectsNS}
\end{figure*}

\subsubsection{Impact on nuclear equation of state}
\label{subsubsec:admixedEoS}

Hidden GeV-mass states charged with baryon number $B$ could explain the long-standing neutron lifetime puzzle~\cite{Fornal:2018eol} and take a crucial part in baryogenesis~(see, {\em e.g.}, Refs.~\cite{McKeen:2015cuz,Alonso-Alvarez:2021oaj}).
One consequential species is the ``{\bf dark neutron}" with $B = 1$ that can mix with the standard neutron, and could arise either as an elementary particle~\cite{Fornal:2018eol} or as a composite in, {\em e.g.}, mirror matter models~\cite{Berezhiani:2018eds}.
The dark neutron could be cosmologically long-lived if its interactions with the visible sector are small enough, in which case it could constitute the dark matter of the universe.
Dark neutrons $\chi$ may be produced in NSs in neutron-nucleon scattering processes $n N \to \chi N$ and neutron decay $n \to \chi + {\rm anything}$.
If the production of $\chi$ occurs on timescales shorter than NS lifetimes, the $\chi$ fluid that is in chemical equilibrium with the nucleonic fluid would generally soften the EoS of NS matter. 
Consequently, the maximum mass of these admixed NSs is reduced compared to standard NSs, giving rise to constraints on dark neutrons (or other dark sectors) from observations of high-mass NSs~\cite{McKeenNelsonReddyZhouNS,SheltonNS,MottaNS,Ivanytskyi:2019wxd,EllisPattavinaNS,Giangrandi:2022wht,Berryman:2022zic,Rutherford:2022xeb}; see Figure~\ref{fig:darkbaryoneffectsNS} for representative limits.
For related studies of millisecond pulsars and NSs as dark matter admixed quark stars, see Refs.~\cite{Lopes:2018oao,Panotopoulos:2018joc,Panotopoulos:2018ipq}.
In the case of neutron decay to dark neutrons, the above arguments regarding neutron decay also apply to a variation of the dark neutron $\chi$ with $B = 1/3$ such that it is produced in the NS via the decay $n \to \chi \chi \chi$~\cite{StrumiaNS:2021ybk}.
These constraints can be evaded in models that introduce repulsive self-interactions between the dark neutrons, which would pre-empt the softening of the EoS~\cite{Narain:2006kx,Cline:2018ami}.
Admixed NSs also exhibit mass-radius relations that could span a 2-dimensional area rather than follow a 1-dimensional sequence~\cite{Hippert:2022snq}.
Another interesting diagnostic of admixed NSs is the tidal Love number impacted by the formation of extended atmospheres~\cite{Collier:2022cpr,Karkevandi:2021ygv}, and the second Love number~\cite{Dengler:2021qcq}.
This is measurable as a phase shift in a binary merger gravitational wave signal at the forthcoming Advanced LIGO and the third-generation observatories Einstein Telescope and Cosmic Explorer. 
Yet another diagnostic, stemming from the modification of NS mass-radius relations, is the NS pulse profile as measured by precision probes such as NICER~\cite{Shakeri:2022dwg}.
In the case where the production of $\chi$ takes longer than NS lifetimes, other effects come into play, chief among which is the overheating of NSs due to formation of holes in the nucleon Fermi sea; we review this in Section~\ref{subsec:nucleonAuger}.

A {\bf sexaquark/hexaquark} electrically neutral state $uuddss$ that is elusive to accelerator searches has been proposed as a dark matter candidate~\cite{Farrar:2017eqq}; to ensure nuclear stability and cosmological lifetimes for the hexaquark, its mass must lie between 1860--1890~MeV~\cite{Farrar:2018hac}.
It was shown that due to rapid thermalization in the early universe, the hexaquark freezes out at an abundance of $10^{-11}$ of the total baryon number, and can thus only be a minuscule relic~\cite{KolbTurnerdibaryons:2018bxv}.
Moreover, hexaquarks in this mass range would be produced within seconds of the birth of a proto-NS during a core-collapse supernova, and the energy released in this process would unbind the proto-NS, strongly disfavoring the existence of this state~\cite{McDermottReddydibaryons:2018ofd}. 
As the latter authors argue, even if the proto-NS somehow survives, since all baryons are converted rapidly to hexaquarks, the EoS of the resulting star would be softer than that of a standard NS, which would run afoul of observations of high NS masses as in the case of dark neutrons.
Nevertheless, the latter limit may be satisfied if the NS undergoes early quark deconfinement such that hexaquarks are not present, or a later deconfinement that leaves a quark core inside a neutron-hexaquark shell~\cite{hexaquarkNS:ShahrbafFarrar:2022upc}.

\subsubsection{More admixed neutron stars}

Further mechanisms to obtain NSs admixed with dark sectors have been proposed; see the review in Ref.~\cite{ReviewAdmixedNSDM:GrippaPoddar:2024ach} for details.
In analogy with dark neutrons, spin-0 states $\phi_\chi$ carrying baryon number may interact with the neutron via the vertex $\mathcal{L} \supset y_{n\nu} n \phi_\chi \bar\nu$, giving rise to $n \to \nu \phi_\chi$ decays producing $\phi_\chi$ that could constitute 1-10\% of the NS mass~\cite{EllisPattavinaNS}.
In a different model with $\mathcal{L} \supset y_{nn} n \phi_\chi \bar n$, nucleon bremsstrahlung $n n \to n n \phi_\chi$ could populate NSs with $\phi_\chi$ for NS internal temperatures $T_{\rm NS}$ satisfying $m_{\phi_\chi}/3 < T_{\rm NS} \lsim m_{\phi_\chi}/2$~\cite{McKeenNelsonReddyZhouNS,EllisPattavinaNS}.
This condition is to ensure that neutron kinetic energies are sufficient to produce $\phi_\chi$ while keeping the products from escaping the NS' gravity.
For $m_{\phi_\chi}$ = 100 MeV one obtains percent-level NS mass fractions of $\phi_\chi$.
Dark compact stars formed from a possibly dissipative dark sector~\cite{Foot:2004pa,Fan:2013yva} could accrete surrounding baryonic matter to form admixed stars~\cite{EllisPattavinaNS}, though the exact mechanism required to obtain comparable amounts of exotica and nucleons in these structures is far from clear.
One probe of admixed NSs orthogonal to EoS effects is the NS cooling curve~\cite{AngelesPerez-Garcia:2022qzs}; in particular, sub-GeV DM that annihilates to neutrino final states and participating in the NS' thermal conduction could have an observable effect on NS cooling.

\subsection{Exotic compact stars} \vspace{0.2cm}
Dark baryons such as mirror neutrons may give rise not only to NSs admixed with them, but also to compact stars composed entirely of them. 
Such exotic ``{\bf mirror neutron stars}" may constitute an appreciable fraction of DM, providing smoking gun signatures in tidal deformability measurements from gravitational waves and binary pulsar observations~\cite{Hippert:2021fch,Hippert:2022snq}.
A mirror-like hidden sector may also give rise to {\bf exotic white dwarfs} that can be detected using their mergers via gravitational waves~\cite{Ryan:2022hku}.
In the presence of kinetic mixing between the photon and its hidden counterpart, mirror NSs may capture interstellar material and emit radiation observable at Gaia~\cite{Howe:2021neq}.
Mirror stars made of mirror anti-neutrons may grow a population of SM anti-matter in their cores which can accrete ISM and produce radiation observable at Fermi-LAT~\cite{Berezhiani:2021src}.

Analogously, asymmetric DM with large self-interactions could form {\bf dark compact stars}~\cite{Kouvaris:2015rea} that could capture ISM protons and electrons which consequently sink to the core and form a hot radiative gas, observable in telescopes as X-ray or gamma-ray point sources~\cite{Kamenetskaia:2022lbf}.
Other exotic compact objects with near-Schwartzschild compactness may exist and constitute DM, {\em e.g.}, some classes of {\bf boson stars}, {\bf Q-balls}, {\bf non-topological solitons}, and {\bf ultra-compact minihalos}.  
For reviews see Ref.~\cite{Cardoso:2019rvt,microlens:erosogle,microlens:subaru,dMACHOS:Bai:2020jfm}.

In the vein of a first-order QCD phase transition giving rise to expanding bubbles of the low-temperature phase compressing regions of the high-temperature phase into dense quark nuggets~\cite{Witten:1984rs}, an excess of ``dark quarks" charged under a confining gauge group and residing in a false vacuum may be compressed by the true vacuum of a first-order transition~\cite{Bai:2018dxf}. 
In the case of fermionic dark quarks heavier than the confinement scale, this process can lead to compact 
``{\bf dark dwarf}" stars supported by Fermi degeneracy pressure {\em \`{a} la} WDs~\cite{Gross:2021qgx}.
The compression may even go on to form primordial black holes, which is the only end state for bosonic dark quarks.

\subsection{Dark sectors leading to internal heating of neutron stars} \vspace{0.2cm}
\label{subsec:nucleonAuger}

In Sec.~\ref{subsec:admixedNS} we mentioned that neutron scattering and decay processes producing dark baryons $n^\prime$ occurring within the lifetime of the NS would give rise to dark neutron-admixed NSs.
The two-state Hamiltonian for the $|n \rangle$-$|n^\prime\rangle$ system with $m_n\simeq m_{n^\prime}$ and mixing amplitude $\epsilon_{nn^\prime}$ is
\beq
\label{eq:H}
H = \begin{pmatrix} 
  m_n + \Delta E & \epsilon_{nn^\prime} \\
  \epsilon_{nn^\prime} & m_{n^\prime}
 \end{pmatrix},
\eeq
where $\Delta E$ is the medium-dependent energy splitting.
The dominant channel of $n^\prime$ production is neutron-nucleon scattering, $n N \ra n^\prime N$, with cross section~\cite{NSheat:Mirror:McKeen:2021jbh}
\begin{equation}
\sigma_{n^\prime N} \simeq  g_N \left(\frac{\epsilon_{nn'}}{\Delta E}\right)^2\sigma_{nN\to nN}~,
\end{equation}
where $\sigma_{nN\to nN}$ is the neutron-nucleon cross section determined experimentally, $g_n (g_p) = 2 (1)$ is a multiplicity factor, and $\epsilon_{nn'}/\Delta E$ an effective in-medium $n$-$n^\prime$ mixing angle.
The typical rate of $n^\prime$ production in an NS may then be computed from the above as
\beq
\Gamma_{n^\prime}=\frac{1}{10^{7}~\rm yr}\left(\frac{\epsilon_{nn'}}{10^{-15}~{\rm eV}}\right)^2\left(\frac{n_{\rm nuc}}{0.3~{\rm fm}^{-3}}\right)~,
\label{eq:prodrateexample}
\eeq
for a total nucleon density $n_{\rm nuc}$.

If the timescale of $n^\prime$ production $\Gamma^{-1}_{n^\prime}$ exceeds NS lifetimes, which typically occurs in parametric regions where $n^\prime$ also constitutes (the cosmologically long-lived) DM, it would give rise to NS overheating via the ``nucleon Auger effect" discussed in the context of DM capture in Sec.~\ref{subsubsec:DMcapturekineticheat}.
When nucleons leave behind holes in their Fermi seas, through either conversion to $\chi$ or upscattering, higher-energy ambient nucleons rapidly fill them in, in the process liberating heat in the form of electromagnetic and kinetic energy.
The total power liberated by $n\to n^\prime$ conversions in the NS is~\cite{NSheat:Mirror:McKeen:2021jbh}
\begin{equation}
    L_{n\to n^\prime} = \int d^3r\, n_n({\bf r})\dot E_{n^\prime}({\bf r})~, \ \ {\rm with} \ \  \dot E_{n^\prime} = \sum_{N=n,p}f_N n_N\left\langle\left(\tilde\mu_{n}-\frac{p_{n^\prime}^2}{2m_{n^\prime}}\right)\sigma_{n^\prime N}v\right\rangle_{p_N>p_{F_{N}}}~~,
    \label{eq:lumiinternalheat}
\end{equation}
where the subscript in the second equation denotes the inclusion only of scattering events that result in spectator nucleons kicked above their Fermi sea. 

This effect arrests the passive cooling of NSs, and very stringent constraints on dark neutrons~\cite{NSheat:DarkBary:McKeen:2020oyr,NSheat:Mirror:McKeen:2021jbh} may be placed using HST observations of the coldest ($<$ 30,000 K) observed pulsar PSR J2144$-$3933~\cite{coldestNSHST}.
The reach on the mixing between neutrons and dark neutrons could be further extended with present and forthcoming ultraviolet, optical and infrared campaigns suited to observe colder NSs: LUVOIR~\cite{theluvoirteam2019luvoir}, Rubin~\cite{Rubin1,Rubin2}, DES~\cite{DES:2019rtl}, Roman~\cite{green2012widefield}, JWST~\cite{JWST:Gardner:2006ky}, TMT~\cite{TMT:2015pvw}, and ELT~\cite{ELT:neichel2018overview}.
This is depicted in the right panel of Fig.~\ref{fig:darkbaryoneffectsNS} in the plane of the off-diagonal mass or transition amplitude versus the surface temperature of various NSs.
The ceiling for these limits arises from the fact that the nucleon Auger effect is effectively non-existent if the timescale of dark neutron production is smaller than the NS age, in which case the Fermi sea of the dark neutron is filled up and thus neutron conversion is Pauli-blocked.
It must also be noted that these limits apply to dark neutron masses in excess of the neutron mass (939.6 MeV) by 10--100 MeV, the values of neutron self-energies from the nuclear potential of the NS medium that effectively raise $m_n$. 
Above this mass conversions of neutrons to dark neutrons are kinematically suppressed.

One scenario in which the above limits may be weakened is when the dark neutron arises as a mirror equivalent of the neutron in mirror world theories, with exact mass degeneracies~\cite{NSheat:Mirror:Goldman:2022rth}.
In this case, mirror electrons produced via mirror beta decay could take away heat by scattering with electrons via a millicharge, and emit mirror bremsstrahlung via mirror photons.
Internal heating of NSs can also occur by baryon number-violating neutron decays, depositing nearly the mass energy of the neutrons, and for some models with long-range potentials this provides the strongest limits from observations of PSR J2144$-$3933~\cite{Davoudiasl:2023peu}.

The heating of NSs from neutron losses via baryon number-conserving and violating processes, and from NS capture of DM in the various ways described in Sec.~\ref{subsec:DMkinannheat}, provide compelling fundamental physics motivations for upcoming astronomical missions to perform systematic measurements of NS luminosities, masses and radii.

\begin{figure*}
\includegraphics[width=0.45\textwidth]{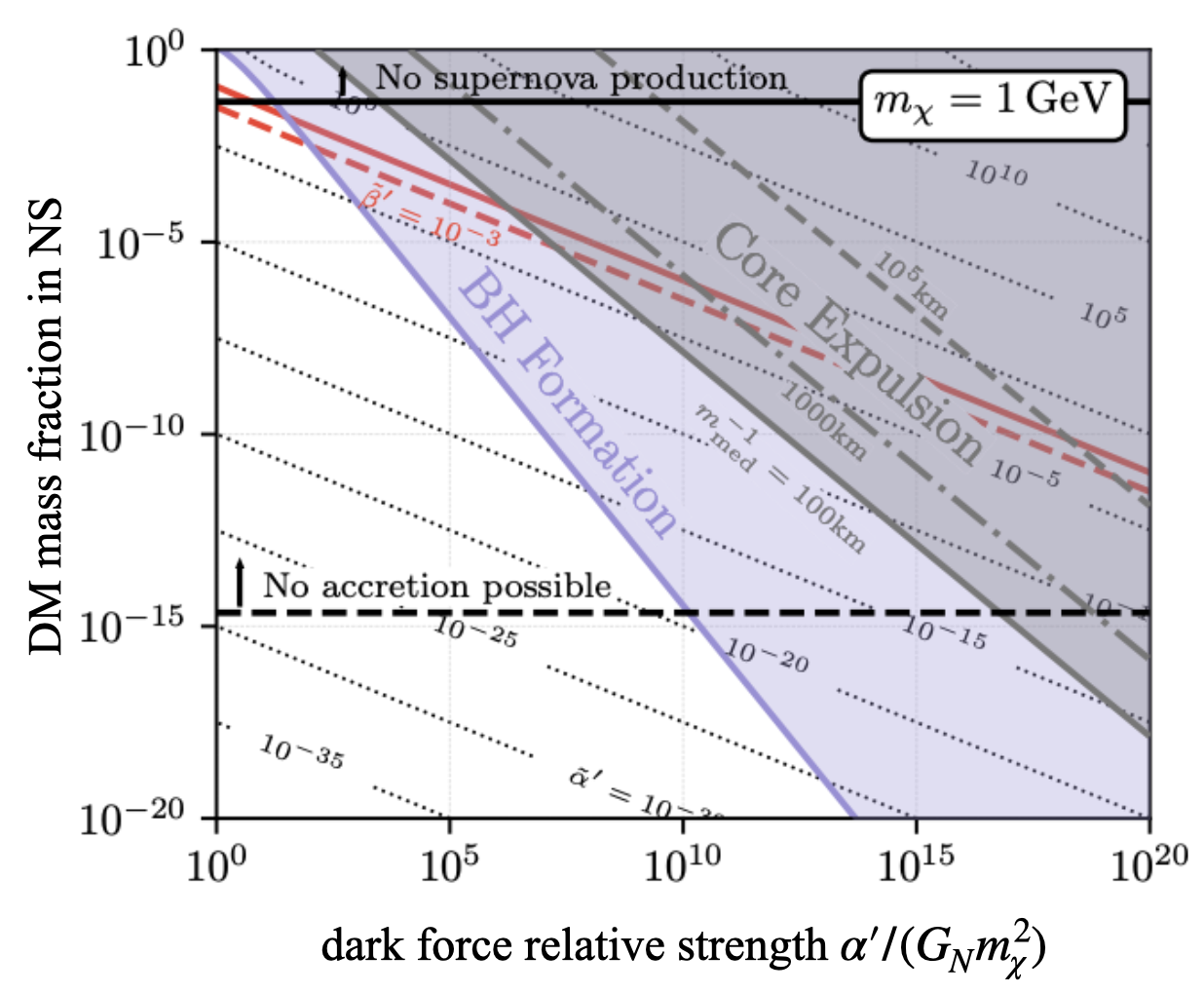} \ \ \ 
\includegraphics[width=0.55\textwidth]{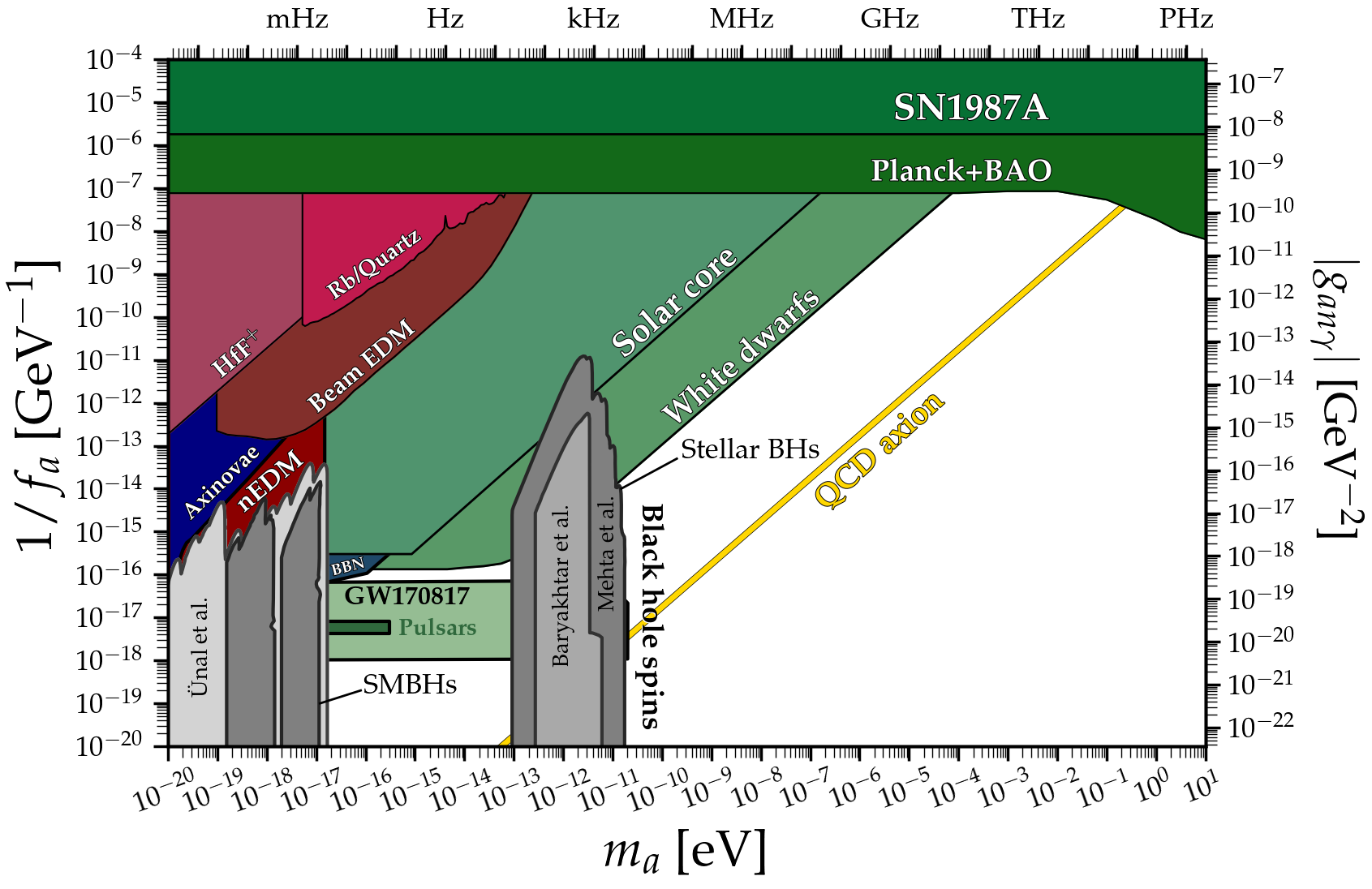} \\
\includegraphics[width=0.5\textwidth]{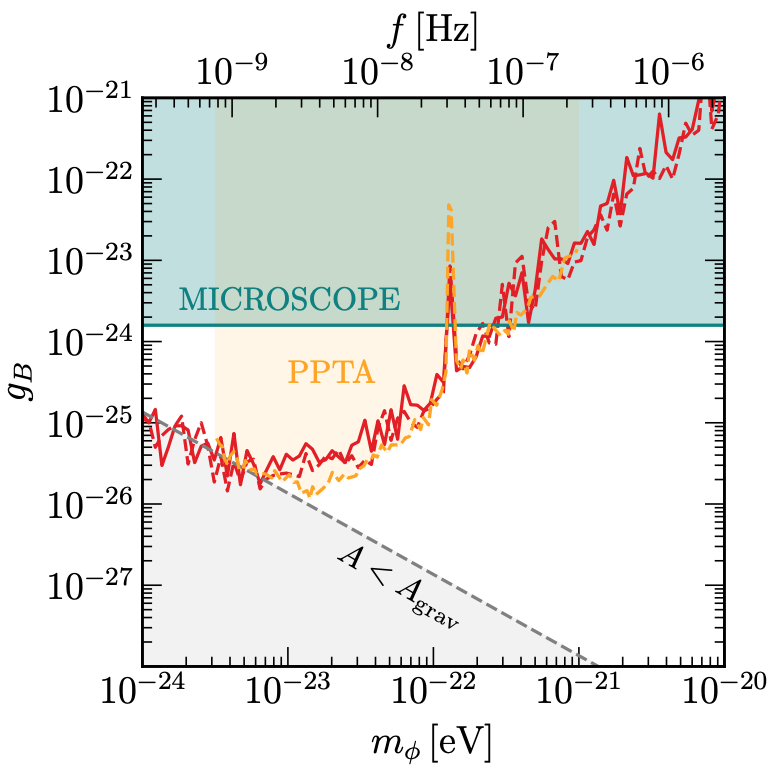}
\includegraphics[width=0.5\textwidth]{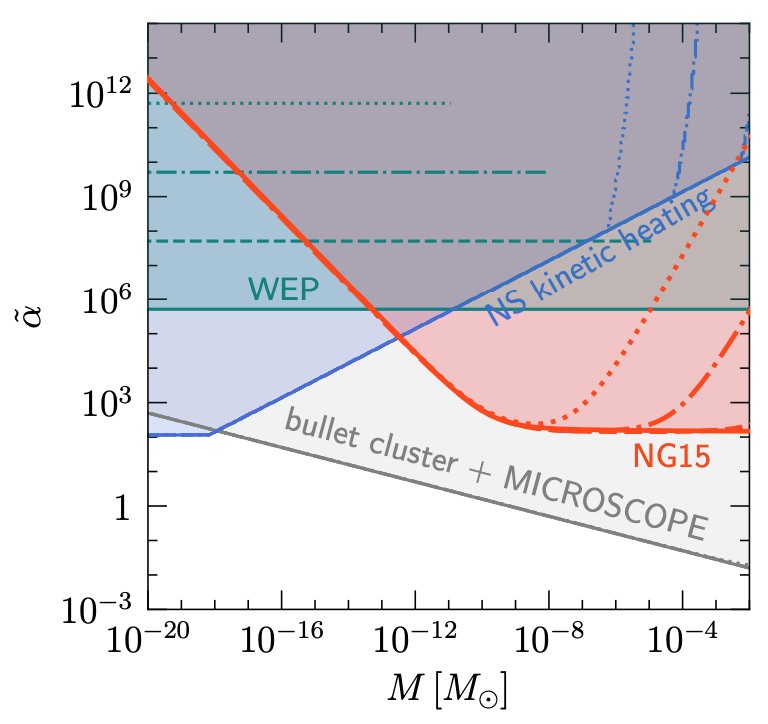}
\caption{
{\bf \em Top left}.
Limits on the mass fraction of dark matter in NSs in a binary as a function of the coupling of an attractive long-range dark force relative to gravity, for DM mass = 1 GeV. 
The solid red contour indicates the minimum value of the effective dark force coupling $\alpha^\prime$ that will have a detectable impact at LIGO; the dashed red contour depicts the value of the coupling when the dark charge resides in only one NS. 
In the purple region, the NSs turns into black holes. 
In the gray region, DM is expelled from the NSs when screening due to the dark force lifts.
See Ref.~\cite{Kopp:2018jom} for further details.
{\bf \em Top right.} Future limits from gravitational waves on an axion-like species sourced by inspiraling binary NSs: the region bounded by $m_a \lsim 10^{-12}$~eV and $f_a \gsim 10^{15}$~GeV can be probed by Advanced LIGO.
For particulars on the other limits shown here, see Ref.~\cite{GWLimitsAxionCAJOHare}.
 {\bf \em Bottom left.} Limits on the coupling of ultralight dark matter to baryons as a function of its mass from the 15-year dataset of the NANOGrav pulsar timing array~\cite{NANOGrav:2023hvm} for correlated (red solid) and uncorrelated (red dashed) signals.
Also shown are erstwhile limits from the Parkes Pulsar Timing Array (PPTA) and MICROSCOPE's equivalence principle test, and the region where the signal amplitude is smaller than a gravitational one. 
{\bf \em Bottom right.} 
Limits on the effective baryonic long-range Yukawa coupling of dark matter subhalos/nuggets versus DM mass using NANOGrav~\cite{NANOGrav:2023hvm} (red region).
The \{solid, dashed, dot-dashed, dotted\} curves correspond to the range of the force $10^{\{0,-1,-2,-3\}}$ pc.
Stronger limits from WD explosions and NS superbursts, not shown here, could come into play for certain additional DM properties~\cite{WDNSboom:Raj:2023azx}.
Also shown are limits from kinetic heating of the coldest observed pulsar PSR J2144$-$3933 (see Sec.~\ref{subsec:DMsubstruct}), weak equivalence principle tests, and the Bullet Cluster.
 }
\label{fig:GWPTA}
\end{figure*}

\subsection{Dark matter signals in gravitational waves from neutron star mergers} \vspace{0.2cm}
\label{subsec:NSmergers}

Admixed NSs containing large quantities of DM could interact with each other via a long-range force (either attractive or repulsive) acting on the dark component. This could leave a distinct signature in the waveforms of the gravitational waves (GWs) picked up during their mergers at detectors such as LIGO and VIRGO~\cite{Croon:2017zcu,Kopp:2018jom,Alexander:2018qzg}. However, for the effects of DM admixed in NSs to be seen, something beyond standard DM accumulation (which results in DM cores with masses $\ll 10^{-10} M_\odot$) is required. As previously noted, one might consider a sizable fraction of the NSs mass converting to dark sector particles, or $e.g.$ some process of DM accumulation through fluid-like accretion in an exceptionally dense region of dark matter~\cite{NSvIR:clumps2021}. 
New GW signatures from highly admixed NSs would be visible via two effects: one, a modification to the measured chirp mass $\mathcal{M} \equiv \mu^{3/5}(M_1+M_2)^{2/5}$ where $M_{1,2}$ are the NS masses and $\mu$ their reduced mass, since the effect of the new force on the evolution of the binary period is degenerate with a shift in NS masses; two, energy loss through dipole radiation of the force-carrier, which would again show up in the binary period evolution.

Other interesting effects are observable in mergers of DM-admixed NSs.
The post-merger GW spectrum could exhibit additional peaks, as was shown by a simple mechanical model in Ref.~\cite{Ellis2018}.

In general, the equation of state and structure of an admixed NSs will depend on DM model properties, including whether the DM is bosonic~\cite{Rutherford:2022xeb,Giangrandi:2022wht,Emma:2022xjs}, and whether such a bosonic core of DM is fully contained with the admixed NS or extends outside the baryonic component of the admixed NS~\cite{Ruter:2023uzc}. In the latter case, it has been shown that the bosonic DM extending beyond the surface of the admixed NSs can lead to a ``mass-shedding" phenomenon in binary systems of DM-admixed NSs~\cite{Emma:2022xjs,Ruter:2023uzc}, resulting from inter-NS tidal forces.

It has also been investigated whether the DM components in a binary admixed NS system might remain gravitationally bound and orbit inside the merger remnant, after the NS component have merged. Assuming the DM components remain intact,~\cite{Bauswein:2020kor} found that orbital separations of typically a few km, resulted in a khz-band GW signal that could be sought in GW searches.

In Fig.~\ref{fig:GWPTA} top left panel we show limits from Ref.~\cite{Kopp:2018jom} on the NS mass fraction of DM as a function of the dark force strength relative to gravity.  
 For sizeable abelian forces, the charge carried by the DM must not be too large lest it unbind the star, limiting 
 the hidden force to be weaker than gravity if gravitational tests are to be made~\cite{Fabbrichesi:2019ema}.
 A long-range dark sector force such as felt by muons in the NS could also induce the effects above~\cite{Dror:2019uea}.
 
Axions may be sourced by NSs due to in-medium corrections to the axion potential, giving rise to inter-NS forces that can be detected during the inspiral~\cite{Hook:2017psm}.
In Fig.~\ref{fig:GWPTA} top right panel we show ensuing limits from Ref.~\cite{Hook:2017psm,GWLimitsAxionCAJOHare} on the decay constant as a function of the mass of an axion-like species.
Ultra-light scalar DM with baryonic interactions can induce time-varying mass shifts in the NS from their coherent background, which could show up in broadband measurements~\cite{Choi:2018axi}. 
Ultra-light DM could also modify the dispersion relation of neutrinos escaping a proto-NS, giving rise to asymmetric emission and hence the natal kick of the NS (Sec.~\ref{sec:physicscompact}) as well as a non-oscillatong permanent strain in the local metric, a ``gravitational memory" signal, that can be picked up at GW detectors~\cite{Lambiase:2023hpq}.
Dark photons~\cite{Diamond:2021ekg} and axions~\cite{Diamond:2023cto} produced in NS mergers could convert to detectably luminous $\gamma$-rays.

\subsection{Dark matter signals in pulsar timing} \vspace{0.2cm}
\label{subsec:pulsartiming} 

Pulsating NSs are incredibly precise celestial metronomes.
As discussed in Sec.~\ref{subsec:Bspindown}, their spin periods slow down at rates smaller than 1 sec per $10^{12}$ sec, affording a pulse regularity second only to atomic clocks.
This precision is exploited to probe fundamental physics in a number of ways, and in this section we collect the implications for dark matter searches.

\subsubsection{Pulsar timing arrays}
\label{subsubsec:PTAs}

Pulsar timing arrays (PTAs), by constraining correlations in the arrival times of pulses emitted by $\Oc(10)$ millisecond pulsars\footnote{Although about 3000 pulsars have been discovered~\cite{ATNF:2004bp}, only these many provide the level of stability and noise-free emission required to achieve interesting precision at PTAs.}, are primarily detectors of nanoHertz to milliHertz gravitational waves~\cite{PTAGW1,PTAGW2}.
These measurements can also be used to constrain DM substructure (including primordial black holes): their transits could induce a shift in the signal phase, 
\beq
\varphi(t) = \varphi_0 + \nu t + \frac{1}{2}\dot \nu t^2 + \frac{1}{6} \ddot \nu t^3 + \Oc(t^4)~,
\label{eq:PTAphase}
\eeq
from a shift in the frequency via two effects:
\bea
\nn \bigg( \frac{\delta \nu}{\nu}\bigg)_{\rm Dopp} &=& \mathbf{\hat d} \cdot \int \nabla \Phi dt~, \\
\bigg( \frac{\delta \nu}{\nu}\bigg)_{\rm Shap} &=& 2 \int \mathbf{v_\chi} \cdot \nabla \Phi dz~,
\label{eq:PTAdelays}
\eea
where $\mathbf{\hat d}$ is the Earth-to-pulsar direction, 
$\mathbf{v_\chi}$ the velocity of the DM structure, 
$\Phi$ the gravitational potential due to it, and $z$ traces the pulsar-to-Earth path of photons.
The first line of Eq.~\eqref{eq:PTAdelays} describes a {\em Doppler time delay} in the observed period of the pulsar brought about by the acceleration of the NS or Earth due to transiting DM~\cite{Seto:2007kj}.
The second line describes a {\em Shapiro time delay} coming from a change in the arriving photon's geodesic due to DM structure's gravitational potential~\cite{Siegel:2007fz}.
Signals of DM substructure could either be {\em dynamic}~\cite{Siegel:2007fz,Seto:2007kj,Baghram:2011is,Seto2012,Kashiyama:2018gsh}, as when the transit times are much smaller than the total observation time, giving rise to blips in the data, or could be {\em static}~\cite{Clark:2015sha,Clark:2015tha,Schutz:2016khr}, as when they are much larger, showing up as a sizeable contribution to the $\ddot \nu$ term in Eq.~\eqref{eq:PTAphase}.
Detailed analyses of dynamic and static signals, treating Doppler and Shapiro time delays, can be found in Refs.~\cite{PTA:Dror:2019twh,Ramani:2020hdo,Lee:2021zqw}, with implications for DM substructure origins explored in Ref.~\cite{Lee:2020wfn}.
Limits from North American Nanohertz Observatory for Gravitational Waves (NANOGrav) data~\cite{NANOGrav:2017wvv,NANOGRAV:2018hou} are derived in Ref.~\cite{NSvIR:tidalfifthforce:Gresham2022,NANOGrav:2023hvm} on dark nuggets (that could range from point-like to $> 10^9$ km in size) interacting with baryons through a long-range Yukawa force; see Fig.~\ref{fig:GWPTA} bottom right panel.

Another scenario amenable to searches at PTAs is that of ultralight ($10^{-24}$-$10^{-22}$~eV) DM, which could induce oscillations of the gravitational potential of the Galactic halo at nanohertz frequencies~\cite{Khmelnitsky:2013lxt,ULDMvPTA:Omiya:2023bio,ULDMvPTA:Armaleo:2020yml,ULDMvPTA:Wu:2023dnp}. 
Various PTAs have constrained these models~\cite{ULDMvNANOGrav:Porayko:2014rfa,ULDMvPPTA:Porayko:2018sfa,ULDMvEPTA:Smarra:2023ljf,NANOGrav:2023hvm}; see Fig.~\ref{fig:GWPTA} bottom left panel.
In addition, a network of topological defects that alter fundamental constants could be uncovered through a variation of pulsar periods across a network of well-timed pulsars~\cite{Stadnik:2014cea}.

PTAs currently operational are 
the European Pulsar Timing Array (EPTA)~\cite{EPTA:Desvignes:2016yex} that uses the Westerbork Synthesis, Effelsberg, Lovell, Nancay and Sardinia radio telescopes, 
Parkes Pulsar Timing Array (PPTA)~\cite{PPTA}, 
NANOGrav~\cite{NANOGrav:Brazier:2019mmu} that uses the Green Bank Telescope, Arecibo Observatory and Very Large Array,
Indian Pulsar Timing Array (InPTA)~\cite{InPTA:Tarafdar:2022toa} that uses the Upgraded Giant Meterwave Radio Telescope (uGMRT)
(these four make up the International Pulsar Timing Array (IPTA)~\cite{IPTA:Perera:2019sca}), 
MeerTime at MeerKAT~\cite{MeerTime:Bailes:2018azh}, 
and the Chinese Pulsar Timing Array (CPTA)~\cite{FASTPTA:Hobbs:2014tqa} that uses the Five-hundred-meter
Aperture Spherical Telescope (FAST).
The future has the Square Kilometer Array (SKA)~\cite{SKAPTA1,SKAPTA2:Rosado:2015epa}.
It would be of significant interest to look for various DM substructure as well as ultra-light DM in the datasets of all these PTAs, an exercise we urge of expert authors.

A recent data release of multiple PTAs announced detection of signals consistent with a stochastic gravitational wave background (SGWB) sourced by supermassive black hole (SMBH) mergers~\cite{NANOGrav:2023hvm,EPTAInPTA:2023fyk,PPTA:2023gzh,CPTA:Xu:2023wog,MeerKAT:Miles:2024seg}.
This has prompted many authors to investigate PTA signatures of dark matter, including
re-interpetation of the data in terms of primordial black holes~\cite{Depta:2023qst,Inomata:2023zup} (which is, however, in tension with the Hubble Space Telescope's measurement of the luminosity function of galaxies~\cite{Gouttenoire:2023nzr}),
a modification of the spectral index of the SGWB due to cosmic DM-induced dynamical friction~\cite{Ghoshal:2023fhh},
a modulation in the amplitude due to a DM spike surrounding the SMBHs~\cite{Shen:2023pan},
an inflationary blue-tilted tensor power spectrum in a setup thermally overproducing WIMPs~\cite{Borah:2023sbc},
a soliton of ultralight DM enclosed by the SMBHs~\cite{Broadhurst:2023tus},
an electroweak phase transition induced by a potential involving DM~\cite{Xiao:2023dbb},
an ultralight radion arising from a fifth spacetime dimension~\cite{Anchordoqui:2023tln},
and dark photon DM produced by the decay of cosmic string loops that may have sourced the SGWB~\cite{Kitajima:2023vre}.

\subsubsection{Binary pulsar timing}
\label{subsubsec:binarypulartiming}

Pulsar timing without the use of a pulsar array can also be a valuable tool to detect DM.
The orbits of inspiraling binary pulsars may undergo seasonal modulation due to dynamical friction caused by DM, which may be used to set bounds on the density of DM in the environment~\cite{Pani:2015qhr}.
Measurements of the orbital period and period decay of binary pulsars also help to set limits on ultralight scalars that could constitute DM via their radiation from NSs~\cite{KumarPoddar:2019jxe} and
their coherent oscillations~\cite{Blas:2016ddr,Blas:2019hxz,DiLuzio:2021pxd}, 
and the rate of mass loss in NSs due to, {\em e.g.}, conversions of neutrons to dark baryons~\cite{Goldman:2019dbq,Berezhiani:2020zck,Berryman:2022zic,Berryman:2023rmh,Gardner:2023wyl}, although these latter limits are superseded by considerations of NS heating via the Auger effect (Sec.~\ref{subsec:nucleonAuger}) as detailed in Ref.~\cite{NSheat:Mirror:McKeen:2021jbh}.

\subsubsection{Pulsar spin-down} 
\label{subsubsec:millichspindown}

Milli-charged DM accreting onto NSs provides surplus charge that must be expelled from the polar caps to maintain charge neutrality. 
Thus an additional electric current (following open magnetic field lines) is induced, contributing to the NS $B$ field, and therefore to the slowing down of its spin as per the discussion in Sec.~\ref{subsec:Bspindown}.
In this way milli-charged DM could explain the observation of pulsar braking indices $n < 3$~\cite{Kouvaris:2014rja,Huang:2015gta}. 

\begin{figure*}
\includegraphics[width=0.45\textwidth]{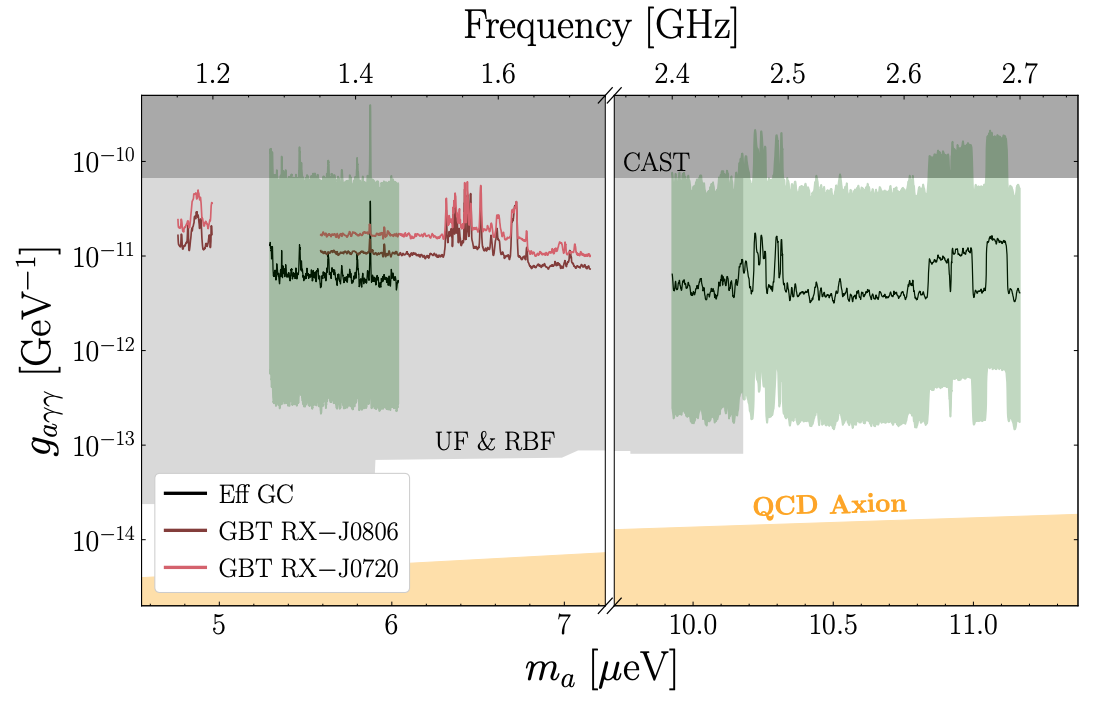}
\includegraphics[width=0.55\textwidth]{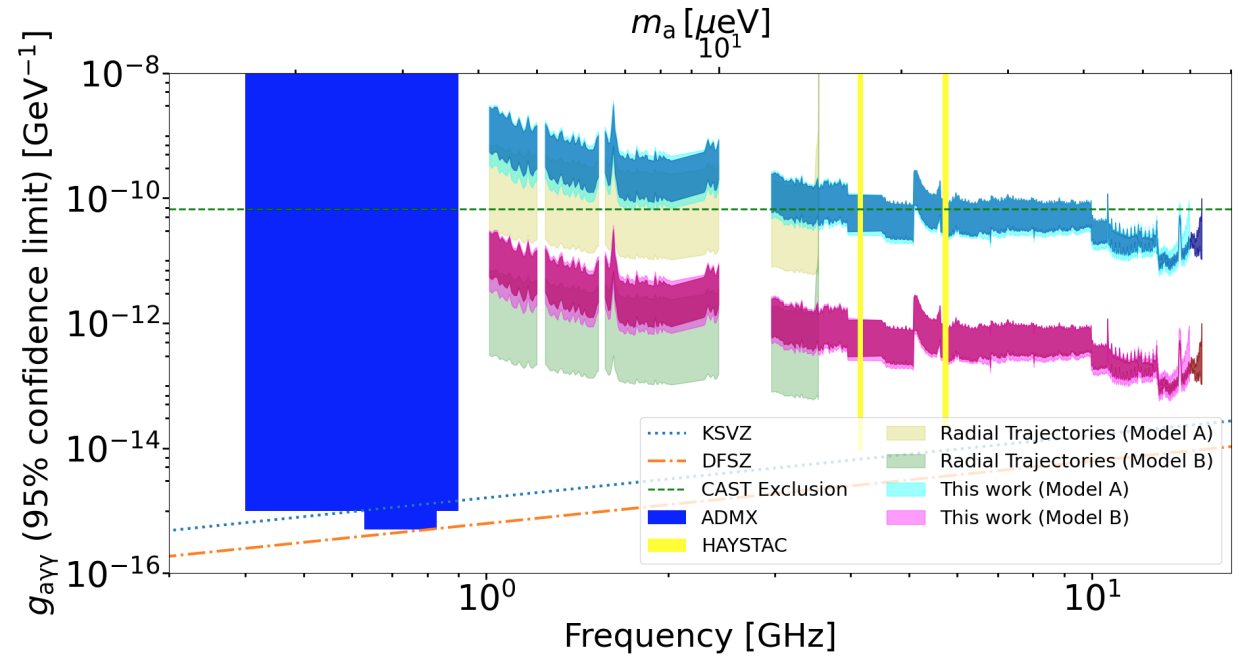}
\caption{{\bf \em Left}. 95\% C.L. bounds on the ALP-photon coupling versus ALP mass from Ref.~\cite{Foster:2020pgt} using the Green Bank and Effelsberg Radio Telescopes' observations of NSs nearby and in the Galactic Center. The green bands span theoretical uncertainties in the Effelsberg analysis. 
Also shown are limits from CAST (helioscope), and UF and RBF (haloscopes), and the region where the ALP could be the QCD axion.
~~ {\bf \em Right.} Same as the left panel, from Ref.~\cite{Battye:2021yue}, using VLA data on the pulsar J1745$-$2900 and ray-tracing models. The bands depict uncertainties in viewing and magnetic angles. Limits from CAST, ADMX and HAYSTAC are also shown. See Sec.~\ref{subsec:NSvALP} for further details.}
\label{fig:alpsvpulsars}
\end{figure*}

\subsection{Axion-like and very light dark matter, and neutron stars} \vspace{0.2cm}
\label{subsec:NSvALP}

The above sub-sections considered particulate DM confronting NS detectors.
Now we will see that very light, wave-like dark matter can also be discovered with NSs.
The poster child for such DM is the QCD axion and its extension, axion-like particles (ALPs).
In this section we will focus on the utility of compact stars in discovering ALP DM, referring the reader to the review in Ref.~\cite{Tinyakov:2021lnt} for a thorough treatment of generic ALPs interacting with compact objects.

ALPs of masses $m_a$ in the range $10^{-7}$--$10^{-4}$ eV, when falling into NSs, convert to photons in their magnetospheres via the Primakoff effect~\cite{Lai:2006af}, {\em i.e.} via the interaction
\begin{equation}
    \mathcal{L} \supset \frac{g_{a\gamma}}{4} a F_{\mu\nu}\tilde{F}^{\mu\nu} \rightarrow - g_{a\gamma} \ a \ \mathbf{E} \cdot \mathbf{B}~.
\end{equation}
The conversion probability is enhanced by the large magnetic field, and resonantly so for ALP masses degenerate with the photon plasma mass~\cite{Pshirkov:2007st,Hook:2018iia,Huang:2018lxq,Safdi:2018oeu,Battye:2019aco,Foster:2022fxn}.
The result is radio emission observable as a monochromatic line signal at a frequency set by the ALP mass: the narrow bandwidth is set by the small dispersion in the speed of the ALP DM. 
Limits exploiting this signature on the ALP-photon coupling and ALP mass  have been set using 
the Green Bank Telescope, 
the Very Large Array, 
and 
the Effelsberg 100-m Radio Telescope~\cite{Foster:2020pgt,Darling:2020plz,Darling:2020uyo,Foster:2022fxn,Battye:2021yue}.
In the near future, observations can be made at
the Parkes Observatory, 
the Sardinia Radio Telescope, 
MeerKAT, 
the Murchison Widefield Array, 
and 
the Hydrogen Epoch of Reionization Array~\cite{ExtremeBaryakhtar:2022hbu}.
Analogous limits have been placed on $\Oc(10^{-5})$ eV mass spin-1 DM converting to photons via kinetic mixing in the plasma of NSs and accreting WDs in the Galactic Center~\cite{Hardy:2022ufh}.
Time-domain data (as opposed to frequency-domain) of PSR J2144$-$3933 at MeerKAT have been used to set limits on ALP emission~\cite{Battye:2023oac}, taking advantage of time variation of the periodic signal due to pulsar rotation and magnetospheric plasma effects.
Resonant conversion of ALP DM can also occur in the corona of a magnetic WD, which could prove more sensitive than NSs to sub-microeV ALPs~\cite{Wang:2021wae}.
In Fig.~\ref{fig:alpsvpulsars} we show recent NS-based constraints on the ALP-photon coupling versus ALP mass.

On the theoretical side, accounting for the semi-relativistic speeds of ALP DM infall and anisotropy of the magnetic field can potentially modify the ALP-photon conversion rate by orders of magnitude~\cite{Millar:2021gzs}.
One can also consider enhanced radio emission from NSs in binary systems with intermediate mass ($10^3$--$10^5 M_\odot$) black holes, taking advantage of a possible spike in the DM density~\cite{Edwards:2019tzf}. 
Ray-tracing simulations accounting for signal photon propagation in the NS gravitational potential are an important direction toward accurate determinations of the outgoing radio flux~\cite{Leroy:2019ghm,Witte:2021arp,Battye:2021xvt}.
Some of the uncertainties addressed by these efforts relate to the absorption of radio waves in the plasma and their refraction and reflection in the magnetosphere, signal broadening due to photon-plasma interactions, and time-dependence of the signal.

As with particulate DM discussed in Sec.~\ref{subsec:NSsubstruct}, ALP DM could also form substructure such as axion mini-clusters and axion stars. 
If large fractions of the DM mass are present in such substructure the sensitivities of laboratory searches for ALP DM may be diluted, and NSs as ALP probes grow in importance. 
Encounters of NSs with ALP substructure can produce distinct transient radio signals, including fast radio bursts~\cite{Kavanagh:2020gcy,Edwards:2020afl,Buckley:2020fmh,Nurmi:2021xds,Bai:2021nrs,Witte:2022cjj}. 
In the bottom right panel of Fig.~\ref{fig:NSvclumps}, taken from Ref.~\cite{Edwards:2020afl}, are shown for two different axion mini-cluster profiles the mean flux densities of the radio signal and the sensitivities of telescope arrays.
These signal rates may be smaller for more conservative estimates of the NS magnetic fields that account for possible mechanisms of their decay~\cite{Witte:2022cjj}.

ALPs that are not necessarily the cosmic DM may also be studied in NS systems. 
Produced with $\mathcal{O}$(100 keV) energies in the core via nucleon bremsstrahlung, ALPs could convert in the magnetosphere to photons that may observed as x-ray and gamma-ray emissions.
As these ALPs carry away energy they also provide an additional cooling mechanism for the NS.
For further details and observation prospects of ALP emissions from within NSs, see Ref.~\cite{ExtremeBaryakhtar:2022hbu} and the references therein. 
ALPs could also overheat NSs by contributing to the NS magnetic energy via a dynamo mechanism generated by axio-electrodynamics~\cite{Anzuini:2022bqd}.

As a reminder, other signatures of ALPs have already been discussed in this review:
their modification of WD EoS (Sec.~\ref{subsec:DMvWDEOS}) and their considerable effects on pulsar timing measurements (Sec.~\ref{subsec:pulsartiming}).

\section{Conclusions and perspective} \vspace{0.2cm}
\label{sec:concs}

In this review we have described how various authors have exploited the unique and extreme properties of compact stars to enable far-reaching searches for a vast assortment of dark matter scenarios\footnote{Although there are some searches for which compact stars hinder discovering dark matter, $e.g.$ searches via microlensing~\cite{Niikura:2019kqi} and searches in globular clusters~\cite{EvansStrigari:2021bsh}!}.
We have, of course, chosen to focus on the commonly accepted and well-observed classes of compact stars: WDs and NSs.
Several adjacent stellar entities may play a role in the discovery of dark matter, {\em e.g.}, black holes, proto-NSs, supernovae and their remnants, and Thorne-\.Zytkow objects; for accounts of some of these, we refer the reader to Ref.~\cite{ExtremeBaryakhtar:2022hbu} and other literature.

Given the richness of physics involving both dark matter and compact stars, it is impossible to exhaust the progress that can be made hereupon.
We mention a few possibilities.
\begin{itemize}
    \item While reseach has gone into detecting wave-like, particulate, macroscopic and black hole dark matter with compact stars,
what happens to DM in the form of topological defects (macroscopic monopoles, cosmic strings, and domain walls)~\cite{topologicaldefect:Murayama:2009nj,topologicaldefect:DereviankoPospelov:2013oaa} in a compact star environment is as yet relatively unexplored, although for some initial work in this direction see Ref.~\cite{Stadnik:2014cea}.
This points to a property of WDs and NSs that could be further exploited -- as sirens of dark matter well-distributed through the galaxy, they have special sensitivity to variations in dark matter properties across the halo and substructure~\cite{NSvIR:clumps2021}.
\item The physics of BEC and BCS states formed by dark matter in NSs, essential to understanding collapse to black holes, is yet to be worked out satisfactorily. Many of the computations have used coarse estimates.
\item Much remains to be studied about thermonuclear explosions in WD cores and NS oceans. 
  Their trigger lengths, which while numerically estimated in Ref.~\cite{TimmesWoosley1992} with a large nuclear reaction network, are only available for a narrow range of densities and compositions.
We believe that the significance of these computations for a scientific question as fundamental as dark matter warrants further exploration of the 32 year-old results of Ref.~\cite{TimmesWoosley1992} by the nuclear astrophysics community.
Rare ``hyperbursts" in NS crusts, 100 times more powerful than superbursts, have physical origins that are even more unclear~\cite{hyperburstPage:2022ikz}.
The impact of WD explosions on the evolution of galactic structure and star formation is yet to be studied in detail.
Some Type Ia SNe suggest WDs of mass exceeding the conventional Chandrasekhar limit~\cite{superChWD:SNLS:2006ics,superChWD:Hachisu2012,superChWD:Das:2013gd,superCh:Zuraiq:2024ypo} if they are globally~\cite{superChWD:instab:Coelho:2013bba} and locally~\cite{superChWD:instab:Chamel:2013tfa} stable.

\end{itemize}
In closing, dark matter may first become manifest through some effect on compact stars detailed in this document. 
On the other hand, since compact stars have the distinction of being the densest objects composed of known particles, it may be that some variety of dark matter, hitherto unexpected, will first become evident through a surprising and as-yet-unforeseen interaction with them. 
In either case, we can look forward to the interplay between our burgeoning understanding of compact stars and the ebullient search for dark matter in the coming decades.

\section*{Acknowledgments} \vspace{0.2cm}

For helpful interactions we thank 
Aryaman Bhutani,
Melissa Diamond,
Michael Fedderke,
Raghuveer Garani,
Bradley Kavanagh,
Ranjan Laha,
and Camellia Sinensis.
The work of J.\,B. is supported by the Natural Sciences and Engineering Research Council of Canada.
N.\,R. acknowledges support from TRIUMF Inc. and the Arthur B. McDonald Canadian Astroparticle Physics Research Institute at Queen's University during the course of this work.
This research was undertaken thanks in part to funding from the Canada First Research Excellence Fund through the Arthur B. McDonald Canadian Astroparticle Research Institute.
Research at Perimeter Institute is supported in part by the Government of Canada through the Department of Innovation, Science and Economic Development Canada and by the Province of Ontario through the Ministry of Colleges and Universities.

\addcontentsline{toc}{section}{References}
\bibliographystyle{elsarticle-num}
\bibliography{refs}

\end{document}

%% file: universalnewcommands.tex
\newcommand{\gsim}{\gtrsim}
\newcommand{\lsim}{\lesssim}
\newcommand{\ra}{\rightarrow}

\def\Oc{\mathcal{O}}

\renewcommand{\tilde}{\widetilde} 

\newcommand{\beq}{\begin{equation}}
\newcommand{\eeq}{\end{equation}}
\newcommand{\bea}{\begin{eqnarray}}
\newcommand{\eea}{\end{eqnarray}}
\newcommand{\nn}{\nonumber}

\definecolor{rosy}{RGB}{230,235,252}
\definecolor{myframetitle}{RGB}{90,89,170}
\definecolor{myblocktitle}{RGB}{140,185,249}
\definecolor{mytitle}{RGB}{10,80,26}

\definecolor{darkgreen}{RGB}{27,130,45}
\definecolor{darkblue}{rgb}{0,0,0.3}
\definecolor{darkred}{rgb}{0.7,0,0}

\definecolor{light gray}{RGB}{220,220,220}
\definecolor{dark purple}{RGB}{108,0,217}
\definecolor{pink}{RGB}{190,20,100}
\definecolor{orang}{RGB}{193,63,0}
\definecolor{green}{RGB}{11,98,17}
\definecolor{darkpink}{RGB}{153,0,76}
\definecolor{bluegreen}{RGB}{0,102,102}
\definecolor{greenlagan}{RGB}{0,102,0}
\definecolor{redgreen}{RGB}{102,102,0}
\definecolor{Redgreen}{RGB}{153,76,0}
\definecolor{vividviolet}{rgb}{0.62, 0.0, 1.0}
\definecolor{amaranth}{rgb}{0.9, 0.17, 0.31}
\definecolor{palatinateblue}{rgb}{0.15, 0.23, 0.89}
\definecolor{brightpink}{rgb}{1.0, 0.0, 0.5}
\definecolor{cornflowerblue}{rgb}{0.39, 0.58, 0.93}
\definecolor{deepcarminepink}{rgb}{0.94, 0.19, 0.22}
\definecolor{radicalred}{rgb}{1.0, 0.21, 0.37}

%% file: main.bbl
\begin{thebibliography}{100}
\expandafter\ifx\csname url\endcsname\relax
  \def\url#1{\texttt{#1}}\fi
\expandafter\ifx\csname urlprefix\endcsname\relax\def\urlprefix{URL }\fi
\expandafter\ifx\csname href\endcsname\relax
  \def\href#1#2{#2} \def\path#1{#1}\fi

\bibitem{DMreview:Cirelli:2024ssz}
M.~Cirelli, A.~Strumia, J.~Zupan, {Dark Matter} (6 2024).
\newblock \href {http://arxiv.org/abs/2406.01705} {\path{arXiv:2406.01705}}.

\bibitem{WDhistory:Fowler:1926}
R.~H. Fowler, {On dense matter}, Mon. Not. Roy. Astron. Soc. 87 (1926)
  114--122.

\bibitem{WDhistory:Anderson:1929}
W.~{Anderson}, {{\"U}ber die Grenzdichte der Materie und der Energie},
  Zeitschrift fur Physik 56~(11-12) (1929) 851--856.
\newblock \href {https://doi.org/10.1007/BF01340146}
  {\path{doi:10.1007/BF01340146}}.

\bibitem{WDhistory:Stoner:1930}
E.~C.~S. Ph.D., \href{https://doi.org/10.1080/14786443008565066}{Lxxxvii. the
  equilibrium of dense stars}, The London, Edinburgh, and Dublin Philosophical
  Magazine and Journal of Science 9~(60) (1930) 944--963.
\newblock \href {https://doi.org/10.1080/14786443008565066}
  {\path{doi:10.1080/14786443008565066}}.
\newline\urlprefix\url{https://doi.org/10.1080/14786443008565066}

\bibitem{WDhistory:Chandra:1931}
S.~{Chandrasekhar}, {The Maximum Mass of Ideal White Dwarfs}, APJ 74 (1931) 81.
\newblock \href {https://doi.org/10.1086/143324} {\path{doi:10.1086/143324}}.

\bibitem{NShistory:BaadeZwicky:1934zex}
W.~Baade, F.~Zwicky, {On Super-Novae}, Proc. Nat. Acad. Sci. 20~(5) (1934)
  254--259.
\newblock \href {https://doi.org/10.1073/pnas.20.5.254}
  {\path{doi:10.1073/pnas.20.5.254}}.

\bibitem{NShistroy:LandauInterpreted2013}
D.~G. {Yakovlev}, P.~{Haensel}, G.~{Baym}, C.~{Pethick}, {Lev Landau and the
  concept of neutron stars}, Physics Uspekhi 56~(3) (2013) 289--295.
\newblock \href {http://arxiv.org/abs/1210.0682} {\path{arXiv:1210.0682}},
  \href {https://doi.org/10.3367/UFNe.0183.201303f.0307}
  {\path{doi:10.3367/UFNe.0183.201303f.0307}}.

\bibitem{WDfundamental:Isern:2022vdx}
J.~Isern, S.~Torres, A.~Rebassa-Mansergas, {White Dwarfs as Physics
  Laboratories: Lights and Shadows}, Front. Astron. Space Sci. 9 (2022) 815517.
\newblock \href {http://arxiv.org/abs/2202.02052} {\path{arXiv:2202.02052}},
  \href {https://doi.org/10.3389/fspas.2022.815517}
  {\path{doi:10.3389/fspas.2022.815517}}.

\bibitem{NSfundamental:Nattila:2022evn}
J.~N\"attil\"a, J.~J.~E. Kajava, {Fundamental physics with neutron stars} (11
  2022).
\newblock \href {http://arxiv.org/abs/2211.15721} {\path{arXiv:2211.15721}},
  \href {https://doi.org/10.1007/978-981-16-4544-0_105-1}
  {\path{doi:10.1007/978-981-16-4544-0_105-1}}.

\bibitem{Press:1985ug}
W.~H. Press, D.~N. Spergel, {Capture by the sun of a galactic population of
  weakly interacting massive particles}, Astrophys. J. 296 (1985) 679--684.
\newblock \href {https://doi.org/10.1086/163485} {\path{doi:10.1086/163485}}.

\bibitem{Gould:1987ir}
A.~Gould, {Resonant Enhancements in WIMP Capture by the Earth}, Astrophys. J.
  321 (1987) 571.
\newblock \href {https://doi.org/10.1086/165653} {\path{doi:10.1086/165653}}.

\bibitem{Goldman:1989nd}
I.~Goldman, S.~Nussinov, {Weakly Interacting Massive Particles and Neutron
  Stars}, Phys. Rev. D40 (1989) 3221--3230.
\newblock \href {https://doi.org/10.1103/PhysRevD.40.3221}
  {\path{doi:10.1103/PhysRevD.40.3221}}.

\bibitem{Bertone:2004pz}
G.~Bertone, D.~Hooper, J.~Silk, {Particle dark matter: Evidence, candidates and
  constraints}, Phys. Rept. 405 (2005) 279--390.
\newblock \href {http://arxiv.org/abs/hep-ph/0404175}
  {\path{arXiv:hep-ph/0404175}}, \href
  {https://doi.org/10.1016/j.physrep.2004.08.031}
  {\path{doi:10.1016/j.physrep.2004.08.031}}.

\bibitem{Leane:2024bvh}
R.~K. Leane, J.~Tong, {Optimal Celestial Bodies for Dark Matter Detection} (5
  2024).
\newblock \href {http://arxiv.org/abs/2405.05312} {\path{arXiv:2405.05312}}.

\bibitem{Saumon:2022gtu}
D.~Saumon, S.~Blouin, P.-E. Tremblay, {Current challenges in the physics of
  white dwarf stars}, Phys. Rept. 988 (2022) 1--63.
\newblock \href {http://arxiv.org/abs/2209.02846} {\path{arXiv:2209.02846}},
  \href {https://doi.org/10.1016/j.physrep.2022.09.001}
  {\path{doi:10.1016/j.physrep.2022.09.001}}.

\bibitem{LattPrakReview}
J.~M. {Lattimer}, M.~{Prakash}, {The Physics of Neutron Stars}, Science
  304~(5670) (2004) 536--542.
\newblock \href {http://arxiv.org/abs/astro-ph/0405262}
  {\path{arXiv:astro-ph/0405262}}, \href
  {https://doi.org/10.1126/science.1090720}
  {\path{doi:10.1126/science.1090720}}.

\bibitem{kick:Lyne:1994az}
A.~G. Lyne, D.~R. Lorimer, {High birth velocities of radio pulsars}, Nature 369
  (1994) 127.
\newblock \href {https://doi.org/10.1038/369127a0}
  {\path{doi:10.1038/369127a0}}.

\bibitem{kick:Scheck:2003rw}
L.~Scheck, T.~Plewa, H.-T. Janka, K.~Kifonidis, E.~Mueller, {Pulsar recoil by
  large scale anisotropies in supernova explosions}, Phys. Rev. Lett. 92 (2004)
  011103.
\newblock \href {http://arxiv.org/abs/astro-ph/0307352}
  {\path{arXiv:astro-ph/0307352}}, \href
  {https://doi.org/10.1103/PhysRevLett.92.011103}
  {\path{doi:10.1103/PhysRevLett.92.011103}}.

\bibitem{kick:Scheck:2006rw}
L.~Scheck, K.~Kifonidis, H.~T. Janka, E.~Mueller, {Multidimensional supernova
  simulations with approximative neutrino transport. 1. neutron star kicks and
  the anisotropy of neutrino-driven explosions in two spatial dimensions},
  Astron. Astrophys. 457 (2006) 963.
\newblock \href {http://arxiv.org/abs/astro-ph/0601302}
  {\path{arXiv:astro-ph/0601302}}, \href
  {https://doi.org/10.1051/0004-6361:20064855}
  {\path{doi:10.1051/0004-6361:20064855}}.

\bibitem{kick:Ng:2007aw}
C.~Y. Ng, R.~W. Romani, {Birth Kick Distributions and the Spin-Kick Correlation
  of Young Pulsars}, Astrophys. J. 660 (2007) 1357--1374.
\newblock \href {http://arxiv.org/abs/astro-ph/0702180}
  {\path{arXiv:astro-ph/0702180}}, \href {https://doi.org/10.1086/513597}
  {\path{doi:10.1086/513597}}.

\bibitem{kick:Nordhaus:2010ub}
J.~Nordhaus, T.~D. Brandt, A.~Burrows, E.~Livne, C.~D. Ott, {Theoretical
  Support for the Hydrodynamic Mechanism of Pulsar Kicks}, Phys. Rev. D 82
  (2010) 103016.
\newblock \href {http://arxiv.org/abs/1010.0674} {\path{arXiv:1010.0674}},
  \href {https://doi.org/10.1103/PhysRevD.82.103016}
  {\path{doi:10.1103/PhysRevD.82.103016}}.

\bibitem{kicknu:Kusenko:1996sr}
A.~Kusenko, G.~Segre, {Velocities of pulsars and neutrino oscillations}, Phys.
  Rev. Lett. 77 (1996) 4872--4875.
\newblock \href {http://arxiv.org/abs/hep-ph/9606428}
  {\path{arXiv:hep-ph/9606428}}, \href
  {https://doi.org/10.1103/PhysRevLett.77.4872}
  {\path{doi:10.1103/PhysRevLett.77.4872}}.

\bibitem{kicknu:Kusenko:1998bk}
A.~Kusenko, G.~Segre, {Pulsar kicks from neutrino oscillations}, Phys. Rev. D
  59 (1999) 061302.
\newblock \href {http://arxiv.org/abs/astro-ph/9811144}
  {\path{arXiv:astro-ph/9811144}}, \href
  {https://doi.org/10.1103/PhysRevD.59.061302}
  {\path{doi:10.1103/PhysRevD.59.061302}}.

\bibitem{kicknu:Barkovich:2002wh}
M.~Barkovich, J.~C. D'Olivo, R.~Montemayor, J.~F. Zanella, {Neutrino
  oscillation mechanism for pulsar kicks revisited}, Phys. Rev. D 66 (2002)
  123005.
\newblock \href {http://arxiv.org/abs/astro-ph/0206471}
  {\path{arXiv:astro-ph/0206471}}, \href
  {https://doi.org/10.1103/PhysRevD.66.123005}
  {\path{doi:10.1103/PhysRevD.66.123005}}.

\bibitem{ReviewKicks:LambiasePoddar:2024cjy}
G.~Lambiase, T.~K. Poddar, {Pulsar Kick: Status and Perspective} (12 2024).
\newblock \href {http://arxiv.org/abs/2412.08446} {\path{arXiv:2412.08446}}.

\bibitem{chandlimit}
S.~Chandrasekhar, \href{https://doi.org/10.1080/14786443109461710}{Xlviii. the
  density of white dwarf stars}, The London, Edinburgh, and Dublin
  Philosophical Magazine and Journal of Science 11~(70) (1931) 592--596.
\newblock \href
  {http://arxiv.org/abs/https://doi.org/10.1080/14786443109461710}
  {\path{arXiv:https://doi.org/10.1080/14786443109461710}}, \href
  {https://doi.org/10.1080/14786443109461710}
  {\path{doi:10.1080/14786443109461710}}.
\newline\urlprefix\url{https://doi.org/10.1080/14786443109461710}

\bibitem{Burrows:2014fla}
A.~Burrows, J.~P. Ostriker, {The Astronomical Reach of Fundamental Physics},
  Proc. Nat. Acad. Sci. 111 (2014) 2409.
\newblock \href {http://arxiv.org/abs/1401.1814} {\path{arXiv:1401.1814}},
  \href {https://doi.org/10.1073/pnas.1318003111}
  {\path{doi:10.1073/pnas.1318003111}}.

\bibitem{ReddySilbar:2003wm}
R.~R. Silbar, S.~Reddy, {Neutron stars for undergraduates}, Am. J. Phys. 72
  (2004) 892--905, [Erratum: Am.J.Phys. 73, 286 (2005)].
\newblock \href {http://arxiv.org/abs/nucl-th/0309041}
  {\path{arXiv:nucl-th/0309041}}, \href {https://doi.org/10.1119/1.1852544}
  {\path{doi:10.1119/1.1852544}}.

\bibitem{ShapiroTeukolsky}
S.~L. {Shapiro}, S.~A. {Teukolsky}, {Black holes, white dwarfs, and neutron
  stars : the physics of compact objects}, 1983.

\bibitem{Haensel:2002cia}
P.~Haensel, J.~L. Zdunik, F.~Douchin, {Equation of state of dense matter and
  the minimum mass of cold neutron stars}, Astron. Astrophys. 385 (2002) 301.
\newblock \href {http://arxiv.org/abs/astro-ph/0201434}
  {\path{arXiv:astro-ph/0201434}}, \href
  {https://doi.org/10.1051/0004-6361:20020131}
  {\path{doi:10.1051/0004-6361:20020131}}.

\bibitem{NSHESS1}
V.~{Doroshenko}, V.~{Suleimanov}, G.~{P{\"u}hlhofer}, A.~{Santangelo}, {A
  strangely light neutron star within a supernova remnant}, Nature Astronomy 6
  (2022) 1444--1451.
\newblock \href {https://doi.org/10.1038/s41550-022-01800-1}
  {\path{doi:10.1038/s41550-022-01800-1}}.

\bibitem{NSHESS:DiClemente:2022wqp}
F.~Di~Clemente, A.~Drago, G.~Pagliara, {Is the compact object associated with
  HESS J1731-347 a strange quark star?} (11 2022).
\newblock \href {http://arxiv.org/abs/2211.07485} {\path{arXiv:2211.07485}}.

\bibitem{NSHESS:Sagun:2023rzp}
V.~Sagun, E.~Giangrandi, T.~Dietrich, O.~Ivanytskyi, R.~Negreiros,
  C.~Provid\^encia, {What is the nature of the HESS J1731-347 compact object?}
  (6 2023).
\newblock \href {http://arxiv.org/abs/2306.12326} {\path{arXiv:2306.12326}}.

\bibitem{Suwa:2018uni}
Y.~Suwa, T.~Yoshida, M.~Shibata, H.~Umeda, K.~Takahashi, {On the minimum mass
  of neutron stars}, Mon. Not. Roy. Astron. Soc. 481~(3) (2018) 3305--3312.
\newblock \href {http://arxiv.org/abs/1808.02328} {\path{arXiv:1808.02328}},
  \href {https://doi.org/10.1093/mnras/sty2460}
  {\path{doi:10.1093/mnras/sty2460}}.

\bibitem{Lattimer:2004pg}
J.~M. Lattimer, M.~Prakash, {The physics of neutron stars}, Science 304 (2004)
  536--542.
\newblock \href {http://arxiv.org/abs/astro-ph/0405262}
  {\path{arXiv:astro-ph/0405262}}, \href
  {https://doi.org/10.1126/science.1090720}
  {\path{doi:10.1126/science.1090720}}.

\bibitem{OzelFreireMRNS:2016oaf}
F.~\"Ozel, P.~Freire, {Masses, Radii, and the Equation of State of Neutron
  Stars}, Ann. Rev. Astron. Astrophys. 54 (2016) 401--440.
\newblock \href {http://arxiv.org/abs/1603.02698} {\path{arXiv:1603.02698}},
  \href {https://doi.org/10.1146/annurev-astro-081915-023322}
  {\path{doi:10.1146/annurev-astro-081915-023322}}.

\bibitem{Lattimer:2021emm}
J.~M. Lattimer, {Neutron Stars and the Nuclear Matter Equation of State}, Ann.
  Rev. Nucl. Part. Sci. 71 (2021) 433--464.
\newblock \href {https://doi.org/10.1146/annurev-nucl-102419-124827}
  {\path{doi:10.1146/annurev-nucl-102419-124827}}.

\bibitem{Rotondo:2011zz}
M.~Rotondo, J.~A. Rueda, R.~Ruffini, S.-S. Xue, {The Relativistic
  Feynman-Metropolis-Teller theory for white dwarfs in general relativity},
  Phys. Rev. D 84 (2011) 084007.
\newblock \href {http://arxiv.org/abs/1012.0154} {\path{arXiv:1012.0154}},
  \href {https://doi.org/10.1103/PhysRevD.84.084007}
  {\path{doi:10.1103/PhysRevD.84.084007}}.

\bibitem{Most:2018hfd}
E.~R. Most, L.~R. Weih, L.~Rezzolla, J.~Schaffner-Bielich, {New constraints on
  radii and tidal deformabilities of neutron stars from GW170817}, Phys. Rev.
  Lett. 120~(26) (2018) 261103.
\newblock \href {http://arxiv.org/abs/1803.00549} {\path{arXiv:1803.00549}},
  \href {https://doi.org/10.1103/PhysRevLett.120.261103}
  {\path{doi:10.1103/PhysRevLett.120.261103}}.

\bibitem{LIGOScientific:2018cki}
B.~P. Abbott, et~al., {GW170817: Measurements of neutron star radii and
  equation of state}, Phys. Rev. Lett. 121~(16) (2018) 161101.
\newblock \href {http://arxiv.org/abs/1805.11581} {\path{arXiv:1805.11581}},
  \href {https://doi.org/10.1103/PhysRevLett.121.161101}
  {\path{doi:10.1103/PhysRevLett.121.161101}}.

\bibitem{Raithel:2018ncd}
C.~Raithel, F.~\"Ozel, D.~Psaltis, {Tidal deformability from GW170817 as a
  direct probe of the neutron star radius}, Astrophys. J. Lett. 857~(2) (2018)
  L23.
\newblock \href {http://arxiv.org/abs/1803.07687} {\path{arXiv:1803.07687}},
  \href {https://doi.org/10.3847/2041-8213/aabcbf}
  {\path{doi:10.3847/2041-8213/aabcbf}}.

\bibitem{NANOGrav:2019jur}
H.~T. Cromartie, et~al., {Relativistic Shapiro delay measurements of an
  extremely massive millisecond pulsar}, Nature Astron. 4~(1) (2019) 72--76.
\newblock \href {http://arxiv.org/abs/1904.06759} {\path{arXiv:1904.06759}},
  \href {https://doi.org/10.1038/s41550-019-0880-2}
  {\path{doi:10.1038/s41550-019-0880-2}}.

\bibitem{Hessels:2006ze}
J.~W.~T. Hessels, S.~M. Ransom, I.~H. Stairs, P.~C.~C. Freire, V.~M. Kaspi,
  F.~Camilo, {A radio pulsar spinning at 716-hz}, Science 311 (2006)
  1901--1904.
\newblock \href {http://arxiv.org/abs/astro-ph/0601337}
  {\path{arXiv:astro-ph/0601337}}, \href
  {https://doi.org/10.1126/science.1123430}
  {\path{doi:10.1126/science.1123430}}.

\bibitem{riley2019nicer}
T.~E. Riley, A.~L. Watts, S.~Bogdanov, P.~S. Ray, R.~M. Ludlam, S.~Guillot,
  Z.~Arzoumanian, C.~L. Baker, A.~V. Bilous, D.~Chakrabarty, et~al., A nicer
  view of psr j0030+ 0451: Millisecond pulsar parameter estimation, The
  Astrophysical Journal Letters 887~(1) (2019) L21.

\bibitem{riley2021nicer}
T.~E. Riley, A.~L. Watts, P.~S. Ray, S.~Bogdanov, S.~Guillot, S.~M. Morsink,
  A.~V. Bilous, Z.~Arzoumanian, D.~Choudhury, J.~S. Deneva, et~al., A nicer
  view of the massive pulsar psr j0740+ 6620 informed by radio timing and
  xmm-newton spectroscopy, The Astrophysical Journal Letters 918~(2) (2021)
  L27.

\bibitem{miller2019psr}
M.~Miller, F.~K. Lamb, A.~Dittmann, S.~Bogdanov, Z.~Arzoumanian, K.~C.
  Gendreau, S.~Guillot, A.~Harding, W.~Ho, J.~Lattimer, et~al., Psr j0030+ 0451
  mass and radius from nicer data and implications for the properties of
  neutron star matter, The Astrophysical Journal Letters 887~(1) (2019) L24.

\bibitem{miller2021radius}
M.~Miller, F.~Lamb, A.~Dittmann, S.~Bogdanov, Z.~Arzoumanian, K.~Gendreau,
  S.~Guillot, W.~Ho, J.~Lattimer, M.~Loewenstein, et~al., The radius of psr
  j0740+ 6620 from nicer and xmm-newton data, The Astrophysical Journal Letters
  918~(2) (2021) L28.

\bibitem{NSvIR:IISc2022}
S.~Chatterjee, R.~Garani, R.~K. Jain, B.~Kanodia, M.~S.~N. Kumar, S.~K.
  Vempati, {Faint light of old neutron stars and detectability at the James
  Webb Space Telescope}, Phys. Rev. D 108~(2) (2023) L021301.
\newblock \href {http://arxiv.org/abs/2205.05048} {\path{arXiv:2205.05048}},
  \href {https://doi.org/10.1103/PhysRevD.108.L021301}
  {\path{doi:10.1103/PhysRevD.108.L021301}}.

\bibitem{orderofmag:2015crq}
A.~Reisenegger, F.~S. Zepeda, {Order-of-magnitude physics of neutron stars},
  Eur. Phys. J. A 52~(3) (2016) 52.
\newblock \href {http://arxiv.org/abs/1511.08813} {\path{arXiv:1511.08813}},
  \href {https://doi.org/10.1140/epja/i2016-16052-y}
  {\path{doi:10.1140/epja/i2016-16052-y}}.

\bibitem{Fedderke:2019jur}
M.~A. Fedderke, P.~W. Graham, S.~Rajendran, {White dwarf bounds on charged
  massive particles}, Phys. Rev. D 101~(11) (2020) 115021.
\newblock \href {http://arxiv.org/abs/1911.08883} {\path{arXiv:1911.08883}},
  \href {https://doi.org/10.1103/PhysRevD.101.115021}
  {\path{doi:10.1103/PhysRevD.101.115021}}.

\bibitem{NSvIR:Pasta}
J.~F. Acevedo, J.~Bramante, R.~K. Leane, N.~Raj, {Warming Nuclear Pasta with
  Dark Matter: Kinetic and Annihilation Heating of Neutron Star Crusts}, JCAP
  03 (2020) 038.
\newblock \href {http://arxiv.org/abs/1911.06334} {\path{arXiv:1911.06334}},
  \href {https://doi.org/10.1088/1475-7516/2020/03/038}
  {\path{doi:10.1088/1475-7516/2020/03/038}}.

\bibitem{Pearson:2018tkr}
J.~M. Pearson, N.~Chamel, A.~Y. Potekhin, A.~F. Fantina, C.~Ducoin, A.~K.
  Dutta, S.~Goriely, {Unified equations of state for cold non-accreting neutron
  stars with Brussels-Montreal functionals: I. Role of symmetry energy}, Mon.
  Not. Roy. Astron. Soc. 481~(3) (2018) 2994--3026, [Erratum: Mon. Not. Roy.
  Astron. Soc.486,no.1,768(2019)].
\newblock \href {http://arxiv.org/abs/1903.04981} {\path{arXiv:1903.04981}},
  \href {https://doi.org/10.1093/mnras/sty2413, 10.1093/mnras/stz800}
  {\path{doi:10.1093/mnras/sty2413, 10.1093/mnras/stz800}}.

\bibitem{Chamel2008}
N.~Chamel, P.~Haensel, \href{https://doi.org/10.12942/lrr-2008-10}{Physics of
  neutron star crusts}, Living Reviews in Relativity 11~(1) (2008) 10.
\newblock \href {https://doi.org/10.12942/lrr-2008-10}
  {\path{doi:10.12942/lrr-2008-10}}.
\newline\urlprefix\url{https://doi.org/10.12942/lrr-2008-10}

\bibitem{Potekhin:2013qqa}
A.~Y. Potekhin, A.~F. Fantina, N.~Chamel, J.~M. Pearson, S.~Goriely,
  {Analytical representations of unified equations of state for neutron-star
  matter}, Astron. Astrophys. 560 (2013) A48.
\newblock \href {http://arxiv.org/abs/1310.0049} {\path{arXiv:1310.0049}},
  \href {https://doi.org/10.1051/0004-6361/201321697}
  {\path{doi:10.1051/0004-6361/201321697}}.

\bibitem{BAYM1971225}
G.~Baym, H.~A. Bethe, C.~J. Pethick,
  \href{http://www.sciencedirect.com/science/article/pii/0375947471902818}{Neutron
  star matter}, Nuclear Physics A 175~(2) (1971) 225 -- 271.
\newblock \href {https://doi.org/https://doi.org/10.1016/0375-9474(71)90281-8}
  {\path{doi:https://doi.org/10.1016/0375-9474(71)90281-8}}.
\newline\urlprefix\url{http://www.sciencedirect.com/science/article/pii/0375947471902818}

\bibitem{LATTIMER1985646}
J.~Lattimer, C.~Pethick, D.~Ravenhall, D.~Lamb,
  \href{http://www.sciencedirect.com/science/article/pii/0375947485900065}{Physical
  properties of hot, dense matter: The general case}, Nuclear Physics A 432~(3)
  (1985) 646 -- 742.
\newblock \href {https://doi.org/https://doi.org/10.1016/0375-9474(85)90006-5}
  {\path{doi:https://doi.org/10.1016/0375-9474(85)90006-5}}.
\newline\urlprefix\url{http://www.sciencedirect.com/science/article/pii/0375947485900065}

\bibitem{Ravenhall:1983uh}
D.~G. Ravenhall, C.~J. Pethick, J.~R. Wilson, {Structure of Matter Below
  Nuclear Saturation Density}, Phys. Rev. Lett. 50 (1983) 2066--2069.
\newblock \href {https://doi.org/10.1103/PhysRevLett.50.2066}
  {\path{doi:10.1103/PhysRevLett.50.2066}}.

\bibitem{10.1143/PTP.71.320}
M.~Hashimoto, H.~Seki, M.~Yamada,
  \href{https://doi.org/10.1143/PTP.71.320}{{Shape of Nuclei in the Crust of
  Neutron Star}}, Progress of Theoretical Physics 71~(2) (1984) 320--326.
\newblock \href
  {http://arxiv.org/abs/http://oup.prod.sis.lan/ptp/article-pdf/71/2/320/5459325/71-2-320.pdf}
  {\path{arXiv:http://oup.prod.sis.lan/ptp/article-pdf/71/2/320/5459325/71-2-320.pdf}},
  \href {https://doi.org/10.1143/PTP.71.320} {\path{doi:10.1143/PTP.71.320}}.
\newline\urlprefix\url{https://doi.org/10.1143/PTP.71.320}

\bibitem{10.1143/PTP.72.373}
K.~Oyamatsu, M.~Hashimoto, M.~Yamada,
  \href{https://doi.org/10.1143/PTP.72.373}{{Further Study of the Nuclear Shape
  in High-Density Matter}}, Progress of Theoretical Physics 72~(2) (1984)
  373--375.
\newblock \href
  {http://arxiv.org/abs/http://oup.prod.sis.lan/ptp/article-pdf/72/2/373/5179866/72-2-373.pdf}
  {\path{arXiv:http://oup.prod.sis.lan/ptp/article-pdf/72/2/373/5179866/72-2-373.pdf}},
  \href {https://doi.org/10.1143/PTP.72.373} {\path{doi:10.1143/PTP.72.373}}.
\newline\urlprefix\url{https://doi.org/10.1143/PTP.72.373}

\bibitem{Williams:1985prf}
R.~D. Williams, S.~E. Koonin, {Sub-saturation phases of nuclear matter}, Nucl.
  Phys. A435 (1985) 844--858.
\newblock \href {https://doi.org/10.1016/0375-9474(85)90191-5}
  {\path{doi:10.1016/0375-9474(85)90191-5}}.

\bibitem{Lorenz:1992zz}
C.~P. Lorenz, D.~G. Ravenhall, C.~J. Pethick, {Neutron star crusts}, Phys. Rev.
  Lett. 70 (1993) 379--382.
\newblock \href {https://doi.org/10.1103/PhysRevLett.70.379}
  {\path{doi:10.1103/PhysRevLett.70.379}}.

\bibitem{Oyamatsu:1993zz}
K.~Oyamatsu, {Nuclear shapes in the inner crust of a neutron star}, Nucl. Phys.
  A561 (1993) 431--452.
\newblock \href {https://doi.org/10.1016/0375-9474(93)90020-X}
  {\path{doi:10.1016/0375-9474(93)90020-X}}.

\bibitem{1975ApJ...199..471H}
J.~B. {Hartle}, R.~F. {Sawyer}, D.~J. {Scalapino}, {Pion condensed matter at
  high densities: equation of state and stellar models.}, Astrophysical Journal
  199 (1975) 471--481.
\newblock \href {https://doi.org/10.1086/153713} {\path{doi:10.1086/153713}}.

\bibitem{1982ApJ...258..306H}
P.~{Haensel}, M.~{Proszynski}, {Pion condensation in cold dense matter and
  neutron stars}, Astrophysical Journal 258 (1982) 306--320.
\newblock \href {https://doi.org/10.1086/160080} {\path{doi:10.1086/160080}}.

\bibitem{BROWN19761}
G.~Brown, W.~Weise,
  \href{http://www.sciencedirect.com/science/article/pii/0370157376900089}{Pion
  condensates}, Physics Reports 27~(1) (1976) 1 -- 34.
\newblock \href {https://doi.org/https://doi.org/10.1016/0370-1573(76)90008-9}
  {\path{doi:https://doi.org/10.1016/0370-1573(76)90008-9}}.
\newline\urlprefix\url{http://www.sciencedirect.com/science/article/pii/0370157376900089}

\bibitem{sfluid:PageReddyReview:2006ud}
D.~Page, S.~Reddy, {Dense Matter in Compact Stars: Theoretical Developments and
  Observational Constraints}, Ann. Rev. Nucl. Part. Sci. 56 (2006) 327--374.
\newblock \href {http://arxiv.org/abs/astro-ph/0608360}
  {\path{arXiv:astro-ph/0608360}}, \href
  {https://doi.org/10.1146/annurev.nucl.56.080805.140600}
  {\path{doi:10.1146/annurev.nucl.56.080805.140600}}.

\bibitem{Holdom:2017gdc}
B.~Holdom, J.~Ren, C.~Zhang, {Quark matter may not be strange}, Phys. Rev.
  Lett. 120~(22) (2018) 222001.
\newblock \href {http://arxiv.org/abs/1707.06610} {\path{arXiv:1707.06610}},
  \href {https://doi.org/10.1103/PhysRevLett.120.222001}
  {\path{doi:10.1103/PhysRevLett.120.222001}}.

\bibitem{Ivanenko1965}
D.~D. Ivanenko, D.~F. Kurdgelaidze,
  \href{https://doi.org/10.1007/BF01042830}{Hypothesis concerning quark stars},
  Astrophysics 1~(4) (1965) 251--252.
\newblock \href {https://doi.org/10.1007/BF01042830}
  {\path{doi:10.1007/BF01042830}}.
\newline\urlprefix\url{https://doi.org/10.1007/BF01042830}

\bibitem{Ivanenko1969}
D.~Ivanenko, D.~F. Kurdgelaidze,
  \href{https://doi.org/10.1007/BF02753988}{Remarks on quark stars}, Lettere al
  Nuovo Cimento (1969-1970) 2~(1) (1969) 13--16.
\newblock \href {https://doi.org/10.1007/BF02753988}
  {\path{doi:10.1007/BF02753988}}.
\newline\urlprefix\url{https://doi.org/10.1007/BF02753988}

\bibitem{Collins:1974ky}
J.~C. Collins, M.~J. Perry, {Superdense Matter: Neutrons Or Asymptotically Free
  Quarks?}, Phys. Rev. Lett. 34 (1975) 1353.
\newblock \href {https://doi.org/10.1103/PhysRevLett.34.1353}
  {\path{doi:10.1103/PhysRevLett.34.1353}}.

\bibitem{Weber:1999qj}
F.~Weber, {Quark matter in neutron stars}, J. Phys. G25 (1999) R195--R229.
\newblock \href {https://doi.org/10.1088/0954-3899/25/9/201}
  {\path{doi:10.1088/0954-3899/25/9/201}}.

\bibitem{Lastowiecki2015}
R.~Lastowiecki, D.~Blaschke, T.~Fischer, T.~Kl{\"a}hn,
  \href{https://doi.org/10.1134/S1063779615050159}{Quark matter in high-mass
  neutron stars?}, Physics of Particles and Nuclei 46~(5) (2015) 843--845.
\newblock \href {https://doi.org/10.1134/S1063779615050159}
  {\path{doi:10.1134/S1063779615050159}}.
\newline\urlprefix\url{https://doi.org/10.1134/S1063779615050159}

\bibitem{Chatterjee:2015pua}
D.~Chatterjee, I.~Vidaña, {Do hyperons exist in the interior of neutron
  stars?}, Eur. Phys. J. A52~(2) (2016) 29.
\newblock \href {http://arxiv.org/abs/1510.06306} {\path{arXiv:1510.06306}},
  \href {https://doi.org/10.1140/epja/i2016-16029-x}
  {\path{doi:10.1140/epja/i2016-16029-x}}.

\bibitem{BARROIS1977390}
B.~C. Barrois,
  \href{https://www.sciencedirect.com/science/article/pii/0550321377901237}{Superconducting
  quark matter}, Nuclear Physics B 129~(3) (1977) 390--396.
\newblock \href {https://doi.org/https://doi.org/10.1016/0550-3213(77)90123-7}
  {\path{doi:https://doi.org/10.1016/0550-3213(77)90123-7}}.
\newline\urlprefix\url{https://www.sciencedirect.com/science/article/pii/0550321377901237}

\bibitem{Frautschi:1978rz}
S.~C. Frautschi, {ASYMPTOTIC FREEDOM AND COLOR SUPERCONDUCTIVITY IN DENSE QUARK
  MATTER}, in: {Workshop on Theoretical Physics: Hadronic Matter at Extreme
  Energy Density}, 1978.

\bibitem{Alford:2007xm}
M.~G. Alford, A.~Schmitt, K.~Rajagopal, T.~Schafer, {Color superconductivity in
  dense quark matter}, Rev. Mod. Phys. 80 (2008) 1455--1515.
\newblock \href {http://arxiv.org/abs/0709.4635} {\path{arXiv:0709.4635}},
  \href {https://doi.org/10.1103/RevModPhys.80.1455}
  {\path{doi:10.1103/RevModPhys.80.1455}}.

\bibitem{Anglani2014}
R.~Anglani, R.~Casalbuoni, M.~Ciminale, N.~Ippolito, R.~Gatto, M.~Mannarelli,
  M.~Ruggieri, \href{http://dx.doi.org/10.1103/RevModPhys.86.509}{Crystalline
  color superconductors}, Reviews of Modern Physics 86~(2) (2014) 509–561.
\newblock \href {https://doi.org/10.1103/revmodphys.86.509}
  {\path{doi:10.1103/revmodphys.86.509}}.
\newline\urlprefix\url{http://dx.doi.org/10.1103/RevModPhys.86.509}

\bibitem{Alford:1998mk}
M.~G. Alford, K.~Rajagopal, F.~Wilczek, {Color flavor locking and chiral
  symmetry breaking in high density QCD}, Nucl. Phys. B537 (1999) 443--458.
\newblock \href {http://arxiv.org/abs/hep-ph/9804403}
  {\path{arXiv:hep-ph/9804403}}, \href
  {https://doi.org/10.1016/S0550-3213(98)00668-3}
  {\path{doi:10.1016/S0550-3213(98)00668-3}}.

\bibitem{Glendenning:1995rd}
N.~K. Glendenning, S.~Pei, {Crystalline structure of the mixed confined -
  deconfined phase in neutron stars}, Phys. Rev. C52 (1995) 2250--2253.
\newblock \href {https://doi.org/10.1103/PhysRevC.52.2250}
  {\path{doi:10.1103/PhysRevC.52.2250}}.

\bibitem{Bedaque:2001je}
P.~F. Bedaque, T.~Schafer, {High density quark matter under stress}, Nucl.
  Phys. A697 (2002) 802--822.
\newblock \href {http://arxiv.org/abs/hep-ph/0105150}
  {\path{arXiv:hep-ph/0105150}}, \href
  {https://doi.org/10.1016/S0375-9474(01)01272-6}
  {\path{doi:10.1016/S0375-9474(01)01272-6}}.

\bibitem{Kaplan:2001qk}
D.~B. Kaplan, S.~Reddy, {Novel phases and transitions in color flavor locked
  matter}, Phys. Rev. D65 (2002) 054042.
\newblock \href {http://arxiv.org/abs/hep-ph/0107265}
  {\path{arXiv:hep-ph/0107265}}, \href
  {https://doi.org/10.1103/PhysRevD.65.054042}
  {\path{doi:10.1103/PhysRevD.65.054042}}.

\bibitem{Maoz:2011iv}
D.~Maoz, F.~Mannucci, {Type-Ia supernova rates and the progenitor problem, a
  review}, Publ. Astron. Soc. Austral. 29 (2012) 447.
\newblock \href {http://arxiv.org/abs/1111.4492} {\path{arXiv:1111.4492}},
  \href {https://doi.org/10.1071/AS11052} {\path{doi:10.1071/AS11052}}.

\bibitem{supburst:Cflashes:Cumming:2001wg}
A.~Cumming, L.~Bildsten, {Carbon flashes in the heavy element ocean on
  accreting neutron stars}, Astrophys. J. Lett. 559 (2001) L127.
\newblock \href {http://arxiv.org/abs/astro-ph/0107213}
  {\path{arXiv:astro-ph/0107213}}, \href {https://doi.org/10.1086/323937}
  {\path{doi:10.1086/323937}}.

\bibitem{supburst:underZanding:2017ugu}
J.~in~'t Zand, {Understanding superbursts}, 2017.
\newblock \href {http://arxiv.org/abs/1702.04899} {\path{arXiv:1702.04899}}.

\bibitem{supburst:MINBARcatalog:2020}
D.~K. Galloway, J.~in~'t Zand, J.~Chenevez, H.~W\"orpel, L.~Keek, L.~Ootes,
  A.~L. Watts, L.~Gisler, C.~Sanchez-Fernandez, E.~Kuulkers, {The
  Multi-INstrument Burst ARchive (MINBAR)}, Astrophys. J. Suppl. 249~(2) (2020)
  32.
\newblock \href {http://arxiv.org/abs/2003.00685} {\path{arXiv:2003.00685}},
  \href {https://doi.org/10.3847/1538-4365/ab9f2e}
  {\path{doi:10.3847/1538-4365/ab9f2e}}.

\bibitem{supburst:catalog2023}
e.~a. Alizai, K, \href{https://doi.org/10.1093/mnras/stad374}{{A catalogue of
  unusually long thermonuclear bursts on neutron stars}}, Monthly Notices of
  the Royal Astronomical Society 521~(3) (2023) 3608--3624.
\newblock \href
  {http://arxiv.org/abs/https://academic.oup.com/mnras/article-pdf/521/3/3608/49630407/stad374.pdf}
  {\path{arXiv:https://academic.oup.com/mnras/article-pdf/521/3/3608/49630407/stad374.pdf}},
  \href {https://doi.org/10.1093/mnras/stad374}
  {\path{doi:10.1093/mnras/stad374}}.
\newline\urlprefix\url{https://doi.org/10.1093/mnras/stad374}

\bibitem{TimmesWoosley1992}
F.~X. {Timmes}, S.~E. {Woosley}, {The Conductive Propagation of Nuclear Flames.
  I. Degenerate C + O and O + NE + MG White Dwarfs}, APJ 396 (1992) 649.
\newblock \href {https://doi.org/10.1086/171746} {\path{doi:10.1086/171746}}.

\bibitem{Bramante:2015cua}
J.~Bramante, {Dark matter ignition of type Ia supernovae}, Phys. Rev. Lett.
  115~(14) (2015) 141301.
\newblock \href {http://arxiv.org/abs/1505.07464} {\path{arXiv:1505.07464}},
  \href {https://doi.org/10.1103/PhysRevLett.115.141301}
  {\path{doi:10.1103/PhysRevLett.115.141301}}.

\bibitem{Acevedo:2019agu}
J.~F. Acevedo, J.~Bramante, R.~K. Leane, N.~Raj, {Cooking Pasta with Dark
  Matter: Kinetic and Annihilation Heating of Neutron Star Crusts} (2019).
\newblock \href {http://arxiv.org/abs/1911.06334} {\path{arXiv:1911.06334}}.

\bibitem{WDcooling:Althaus2010}
L.~G. Althaus, A.~H. C{\'{o}}rsico, J.~Isern, E.~Garc{\'{\i}}a-Berro,
  Evolutionary and pulsational properties of white dwarf stars, The Astronomy
  and Astrophysics Review 18~(4) (2010) 471--566.
\newblock \href {https://doi.org/10.1007/s00159-010-0033-1}
  {\path{doi:10.1007/s00159-010-0033-1}}.

\bibitem{heatcapacity:ReddyPageHorowitz:2016weq}
A.~Cumming, E.~F. Brown, F.~J. Fattoyev, C.~J. Horowitz, D.~Page, S.~Reddy, {A
  lower limit on the heat capacity of the neutron star core}, Phys. Rev. C
  95~(2) (2017) 025806.
\newblock \href {http://arxiv.org/abs/1608.07532} {\path{arXiv:1608.07532}},
  \href {https://doi.org/10.1103/PhysRevC.95.025806}
  {\path{doi:10.1103/PhysRevC.95.025806}}.

\bibitem{coolingminimal:Page:2004fy}
D.~Page, J.~M. Lattimer, M.~Prakash, A.~W. Steiner, {Minimal cooling of neutron
  stars: A New paradigm}, Astrophys. J. Suppl. 155 (2004) 623--650.
\newblock \href {http://arxiv.org/abs/astro-ph/0403657}
  {\path{arXiv:astro-ph/0403657}}, \href {https://doi.org/10.1086/424844}
  {\path{doi:10.1086/424844}}.

\bibitem{cooling:Yakovlev:2004iq}
D.~G. Yakovlev, C.~J. Pethick, {Neutron star cooling}, Ann. Rev. Astron.
  Astrophys. 42 (2004) 169--210.
\newblock \href {http://arxiv.org/abs/astro-ph/0402143}
  {\path{arXiv:astro-ph/0402143}}, \href
  {https://doi.org/10.1146/annurev.astro.42.053102.134013}
  {\path{doi:10.1146/annurev.astro.42.053102.134013}}.

\bibitem{Harvey:2007rd}
J.~A. Harvey, C.~T. Hill, R.~J. Hill, {Anomaly mediated neutrino-photon
  interactions at finite baryon density}, Phys. Rev. Lett. 99 (2007) 261601.
\newblock \href {http://arxiv.org/abs/0708.1281} {\path{arXiv:0708.1281}},
  \href {https://doi.org/10.1103/PhysRevLett.99.261601}
  {\path{doi:10.1103/PhysRevLett.99.261601}}.

\bibitem{Chakraborty:2023wgl}
S.~Chakraborty, A.~Gupta, M.~Vanvlasselaer, {Anomaly induced cooling of Neutron
  Stars: A Standard Model contribution} (6 2023).
\newblock \href {http://arxiv.org/abs/2306.15872} {\path{arXiv:2306.15872}}.

\bibitem{cvanalytic:Ofengeim:2017xxr}
D.~D. Ofengeim, M.~Fortin, P.~Haensel, D.~G. Yakovlev, J.~L. Zdunik, {Neutrino
  luminosities and heat capacities of neutron stars in analytic form}, Phys.
  Rev. D 96~(4) (2017) 043002.
\newblock \href {http://arxiv.org/abs/1708.08272} {\path{arXiv:1708.08272}},
  \href {https://doi.org/10.1103/PhysRevD.96.043002}
  {\path{doi:10.1103/PhysRevD.96.043002}}.

\bibitem{coolingcatalogue:Potekhin:2020ttj}
A.~Y. Potekhin, D.~A. Zyuzin, D.~G. Yakovlev, M.~V. Beznogov, Y.~A. Shibanov,
  {Thermal luminosities of cooling neutron stars}, Mon. Not. Roy. Astron. Soc.
  496~(4) (2020) 5052--5071.
\newblock \href {http://arxiv.org/abs/2006.15004} {\path{arXiv:2006.15004}},
  \href {https://doi.org/10.1093/mnras/staa1871}
  {\path{doi:10.1093/mnras/staa1871}}.

\bibitem{TbTs-Fe-Pethick1983}
E.~H. {Gudmundsson}, C.~J. {Pethick}, R.~I. {Epstein}, {Structure of neutron
  star envelopes}, apj 272 (1983) 286--300.
\newblock \href {https://doi.org/10.1086/161292} {\path{doi:10.1086/161292}}.

\bibitem{NSenvelope:Beznogov:2021ijc}
M.~V. Beznogov, A.~Y. Potekhin, D.~G. Yakovlev, {Heat blanketing envelopes of
  neutron stars}, Phys. Rept. 919 (2021) 1--68.
\newblock \href {http://arxiv.org/abs/2103.12422} {\path{arXiv:2103.12422}},
  \href {https://doi.org/10.1016/j.physrep.2021.03.004}
  {\path{doi:10.1016/j.physrep.2021.03.004}}.

\bibitem{coolinganalytic:Ofengeim:2017cum}
D.~D. Ofengeim, D.~G. Yakovlev, {Analytic description of neutron star cooling},
  Mon. Not. Roy. Astron. Soc. 467~(3) (2017) 3598--3603.
\newblock \href {https://doi.org/10.1093/mnras/stx366}
  {\path{doi:10.1093/mnras/stx366}}.

\bibitem{sfluid:Migdal1959}
A.~B. {Migdal}, {Superfluidity and the moments of inertia of nuclei}, Nuclear
  Physics 13~(5) (1959) 655--674.
\newblock \href {https://doi.org/10.1016/0029-5582(59)90264-0}
  {\path{doi:10.1016/0029-5582(59)90264-0}}.

\bibitem{sfluid:Yakovlev:1999sk}
D.~G. Yakovlev, K.~P. Levenfish, Y.~A. Shibanov, {Cooling neutron stars and
  superfluidity in their interiors}, Phys. Usp. 42 (1999) 737--778.
\newblock \href {http://arxiv.org/abs/astro-ph/9906456}
  {\path{arXiv:astro-ph/9906456}}, \href
  {https://doi.org/10.1070/PU1999v042n08ABEH000556}
  {\path{doi:10.1070/PU1999v042n08ABEH000556}}.

\bibitem{sfluid:Haskell:2017lkl}
B.~Haskell, A.~Sedrakian, {Superfluidity and Superconductivity in Neutron
  Stars}, Astrophys. Space Sci. Libr. 457 (2018) 401--454.
\newblock \href {http://arxiv.org/abs/1709.10340} {\path{arXiv:1709.10340}},
  \href {https://doi.org/10.1007/978-3-319-97616-7-8}
  {\path{doi:10.1007/978-3-319-97616-7-8}}.

\bibitem{sfluid:colorAlfordRajagopalWilczek:1997zt}
M.~G. Alford, K.~Rajagopal, F.~Wilczek, {QCD at finite baryon density: Nucleon
  droplets and color superconductivity}, Phys. Lett. B 422 (1998) 247--256.
\newblock \href {http://arxiv.org/abs/hep-ph/9711395}
  {\path{arXiv:hep-ph/9711395}}, \href
  {https://doi.org/10.1016/S0370-2693(98)00051-3}
  {\path{doi:10.1016/S0370-2693(98)00051-3}}.

\bibitem{glitch:Anderson:1975zze}
P.~W. Anderson, N.~Itoh, {Pulsar glitches and restlessness as a hard
  superfluidity phenomenon}, Nature 256 (1975) 25--27.
\newblock \href {https://doi.org/10.1038/256025a0}
  {\path{doi:10.1038/256025a0}}.

\bibitem{glitch:Haskell:2015jra}
B.~Haskell, A.~Melatos, {Models of Pulsar Glitches}, Int. J. Mod. Phys. D
  24~(03) (2015) 1530008.
\newblock \href {http://arxiv.org/abs/1502.07062} {\path{arXiv:1502.07062}},
  \href {https://doi.org/10.1142/S0218271815300086}
  {\path{doi:10.1142/S0218271815300086}}.

\bibitem{glitchreview:Zhou:2022cyp}
S.~Zhou, E.~G\"ugercino\u{g}lu, J.~Yuan, M.~Ge, C.~Yu, {Pulsar Glitches: A
  Review}, Universe 8~(12) (2022) 641.
\newblock \href {http://arxiv.org/abs/2211.13885} {\path{arXiv:2211.13885}},
  \href {https://doi.org/10.3390/universe8120641}
  {\path{doi:10.3390/universe8120641}}.

\bibitem{glitch:Antonelli:2023vpd}
M.~Antonelli, A.~Montoli, P.~Pizzochero, {Insights into the physics of neutron
  star interiors from pulsar glitches}, 2023.
\newblock \href {http://arxiv.org/abs/2301.12769} {\path{arXiv:2301.12769}},
  \href {https://doi.org/10.1142/9789811220944_0007}
  {\path{doi:10.1142/9789811220944_0007}}.

\bibitem{glitch:Manchester2017}
R.~N. Manchester, Pulsar glitches, Proceedings of the International
  Astronomical Union 13~(S337) (2017) 197–202.
\newblock \href {https://doi.org/10.1017/S1743921317009607}
  {\path{doi:10.1017/S1743921317009607}}.

\bibitem{NSBfieldspindown}
\url{https://www.cv.nrao.edu/~sransom/web/Ch6.html/}.

\bibitem{PPdot:Caleb:2022xyo}
M.~Caleb, et~al., {Discovery of a radio-emitting neutron star with an
  ultra-long spin period of 76\,s}, Nature Astron. 6~(7) (2022) 828--836.
\newblock \href {http://arxiv.org/abs/2206.01346} {\path{arXiv:2206.01346}},
  \href {https://doi.org/10.1038/s41550-022-01688-x}
  {\path{doi:10.1038/s41550-022-01688-x}}.

\bibitem{ReiseneggerBfield2009}
A.~{Reisenegger}, {Stable magnetic equilibria and their evolution in the upper
  main sequence, white dwarfs, and neutron stars}, AAP 499~(2) (2009) 557--566.
\newblock \href {http://arxiv.org/abs/0809.0361} {\path{arXiv:0809.0361}},
  \href {https://doi.org/10.1051/0004-6361/200810895}
  {\path{doi:10.1051/0004-6361/200810895}}.

\bibitem{PulsarB:1975RudermanSutherland}
M.~A. {Ruderman}, P.~G. {Sutherland}, {Theory of pulsars: polar gaps, sparks,
  and coherent microwave radiation.}, APJ 196 (1975) 51--72.
\newblock \href {https://doi.org/10.1086/153393} {\path{doi:10.1086/153393}}.

\bibitem{PulsarB:reviewMichel:1982fj}
F.~C. Michel, {Theory of pulsar magnetospheres}, Rev. Mod. Phys. 54 (1982)
  1--66.
\newblock \href {https://doi.org/10.1103/RevModPhys.54.1}
  {\path{doi:10.1103/RevModPhys.54.1}}.

\bibitem{WDpulsar:Rea:2023stb}
N.~Rea, et~al., {Long-period Radio Pulsars: Population Study in the Neutron
  Star and White Dwarf Rotating Dipole Scenarios}, Astrophys. J. 961~(2) (2024)
  214.
\newblock \href {http://arxiv.org/abs/2307.10351} {\path{arXiv:2307.10351}},
  \href {https://doi.org/10.3847/1538-4357/ad165d}
  {\path{doi:10.3847/1538-4357/ad165d}}.

\bibitem{ATNF:2004bp}
R.~N. Manchester, G.~B. Hobbs, A.~Teoh, M.~Hobbs, {The Australia Telescope
  National Facility pulsar catalogue}, Astron. J. 129 (2005) 1993.
\newblock \href {http://arxiv.org/abs/astro-ph/0412641}
  {\path{arXiv:astro-ph/0412641}}, \href {https://doi.org/10.1086/428488}
  {\path{doi:10.1086/428488}}.

\bibitem{Tauris}
T.~M. {Tauris}, {Spin-Down of Radio Millisecond Pulsars at Genesis}, Science
  335~(6068) (2012) 561.
\newblock \href {http://arxiv.org/abs/1202.0551} {\path{arXiv:1202.0551}},
  \href {https://doi.org/10.1126/science.1216355}
  {\path{doi:10.1126/science.1216355}}.

\bibitem{Planck:2018vyg}
N.~Aghanim, et~al., {Planck 2018 results. VI. Cosmological parameters}, Astron.
  Astrophys. 641 (2020) A6, [Erratum: Astron.Astrophys. 652, C4 (2021)].
\newblock \href {http://arxiv.org/abs/1807.06209} {\path{arXiv:1807.06209}},
  \href {https://doi.org/10.1051/0004-6361/201833910}
  {\path{doi:10.1051/0004-6361/201833910}}.

\bibitem{NSkinematicageMag7}
N.~{Tetzlaff}, R.~{Neuh{\"a}user}, M.~M. {Hohle}, G.~{Maciejewski},
  {Identifying birth places of young isolated neutron stars}, MNRAS 402~(4)
  (2010) 2369--2387.
\newblock \href {http://arxiv.org/abs/0911.4441} {\path{arXiv:0911.4441}},
  \href {https://doi.org/10.1111/j.1365-2966.2009.16093.x}
  {\path{doi:10.1111/j.1365-2966.2009.16093.x}}.

\bibitem{PulsarDeathLineAnomaly:PSRJ0250+5854}
F.~F. {Kou}, H.~{Tong}, R.~X. {Xu}, X.~{Zhou}, {Rotational Evolution of the
  Slowest Radio Pulsar, PSR J0250+5854}, APJ 876~(2) (2019) 131.
\newblock \href {http://arxiv.org/abs/1901.00300} {\path{arXiv:1901.00300}},
  \href {https://doi.org/10.3847/1538-4357/ab17da}
  {\path{doi:10.3847/1538-4357/ab17da}}.

\bibitem{PulsarDeathLineAnomaly:PSRJ0901-4046}
M.~Caleb, et~al., {Discovery of a radio-emitting neutron star with an
  ultra-long spin period of 76\,s}, Nature Astron. 6~(7) (2022) 828--836.
\newblock \href {http://arxiv.org/abs/2206.01346} {\path{arXiv:2206.01346}},
  \href {https://doi.org/10.1038/s41550-022-01688-x}
  {\path{doi:10.1038/s41550-022-01688-x}}.

\bibitem{PulsarDeathLineAnomaly:PSRJ2144-3933}
M.~D. {Young}, R.~N. {Manchester}, S.~{Johnston}, {A radio pulsar with an
  8.5-second period that challenges emission models}, Nature 400~(6747) (1999)
  848--849.
\newblock \href {https://doi.org/10.1038/23650} {\path{doi:10.1038/23650}}.

\bibitem{PulsarDeathLineAnomaly:PSRJ2251-3711}
F.~F. Kou, H.~Tong,
  \href{https://dx.doi.org/10.3847/2515-5172/ab595f}{Rotational evolution of
  psr j2251-3711 (the second slowest radio pulsar)}, Research Notes of the AAS
  3~(11) (2019) 175.
\newblock \href {https://doi.org/10.387/2515-5172/ab595f}
  {\path{doi:10.387/2515-5172/ab595f}}.
\newline\urlprefix\url{https://dx.doi.org/10.3847/2515-5172/ab595f}

\bibitem{Wang:2021wae}
J.-W. Wang, X.-J. Bi, R.-M. Yao, P.-F. Yin, {Exploring axion dark matter
  through radio signals from magnetic white dwarf stars}, Phys. Rev. D 103~(11)
  (2021) 115021.
\newblock \href {http://arxiv.org/abs/2101.02585} {\path{arXiv:2101.02585}},
  \href {https://doi.org/10.1103/PhysRevD.103.115021}
  {\path{doi:10.1103/PhysRevD.103.115021}}.

\bibitem{Garani:2023esk}
R.~Garani, N.~Raj, J.~Reynoso-Cordova, {Could compact stars in globular
  clusters constrain dark matter?} (3 2023).
\newblock \href {http://arxiv.org/abs/2303.18009} {\path{arXiv:2303.18009}}.

\bibitem{McCullough:2010ai}
M.~McCullough, M.~Fairbairn, {Capture of Inelastic Dark Matter in White
  Dwarves}, Phys. Rev. D81 (2010) 083520.
\newblock \href {http://arxiv.org/abs/1001.2737} {\path{arXiv:1001.2737}},
  \href {https://doi.org/10.1103/PhysRevD.81.083520}
  {\path{doi:10.1103/PhysRevD.81.083520}}.

\bibitem{Bertone:2007ae}
G.~Bertone, M.~Fairbairn, {Compact Stars as Dark Matter Probes}, Phys. Rev. D77
  (2008) 043515.
\newblock \href {http://arxiv.org/abs/0709.1485} {\path{arXiv:0709.1485}},
  \href {https://doi.org/10.1103/PhysRevD.77.043515}
  {\path{doi:10.1103/PhysRevD.77.043515}}.

\bibitem{Hooper:2010es}
D.~Hooper, D.~Spolyar, A.~Vallinotto, N.~Y. Gnedin, {Inelastic Dark Matter As
  An Efficient Fuel For Compact Stars}, Phys. Rev. D 81 (2010) 103531.
\newblock \href {http://arxiv.org/abs/1002.0005} {\path{arXiv:1002.0005}},
  \href {https://doi.org/10.1103/PhysRevD.81.103531}
  {\path{doi:10.1103/PhysRevD.81.103531}}.

\bibitem{Horowitz:2020axx}
C.~Horowitz, {Nuclear and dark matter heating in massive white dwarf stars} (8
  2020).
\newblock \href {http://arxiv.org/abs/2008.03291} {\path{arXiv:2008.03291}}.

\bibitem{Bramante:2017xlb}
J.~Bramante, A.~Delgado, A.~Martin, {Multiscatter stellar capture of dark
  matter}, Phys. Rev. D96~(6) (2017) 063002.
\newblock \href {http://arxiv.org/abs/1703.04043} {\path{arXiv:1703.04043}},
  \href {https://doi.org/10.1103/PhysRevD.96.063002}
  {\path{doi:10.1103/PhysRevD.96.063002}}.

\bibitem{Kouvaris:2010jy}
C.~Kouvaris, P.~Tinyakov, {Constraining Asymmetric Dark Matter through
  observations of compact stars}, Phys. Rev. D83 (2011) 083512.
\newblock \href {http://arxiv.org/abs/1012.2039} {\path{arXiv:1012.2039}},
  \href {https://doi.org/10.1103/PhysRevD.83.083512}
  {\path{doi:10.1103/PhysRevD.83.083512}}.

\bibitem{Cermeno:2018qgu}
M.~Cermeno, M.~Perez-Garcia, {Gamma rays from dark mediators in white dwarfs},
  Phys. Rev. D98~(6) (2018) 063002.
\newblock \href {http://arxiv.org/abs/1807.03318} {\path{arXiv:1807.03318}},
  \href {https://doi.org/10.1103/PhysRevD.98.063002}
  {\path{doi:10.1103/PhysRevD.98.063002}}.

\bibitem{Dasgupta:2019juq}
B.~Dasgupta, A.~Gupta, A.~Ray, {Dark matter capture in celestial objects:
  Improved treatment of multiple scattering and updated constraints from white
  dwarfs}, JCAP 08 (2019) 018.
\newblock \href {http://arxiv.org/abs/1906.04204} {\path{arXiv:1906.04204}},
  \href {https://doi.org/10.1088/1475-7516/2019/08/018}
  {\path{doi:10.1088/1475-7516/2019/08/018}}.

\bibitem{Panotopoulos_2020}
G.~Panotopoulos, I.~Lopes, Constraints on light dark matter particles using
  white dwarf stars, International Journal of Modern Physics D 29~(08) (2020)
  2050058.
\newblock \href {https://doi.org/10.1142/s0218271820500583}
  {\path{doi:10.1142/s0218271820500583}}.

\bibitem{LongLivedMedNS:Leane:2021ihh}
R.~K. Leane, T.~Linden, P.~Mukhopadhyay, N.~Toro, {Celestial-Body Focused Dark
  Matter Annihilation Throughout the Galaxy}, Phys. Rev. D 103~(7) (2021)
  075030.
\newblock \href {http://arxiv.org/abs/2101.12213} {\path{arXiv:2101.12213}},
  \href {https://doi.org/10.1103/PhysRevD.103.075030}
  {\path{doi:10.1103/PhysRevD.103.075030}}.

\bibitem{Bell:2021fye}
N.~F. Bell, G.~Busoni, M.~E. Ramirez-Quezada, S.~Robles, M.~Virgato, {Improved
  treatment of dark matter capture in white dwarfs}, JCAP 10 (2021) 083.
\newblock \href {http://arxiv.org/abs/2104.14367} {\path{arXiv:2104.14367}},
  \href {https://doi.org/10.1088/1475-7516/2021/10/083}
  {\path{doi:10.1088/1475-7516/2021/10/083}}.

\bibitem{Biswas:2022cyh}
A.~Biswas, A.~Kar, H.~Kim, S.~Scopel, L.~Velasco-Sevilla, {Improved white
  dwarves constraints on inelastic dark matter and left-right symmetric
  models}, Phys. Rev. D 106~(8) (2022) 083012.
\newblock \href {http://arxiv.org/abs/2206.06667} {\path{arXiv:2206.06667}},
  \href {https://doi.org/10.1103/PhysRevD.106.083012}
  {\path{doi:10.1103/PhysRevD.106.083012}}.

\bibitem{Ramirez-Quezada:2022uou}
M.~E. Ramirez-Quezada, {Constraining dark matter interactions mediated by a
  light scalar with white dwarfs} (12 2022).
\newblock \href {http://arxiv.org/abs/2212.09785} {\path{arXiv:2212.09785}}.

\bibitem{SearleZinn:1978ApJ}
L.~{Searle}, R.~{Zinn}, {Composition of halo clusters and the formation of the
  galactic halo.}, apj 225 (1978) 357--379.
\newblock \href {https://doi.org/10.1086/156499} {\path{doi:10.1086/156499}}.

\bibitem{Peebles:1984ApJ}
P.~J.~E. {Peebles}, {Dark matter and the origin of galaxies and globular star
  clusters}, apj 277 (1984) 470--477.
\newblock \href {https://doi.org/10.1086/161714} {\path{doi:10.1086/161714}}.

\bibitem{Diemand:2005MNRAS}
J.~{Diemand}, P.~{Madau}, B.~{Moore}, {The distribution and kinematics of early
  high-{\ensuremath{\sigma}} peaks in present-day haloes: implications for rare
  objects and old stellar populations}, MNRAS 364~(2) (2005) 367--383.
\newblock \href {http://arxiv.org/abs/astro-ph/0506615}
  {\path{arXiv:astro-ph/0506615}}, \href
  {https://doi.org/10.1111/j.1365-2966.2005.09604.x}
  {\path{doi:10.1111/j.1365-2966.2005.09604.x}}.

\bibitem{Kravtsov:2003sm}
A.~V. Kravtsov, O.~Y. Gnedin, {Formation of globular clusters in hierarchical
  cosmology}, Astrophys. J. 623 (2005) 650--665.
\newblock \href {http://arxiv.org/abs/astro-ph/0305199}
  {\path{arXiv:astro-ph/0305199}}, \href {https://doi.org/10.1086/428636}
  {\path{doi:10.1086/428636}}.

\bibitem{Claydon_2019}
I.~Claydon, M.~Gieles, A.~L. Varri, D.~C. Heggie, A.~Zocchi, Spherical models
  of star clusters with potential escapers, Monthly Notices of the Royal
  Astronomical Society 487~(1) (2019) 147--160.
\newblock \href {https://doi.org/10.1093/mnras/stz1109}
  {\path{doi:10.1093/mnras/stz1109}}.

\bibitem{Ashman2001}
K.~M. Ashman, S.~E. Zepf, Some constraints on the formation of globular
  clusters, The Astronomical Journal 122~(4) (2001) 1888--1895.
\newblock \href {https://doi.org/10.1086/323133} {\path{doi:10.1086/323133}}.

\bibitem{vandenBergh2001}
S.~van~den Bergh, How did globular clusters form?, The Astrophysical Journal
  559~(2) (2001) L113--L114.
\newblock \href {https://doi.org/10.1086/323754} {\path{doi:10.1086/323754}}.

\bibitem{clustervgalaxyWillman:2012uj}
B.~Willman, J.~Strader, {'Galaxy,' Defined}, Astron. J. 144 (2012) 76.
\newblock \href {http://arxiv.org/abs/1203.2608} {\path{arXiv:1203.2608}},
  \href {https://doi.org/10.1088/0004-6256/144/3/76}
  {\path{doi:10.1088/0004-6256/144/3/76}}.

\bibitem{EvansStrigari:2021bsh}
A.~J. Evans, L.~E. Strigari, P.~Zivick, {Dark and luminous mass components of
  Omega Centauri from stellar kinematics}, Mon. Not. Roy. Astron. Soc. 511~(3)
  (2022) 4251--4264.
\newblock \href {http://arxiv.org/abs/2109.10998} {\path{arXiv:2109.10998}},
  \href {https://doi.org/10.1093/mnras/stac261}
  {\path{doi:10.1093/mnras/stac261}}.

\bibitem{Ibata2012}
R.~Ibata, C.~Nipoti, A.~Sollima, M.~Bellazzini, S.~C. Chapman, E.~Dalessandro,
  \href{https://doi.org/10.1093%2Fmnras%2Fsts302}{Do globular clusters possess
  dark matter haloes? a case study in {NGC} 2419}, Monthly Notices of the Royal
  Astronomical Society 428~(4) (2012) 3648--3659.
\newblock \href {https://doi.org/10.1093/mnras/sts302}
  {\path{doi:10.1093/mnras/sts302}}.
\newline\urlprefix\url{https://doi.org/10.1093%2Fmnras%2Fsts302}

\bibitem{Krall:2017xij}
R.~Krall, M.~Reece, {Last Electroweak WIMP Standing: Pseudo-Dirac Higgsino
  Status and Compact Stars as Future Probes}, Chin. Phys. C42~(4) (2018)
  043105.
\newblock \href {http://arxiv.org/abs/1705.04843} {\path{arXiv:1705.04843}},
  \href {https://doi.org/10.1088/1674-1137/42/4/043105}
  {\path{doi:10.1088/1674-1137/42/4/043105}}.

\bibitem{Janish:2019nkk}
R.~Janish, V.~Narayan, P.~Riggins, {Type Ia supernovae from dark matter core
  collapse}, Phys. Rev. D100~(3) (2019) 035008.
\newblock \href {http://arxiv.org/abs/1905.00395} {\path{arXiv:1905.00395}},
  \href {https://doi.org/10.1103/PhysRevD.100.035008}
  {\path{doi:10.1103/PhysRevD.100.035008}}.

\bibitem{Petraki:2013wwa}
K.~Petraki, R.~R. Volkas, {Review of asymmetric dark matter}, Int. J. Mod.
  Phys. A 28 (2013) 1330028.
\newblock \href {http://arxiv.org/abs/1305.4939} {\path{arXiv:1305.4939}},
  \href {https://doi.org/10.1142/S0217751X13300287}
  {\path{doi:10.1142/S0217751X13300287}}.

\bibitem{Acevedo:2020gro}
J.~F. Acevedo, J.~Bramante, A.~Goodman, J.~Kopp, T.~Opferkuch, {Dark Matter,
  Destroyer of Worlds: Neutrino, Thermal, and Existential Signatures from Black
  Holes in the Sun and Earth}, JCAP 04 (2021) 026.
\newblock \href {http://arxiv.org/abs/2012.09176} {\path{arXiv:2012.09176}},
  \href {https://doi.org/10.1088/1475-7516/2021/04/026}
  {\path{doi:10.1088/1475-7516/2021/04/026}}.

\bibitem{Ellis:2021ztw}
S.~A.~R. Ellis, {Premature black hole death of Population III stars by dark
  matter}, JCAP 05~(05) (2022) 025.
\newblock \href {http://arxiv.org/abs/2111.02414} {\path{arXiv:2111.02414}},
  \href {https://doi.org/10.1088/1475-7516/2022/05/025}
  {\path{doi:10.1088/1475-7516/2022/05/025}}.

\bibitem{Diamond:2021scl}
M.~D. Diamond, D.~E. Kaplan, {Constraints on relic magnetic black holes}, JHEP
  03 (2022) 157.
\newblock \href {http://arxiv.org/abs/2103.01850} {\path{arXiv:2103.01850}},
  \href {https://doi.org/10.1007/JHEP03(2022)157}
  {\path{doi:10.1007/JHEP03(2022)157}}.

\bibitem{BramanteRajSelfReplic:2024idl}
J.~Bramante, N.~Raj, {Cosmology of self-replicating universes in the interior
  of black holes formed by dark matter-seeded stellar collapse}, Phys. Rev. D
  110~(4) (2024) 043537.
\newblock \href {http://arxiv.org/abs/2405.12277} {\path{arXiv:2405.12277}},
  \href {https://doi.org/10.1103/PhysRevD.110.043537}
  {\path{doi:10.1103/PhysRevD.110.043537}}.

\bibitem{Graham:2015apa}
P.~W. Graham, S.~Rajendran, J.~Varela, {Dark Matter Triggers of Supernovae},
  Phys. Rev. D92~(6) (2015) 063007.
\newblock \href {http://arxiv.org/abs/1505.04444} {\path{arXiv:1505.04444}},
  \href {https://doi.org/10.1103/PhysRevD.92.063007}
  {\path{doi:10.1103/PhysRevD.92.063007}}.

\bibitem{Leung:2013pra}
S.~C. Leung, M.~C. Chu, L.~M. Lin, K.~W. Wong, {Dark-matter admixed white
  dwarfs}, Phys. Rev. D87~(12) (2013) 123506.
\newblock \href {http://arxiv.org/abs/1305.6142} {\path{arXiv:1305.6142}},
  \href {https://doi.org/10.1103/PhysRevD.87.123506}
  {\path{doi:10.1103/PhysRevD.87.123506}}.

\bibitem{Acevedo:2019gre}
J.~F. Acevedo, J.~Bramante, {Supernovae Sparked By Dark Matter in White
  Dwarfs}, Phys. Rev. D100~(4) (2019) 043020.
\newblock \href {http://arxiv.org/abs/1904.11993} {\path{arXiv:1904.11993}},
  \href {https://doi.org/10.1103/PhysRevD.100.043020}
  {\path{doi:10.1103/PhysRevD.100.043020}}.

\bibitem{Graham:2018efk}
P.~W. Graham, R.~Janish, V.~Narayan, S.~Rajendran, P.~Riggins, {White Dwarfs as
  Dark Matter Detectors}, Phys. Rev. D98~(11) (2018) 115027.
\newblock \href {http://arxiv.org/abs/1805.07381} {\path{arXiv:1805.07381}},
  \href {https://doi.org/10.1103/PhysRevD.98.115027}
  {\path{doi:10.1103/PhysRevD.98.115027}}.

\bibitem{Acevedo:2020avd}
J.~F. Acevedo, J.~Bramante, A.~Goodman, {Nuclear fusion inside dark matter},
  Phys. Rev. D 103~(12) (2021) 123022.
\newblock \href {http://arxiv.org/abs/2012.10998} {\path{arXiv:2012.10998}},
  \href {https://doi.org/10.1103/PhysRevD.103.123022}
  {\path{doi:10.1103/PhysRevD.103.123022}}.

\bibitem{Montero-Camacho:2019jte}
P.~Montero-Camacho, X.~Fang, G.~Vasquez, M.~Silva, C.~M. Hirata, {Revisiting
  constraints on asteroid-mass primordial black holes as dark matter
  candidates}, JCAP 08 (2019) 031.
\newblock \href {http://arxiv.org/abs/1906.05950} {\path{arXiv:1906.05950}},
  \href {https://doi.org/10.1088/1475-7516/2019/08/031}
  {\path{doi:10.1088/1475-7516/2019/08/031}}.

\bibitem{MACROSidhu2020}
J.~S. Sidhu, G.~D. Starkman, Reconsidering astrophysical constraints on
  macroscopic dark matter, Physical Review D 101~(8) (apr 2020).
\newblock \href {https://doi.org/10.1103/physrevd.101.083503}
  {\path{doi:10.1103/physrevd.101.083503}}.

\bibitem{WDNSboom:Raj:2023azx}
N.~Raj, {Supernovae and superbursts by dark matter clumps} (6 2023).
\newblock \href {http://arxiv.org/abs/2306.14981} {\path{arXiv:2306.14981}}.

\bibitem{DasEllisSchusterZhou:2021drz}
A.~Das, S.~A.~R. Ellis, P.~C. Schuster, K.~Zhou, {Stellar Shocks From Dark
  Matter} (6 2021).
\newblock \href {http://arxiv.org/abs/2106.09033} {\path{arXiv:2106.09033}}.

\bibitem{Steigerwald:2021vgi}
H.~Steigerwald, E.~Tejeda, {Bondi-Hoyle-Lyttleton Accretion in a Reactive
  Medium: Detonation Ignition and a Mechanism for Type Ia Supernovae}, Phys.
  Rev. Lett. 127~(1) (2021) 011101.
\newblock \href {http://arxiv.org/abs/2104.07066} {\path{arXiv:2104.07066}},
  \href {https://doi.org/10.1103/PhysRevLett.127.011101}
  {\path{doi:10.1103/PhysRevLett.127.011101}}.

\bibitem{Steigerwald:2022pjo}
H.~Steigerwald, V.~Marra, S.~Profumo, {Revisiting constraints on asymmetric
  dark matter from collapse in white dwarf stars}, Phys. Rev. D 105~(8) (2022)
  083507.
\newblock \href {http://arxiv.org/abs/2203.09054} {\path{arXiv:2203.09054}},
  \href {https://doi.org/10.1103/PhysRevD.105.083507}
  {\path{doi:10.1103/PhysRevD.105.083507}}.

\bibitem{Scalzo:2014wxa}
R.~A. Scalzo, A.~J. Ruiter, S.~A. Sim, {The ejected mass distribution of type
  Ia supernovae: A significant rate of non-Chandrasekhar-mass progenitors},
  Mon. Not. Roy. Astron. Soc. 445~(3) (2014) 2535--2544.
\newblock \href {http://arxiv.org/abs/1408.6601} {\path{arXiv:1408.6601}},
  \href {https://doi.org/10.1093/mnras/stu1808}
  {\path{doi:10.1093/mnras/stu1808}}.

\bibitem{Pan:2013cva}
Y.~C. Pan, et~al., {The Host Galaxies of Type Ia Supernovae Discovered by the
  Palomar Transient Factory}, Mon. Not. Roy. Astron. Soc. 438~(2) (2014)
  1391--1416.
\newblock \href {http://arxiv.org/abs/1311.6344} {\path{arXiv:1311.6344}},
  \href {https://doi.org/10.1093/mnras/stt2287}
  {\path{doi:10.1093/mnras/stt2287}}.

\bibitem{Acevedo:2023cab}
J.~F. Acevedo, H.~An, Y.~Boukhtouchen, J.~Bramante, M.~L.~A. Richardson,
  L.~Sansom, {Dark matter induced baryonic feedback in galaxies}, Phys. Rev. D
  110~(8) (2024) 083004.
\newblock \href {http://arxiv.org/abs/2309.08661} {\path{arXiv:2309.08661}},
  \href {https://doi.org/10.1103/PhysRevD.110.083004}
  {\path{doi:10.1103/PhysRevD.110.083004}}.

\bibitem{Kasliwal:2011se}
M.~M. Kasliwal, et~al., {Calcium-rich gap transients in the remote outskirts of
  galaxies}, Astrophys. J. 755 (2012) 161.
\newblock \href {http://arxiv.org/abs/1111.6109} {\path{arXiv:1111.6109}},
  \href {https://doi.org/10.1088/0004-637X/755/2/161}
  {\path{doi:10.1088/0004-637X/755/2/161}}.

\bibitem{Steigerwald:2021bro}
H.~Steigerwald, D.~Rodrigues, S.~Profumo, V.~Marra, {Type Ia supernova
  magnitude step from the local dark matter environment}, Mon. Not. Roy.
  Astron. Soc. 510~(4) (2022) 4779--4795.
\newblock \href {http://arxiv.org/abs/2112.09739} {\path{arXiv:2112.09739}},
  \href {https://doi.org/10.1093/mnras/stab3747}
  {\path{doi:10.1093/mnras/stab3747}}.

\bibitem{Smirnov:2022zip}
J.~Smirnov, A.~Goobar, T.~Linden, E.~M\"ortsell, {White Dwarfs in Dwarf
  Spheroidal Galaxies: A New Class of Compact-Dark-Matter Detectors} (10 2022).
\newblock \href {http://arxiv.org/abs/2211.00013} {\path{arXiv:2211.00013}}.

\bibitem{Leung:2019ctw}
S.-C. Leung, S.~Zha, M.-C. Chu, L.-M. Lin, K.~Nomoto, {Accretion-Induced
  Collapse of Dark Matter Admixed White Dwarfs -- I: Formation of Low-mass
  Neutron Stars}, Astrophys. J. 884 (2019) 9.
\newblock \href {http://arxiv.org/abs/1908.05102} {\path{arXiv:1908.05102}},
  \href {https://doi.org/10.3847/1538-4357/ab3b5e}
  {\path{doi:10.3847/1538-4357/ab3b5e}}.

\bibitem{Zha:2019lxw}
S.~Zha, M.-C. Chu, S.-C. Leung, L.-M. Lin, {Accretion-Induced Collapse of Dark
  Matter Admixed White Dwarfs -- II: Rotation and Gravitational-wave Signals}
  (8 2019).
\newblock \href {http://arxiv.org/abs/1908.05150} {\path{arXiv:1908.05150}},
  \href {https://doi.org/10.3847/1538-4357/ab3640}
  {\path{doi:10.3847/1538-4357/ab3640}}.

\bibitem{NSvIR:clumps2021}
J.~Bramante, B.~J. Kavanagh, N.~Raj, {Scattering Searches for Dark Matter in
  Subhalos: Neutron Stars, Cosmic Rays, and Old Rocks}, Phys. Rev. Lett.
  128~(23) (2022) 231801.
\newblock \href {http://arxiv.org/abs/2109.04582} {\path{arXiv:2109.04582}},
  \href {https://doi.org/10.1103/PhysRevLett.128.231801}
  {\path{doi:10.1103/PhysRevLett.128.231801}}.

\bibitem{Balkin:2022qer}
R.~Balkin, J.~Serra, K.~Springmann, S.~Stelzl, A.~Weiler, {White dwarfs as a
  probe of light QCD axions} (11 2022).
\newblock \href {http://arxiv.org/abs/2211.02661} {\path{arXiv:2211.02661}}.

\bibitem{NSvIR:Riverside:LeptophilicLong}
A.~Joglekar, N.~Raj, P.~Tanedo, H.-B. Yu, {Kinetic Heating from Contact
  Interactions with Relativistic Targets: Electrons Capture Dark Matter in
  Neutron Stars} (4 2020).
\newblock \href {http://arxiv.org/abs/2004.09539} {\path{arXiv:2004.09539}}.

\bibitem{NSvIR:GaraniGenoliniHambye}
R.~Garani, Y.~Genolini, T.~Hambye, {New Analysis of Neutron Star Constraints on
  Asymmetric Dark Matter}, JCAP 1905~(05) (2019) 035.
\newblock \href {http://arxiv.org/abs/1812.08773} {\path{arXiv:1812.08773}},
  \href {https://doi.org/10.1088/1475-7516/2019/05/035}
  {\path{doi:10.1088/1475-7516/2019/05/035}}.

\bibitem{NSvIR:anzuiniBell2021improved}
F.~Anzuini, N.~F. Bell, G.~Busoni, T.~F. Motta, S.~Robles, A.~W. Thomas,
  M.~Virgato, {Improved Treatment of Dark Matter Capture in Neutron Stars III:
  Nucleon and Exotic Targets} (8 2021).
\newblock \href {http://arxiv.org/abs/2108.02525} {\path{arXiv:2108.02525}}.

\bibitem{NSvIR:Bell:Improved}
N.~F. Bell, G.~Busoni, S.~Robles, M.~Virgato, {Improved Treatment of Dark
  Matter Capture in Neutron Stars} (4 2020).
\newblock \href {http://arxiv.org/abs/2004.14888} {\path{arXiv:2004.14888}}.

\bibitem{NSvIR:Bell2020improved}
N.~F. Bell, G.~Busoni, T.~F. Motta, S.~Robles, A.~W. Thomas, M.~Virgato,
  {Nucleon Structure and Strong Interactions in Dark Matter Capture in Neutron
  Stars} (12 2020).
\newblock \href {http://arxiv.org/abs/2012.08918} {\path{arXiv:2012.08918}}.

\bibitem{NSvIR:DIS:Su:2024flx}
L.~Su, L.~Wu, M.~Yang, {Deep inelastic scattering in the capture of dark matter
  by neutron stars}, Phys. Rev. D 110~(5) (2024) 055014.
\newblock \href {http://arxiv.org/abs/2408.03759} {\path{arXiv:2408.03759}},
  \href {https://doi.org/10.1103/PhysRevD.110.055014}
  {\path{doi:10.1103/PhysRevD.110.055014}}.

\bibitem{collectiveDeRoccoLasenby:2022rze}
W.~DeRocco, M.~Galanis, R.~Lasenby, {Dark matter scattering in astrophysical
  media: collective effects}, JCAP 05~(05) (2022) 015.
\newblock \href {http://arxiv.org/abs/2201.05167} {\path{arXiv:2201.05167}},
  \href {https://doi.org/10.1088/1475-7516/2022/05/015}
  {\path{doi:10.1088/1475-7516/2022/05/015}}.

\bibitem{NSvIR:DasguptaGuptaRay:LightMed}
B.~Dasgupta, A.~Gupta, A.~Ray, {Dark matter capture in celestial objects: light
  mediators, self-interactions, and complementarity with direct detection},
  JCAP 10 (2020) 023.
\newblock \href {http://arxiv.org/abs/2006.10773} {\path{arXiv:2006.10773}},
  \href {https://doi.org/10.1088/1475-7516/2020/10/023}
  {\path{doi:10.1088/1475-7516/2020/10/023}}.

\bibitem{Bose:2022ola}
D.~Bose, S.~Sarkar, {Impact of galactic distributions in celestial capture of
  dark matter}, Phys. Rev. D 107~(6) (2023) 063010.
\newblock \href {http://arxiv.org/abs/2211.16982} {\path{arXiv:2211.16982}},
  \href {https://doi.org/10.1103/PhysRevD.107.063010}
  {\path{doi:10.1103/PhysRevD.107.063010}}.

\bibitem{NSvIR:Baryakhtar:DKHNS}
M.~Baryakhtar, J.~Bramante, S.~W. Li, T.~Linden, N.~Raj, {Dark Kinetic Heating
  of Neutron Stars and An Infrared Window On WIMPs, SIMPs, and Pure Higgsinos},
  Phys. Rev. Lett. 119~(13) (2017) 131801.
\newblock \href {http://arxiv.org/abs/1704.01577} {\path{arXiv:1704.01577}},
  \href {https://doi.org/10.1103/PhysRevLett.119.131801}
  {\path{doi:10.1103/PhysRevLett.119.131801}}.

\bibitem{TMT:2015pvw}
W.~Skidmore, et~al., {Thirty Meter Telescope Detailed Science Case: 2015}, Res.
  Astron. Astrophys. 15~(12) (2015) 1945--2140.
\newblock \href {http://arxiv.org/abs/1505.01195} {\path{arXiv:1505.01195}},
  \href {https://doi.org/10.1088/1674-4527/15/12/001}
  {\path{doi:10.1088/1674-4527/15/12/001}}.

\bibitem{ELT:neichel2018overview}
B.~Neichel, D.~Mouillet, E.~Gendron, C.~Correia, J.~F. Sauvage, T.~Fusco,
  Overview of the european extremely large telescope and its instrument suite
  (2018).
\newblock \href {http://arxiv.org/abs/1812.06639} {\path{arXiv:1812.06639}}.

\bibitem{JWST:Gardner:2006ky}
J.~P. Gardner, et~al., {The James Webb Space Telescope}, Space Sci. Rev. 123
  (2006) 485.
\newblock \href {http://arxiv.org/abs/astro-ph/0606175}
  {\path{arXiv:astro-ph/0606175}}, \href
  {https://doi.org/10.1007/s11214-006-8315-7}
  {\path{doi:10.1007/s11214-006-8315-7}}.

\bibitem{FAST2011}
R.~Nan, D.~Li, C.~Jin, Q.~Wang, L.~Zhu, W.~Zhu, H.~Zhang, Y.~Yue, L.~Qian,
  \href{http://dx.doi.org/10.1142/S0218271811019335}{The five-hundred-meter
  aperture spherical radio telescope (fast) project}, International Journal of
  Modern Physics D 20~(06) (2011) 989–1024.
\newblock \href {https://doi.org/10.1142/s0218271811019335}
  {\path{doi:10.1142/s0218271811019335}}.
\newline\urlprefix\url{http://dx.doi.org/10.1142/S0218271811019335}

\bibitem{CHIME2021}
M.~Amiri, K.~M. Bandura, P.~J. Boyle, C.~Brar, J.-F. Cliche, K.~Crowter,
  D.~Cubranic, P.~B. Demorest, N.~T. Denman, M.~Dobbs, et~al.,
  \href{http://dx.doi.org/10.3847/1538-4365/abfdcb}{The chime pulsar project:
  System overview}, The Astrophysical Journal Supplement Series 255~(1) (2021)
  5.
\newblock \href {https://doi.org/10.3847/1538-4365/abfdcb}
  {\path{doi:10.3847/1538-4365/abfdcb}}.
\newline\urlprefix\url{http://dx.doi.org/10.3847/1538-4365/abfdcb}

\bibitem{SKA:2004nx}
C.~L. Carilli, S.~Rawlings, {Science with the Square Kilometer Array:
  Motivation, key science projects, standards and assumptions}, New Astron.
  Rev. 48 (2004) 979.
\newblock \href {http://arxiv.org/abs/astro-ph/0409274}
  {\path{arXiv:astro-ph/0409274}}, \href
  {https://doi.org/10.1016/j.newar.2004.09.001}
  {\path{doi:10.1016/j.newar.2004.09.001}}.

\bibitem{vcircMW}
A.-C. Eilers, D.~W. Hogg, H.-W. Rix, M.~K. Ness,
  \href{http://dx.doi.org/10.3847/1538-4357/aaf648}{The circular velocity curve
  of the milky way from 5 to 25 kpc}, The Astrophysical Journal 871~(1) (2019)
  120.
\newblock \href {https://doi.org/10.3847/1538-4357/aaf648}
  {\path{doi:10.3847/1538-4357/aaf648}}.
\newline\urlprefix\url{http://dx.doi.org/10.3847/1538-4357/aaf648}

\bibitem{BrayeurTinyakovBinaryCap:2011yw}
L.~Brayeur, P.~Tinyakov, {Enhancement of dark matter capture by neutron stars
  in binary systems}, Phys. Rev. Lett. 109 (2012) 061301.
\newblock \href {http://arxiv.org/abs/1111.3205} {\path{arXiv:1111.3205}},
  \href {https://doi.org/10.1103/PhysRevLett.109.061301}
  {\path{doi:10.1103/PhysRevLett.109.061301}}.

\bibitem{EvaporationGaraniSergio:2021feo}
R.~Garani, S.~Palomares-Ruiz, {Evaporation of dark matter from celestial
  bodies}, JCAP 05~(05) (2022) 042.
\newblock \href {http://arxiv.org/abs/2104.12757} {\path{arXiv:2104.12757}},
  \href {https://doi.org/10.1088/1475-7516/2022/05/042}
  {\path{doi:10.1088/1475-7516/2022/05/042}}.

\bibitem{Kouvaris:2007ay}
C.~Kouvaris, {WIMP Annihilation and Cooling of Neutron Stars}, Phys. Rev. D 77
  (2008) 023006.
\newblock \href {http://arxiv.org/abs/0708.2362} {\path{arXiv:0708.2362}},
  \href {https://doi.org/10.1103/PhysRevD.77.023006}
  {\path{doi:10.1103/PhysRevD.77.023006}}.

\bibitem{NSvIR:SelfIntDM}
C.-S. Chen, Y.-H. Lin, {Reheating neutron stars with the annihilation of
  self-interacting dark matter}, JHEP 08 (2018) 069.
\newblock \href {http://arxiv.org/abs/1804.03409} {\path{arXiv:1804.03409}},
  \href {https://doi.org/10.1007/JHEP08(2018)069}
  {\path{doi:10.1007/JHEP08(2018)069}}.

\bibitem{NSvIR:Raj:DKHNSOps}
N.~Raj, P.~Tanedo, H.-B. Yu, {Neutron stars at the dark matter direct detection
  frontier}, Phys. Rev. D 97~(4) (2018) 043006.
\newblock \href {http://arxiv.org/abs/1707.09442} {\path{arXiv:1707.09442}},
  \href {https://doi.org/10.1103/PhysRevD.97.043006}
  {\path{doi:10.1103/PhysRevD.97.043006}}.

\bibitem{NSdistribs:Ofek2009}
E.~O. {Ofek}, {Space and Velocity Distributions of Galactic Isolated Old
  Neutron Stars}, pasp 121~(882) (2009) 814.
\newblock \href {http://arxiv.org/abs/0910.3684} {\path{arXiv:0910.3684}},
  \href {https://doi.org/10.1086/605389} {\path{doi:10.1086/605389}}.

\bibitem{NSdistribs:Sartore2010}
N.~Sartore, E.~Ripamonti, A.~Treves, R.~Turolla,
  \href{http://dx.doi.org/10.1051/0004-6361/200912222}{Galactic neutron stars},
  Astronomy and Astrophysics 510 (2010) A23.
\newblock \href {https://doi.org/10.1051/0004-6361/200912222}
  {\path{doi:10.1051/0004-6361/200912222}}.
\newline\urlprefix\url{http://dx.doi.org/10.1051/0004-6361/200912222}

\bibitem{NSHeatObs:Raj:2024kjq}
N.~Raj, P.~Shivanna, G.~N. Rachh, {Exploring reheated sub-40000 Kelvin neutron
  stars with JWST, ELT, and TMT}, Phys. Rev. D 109~(12) (2024) 123040.
\newblock \href {http://arxiv.org/abs/2403.07496} {\path{arXiv:2403.07496}},
  \href {https://doi.org/10.1103/PhysRevD.109.123040}
  {\path{doi:10.1103/PhysRevD.109.123040}}.

\bibitem{NSHeatObs:Bramante:2024ikc}
J.~Bramante, K.~Mack, N.~Raj, L.~Shao, N.~Tyagi, {Seeking the nearest neutron
  stars using a new local electron density map} (11 2024).
\newblock \href {http://arxiv.org/abs/2411.18647} {\path{arXiv:2411.18647}}.

\bibitem{Bertoni:2013bsa}
B.~Bertoni, A.~E. Nelson, S.~Reddy, {Dark Matter Thermalization in Neutron
  Stars}, Phys. Rev. D88 (2013) 123505.
\newblock \href {http://arxiv.org/abs/1309.1721} {\path{arXiv:1309.1721}},
  \href {https://doi.org/10.1103/PhysRevD.88.123505}
  {\path{doi:10.1103/PhysRevD.88.123505}}.

\bibitem{NSvIR:Bell2019:Leptophilic}
N.~F. Bell, G.~Busoni, S.~Robles, {Capture of Leptophilic Dark Matter in
  Neutron Stars}, JCAP 1906~(06) (2019) 054.
\newblock \href {http://arxiv.org/abs/1904.09803} {\path{arXiv:1904.09803}},
  \href {https://doi.org/10.1088/1475-7516/2019/06/054}
  {\path{doi:10.1088/1475-7516/2019/06/054}}.

\bibitem{Garani:2019fpa}
R.~Garani, J.~Heeck, {Dark matter interactions with muons in neutron stars},
  Phys. Rev. D100~(3) (2019) 035039.
\newblock \href {http://arxiv.org/abs/1906.10145} {\path{arXiv:1906.10145}},
  \href {https://doi.org/10.1103/PhysRevD.100.035039}
  {\path{doi:10.1103/PhysRevD.100.035039}}.

\bibitem{NSvIR:Riverside:LeptophilicShort}
A.~Joglekar, N.~Raj, P.~Tanedo, H.-B. Yu, {Relativistic capture of dark matter
  by electrons in neutron stars} (11 2019).
\newblock \href {http://arxiv.org/abs/1911.13293} {\path{arXiv:1911.13293}}.

\bibitem{NSvIR:Bell:ImprovedLepton}
N.~F. Bell, G.~Busoni, S.~Robles, M.~Virgato, {Improved Treatment of Dark
  Matter Capture in Neutron Stars II: Leptonic Targets}, JCAP 03 (2021) 086.
\newblock \href {http://arxiv.org/abs/2010.13257} {\path{arXiv:2010.13257}},
  \href {https://doi.org/10.1088/1475-7516/2021/03/086}
  {\path{doi:10.1088/1475-7516/2021/03/086}}.

\bibitem{NSvIR:GaraniGuptaRaj:Thermalizn}
R.~Garani, A.~Gupta, N.~Raj, {Observing the thermalization of dark matter in
  neutron stars}, Phys. Rev. D 103~(4) (2021) 043019.
\newblock \href {http://arxiv.org/abs/2009.10728} {\path{arXiv:2009.10728}},
  \href {https://doi.org/10.1103/PhysRevD.103.043019}
  {\path{doi:10.1103/PhysRevD.103.043019}}.

\bibitem{NSvIR:PseudoscaTRIUMF:2022eav}
J.~Coffey, D.~McKeen, D.~E. Morrissey, N.~Raj, {Neutron star observations of
  pseudoscalar-mediated dark matter}, Phys. Rev. D 106~(11) (2022) 115019.
\newblock \href {http://arxiv.org/abs/2207.02221} {\path{arXiv:2207.02221}},
  \href {https://doi.org/10.1103/PhysRevD.106.115019}
  {\path{doi:10.1103/PhysRevD.106.115019}}.

\bibitem{NSvIR:Bell2018:Inelastic}
N.~F. Bell, G.~Busoni, S.~Robles, {Heating up Neutron Stars with Inelastic Dark
  Matter}, JCAP 1809~(09) (2018) 018.
\newblock \href {http://arxiv.org/abs/1807.02840} {\path{arXiv:1807.02840}},
  \href {https://doi.org/10.1088/1475-7516/2018/09/018}
  {\path{doi:10.1088/1475-7516/2018/09/018}}.

\bibitem{NSvIR:InelasticJoglekarYu:2023fjj}
G.~Alvarez, A.~Joglekar, M.~Phoroutan-Mehr, H.-B. Yu, {Heating Neutron Stars
  with Inelastic Dark Matter and Relativistic Targets} (1 2023).
\newblock \href {http://arxiv.org/abs/2301.08767} {\path{arXiv:2301.08767}}.

\bibitem{NSMultiscat:Bramante:2017xlb}
J.~Bramante, A.~Delgado, A.~Martin, {Multiscatter stellar capture of dark
  matter}, Phys. Rev. D 96~(6) (2017) 063002.
\newblock \href {http://arxiv.org/abs/1703.04043} {\path{arXiv:1703.04043}},
  \href {https://doi.org/10.1103/PhysRevD.96.063002}
  {\path{doi:10.1103/PhysRevD.96.063002}}.

\bibitem{NSMultiscat:Ilie:2020vec}
C.~Ilie, J.~Pilawa, S.~Zhang, {Comment on \textquotedblleft{}Multiscatter
  stellar capture of dark matter\textquotedblright{}}, Phys. Rev. D 102~(4)
  (2020) 048301.
\newblock \href {http://arxiv.org/abs/2005.05946} {\path{arXiv:2005.05946}},
  \href {https://doi.org/10.1103/PhysRevD.102.048301}
  {\path{doi:10.1103/PhysRevD.102.048301}}.

\bibitem{NSMultiscat:Ilie:2024sos}
C.~Ilie, {Closed-form Expressions for Multiscatter Dark Matter Capture Rates},
  Astrophys. J. 970~(2) (2024) 159.
\newblock \href {http://arxiv.org/abs/2402.07713} {\path{arXiv:2402.07713}},
  \href {https://doi.org/10.3847/1538-4357/ad5556}
  {\path{doi:10.3847/1538-4357/ad5556}}.

\bibitem{deLavallaz:2010wp}
A.~de~Lavallaz, M.~Fairbairn, {Neutron Stars as Dark Matter Probes}, Phys. Rev.
  D81 (2010) 123521.
\newblock \href {http://arxiv.org/abs/1004.0629} {\path{arXiv:1004.0629}},
  \href {https://doi.org/10.1103/PhysRevD.81.123521}
  {\path{doi:10.1103/PhysRevD.81.123521}}.

\bibitem{Reddy:1997yr}
S.~Reddy, M.~Prakash, J.~M. Lattimer, {Neutrino interactions in hot and dense
  matter}, Phys. Rev. D58 (1998) 013009.
\newblock \href {http://arxiv.org/abs/astro-ph/9710115}
  {\path{arXiv:astro-ph/9710115}}, \href
  {https://doi.org/10.1103/PhysRevD.58.013009}
  {\path{doi:10.1103/PhysRevD.58.013009}}.

\bibitem{NSvIR:Bell:Thermalization:2023ysh}
N.~F. Bell, G.~Busoni, S.~Robles, M.~Virgato, {Thermalization and annihilation
  of dark matter in neutron stars}, JCAP 04 (2024) 006.
\newblock \href {http://arxiv.org/abs/2312.11892} {\path{arXiv:2312.11892}},
  \href {https://doi.org/10.1088/1475-7516/2024/04/006}
  {\path{doi:10.1088/1475-7516/2024/04/006}}.

\bibitem{UnthermalizedHiggsinoAcevedo:2024ttq}
J.~F. Acevedo, J.~Bramante, Q.~Liu, N.~Tyagi, {Neutrino and Gamma-Ray
  Signatures of Inelastic Dark Matter Annihilating outside Neutron Stars} (4
  2024).
\newblock \href {http://arxiv.org/abs/2404.10039} {\path{arXiv:2404.10039}}.

\bibitem{LongLivedMedNS:BoseMaity:2021yhz}
D.~Bose, T.~N. Maity, T.~S. Ray, {Neutrinos from captured dark matter
  annihilation in a galactic population of neutron stars}, JCAP 05~(05) (2022)
  001.
\newblock \href {http://arxiv.org/abs/2108.12420} {\path{arXiv:2108.12420}},
  \href {https://doi.org/10.1088/1475-7516/2022/05/001}
  {\path{doi:10.1088/1475-7516/2022/05/001}}.

\bibitem{LongLivedMedNS:NguyenTait:2022zwb}
T.~T.~Q. Nguyen, T.~M.~P. Tait, {Bounds on Long-lived Dark Matter Mediators
  from Neutron Stars} (12 2022).
\newblock \href {http://arxiv.org/abs/2212.12547} {\path{arXiv:2212.12547}}.

\bibitem{NSvIR:hylogenenesis1:2010}
H.~Davoudiasl, D.~E. Morrissey, K.~Sigurdson, S.~Tulin, {Hylogenesis: A Unified
  Origin for Baryonic Visible Matter and Antibaryonic Dark Matter}, Phys. Rev.
  Lett. 105 (2010) 211304.
\newblock \href {http://arxiv.org/abs/1008.2399} {\path{arXiv:1008.2399}},
  \href {https://doi.org/10.1103/PhysRevLett.105.211304}
  {\path{doi:10.1103/PhysRevLett.105.211304}}.

\bibitem{NSvIR:hylogenenesis2:2011}
H.~Davoudiasl, D.~E. Morrissey, K.~Sigurdson, S.~Tulin, {Baryon Destruction by
  Asymmetric Dark Matter}, Phys. Rev. D 84 (2011) 096008.
\newblock \href {http://arxiv.org/abs/1106.4320} {\path{arXiv:1106.4320}},
  \href {https://doi.org/10.1103/PhysRevD.84.096008}
  {\path{doi:10.1103/PhysRevD.84.096008}}.

\bibitem{NSvIR:coann-nucleons:JinGao2018moh}
M.~Jin, Y.~Gao, {Nucleon - Light Dark Matter Annihilation through Baryon Number
  Violation}, Phys. Rev. D 98~(7) (2018) 075026.
\newblock \href {http://arxiv.org/abs/1808.10644} {\path{arXiv:1808.10644}},
  \href {https://doi.org/10.1103/PhysRevD.98.075026}
  {\path{doi:10.1103/PhysRevD.98.075026}}.

\bibitem{NSvIR:Marfatia:DarkBaryon}
W.-Y. Keung, D.~Marfatia, P.-Y. Tseng, {Heating neutron stars with GeV dark
  matter}, JHEP 07 (2020) 181.
\newblock \href {http://arxiv.org/abs/2001.09140} {\path{arXiv:2001.09140}},
  \href {https://doi.org/10.1007/JHEP07(2020)181}
  {\path{doi:10.1007/JHEP07(2020)181}}.

\bibitem{NSvIR:baryodestruct:Ema:2024wqr}
Y.~Ema, R.~McGehee, M.~Pospelov, A.~Ray, {Dark Matter Catalyzed Baryon
  Destruction} (5 2024).
\newblock \href {http://arxiv.org/abs/2405.18472} {\path{arXiv:2405.18472}}.

\bibitem{NSvIR:magneticBH:2020}
Y.~Bai, J.~Berger, M.~Korwar, N.~Orlofsky, {Phenomenology of magnetic black
  holes with electroweak-symmetric coronas}, JHEP 10 (2020) 210.
\newblock \href {http://arxiv.org/abs/2007.03703} {\path{arXiv:2007.03703}},
  \href {https://doi.org/10.1007/JHEP10(2020)210}
  {\path{doi:10.1007/JHEP10(2020)210}}.

\bibitem{DMann:strangelet:Silk2010xlt}
M.~A. Perez-Garcia, J.~Silk, J.~R. Stone, {Dark matter, neutron stars and
  strange quark matter}, Phys. Rev. Lett. 105 (2010) 141101.
\newblock \href {http://arxiv.org/abs/1007.1421} {\path{arXiv:1007.1421}},
  \href {https://doi.org/10.1103/PhysRevLett.105.141101}
  {\path{doi:10.1103/PhysRevLett.105.141101}}.

\bibitem{DMdecayNS:Silk2014dra}
M.~Angeles Perez-Garcia, J.~Silk, {Constraining decaying dark matter with
  neutron stars}, Phys. Lett. B744 (2015) 13--17.
\newblock \href {http://arxiv.org/abs/1403.6111} {\path{arXiv:1403.6111}},
  \href {https://doi.org/10.1016/j.physletb.2015.03.026}
  {\path{doi:10.1016/j.physletb.2015.03.026}}.

\bibitem{DMann:bubblenucleation:Silk2019}
A.~Herrero, M.~A. P\'erez-Garc\'\i{}a, J.~Silk, C.~Albertus, {Dark matter and
  bubble nucleation in old neutron stars}, Phys. Rev. D 100~(10) (2019) 103019.
\newblock \href {http://arxiv.org/abs/1905.00893} {\path{arXiv:1905.00893}},
  \href {https://doi.org/10.1103/PhysRevD.100.103019}
  {\path{doi:10.1103/PhysRevD.100.103019}}.

\bibitem{Kolb:1982si}
E.~W. Kolb, S.~A. Colgate, J.~A. Harvey, {Monopole Catalysis of Nucleon Decay
  in Neutron Stars}, Phys. Rev. Lett. 49 (1982) 1373.
\newblock \href {https://doi.org/10.1103/PhysRevLett.49.1373}
  {\path{doi:10.1103/PhysRevLett.49.1373}}.

\bibitem{Dimopoulos:1982cz}
S.~Dimopoulos, J.~Preskill, F.~Wilczek, {Catalyzed Nucleon Decay in Neutron
  Stars}, Phys. Lett. B 119 (1982) 320.
\newblock \href {https://doi.org/10.1016/0370-2693(82)90679-7}
  {\path{doi:10.1016/0370-2693(82)90679-7}}.

\bibitem{NSvIR:GaraniHeeck:Muophilic}
R.~Garani, J.~Heeck, {Dark Matter Interactions with Muons in Neutron Stars},
  Phys. Rev. D100~(3) (2019) 035039.
\newblock \href {http://arxiv.org/abs/1906.10145} {\path{arXiv:1906.10145}},
  \href {https://doi.org/10.1103/PhysRevD.100.035039}
  {\path{doi:10.1103/PhysRevD.100.035039}}.

\bibitem{NSvIR:HamaguchiEWmultiplet:2022uiq}
M.~Fujiwara, K.~Hamaguchi, N.~Nagata, J.~Zheng, {Capture of Electroweak
  Multiplet Dark Matter in Neutron Stars} (4 2022).
\newblock \href {http://arxiv.org/abs/2204.02238} {\path{arXiv:2204.02238}}.

\bibitem{NSvIR:Queiroz:Spectroscopy}
D.~A. Camargo, F.~S. Queiroz, R.~Sturani, {Detecting Dark Matter with Neutron
  Star Spectroscopy}, JCAP 1909~(09) (2019) 051.
\newblock \href {http://arxiv.org/abs/1901.05474} {\path{arXiv:1901.05474}},
  \href {https://doi.org/10.1088/1475-7516/2019/09/051}
  {\path{doi:10.1088/1475-7516/2019/09/051}}.

\bibitem{NSvIR:Queiroz:BosonDM}
T.~N. Maity, F.~S. Queiroz, Detecting bosonic dark matter with neutron stars
  (2021).
\newblock \href {http://arxiv.org/abs/2104.02700} {\path{arXiv:2104.02700}}.

\bibitem{NSvIR:Lin2021:spin1med}
G.-L. Lin, Y.-H. Lin, {Exploring dark sector parameters in light of neutron
  star temperatures}, Phys. Rev. D 104~(6) (2021) 063021.
\newblock \href {http://arxiv.org/abs/2102.11151} {\path{arXiv:2102.11151}},
  \href {https://doi.org/10.1103/PhysRevD.104.063021}
  {\path{doi:10.1103/PhysRevD.104.063021}}.

\bibitem{NSvIR:zeng2021PNGBDM}
Y.-P. Zeng, X.~Xiao, W.~Wang, {Constraints on Pseudo-Nambu-Goldstone dark
  matter from direct detection experiment and neutron star reheating
  temperature}, Phys. Lett. B 824 (2022) 136822.
\newblock \href {http://arxiv.org/abs/2108.11381} {\path{arXiv:2108.11381}},
  \href {https://doi.org/10.1016/j.physletb.2021.136822}
  {\path{doi:10.1016/j.physletb.2021.136822}}.

\bibitem{NsvIR:HamaguchiMug-2:2022wpz}
K.~Hamaguchi, N.~Nagata, M.~E. Ramirez-Quezada, {Neutron Star Heating in Dark
  Matter Models for the Muon $g-2$ Discrepancy} (4 2022).
\newblock \href {http://arxiv.org/abs/2204.02413} {\path{arXiv:2204.02413}}.

\bibitem{NSvIR:sterilenu:Das:2024thc}
S.~Das, P.~S.~B. Dev, T.~Okawa, A.~Soni, {Old neutron stars as a new probe of
  relic neutrinos and sterile neutrino dark matter} (8 2024).
\newblock \href {http://arxiv.org/abs/2408.01484} {\path{arXiv:2408.01484}}.

\bibitem{NSvIR:ISMaccretionphases:Treves:1999ne}
A.~Treves, R.~Turolla, S.~Zane, M.~Colpi, {Isolated neutron stars: accretors
  and coolers}, Publ. Astron. Soc. Pac. 112 (2000) 297.
\newblock \href {http://arxiv.org/abs/astro-ph/9911430}
  {\path{arXiv:astro-ph/9911430}}, \href {https://doi.org/10.1086/316529}
  {\path{doi:10.1086/316529}}.

\bibitem{ISMBubble:Jenkins2009}
E.~B. Jenkins, \href{https://dx.doi.org/10.1088/0004-637X/700/2/1299}{A unified
  representation of gas-phase element depletions in the interstellar medium*},
  The Astrophysical Journal 700~(2) (2009) 1299.
\newblock \href {https://doi.org/10.1088/0004-637X/700/2/1299}
  {\path{doi:10.1088/0004-637X/700/2/1299}}.
\newline\urlprefix\url{https://dx.doi.org/10.1088/0004-637X/700/2/1299}

\bibitem{Rotochem:FernandezReisenegger:2005cg}
R.~Fernandez, A.~Reisenegger, {Rotochemical heating in millisecond pulsars.
  Formalism and non-superfluid case}, Astrophys. J. 625 (2005) 291--306.
\newblock \href {http://arxiv.org/abs/astro-ph/0502116}
  {\path{arXiv:astro-ph/0502116}}, \href {https://doi.org/10.1086/429551}
  {\path{doi:10.1086/429551}}.

\bibitem{Rotochem:Reisenegger:2006ky}
A.~Reisenegger, P.~Jofre, R.~Fernandez, E.~Kantor, {Rotochemical Heating of
  Neutron Stars: Rigorous Formalism with Electrostatic Potential
  Perturbations}, Astrophys. J. 653 (2006) 568--572.
\newblock \href {http://arxiv.org/abs/astro-ph/0606322}
  {\path{arXiv:astro-ph/0606322}}, \href {https://doi.org/10.1086/506601}
  {\path{doi:10.1086/506601}}.

\bibitem{Rotochem:PetrovichReis:2009yh}
C.~Petrovich, A.~Reisenegger, {Rotochemical heating in millisecond pulsars:
  modified Urca reactions with uniform Cooper pairing gaps}, Astron. Astrophys.
  521 (2010) A77.
\newblock \href {http://arxiv.org/abs/0912.2564} {\path{arXiv:0912.2564}},
  \href {https://doi.org/10.1051/0004-6361/200913861}
  {\path{doi:10.1051/0004-6361/200913861}}.

\bibitem{Rotochem:GonzalezJimenezReis:2014iia}
N.~Gonz\'alez-Jim\'enez, C.~Petrovich, A.~Reisenegger, {Rotochemical heating of
  millisecond and classical pulsars with anisotropic and density-dependent
  superfluid gap models}, Mon. Not. Roy. Astron. Soc. 447 (2015) 2073.
\newblock \href {http://arxiv.org/abs/1411.6500} {\path{arXiv:1411.6500}},
  \href {https://doi.org/10.1093/mnras/stu2558}
  {\path{doi:10.1093/mnras/stu2558}}.

\bibitem{Rotochem:GusakovReisenegger2015:CrustNSE}
M.~E. {Gusakov}, E.~M. {Kantor}, A.~{Reisenegger}, {Rotation-induced deep
  crustal heating of millisecond pulsars}, MNRAS 453~(1) (2015) L36--L40.
\newblock \href {http://arxiv.org/abs/1507.04586} {\path{arXiv:1507.04586}},
  \href {https://doi.org/10.1093/mnrasl/slv095}
  {\path{doi:10.1093/mnrasl/slv095}}.

\bibitem{Rotochem:Gusakov2021:Refined}
E.~M. {Kantor}, M.~E. {Gusakov}, {Long-lasting accretion-powered chemical
  heating of millisecond pulsars}, MNRAS 508~(4) (2021) 6118--6127.
\newblock \href {http://arxiv.org/abs/2110.02881} {\path{arXiv:2110.02881}},
  \href {https://doi.org/10.1093/mnras/stab2922}
  {\path{doi:10.1093/mnras/stab2922}}.

\bibitem{NSvIR:Hamaguchi:RotochemicalvDM2019}
K.~Hamaguchi, N.~Nagata, K.~Yanagi, {Dark Matter Heating vs. Rotochemical
  Heating in Old Neutron Stars}, Phys. Lett. B795 (2019) 484--489.
\newblock \href {http://arxiv.org/abs/1905.02991} {\path{arXiv:1905.02991}},
  \href {https://doi.org/10.1016/j.physletb.2019.06.060}
  {\path{doi:10.1016/j.physletb.2019.06.060}}.

\bibitem{NSvIR:Hamaguchi:RotochemicalPure2019}
K.~Yanagi, N.~Nagata, K.~Hamaguchi, {Cooling Theory Faced with Old Warm Neutron
  Stars: Role of Non-Equilibrium Processes with Proton and Neutron Gaps}, Mon.
  Not. Roy. Astron. Soc. 492~(4) (2020) 5508--5523.
\newblock \href {http://arxiv.org/abs/1904.04667} {\path{arXiv:1904.04667}},
  \href {https://doi.org/10.1093/mnras/staa076}
  {\path{doi:10.1093/mnras/staa076}}.

\bibitem{NSvIR:Fujiwara:VortexCreepPure}
M.~Fujiwara, K.~Hamaguchi, N.~Nagata, M.~E. Ramirez-Quezada, {Vortex creep
  heating in neutron stars}, JCAP 03 (2024) 051.
\newblock \href {http://arxiv.org/abs/2308.16066} {\path{arXiv:2308.16066}},
  \href {https://doi.org/10.1088/1475-7516/2024/03/051}
  {\path{doi:10.1088/1475-7516/2024/03/051}}.

\bibitem{NSvIR:Fujiwara:VortexCreepvDM2023}
M.~Fujiwara, K.~Hamaguchi, N.~Nagata, M.~E. Ramirez-Quezada, {Vortex creep
  heating vs. dark matter heating in neutron stars}, Phys. Lett. B 848 (2024)
  138341.
\newblock \href {http://arxiv.org/abs/2309.02633} {\path{arXiv:2309.02633}},
  \href {https://doi.org/10.1016/j.physletb.2023.138341}
  {\path{doi:10.1016/j.physletb.2023.138341}}.

\bibitem{NsvIR:otherinternalheatings:Reisenegger}
D.~{Gonzalez}, A.~{Reisenegger}, {Internal heating of old neutron stars:
  contrasting different mechanisms}, AAP 522 (2010) A16.
\newblock \href {http://arxiv.org/abs/1005.5699} {\path{arXiv:1005.5699}},
  \href {https://doi.org/10.1051/0004-6361/201015084}
  {\path{doi:10.1051/0004-6361/201015084}}.

\bibitem{NSvIR:otherinternalheatings:2022}
F.~K\"opp, J.~E. Horvath, D.~Hadjimichef, C.~A.~Z. Vasconcellos, P.~O. Hess,
  {Internal heating mechanisms in neutron stars} (8 2022).
\newblock \href {http://arxiv.org/abs/2208.07770} {\path{arXiv:2208.07770}}.

\bibitem{RotationPowervXRayLumi:PrinzWerner2015}
T.~Prinz, W.~Becker, {A Search for X-ray Counterparts of Radio Pulsars} (11
  2015).
\newblock \href {http://arxiv.org/abs/1511.07713} {\path{arXiv:1511.07713}}.

\bibitem{magnetothermalheating:Pons2008fd}
J.~A. Pons, J.~A. Miralles, U.~Geppert, {Magneto--thermal evolution of neutron
  stars}, Astron. Astrophys. 496 (2009) 207--216.
\newblock \href {http://arxiv.org/abs/0812.3018} {\path{arXiv:0812.3018}},
  \href {https://doi.org/10.1051/0004-6361:200811229}
  {\path{doi:10.1051/0004-6361:200811229}}.

\bibitem{Edwards:2020afl}
T.~D.~P. Edwards, B.~J. Kavanagh, L.~Visinelli, C.~Weniger, {Transient Radio
  Signatures from Neutron Star Encounters with QCD Axion Miniclusters} (11
  2020).
\newblock \href {http://arxiv.org/abs/2011.05378} {\path{arXiv:2011.05378}}.

\bibitem{ErickcekSigurdson}
A.~L. Erickcek, K.~Sigurdson,
  \href{https://link.aps.org/doi/10.1103/PhysRevD.84.083503}{Reheating effects
  in the matter power spectrum and implications for substructure}, Phys. Rev. D
  84 (2011) 083503.
\newblock \href {https://doi.org/10.1103/PhysRevD.84.083503}
  {\path{doi:10.1103/PhysRevD.84.083503}}.
\newline\urlprefix\url{https://link.aps.org/doi/10.1103/PhysRevD.84.083503}

\bibitem{Barenboim:2013gya}
G.~Barenboim, J.~Rasero, {Structure Formation during an early period of matter
  domination}, JHEP 04 (2014) 138.
\newblock \href {http://arxiv.org/abs/1311.4034} {\path{arXiv:1311.4034}},
  \href {https://doi.org/10.1007/JHEP04(2014)138}
  {\path{doi:10.1007/JHEP04(2014)138}}.

\bibitem{FanWatson}
J.~Fan, O.~\"Ozsoy, S.~Watson,
  \href{https://link.aps.org/doi/10.1103/PhysRevD.90.043536}{Nonthermal
  histories and implications for structure formation}, Phys. Rev. D 90 (2014)
  043536.
\newblock \href {https://doi.org/10.1103/PhysRevD.90.043536}
  {\path{doi:10.1103/PhysRevD.90.043536}}.
\newline\urlprefix\url{https://link.aps.org/doi/10.1103/PhysRevD.90.043536}

\bibitem{drorcodecay}
J.~A. Dror, E.~Kuflik, B.~Melcher, S.~Watson,
  \href{https://link.aps.org/doi/10.1103/PhysRevD.97.063524}{Concentrated dark
  matter: Enhanced small-scale structure from codecaying dark matter}, Phys.
  Rev. D 97 (2018) 063524.
\newblock \href {https://doi.org/10.1103/PhysRevD.97.063524}
  {\path{doi:10.1103/PhysRevD.97.063524}}.
\newline\urlprefix\url{https://link.aps.org/doi/10.1103/PhysRevD.97.063524}

\bibitem{inflatflucs}
P.~W. Graham, J.~Mardon, S.~Rajendran,
  \href{https://link.aps.org/doi/10.1103/PhysRevD.93.103520}{Vector dark matter
  from inflationary fluctuations}, Phys. Rev. D 93 (2016) 103520.
\newblock \href {https://doi.org/10.1103/PhysRevD.93.103520}
  {\path{doi:10.1103/PhysRevD.93.103520}}.
\newline\urlprefix\url{https://link.aps.org/doi/10.1103/PhysRevD.93.103520}

\bibitem{Buckley:2017ttd}
M.~R. Buckley, A.~DiFranzo, {Collapsed Dark Matter Structures}, Phys. Rev.
  Lett. 120~(5) (2018) 051102.
\newblock \href {http://arxiv.org/abs/1707.03829} {\path{arXiv:1707.03829}},
  \href {https://doi.org/10.1103/PhysRevLett.120.051102}
  {\path{doi:10.1103/PhysRevLett.120.051102}}.

\bibitem{nussinovcluster}
S.~Nussinov, Y.~Zhang, {Dark Matter Clusters and Time Correlations in Direct
  Detection Experiments}, JHEP 03 (2020) 133.
\newblock \href {http://arxiv.org/abs/1807.00846} {\path{arXiv:1807.00846}},
  \href {https://doi.org/10.1007/JHEP03(2020)133}
  {\path{doi:10.1007/JHEP03(2020)133}}.

\bibitem{Barenboim:2021swl}
G.~Barenboim, N.~Blinov, A.~Stebbins, {Smallest Remnants of Early Matter
  Domination} (7 2021).
\newblock \href {http://arxiv.org/abs/2107.10293} {\path{arXiv:2107.10293}}.

\bibitem{coldestNSHST}
S.~Guillot, G.~G. Pavlov, C.~Reyes, A.~Reisenegger, L.~Rodriguez, B.~Rangelov,
  O.~Kargaltsev, {Hubble Space Telescope Nondetection of PSR
  J2144\textendash{}3933: The Coldest Known Neutron Star}, Astrophys. J.
  874~(2) (2019) 175.
\newblock \href {http://arxiv.org/abs/1901.07998} {\path{arXiv:1901.07998}},
  \href {https://doi.org/10.3847/1538-4357/ab0f38}
  {\path{doi:10.3847/1538-4357/ab0f38}}.

\bibitem{BondiHoyle1944}
H.~{Bondi}, F.~{Hoyle}, {On the mechanism of accretion by stars}, Mon. Not.
  Roy. Astron. Soc. 104 (1944) 273.
\newblock \href {https://doi.org/10.1093/mnras/104.5.273}
  {\path{doi:10.1093/mnras/104.5.273}}.

\bibitem{Bondi1952}
H.~{Bondi}, {On spherically symmetrical accretion}, Mon. Not. Roy. Astron. Soc.
  112 (1952) 195.
\newblock \href {https://doi.org/10.1093/mnras/112.2.195}
  {\path{doi:10.1093/mnras/112.2.195}}.

\bibitem{Begelman1977}
M.~C. {Begelman}, {Nearly collisionless spherical accretion.}, Mon. Not. Roy.
  Astron. Soc. 181 (1977) 347--363.
\newblock \href {https://doi.org/10.1093/mnras/181.2.347}
  {\path{doi:10.1093/mnras/181.2.347}}.

\bibitem{NANOGrav:2023hvm}
A.~Afzal, et~al., {The NANOGrav 15 yr Data Set: Search for Signals from New
  Physics}, Astrophys. J. Lett. 951~(1) (2023) L11.
\newblock \href {http://arxiv.org/abs/2306.16219} {\path{arXiv:2306.16219}},
  \href {https://doi.org/10.3847/2041-8213/acdc91}
  {\path{doi:10.3847/2041-8213/acdc91}}.

\bibitem{NSvIR:tidalfifthforce:Gresham2022}
M.~I. Gresham, V.~S.~H. Lee, K.~M. Zurek, {Astrophysical observations of a dark
  matter-Baryon fifth force}, JCAP 02 (2023) 048.
\newblock \href {http://arxiv.org/abs/2209.03963} {\path{arXiv:2209.03963}},
  \href {https://doi.org/10.1088/1475-7516/2023/02/048}
  {\path{doi:10.1088/1475-7516/2023/02/048}}.

\bibitem{Gould:1989gw}
A.~Gould, B.~T. Draine, R.~W. Romani, S.~Nussinov, {Neutron Stars: Graveyard of
  Charged Dark Matter}, Phys. Lett. B238 (1990) 337--343.
\newblock \href {https://doi.org/10.1016/0370-2693(90)91745-W}
  {\path{doi:10.1016/0370-2693(90)91745-W}}.

\bibitem{Kouvaris:2010vv}
C.~Kouvaris, P.~Tinyakov, {Can Neutron stars constrain Dark Matter?}, Phys.
  Rev. D82 (2010) 063531.
\newblock \href {http://arxiv.org/abs/1004.0586} {\path{arXiv:1004.0586}},
  \href {https://doi.org/10.1103/PhysRevD.82.063531}
  {\path{doi:10.1103/PhysRevD.82.063531}}.

\bibitem{McDermott:2011jp}
S.~D. McDermott, H.-B. Yu, K.~M. Zurek, {Constraints on Scalar Asymmetric Dark
  Matter from Black Hole Formation in Neutron Stars}, Phys. Rev. D85 (2012)
  023519.
\newblock \href {http://arxiv.org/abs/1103.5472} {\path{arXiv:1103.5472}},
  \href {https://doi.org/10.1103/PhysRevD.85.023519}
  {\path{doi:10.1103/PhysRevD.85.023519}}.

\bibitem{Kouvaris:2011fi}
C.~Kouvaris, P.~Tinyakov, {Excluding Light Asymmetric Bosonic Dark Matter},
  Phys. Rev. Lett. 107 (2011) 091301.
\newblock \href {http://arxiv.org/abs/1104.0382} {\path{arXiv:1104.0382}},
  \href {https://doi.org/10.1103/PhysRevLett.107.091301}
  {\path{doi:10.1103/PhysRevLett.107.091301}}.

\bibitem{Kouvaris:2011gb}
C.~Kouvaris, {Limits on Self-Interacting Dark Matter}, Phys. Rev. Lett. 108
  (2012) 191301.
\newblock \href {http://arxiv.org/abs/1111.4364} {\path{arXiv:1111.4364}},
  \href {https://doi.org/10.1103/PhysRevLett.108.191301}
  {\path{doi:10.1103/PhysRevLett.108.191301}}.

\bibitem{Bramante:2013hn}
J.~Bramante, K.~Fukushima, J.~Kumar, {Constraints on bosonic dark matter from
  observation of old neutron stars}, Phys. Rev. D87~(5) (2013) 055012.
\newblock \href {http://arxiv.org/abs/1301.0036} {\path{arXiv:1301.0036}},
  \href {https://doi.org/10.1103/PhysRevD.87.055012}
  {\path{doi:10.1103/PhysRevD.87.055012}}.

\bibitem{Bell:2013xk}
N.~F. Bell, A.~Melatos, K.~Petraki, {Realistic neutron star constraints on
  bosonic asymmetric dark matter}, Phys. Rev. D87~(12) (2013) 123507.
\newblock \href {http://arxiv.org/abs/1301.6811} {\path{arXiv:1301.6811}},
  \href {https://doi.org/10.1103/PhysRevD.87.123507}
  {\path{doi:10.1103/PhysRevD.87.123507}}.

\bibitem{Bramante:2014zca}
J.~Bramante, T.~Linden, {Detecting Dark Matter with Imploding Pulsars in the
  Galactic Center}, Phys. Rev. Lett. 113~(19) (2014) 191301.
\newblock \href {http://arxiv.org/abs/1405.1031} {\path{arXiv:1405.1031}},
  \href {https://doi.org/10.1103/PhysRevLett.113.191301}
  {\path{doi:10.1103/PhysRevLett.113.191301}}.

\bibitem{Bramante:2016mzo}
J.~Bramante, T.~Linden, {On the $r$-Process Enrichment of Dwarf Spheroidal
  Galaxies}, Astrophys. J. 826~(1) (2016) 57.
\newblock \href {http://arxiv.org/abs/1601.06784} {\path{arXiv:1601.06784}},
  \href {https://doi.org/10.3847/0004-637X/826/1/57}
  {\path{doi:10.3847/0004-637X/826/1/57}}.

\bibitem{Bramante:2017ulk}
J.~Bramante, T.~Linden, Y.-D. Tsai, {Searching for dark matter with neutron
  star mergers and quiet kilonovae}, Phys. Rev. D 97~(5) (2018) 055016.
\newblock \href {http://arxiv.org/abs/1706.00001} {\path{arXiv:1706.00001}},
  \href {https://doi.org/10.1103/PhysRevD.97.055016}
  {\path{doi:10.1103/PhysRevD.97.055016}}.

\bibitem{Garani:2018kkd}
R.~Garani, Y.~Genolini, T.~Hambye, {New Analysis of Neutron Star Constraints on
  Asymmetric Dark Matter}, JCAP 1905~(05) (2019) 035.
\newblock \href {http://arxiv.org/abs/1812.08773} {\path{arXiv:1812.08773}},
  \href {https://doi.org/10.1088/1475-7516/2019/05/035}
  {\path{doi:10.1088/1475-7516/2019/05/035}}.

\bibitem{Kouvaris:2018wnh}
C.~Kouvaris, P.~Tinyakov, M.~H. Tytgat, {NonPrimordial Solar Mass Black Holes},
  Phys. Rev. Lett. 121~(22) (2018) 221102.
\newblock \href {http://arxiv.org/abs/1804.06740} {\path{arXiv:1804.06740}},
  \href {https://doi.org/10.1103/PhysRevLett.121.221102}
  {\path{doi:10.1103/PhysRevLett.121.221102}}.

\bibitem{Kopp:2018jom}
J.~Kopp, R.~Laha, T.~Opferkuch, W.~Shepherd, {Cuckoo’s eggs in neutron stars:
  can LIGO hear chirps from the dark sector?}, JHEP 11 (2018) 096.
\newblock \href {http://arxiv.org/abs/1807.02527} {\path{arXiv:1807.02527}},
  \href {https://doi.org/10.1007/JHEP11(2018)096}
  {\path{doi:10.1007/JHEP11(2018)096}}.

\bibitem{East:2019dxt}
W.~E. East, L.~Lehner, {Fate of a neutron star with an endoparasitic black hole
  and implications for dark matter}, Phys. Rev. D100~(12) (2019) 124026.
\newblock \href {http://arxiv.org/abs/1909.07968} {\path{arXiv:1909.07968}},
  \href {https://doi.org/10.1103/PhysRevD.100.124026}
  {\path{doi:10.1103/PhysRevD.100.124026}}.

\bibitem{Tsai:2020hpi}
Y.-D. Tsai, A.~Palmese, S.~Profumo, T.~Jeltema, {Is GW170817 a Multimessenger
  Neutron Star-Primordial Black Hole Merger?}, JCAP 10 (2021) 019.
\newblock \href {http://arxiv.org/abs/2007.03686} {\path{arXiv:2007.03686}},
  \href {https://doi.org/10.1088/1475-7516/2021/10/019}
  {\path{doi:10.1088/1475-7516/2021/10/019}}.

\bibitem{Takhistov:2020vxs}
V.~Takhistov, G.~M. Fuller, A.~Kusenko, {Test for the Origin of Solar Mass
  Black Holes}, Phys. Rev. Lett. 126~(7) (2021) 071101.
\newblock \href {http://arxiv.org/abs/2008.12780} {\path{arXiv:2008.12780}},
  \href {https://doi.org/10.1103/PhysRevLett.126.071101}
  {\path{doi:10.1103/PhysRevLett.126.071101}}.

\bibitem{Dasgupta:2020mqg}
B.~Dasgupta, R.~Laha, A.~Ray, {Low Mass Black Holes from Dark Core Collapse} (9
  2020).
\newblock \href {http://arxiv.org/abs/2009.01825} {\path{arXiv:2009.01825}}.

\bibitem{Garani:2021gvc}
R.~Garani, D.~Levkov, P.~Tinyakov, {Solar mass black holes from neutron stars
  and bosonic dark matter}, Phys. Rev. D 105~(6) (2022) 063019.
\newblock \href {http://arxiv.org/abs/2112.09716} {\path{arXiv:2112.09716}},
  \href {https://doi.org/10.1103/PhysRevD.105.063019}
  {\path{doi:10.1103/PhysRevD.105.063019}}.

\bibitem{Giffin:2021kgb}
P.~Giffin, J.~Lloyd, S.~D. McDermott, S.~Profumo, {Neutron star quantum death
  by small black holes}, Phys. Rev. D 105~(12) (2022) 123030.
\newblock \href {http://arxiv.org/abs/2105.06504} {\path{arXiv:2105.06504}},
  \href {https://doi.org/10.1103/PhysRevD.105.123030}
  {\path{doi:10.1103/PhysRevD.105.123030}}.

\bibitem{Ray:2023auh}
A.~Ray, {Celestial objects as strongly-interacting nonannihilating dark matter
  detectors}, Phys. Rev. D 107~(8) (2023) 083012.
\newblock \href {http://arxiv.org/abs/2301.03625} {\path{arXiv:2301.03625}},
  \href {https://doi.org/10.1103/PhysRevD.107.083012}
  {\path{doi:10.1103/PhysRevD.107.083012}}.

\bibitem{Liu:2024qbe}
N.~Liu, A.~K. Mishra, {Neutron star collapse from accretion: A probe of massive
  dark matter particles}, Phys. Dark Univ. 47 (2025) 101740.
\newblock \href {http://arxiv.org/abs/2408.00594} {\path{arXiv:2408.00594}},
  \href {https://doi.org/10.1016/j.dark.2024.101740}
  {\path{doi:10.1016/j.dark.2024.101740}}.

\bibitem{Dutta:2024vzw}
K.~Dutta, D.~Ghosh, B.~Mukhopadhyaya, {Improved Treatment of Bosonic Dark
  Matter Dynamics in Neutron Stars: Consequences and Constraints} (8 2024).
\newblock \href {http://arxiv.org/abs/2408.16091} {\path{arXiv:2408.16091}}.

\bibitem{Diks:2024cww}
J.~Diks, C.~Ilie, {Constraining Asymmetric DM Properties by Black Hole
  Formation in Neutron Stars and Population III Stars} (12 2024).
\newblock \href {http://arxiv.org/abs/2412.07953} {\path{arXiv:2412.07953}}.

\bibitem{Fuller:2014rza}
J.~Fuller, C.~Ott, {Dark Matter-induced Collapse of Neutron Stars: A Possible
  Link Between Fast Radio Bursts and the Missing Pulsar Problem}, Mon. Not.
  Roy. Astron. Soc. 450~(1) (2015) L71--L75.
\newblock \href {http://arxiv.org/abs/1412.6119} {\path{arXiv:1412.6119}},
  \href {https://doi.org/10.1093/mnrasl/slv049}
  {\path{doi:10.1093/mnrasl/slv049}}.

\bibitem{Zurek:2013wia}
K.~M. Zurek, {Asymmetric Dark Matter: Theories, Signatures, and Constraints},
  Phys. Rept. 537 (2014) 91--121.
\newblock \href {http://arxiv.org/abs/1308.0338} {\path{arXiv:1308.0338}},
  \href {https://doi.org/10.1016/j.physrep.2013.12.001}
  {\path{doi:10.1016/j.physrep.2013.12.001}}.

\bibitem{asymmetricannih:Kumar:2013vba}
J.~Kumar, {Asymmetric Dark Matter}, AIP Conf. Proc. 1604~(1) (2015) 389--396.
\newblock \href {http://arxiv.org/abs/1308.4513} {\path{arXiv:1308.4513}},
  \href {https://doi.org/10.1063/1.4883455} {\path{doi:10.1063/1.4883455}}.

\bibitem{Bramante:2013nma}
J.~Bramante, K.~Fukushima, J.~Kumar, E.~Stopnitzky, {Bounds on self-interacting
  fermion dark matter from observations of old neutron stars}, Phys. Rev.
  D89~(1) (2014) 015010.
\newblock \href {http://arxiv.org/abs/1310.3509} {\path{arXiv:1310.3509}},
  \href {https://doi.org/10.1103/PhysRevD.89.015010}
  {\path{doi:10.1103/PhysRevD.89.015010}}.

\bibitem{PBHreviewGreenKavanagh:2020jor}
A.~M. Green, B.~J. Kavanagh, {Primordial Black Holes as a dark matter
  candidate}, J. Phys. G 48~(4) (2021) 043001.
\newblock \href {http://arxiv.org/abs/2007.10722} {\path{arXiv:2007.10722}},
  \href {https://doi.org/10.1088/1361-6471/abc534}
  {\path{doi:10.1088/1361-6471/abc534}}.

\bibitem{PBHreviewCarr:2016drx}
B.~Carr, F.~Kuhnel, M.~Sandstad, {Primordial Black Holes as Dark Matter}, Phys.
  Rev. D 94~(8) (2016) 083504.
\newblock \href {http://arxiv.org/abs/1607.06077} {\path{arXiv:1607.06077}},
  \href {https://doi.org/10.1103/PhysRevD.94.083504}
  {\path{doi:10.1103/PhysRevD.94.083504}}.

\bibitem{Capela:2012jz}
F.~Capela, M.~Pshirkov, P.~Tinyakov, {Constraints on Primordial Black Holes as
  Dark Matter Candidates from Star Formation}, Phys. Rev. D 87~(2) (2013)
  023507.
\newblock \href {http://arxiv.org/abs/1209.6021} {\path{arXiv:1209.6021}},
  \href {https://doi.org/10.1103/PhysRevD.87.023507}
  {\path{doi:10.1103/PhysRevD.87.023507}}.

\bibitem{Capela:2013yf}
F.~Capela, M.~Pshirkov, P.~Tinyakov, {Constraints on primordial black holes as
  dark matter candidates from capture by neutron stars}, Phys. Rev. D 87~(12)
  (2013) 123524.
\newblock \href {http://arxiv.org/abs/1301.4984} {\path{arXiv:1301.4984}},
  \href {https://doi.org/10.1103/PhysRevD.87.123524}
  {\path{doi:10.1103/PhysRevD.87.123524}}.

\bibitem{Pani:2014rca}
P.~Pani, A.~Loeb, {Tidal capture of a primordial black hole by a neutron star:
  implications for constraints on dark matter}, JCAP 06 (2014) 026.
\newblock \href {http://arxiv.org/abs/1401.3025} {\path{arXiv:1401.3025}},
  \href {https://doi.org/10.1088/1475-7516/2014/06/026}
  {\path{doi:10.1088/1475-7516/2014/06/026}}.

\bibitem{Kurita:2015vga}
Y.~Kurita, H.~Nakano, {Gravitational waves from dark matter collapse in a
  star}, Phys. Rev. D93~(2) (2016) 023508.
\newblock \href {http://arxiv.org/abs/1510.00893} {\path{arXiv:1510.00893}},
  \href {https://doi.org/10.1103/PhysRevD.93.023508}
  {\path{doi:10.1103/PhysRevD.93.023508}}.

\bibitem{Abramowicz:2017zbp}
M.~A. Abramowicz, M.~Bejger, M.~Wielgus, {Collisions of neutron stars with
  primordial black holes as fast radio bursts engines}, Astrophys. J. 868~(1)
  (2018) 17.
\newblock \href {http://arxiv.org/abs/1704.05931} {\path{arXiv:1704.05931}},
  \href {https://doi.org/10.3847/1538-4357/aae64a}
  {\path{doi:10.3847/1538-4357/aae64a}}.

\bibitem{Takhistov:2017nmt}
V.~Takhistov, {Positrons from Primordial Black Hole Microquasars and Gamma-ray
  Bursts}, Phys. Lett. B 789 (2019) 538--544.
\newblock \href {http://arxiv.org/abs/1710.09458} {\path{arXiv:1710.09458}},
  \href {https://doi.org/10.1016/j.physletb.2018.12.043}
  {\path{doi:10.1016/j.physletb.2018.12.043}}.

\bibitem{Takhistov:2017bpt}
V.~Takhistov, {Transmuted Gravity Wave Signals from Primordial Black Holes},
  Phys. Lett. B 782 (2018) 77--82.
\newblock \href {http://arxiv.org/abs/1707.05849} {\path{arXiv:1707.05849}},
  \href {https://doi.org/10.1016/j.physletb.2018.05.026}
  {\path{doi:10.1016/j.physletb.2018.05.026}}.

\bibitem{PBHvsNSGenoliniTinyakov:2020ejw}
Y.~Genolini, P.~Serpico, P.~Tinyakov, {Revisiting primordial black hole capture
  into neutron stars}, Phys. Rev. D 102~(8) (2020) 083004.
\newblock \href {http://arxiv.org/abs/2006.16975} {\path{arXiv:2006.16975}},
  \href {https://doi.org/10.1103/PhysRevD.102.083004}
  {\path{doi:10.1103/PhysRevD.102.083004}}.

\bibitem{Baumgarte:2021thx}
T.~W. Baumgarte, S.~L. Shapiro, {Neutron Stars Harboring a Primordial Black
  Hole: Maximum Survival Time}, Phys. Rev. D 103~(8) (2021) L081303.
\newblock \href {http://arxiv.org/abs/2101.12220} {\path{arXiv:2101.12220}},
  \href {https://doi.org/10.1103/PhysRevD.103.L081303}
  {\path{doi:10.1103/PhysRevD.103.L081303}}.

\bibitem{Estes:2022buj}
J.~Estes, M.~Kavic, S.~L. Liebling, M.~Lippert, J.~H. Simonetti, {Stability and
  observability of magnetic primordial black hole-neutron star collisions},
  JCAP 06 (2023) 017.
\newblock \href {http://arxiv.org/abs/2209.06060} {\path{arXiv:2209.06060}},
  \href {https://doi.org/10.1088/1475-7516/2023/06/017}
  {\path{doi:10.1088/1475-7516/2023/06/017}}.

\bibitem{Zenati:2023ckz}
Y.~Zenati, C.~Albertus, M.~A. P\'erez-Garc\'\i{}a, J.~Silk, {Neutrino signals
  from Neutron Star implosions to Black Holes} (4 2023).
\newblock \href {http://arxiv.org/abs/2304.06746} {\path{arXiv:2304.06746}}.

\bibitem{Bramante:2015dfa}
J.~Bramante, F.~Elahi, {Higgs portals to pulsar collapse}, Phys. Rev. D91~(11)
  (2015) 115001.
\newblock \href {http://arxiv.org/abs/1504.04019} {\path{arXiv:1504.04019}},
  \href {https://doi.org/10.1103/PhysRevD.91.115001}
  {\path{doi:10.1103/PhysRevD.91.115001}}.

\bibitem{Gresham:2018rqo}
M.~I. Gresham, K.~M. Zurek, {Asymmetric Dark Stars and Neutron Star Stability},
  Phys. Rev. D99~(8) (2019) 083008.
\newblock \href {http://arxiv.org/abs/1809.08254} {\path{arXiv:1809.08254}},
  \href {https://doi.org/10.1103/PhysRevD.99.083008}
  {\path{doi:10.1103/PhysRevD.99.083008}}.

\bibitem{Kouvaris:2012dz}
C.~Kouvaris, P.~Tinyakov, {(Not)-constraining heavy asymmetric bosonic dark
  matter}, Phys. Rev. D 87~(12) (2013) 123537.
\newblock \href {http://arxiv.org/abs/1212.4075} {\path{arXiv:1212.4075}},
  \href {https://doi.org/10.1103/PhysRevD.87.123537}
  {\path{doi:10.1103/PhysRevD.87.123537}}.

\bibitem{Garani:2022quc}
R.~Garani, M.~H.~G. Tytgat, J.~Vandecasteele, {Condensed dark matter with a
  Yukawa interaction}, Phys. Rev. D 106~(11) (2022) 116003.
\newblock \href {http://arxiv.org/abs/2207.06928} {\path{arXiv:2207.06928}},
  \href {https://doi.org/10.1103/PhysRevD.106.116003}
  {\path{doi:10.1103/PhysRevD.106.116003}}.

\bibitem{Kouvaris:2013kra}
C.~Kouvaris, P.~Tinyakov, {Growth of Black Holes in the interior of Rotating
  Neutron Stars}, Phys. Rev. D 90~(4) (2014) 043512.
\newblock \href {http://arxiv.org/abs/1312.3764} {\path{arXiv:1312.3764}},
  \href {https://doi.org/10.1103/PhysRevD.90.043512}
  {\path{doi:10.1103/PhysRevD.90.043512}}.

\bibitem{Autzen:2014tza}
M.~Autzen, C.~Kouvaris, {Blocking the Hawking Radiation}, Phys. Rev. D 89~(12)
  (2014) 123519.
\newblock \href {http://arxiv.org/abs/1403.1072} {\path{arXiv:1403.1072}},
  \href {https://doi.org/10.1103/PhysRevD.89.123519}
  {\path{doi:10.1103/PhysRevD.89.123519}}.

\bibitem{Basumatary:2024uwo}
U.~Basumatary, N.~Raj, A.~Ray, {Beyond Hawking evaporation of black holes
  formed by dark matter in compact stars} (10 2024).
\newblock \href {http://arxiv.org/abs/2410.22702} {\path{arXiv:2410.22702}}.

\bibitem{Liang:2023nvo}
D.~Liang, L.~Shao, {Improved bounds on the bosonic dark matter with pulsars in
  the Milky Way} (3 2023).
\newblock \href {http://arxiv.org/abs/2303.05107} {\path{arXiv:2303.05107}}.

\bibitem{Dexter:2013xga}
J.~Dexter, R.~M. O'Leary, {The Peculiar Pulsar Population of the Central
  Parsec}, Astrophys. J. Lett. 783 (2014) L7.
\newblock \href {http://arxiv.org/abs/1310.7022} {\path{arXiv:1310.7022}},
  \href {https://doi.org/10.1088/2041-8205/783/1/L7}
  {\path{doi:10.1088/2041-8205/783/1/L7}}.

\bibitem{Suresh:2022vmf}
A.~Suresh, J.~M. Cordes, S.~Chatterjee, V.~Gajjar, K.~I. Perez, A.~P.~V.
  Siemion, M.~Lebofsky, D.~H.~E. MacMahon, C.~Ng, {4\textendash{}8 GHz
  Fourier-domain Searches for Galactic Center Pulsars}, Astrophys. J. 933~(2)
  (2022) 121.
\newblock \href {http://arxiv.org/abs/2203.00036} {\path{arXiv:2203.00036}},
  \href {https://doi.org/10.3847/1538-4357/ac74c0}
  {\path{doi:10.3847/1538-4357/ac74c0}}.

\bibitem{Fuller:2017uyd}
G.~M. Fuller, A.~Kusenko, V.~Takhistov, {Primordial Black Holes and r-Process
  Nucleosynthesis}, Phys. Rev. Lett. 119~(6) (2017) 061101.
\newblock \href {http://arxiv.org/abs/1704.01129} {\path{arXiv:1704.01129}},
  \href {https://doi.org/10.1103/PhysRevLett.119.061101}
  {\path{doi:10.1103/PhysRevLett.119.061101}}.

\bibitem{Richards:2021upu}
C.~B. Richards, T.~W. Baumgarte, S.~L. Shapiro, {Accretion onto a small black
  hole at the center of a neutron star}, Phys. Rev. D 103~(10) (2021) 104009.
\newblock \href {http://arxiv.org/abs/2102.09574} {\path{arXiv:2102.09574}},
  \href {https://doi.org/10.1103/PhysRevD.103.104009}
  {\path{doi:10.1103/PhysRevD.103.104009}}.

\bibitem{NeutrinosNSBHZenatiSilk:2023ckz}
Y.~Zenati, C.~Albertus, M.~A. P\'erez-Garc\'\i{}a, J.~Silk, {Neutrino signals
  from Neutron Star implosions to Black Holes} (4 2023).
\newblock \href {http://arxiv.org/abs/2304.06746} {\path{arXiv:2304.06746}}.

\bibitem{Bhattacharya:2023stq}
S.~Bhattacharya, B.~Dasgupta, R.~Laha, A.~Ray, {Can LIGO Detect Asymmetric Dark
  Matter?} (2 2023).
\newblock \href {http://arxiv.org/abs/2302.07898} {\path{arXiv:2302.07898}}.

\bibitem{Bhattacharya:2024pmp}
S.~Bhattacharya, A.~L. Miller, A.~Ray, {Continuous gravitational waves: A new
  window to look for heavy nonannihilating dark matter}, Phys. Rev. D 110
  (2024) 043006.
\newblock \href {http://arxiv.org/abs/2403.13886} {\path{arXiv:2403.13886}},
  \href {https://doi.org/10.1103/PhysRevD.110.043006}
  {\path{doi:10.1103/PhysRevD.110.043006}}.

\bibitem{Yang:2017gfb}
H.~Yang, W.~E. East, L.~Lehner, {Can we distinguish low mass black holes in
  neutron star binaries?}, Astrophys. J. 856~(2) (2018) 110, [Erratum:
  Astrophys.J. 870, 139 (2019)].
\newblock \href {http://arxiv.org/abs/1710.05891} {\path{arXiv:1710.05891}},
  \href {https://doi.org/10.3847/1538-4357/aab2b0}
  {\path{doi:10.3847/1538-4357/aab2b0}}.

\bibitem{PBHvsNSCapelaTinyakov:2013yf}
F.~Capela, M.~Pshirkov, P.~Tinyakov, {Constraints on primordial black holes as
  dark matter candidates from capture by neutron stars}, Phys. Rev. D 87~(12)
  (2013) 123524.
\newblock \href {http://arxiv.org/abs/1301.4984} {\path{arXiv:1301.4984}},
  \href {https://doi.org/10.1103/PhysRevD.87.123524}
  {\path{doi:10.1103/PhysRevD.87.123524}}.

\bibitem{PBHvsNSLoebPani:2014rca}
P.~Pani, A.~Loeb, {Tidal capture of a primordial black hole by a neutron star:
  implications for constraints on dark matter}, JCAP 06 (2014) 026.
\newblock \href {http://arxiv.org/abs/1401.3025} {\path{arXiv:1401.3025}},
  \href {https://doi.org/10.1088/1475-7516/2014/06/026}
  {\path{doi:10.1088/1475-7516/2014/06/026}}.

\bibitem{PBHvsNSDefillonTinyakov:2014wla}
G.~Defillon, E.~Granet, P.~Tinyakov, M.~H.~G. Tytgat, {Tidal capture of
  primordial black holes by neutron stars}, Phys. Rev. D 90~(10) (2014) 103522.
\newblock \href {http://arxiv.org/abs/1409.0469} {\path{arXiv:1409.0469}},
  \href {https://doi.org/10.1103/PhysRevD.90.103522}
  {\path{doi:10.1103/PhysRevD.90.103522}}.

\bibitem{PBHvsNS-Bondi-Richards:2021upu}
C.~B. Richards, T.~W. Baumgarte, S.~L. Shapiro, {Accretion onto a small black
  hole at the center of a neutron star}, Phys. Rev. D 103~(10) (2021) 104009.
\newblock \href {http://arxiv.org/abs/2102.09574} {\path{arXiv:2102.09574}},
  \href {https://doi.org/10.1103/PhysRevD.103.104009}
  {\path{doi:10.1103/PhysRevD.103.104009}}.

\bibitem{PBHvsNS-survivaltime-Baumgarte:2021thx}
T.~W. Baumgarte, S.~L. Shapiro, {Neutron Stars Harboring a Primordial Black
  Hole: Maximum Survival Time}, Phys. Rev. D 103~(8) (2021) L081303.
\newblock \href {http://arxiv.org/abs/2101.12220} {\path{arXiv:2101.12220}},
  \href {https://doi.org/10.1103/PhysRevD.103.L081303}
  {\path{doi:10.1103/PhysRevD.103.L081303}}.

\bibitem{PBHvNS-FRB-Kainulainen:2021rbg}
K.~Kainulainen, S.~Nurmi, E.~D. Schiappacasse, T.~T. Yanagida, {Can primordial
  black holes as all dark matter explain fast radio bursts?}, Phys. Rev. D
  104~(12) (2021) 123033.
\newblock \href {http://arxiv.org/abs/2108.08717} {\path{arXiv:2108.08717}},
  \href {https://doi.org/10.1103/PhysRevD.104.123033}
  {\path{doi:10.1103/PhysRevD.104.123033}}.

\bibitem{microlens:NarayanBartelmann}
R.~Narayan, M.~Bartelmann, {Lectures on gravitational lensing}, in: {13th
  Jerusalem Winter School in Theoretical Physics: Formation of Structure in the
  Universe}, 1996.
\newblock \href {http://arxiv.org/abs/astro-ph/9606001}
  {\path{arXiv:astro-ph/9606001}}.

\bibitem{microlens:erosogle}
D.~Croon, D.~McKeen, N.~Raj, {Gravitational microlensing by dark matter in
  extended structures}, Phys. Rev. D 101~(8) (2020) 083013.
\newblock \href {http://arxiv.org/abs/2002.08962} {\path{arXiv:2002.08962}},
  \href {https://doi.org/10.1103/PhysRevD.101.083013}
  {\path{doi:10.1103/PhysRevD.101.083013}}.

\bibitem{microlens:subaru}
D.~Croon, D.~McKeen, N.~Raj, Z.~Wang, {Subaru-HSC through a different lens:
  Microlensing by extended dark matter structures}, Phys. Rev. D 102~(8) (2020)
  083021.
\newblock \href {http://arxiv.org/abs/2007.12697} {\path{arXiv:2007.12697}},
  \href {https://doi.org/10.1103/PhysRevD.102.083021}
  {\path{doi:10.1103/PhysRevD.102.083021}}.

\bibitem{Nakamura:1999:waveoptix}
T.~T. Nakamura, S.~Deguchi, {Wave Optics in Gravitational Lensing}, Prog.
  Theor. Phys. Suppl. 133 (1999) 137--153.
\newblock \href {https://doi.org/10.1143/ptps.133.137}
  {\path{doi:10.1143/ptps.133.137}}.

\bibitem{Matsunaga2006FiniteSourceWaveOptix}
N.~Matsunaga, K.~Yamamoto,
  \href{http://dx.doi.org/10.1088/1475-7516/2006/01/023}{The finite source size
  effect and wave optics in gravitational lensing}, Journal of Cosmology and
  Astroparticle Physics 2006~(01) (2006) 023–023.
\newblock \href {https://doi.org/10.1088/1475-7516/2006/01/023}
  {\path{doi:10.1088/1475-7516/2006/01/023}}.
\newline\urlprefix\url{http://dx.doi.org/10.1088/1475-7516/2006/01/023}

\bibitem{SugiyamaWaveptics:2019dgt}
S.~Sugiyama, T.~Kurita, M.~Takada, {On the wave optics effect on primordial
  black hole constraints from optical microlensing search}, Mon. Not. Roy.
  Astron. Soc. 493~(3) (2020) 3632--3641.
\newblock \href {http://arxiv.org/abs/1905.06066} {\path{arXiv:1905.06066}},
  \href {https://doi.org/10.1093/mnras/staa407}
  {\path{doi:10.1093/mnras/staa407}}.

\bibitem{microlens:xray:Bai:2018bej}
Y.~Bai, N.~Orlofsky, {Microlensing of X-ray Pulsars: a Method to Detect
  Primordial Black Hole Dark Matter}, Phys. Rev. D 99~(12) (2019) 123019.
\newblock \href {http://arxiv.org/abs/1812.01427} {\path{arXiv:1812.01427}},
  \href {https://doi.org/10.1103/PhysRevD.99.123019}
  {\path{doi:10.1103/PhysRevD.99.123019}}.

\bibitem{microlens:xray:Tamta:2024pow}
M.~Tamta, N.~Raj, P.~Sharma, {Breaking into the window of primordial black hole
  dark matter with x-ray microlensing} (5 2024).
\newblock \href {http://arxiv.org/abs/2405.20365} {\path{arXiv:2405.20365}}.

\bibitem{NICERDesign2016}
K.~C. {Gendreau}, et~al., {The Neutron star Interior Composition Explorer
  (NICER): design and development}, in: J.-W.~A. {den Herder}, T.~{Takahashi},
  M.~{Bautz} (Eds.), Space Telescopes and Instrumentation 2016: Ultraviolet to
  Gamma Ray, Vol. 9905 of Society of Photo-Optical Instrumentation Engineers
  (SPIE) Conference Series, 2016, p. 99051H.
\newblock \href {https://doi.org/10.1117/12.2231304}
  {\path{doi:10.1117/12.2231304}}.

\bibitem{STROBE-XScienceWorkingGroup:2019cyd}
P.~S. Ray, et~al., {STROBE-X: X-ray Timing and Spectroscopy on Dynamical
  Timescales from Microseconds to Years} (3 2019).
\newblock \href {http://arxiv.org/abs/1903.03035} {\path{arXiv:1903.03035}}.

\bibitem{Parallaxmicrolens1995}
R.~J. {Nemiroff}, A.~{Gould}, {Probing for MACHOs of Mass 10 -15 M$_{sun}$ to
  10 -7 M$_{sun}$ with Gamma-Ray Burst Parallax Spacecraft}, APJL 452 (1995)
  L111.
\newblock \href {http://arxiv.org/abs/astro-ph/9505019}
  {\path{arXiv:astro-ph/9505019}}, \href {https://doi.org/10.1086/309722}
  {\path{doi:10.1086/309722}}.

\bibitem{Parallaxmicrolens1998}
R.~J. {Nemiroff}, {Attributes of Gravitational Lensing Parallax}, APSS 259~(3)
  (1998) 309--315.
\newblock \href {http://arxiv.org/abs/astro-ph/9806012}
  {\path{arXiv:astro-ph/9806012}}, \href
  {https://doi.org/10.1023/A:1001769807077}
  {\path{doi:10.1023/A:1001769807077}}.

\bibitem{dakshabhalerao2022science}
V.~Bhalerao, et~al., Science with the daksha high energy transients mission,
  arXiv preprint arXiv:2211.12052 (2022).

\bibitem{Gawade:2023gmt}
P.~Gawade, S.~More, V.~Bhalerao, {On the feasibility of primordial black hole
  abundance constraints using lensing parallax of GRBs}, Mon. Not. Roy. Astron.
  Soc. 527~(2) (2023) 3306--3314.
\newblock \href {http://arxiv.org/abs/2308.01775} {\path{arXiv:2308.01775}},
  \href {https://doi.org/10.1093/mnras/stad3336}
  {\path{doi:10.1093/mnras/stad3336}}.

\bibitem{Fedderke:2024wpy}
M.~A. Fedderke, S.~Sibiryakov, {Kicking the Tires on Picolensing as a Probe of
  Primordial Black Hole Dark Matter} (11 2024).
\newblock \href {http://arxiv.org/abs/2411.12947} {\path{arXiv:2411.12947}}.

\bibitem{femtolensing:Katz:2018zrn}
A.~Katz, J.~Kopp, S.~Sibiryakov, W.~Xue, {Femtolensing by Dark Matter
  Revisited}, JCAP 12 (2018) 005.
\newblock \href {http://arxiv.org/abs/1807.11495} {\path{arXiv:1807.11495}},
  \href {https://doi.org/10.1088/1475-7516/2018/12/005}
  {\path{doi:10.1088/1475-7516/2018/12/005}}.

\bibitem{McKeenNelsonReddyZhouNS}
D.~McKeen, A.~E. Nelson, S.~Reddy, D.~Zhou, {Neutron stars exclude light dark
  baryons}, Phys. Rev. Lett. 121~(6) (2018) 061802.
\newblock \href {http://arxiv.org/abs/1802.08244} {\path{arXiv:1802.08244}},
  \href {https://doi.org/10.1103/PhysRevLett.121.061802}
  {\path{doi:10.1103/PhysRevLett.121.061802}}.

\bibitem{NSFastestHeaviest:Romani:2022jhd}
R.~W. Romani, D.~Kandel, A.~V. Filippenko, T.~G. Brink, W.~Zheng, {PSR
  J0952\ensuremath{-}0607: The Fastest and Heaviest Known Galactic Neutron
  Star}, Astrophys. J. Lett. 934~(2) (2022) L17.
\newblock \href {http://arxiv.org/abs/2207.05124} {\path{arXiv:2207.05124}},
  \href {https://doi.org/10.3847/2041-8213/ac8007}
  {\path{doi:10.3847/2041-8213/ac8007}}.

\bibitem{McDermottReddydibaryons:2018ofd}
S.~D. McDermott, S.~Reddy, S.~Sen, {Deeply bound dibaryon is incompatible with
  neutron stars and supernovae}, Phys. Rev. D 99~(3) (2019) 035013.
\newblock \href {http://arxiv.org/abs/1809.06765} {\path{arXiv:1809.06765}},
  \href {https://doi.org/10.1103/PhysRevD.99.035013}
  {\path{doi:10.1103/PhysRevD.99.035013}}.

\bibitem{NSheat:Mirror:McKeen:2021jbh}
D.~McKeen, M.~Pospelov, N.~Raj, {Neutron Star Internal Heating Constraints on
  Mirror Matter}, Phys. Rev. Lett. 127~(6) (2021) 061805.
\newblock \href {http://arxiv.org/abs/2105.09951} {\path{arXiv:2105.09951}},
  \href {https://doi.org/10.1103/PhysRevLett.127.061805}
  {\path{doi:10.1103/PhysRevLett.127.061805}}.

\bibitem{Goldman:2019dbq}
I.~Goldman, R.~N. Mohapatra, S.~Nussinov, {Bounds on neutron-mirror neutron
  mixing from pulsar timing}, Phys. Rev. D 100~(12) (2019) 123021.
\newblock \href {http://arxiv.org/abs/1901.07077} {\path{arXiv:1901.07077}},
  \href {https://doi.org/10.1103/PhysRevD.100.123021}
  {\path{doi:10.1103/PhysRevD.100.123021}}.

\bibitem{Fornal:2018eol}
B.~Fornal, B.~Grinstein, {Dark Matter Interpretation of the Neutron Decay
  Anomaly}, Phys. Rev. Lett. 120~(19) (2018) 191801, [Erratum: Phys.Rev.Lett.
  124, 219901 (2020)].
\newblock \href {http://arxiv.org/abs/1801.01124} {\path{arXiv:1801.01124}},
  \href {https://doi.org/10.1103/PhysRevLett.120.191801}
  {\path{doi:10.1103/PhysRevLett.120.191801}}.

\bibitem{McKeen:2015cuz}
D.~McKeen, A.~E. Nelson, {CP Violating Baryon Oscillations}, Phys. Rev. D
  94~(7) (2016) 076002.
\newblock \href {http://arxiv.org/abs/1512.05359} {\path{arXiv:1512.05359}},
  \href {https://doi.org/10.1103/PhysRevD.94.076002}
  {\path{doi:10.1103/PhysRevD.94.076002}}.

\bibitem{Alonso-Alvarez:2021oaj}
G.~Alonso-\'Alvarez, G.~Elor, M.~Escudero, B.~Fornal, B.~Grinstein, J.~M.
  Camalich, {The Strange Physics of Dark Baryons} (11 2021).
\newblock \href {http://arxiv.org/abs/2111.12712} {\path{arXiv:2111.12712}}.

\bibitem{Berezhiani:2018eds}
Z.~Berezhiani, {Neutron lifetime puzzle and neutron\textendash{}mirror neutron
  oscillation}, Eur. Phys. J. C 79~(6) (2019) 484.
\newblock \href {http://arxiv.org/abs/1807.07906} {\path{arXiv:1807.07906}},
  \href {https://doi.org/10.1140/epjc/s10052-019-6995-x}
  {\path{doi:10.1140/epjc/s10052-019-6995-x}}.

\bibitem{SheltonNS}
G.~Baym, D.~Beck, P.~Geltenbort, J.~Shelton, {Testing dark decays of baryons in
  neutron stars}, Phys. Rev. Lett. 121~(6) (2018) 061801.
\newblock \href {http://arxiv.org/abs/1802.08282} {\path{arXiv:1802.08282}},
  \href {https://doi.org/10.1103/PhysRevLett.121.061801}
  {\path{doi:10.1103/PhysRevLett.121.061801}}.

\bibitem{MottaNS}
T.~Motta, P.~Guichon, A.~Thomas, {Implications of Neutron Star Properties for
  the Existence of Light Dark Matter}, J. Phys. G 45~(5) (2018) 05LT01.
\newblock \href {http://arxiv.org/abs/1802.08427} {\path{arXiv:1802.08427}},
  \href {https://doi.org/10.1088/1361-6471/aab689}
  {\path{doi:10.1088/1361-6471/aab689}}.

\bibitem{Ivanytskyi:2019wxd}
O.~Ivanytskyi, V.~Sagun, I.~Lopes, {Neutron stars: New constraints on
  asymmetric dark matter}, Phys. Rev. D 102~(6) (2020) 063028.
\newblock \href {http://arxiv.org/abs/1910.09925} {\path{arXiv:1910.09925}},
  \href {https://doi.org/10.1103/PhysRevD.102.063028}
  {\path{doi:10.1103/PhysRevD.102.063028}}.

\bibitem{EllisPattavinaNS}
J.~Ellis, G.~H\"utsi, K.~Kannike, L.~Marzola, M.~Raidal, V.~Vaskonen, {Dark
  Matter Effects On Neutron Star Properties}, Phys. Rev. D 97~(12) (2018)
  123007.
\newblock \href {http://arxiv.org/abs/1804.01418} {\path{arXiv:1804.01418}},
  \href {https://doi.org/10.1103/PhysRevD.97.123007}
  {\path{doi:10.1103/PhysRevD.97.123007}}.

\bibitem{Giangrandi:2022wht}
E.~Giangrandi, V.~Sagun, O.~Ivanytskyi, C.~Provid\^encia, T.~Dietrich, {The
  Effects of Self-interacting Bosonic Dark Matter on Neutron Star Properties},
  Astrophys. J. 953~(1) (2023) 115.
\newblock \href {http://arxiv.org/abs/2209.10905} {\path{arXiv:2209.10905}},
  \href {https://doi.org/10.3847/1538-4357/ace104}
  {\path{doi:10.3847/1538-4357/ace104}}.

\bibitem{Berryman:2022zic}
J.~M. Berryman, S.~Gardner, M.~Zakeri, {Neutron Stars with Baryon Number
  Violation, Probing Dark Sectors}, Symmetry 14~(3) (2022) 518.
\newblock \href {http://arxiv.org/abs/2201.02637} {\path{arXiv:2201.02637}},
  \href {https://doi.org/10.3390/sym14030518} {\path{doi:10.3390/sym14030518}}.

\bibitem{Rutherford:2022xeb}
N.~Rutherford, G.~Raaijmakers, C.~Prescod-Weinstein, A.~Watts, {Constraining
  bosonic asymmetric dark matter with neutron star mass-radius measurements},
  Phys. Rev. D 107~(10) (2023) 103051.
\newblock \href {http://arxiv.org/abs/2208.03282} {\path{arXiv:2208.03282}},
  \href {https://doi.org/10.1103/PhysRevD.107.103051}
  {\path{doi:10.1103/PhysRevD.107.103051}}.

\bibitem{Lopes:2018oao}
I.~Lopes, G.~Panotopoulos, {Dark matter admixed strange quark stars in the
  Starobinsky model}, Phys. Rev. D 97~(2) (2018) 024030.
\newblock \href {http://arxiv.org/abs/1801.05031} {\path{arXiv:1801.05031}},
  \href {https://doi.org/10.1103/PhysRevD.97.024030}
  {\path{doi:10.1103/PhysRevD.97.024030}}.

\bibitem{Panotopoulos:2018joc}
G.~Panotopoulos, I.~Lopes, {Millisecond pulsars modeled as strange quark stars
  admixed with condensed dark matter}, Int. J. Mod. Phys. D 27~(09) (2018)
  1850093.
\newblock \href {http://arxiv.org/abs/1804.05023} {\path{arXiv:1804.05023}},
  \href {https://doi.org/10.1142/S0218271818500931}
  {\path{doi:10.1142/S0218271818500931}}.

\bibitem{Panotopoulos:2018ipq}
G.~Panotopoulos, I.~Lopes, {Radial oscillations of strange quark stars admixed
  with fermionic dark matter}, Phys. Rev. D 98~(8) (2018) 083001.
\newblock \href {https://doi.org/10.1103/PhysRevD.98.083001}
  {\path{doi:10.1103/PhysRevD.98.083001}}.

\bibitem{StrumiaNS:2021ybk}
A.~Strumia, {Dark Matter interpretation of the neutron decay anomaly}, JHEP 02
  (2022) 067.
\newblock \href {http://arxiv.org/abs/2112.09111} {\path{arXiv:2112.09111}},
  \href {https://doi.org/10.1007/JHEP02(2022)067}
  {\path{doi:10.1007/JHEP02(2022)067}}.

\bibitem{Narain:2006kx}
G.~Narain, J.~Schaffner-Bielich, I.~N. Mishustin, {Compact stars made of
  fermionic dark matter}, Phys. Rev. D 74 (2006) 063003.
\newblock \href {http://arxiv.org/abs/astro-ph/0605724}
  {\path{arXiv:astro-ph/0605724}}, \href
  {https://doi.org/10.1103/PhysRevD.74.063003}
  {\path{doi:10.1103/PhysRevD.74.063003}}.

\bibitem{Cline:2018ami}
J.~M. Cline, J.~M. Cornell, {Dark decay of the neutron}, JHEP 07 (2018) 081.
\newblock \href {http://arxiv.org/abs/1803.04961} {\path{arXiv:1803.04961}},
  \href {https://doi.org/10.1007/JHEP07(2018)081}
  {\path{doi:10.1007/JHEP07(2018)081}}.

\bibitem{Hippert:2022snq}
M.~Hippert, E.~Dillingham, H.~Tan, D.~Curtin, J.~Noronha-Hostler, N.~Yunes,
  {Dark Matter or Regular Matter in Neutron Stars? How to tell the difference
  from the coalescence of compact objects} (11 2022).
\newblock \href {http://arxiv.org/abs/2211.08590} {\path{arXiv:2211.08590}}.

\bibitem{Collier:2022cpr}
M.~Collier, D.~Croon, R.~K. Leane, {Tidal Love numbers of novel and admixed
  celestial objects}, Phys. Rev. D 106~(12) (2022) 123027.
\newblock \href {http://arxiv.org/abs/2205.15337} {\path{arXiv:2205.15337}},
  \href {https://doi.org/10.1103/PhysRevD.106.123027}
  {\path{doi:10.1103/PhysRevD.106.123027}}.

\bibitem{Karkevandi:2021ygv}
D.~R. Karkevandi, S.~Shakeri, V.~Sagun, O.~Ivanytskyi, {Bosonic dark matter in
  neutron stars and its effect on gravitational wave signal}, Phys. Rev. D
  105~(2) (2022) 023001.
\newblock \href {http://arxiv.org/abs/2109.03801} {\path{arXiv:2109.03801}},
  \href {https://doi.org/10.1103/PhysRevD.105.023001}
  {\path{doi:10.1103/PhysRevD.105.023001}}.

\bibitem{Dengler:2021qcq}
Y.~Dengler, J.~Schaffner-Bielich, L.~Tolos, {Second Love number of dark compact
  planets and neutron stars with dark matter}, Phys. Rev. D 105~(4) (2022)
  043013.
\newblock \href {http://arxiv.org/abs/2111.06197} {\path{arXiv:2111.06197}},
  \href {https://doi.org/10.1103/PhysRevD.105.043013}
  {\path{doi:10.1103/PhysRevD.105.043013}}.

\bibitem{Shakeri:2022dwg}
S.~Shakeri, D.~R. Karkevandi, {Bosonic Dark Matter in Light of the NICER
  Precise Mass-Radius Measurements} (10 2022).
\newblock \href {http://arxiv.org/abs/2210.17308} {\path{arXiv:2210.17308}}.

\bibitem{Farrar:2017eqq}
G.~R. Farrar, {Stable Sexaquark} (8 2017).
\newblock \href {http://arxiv.org/abs/1708.08951} {\path{arXiv:1708.08951}}.

\bibitem{Farrar:2018hac}
G.~R. Farrar, {A precision test of the nature of Dark Matter and a probe of the
  QCD phase transition} (5 2018).
\newblock \href {http://arxiv.org/abs/1805.03723} {\path{arXiv:1805.03723}}.

\bibitem{KolbTurnerdibaryons:2018bxv}
E.~W. Kolb, M.~S. Turner, {Dibaryons cannot be the dark matter}, Phys. Rev. D
  99~(6) (2019) 063519.
\newblock \href {http://arxiv.org/abs/1809.06003} {\path{arXiv:1809.06003}},
  \href {https://doi.org/10.1103/PhysRevD.99.063519}
  {\path{doi:10.1103/PhysRevD.99.063519}}.

\bibitem{hexaquarkNS:ShahrbafFarrar:2022upc}
M.~Shahrbaf, D.~Blaschke, S.~Typel, G.~R. Farrar, D.~E. Alvarez-Castillo,
  {Sexaquark dilemma in neutron stars and its solution by quark deconfinement},
  Phys. Rev. D 105~(10) (2022) 103005.
\newblock \href {http://arxiv.org/abs/2202.00652} {\path{arXiv:2202.00652}},
  \href {https://doi.org/10.1103/PhysRevD.105.103005}
  {\path{doi:10.1103/PhysRevD.105.103005}}.

\bibitem{ReviewAdmixedNSDM:GrippaPoddar:2024ach}
F.~Grippa, G.~Lambiase, T.~K. Poddar, {Searching for New Physics in Ultradense
  Environment: a Review on Dark Matter Admixed Neutron Stars} (12 2024).
\newblock \href {http://arxiv.org/abs/2412.09381} {\path{arXiv:2412.09381}}.

\bibitem{Foot:2004pa}
R.~Foot, {Mirror matter-type dark matter}, Int. J. Mod. Phys. D 13 (2004)
  2161--2192.
\newblock \href {http://arxiv.org/abs/astro-ph/0407623}
  {\path{arXiv:astro-ph/0407623}}, \href
  {https://doi.org/10.1142/S0218271804006449}
  {\path{doi:10.1142/S0218271804006449}}.

\bibitem{Fan:2013yva}
J.~Fan, A.~Katz, L.~Randall, M.~Reece, {Double-Disk Dark Matter}, Phys. Dark
  Univ. 2 (2013) 139--156.
\newblock \href {http://arxiv.org/abs/1303.1521} {\path{arXiv:1303.1521}},
  \href {https://doi.org/10.1016/j.dark.2013.07.001}
  {\path{doi:10.1016/j.dark.2013.07.001}}.

\bibitem{AngelesPerez-Garcia:2022qzs}
M.~\'Angeles P\'erez-Garc\'\i{}a, H.~Grigorian, C.~Albertus, D.~Barba, J.~Silk,
  {Cooling of Neutron Stars admixed with light dark matter: A case study},
  Phys. Lett. B 827 (2022) 136937.
\newblock \href {http://arxiv.org/abs/2202.00702} {\path{arXiv:2202.00702}},
  \href {https://doi.org/10.1016/j.physletb.2022.136937}
  {\path{doi:10.1016/j.physletb.2022.136937}}.

\bibitem{Hippert:2021fch}
M.~Hippert, J.~Setford, H.~Tan, D.~Curtin, J.~Noronha-Hostler, N.~Yunes,
  {Mirror neutron stars}, Phys. Rev. D 106~(3) (2022) 035025.
\newblock \href {http://arxiv.org/abs/2103.01965} {\path{arXiv:2103.01965}},
  \href {https://doi.org/10.1103/PhysRevD.106.035025}
  {\path{doi:10.1103/PhysRevD.106.035025}}.

\bibitem{Ryan:2022hku}
M.~Ryan, D.~Radice, {Exotic compact objects: The dark white dwarf}, Phys. Rev.
  D 105~(11) (2022) 115034.
\newblock \href {http://arxiv.org/abs/2201.05626} {\path{arXiv:2201.05626}},
  \href {https://doi.org/10.1103/PhysRevD.105.115034}
  {\path{doi:10.1103/PhysRevD.105.115034}}.

\bibitem{Howe:2021neq}
A.~Howe, J.~Setford, D.~Curtin, C.~D. Matzner, {How to search for mirror stars
  with Gaia}, JHEP 07 (2022) 059.
\newblock \href {http://arxiv.org/abs/2112.05766} {\path{arXiv:2112.05766}},
  \href {https://doi.org/10.1007/JHEP07(2022)059}
  {\path{doi:10.1007/JHEP07(2022)059}}.

\bibitem{Berezhiani:2021src}
Z.~Berezhiani, {Antistars or Antimatter Cores in Mirror Neutron Stars?},
  Universe 8~(6) (2022) 313.
\newblock \href {http://arxiv.org/abs/2106.11203} {\path{arXiv:2106.11203}},
  \href {https://doi.org/10.3390/universe8060313}
  {\path{doi:10.3390/universe8060313}}.

\bibitem{Kouvaris:2015rea}
C.~Kouvaris, N.~G. Nielsen, {Asymmetric Dark Matter Stars}, Phys. Rev. D 92~(6)
  (2015) 063526.
\newblock \href {http://arxiv.org/abs/1507.00959} {\path{arXiv:1507.00959}},
  \href {https://doi.org/10.1103/PhysRevD.92.063526}
  {\path{doi:10.1103/PhysRevD.92.063526}}.

\bibitem{Kamenetskaia:2022lbf}
B.~B. Kamenetskaia, A.~Brenner, A.~Ibarra, C.~Kouvaris, {Proton Capture in
  Compact Dark Stars and Observable Implications} (11 2022).
\newblock \href {http://arxiv.org/abs/2211.05845} {\path{arXiv:2211.05845}}.

\bibitem{Cardoso:2019rvt}
V.~Cardoso, P.~Pani, {Testing the nature of dark compact objects: a status
  report}, Living Rev. Rel. 22~(1) (2019) 4.
\newblock \href {http://arxiv.org/abs/1904.05363} {\path{arXiv:1904.05363}},
  \href {https://doi.org/10.1007/s41114-019-0020-4}
  {\path{doi:10.1007/s41114-019-0020-4}}.

\bibitem{dMACHOS:Bai:2020jfm}
Y.~Bai, A.~J. Long, S.~Lu, {Tests of Dark MACHOs: Lensing, Accretion, and
  Glow}, JCAP 09 (2020) 044.
\newblock \href {http://arxiv.org/abs/2003.13182} {\path{arXiv:2003.13182}},
  \href {https://doi.org/10.1088/1475-7516/2020/09/044}
  {\path{doi:10.1088/1475-7516/2020/09/044}}.

\bibitem{Witten:1984rs}
E.~Witten, {Cosmic Separation of Phases}, Phys. Rev. D30 (1984) 272--285.
\newblock \href {https://doi.org/10.1103/PhysRevD.30.272}
  {\path{doi:10.1103/PhysRevD.30.272}}.

\bibitem{Bai:2018dxf}
Y.~Bai, A.~J. Long, S.~Lu, {Dark Quark Nuggets}, Phys. Rev. D 99~(5) (2019)
  055047.
\newblock \href {http://arxiv.org/abs/1810.04360} {\path{arXiv:1810.04360}},
  \href {https://doi.org/10.1103/PhysRevD.99.055047}
  {\path{doi:10.1103/PhysRevD.99.055047}}.

\bibitem{Gross:2021qgx}
C.~Gross, G.~Landini, A.~Strumia, D.~Teresi, {Dark Matter as dark dwarfs and
  other macroscopic objects: multiverse relics?}, JHEP 09 (2021) 033.
\newblock \href {http://arxiv.org/abs/2105.02840} {\path{arXiv:2105.02840}},
  \href {https://doi.org/10.1007/JHEP09(2021)033}
  {\path{doi:10.1007/JHEP09(2021)033}}.

\bibitem{NSheat:DarkBary:McKeen:2020oyr}
D.~McKeen, M.~Pospelov, N.~Raj, {Cosmological and astrophysical probes of dark
  baryons}, Phys. Rev. D 103~(11) (2021) 115002.
\newblock \href {http://arxiv.org/abs/2012.09865} {\path{arXiv:2012.09865}},
  \href {https://doi.org/10.1103/PhysRevD.103.115002}
  {\path{doi:10.1103/PhysRevD.103.115002}}.

\bibitem{theluvoirteam2019luvoir}
T.~L. Team, The luvoir mission concept study final report (2019).
\newblock \href {http://arxiv.org/abs/1912.06219} {\path{arXiv:1912.06219}}.

\bibitem{Rubin1}
J.~R. Peterson, \href{https://www.osti.gov/biblio/1272167}{Dark energy studies
  with lsst image simulations, final report} (7 2016).
\newblock \href {https://doi.org/10.2172/1272167} {\path{doi:10.2172/1272167}}.
\newline\urlprefix\url{https://www.osti.gov/biblio/1272167}

\bibitem{Rubin2}
v.~Ivezi\'c, et~al., {LSST: from Science Drivers to Reference Design and
  Anticipated Data Products}, Astrophys. J. 873~(2) (2019) 111.
\newblock \href {http://arxiv.org/abs/0805.2366} {\path{arXiv:0805.2366}},
  \href {https://doi.org/10.3847/1538-4357/ab042c}
  {\path{doi:10.3847/1538-4357/ab042c}}.

\bibitem{DES:2019rtl}
H.~T. Diehl, et~al., {The Dark Energy Survey and Operations: Year 6
  \textendash{} The Finale} (10 2019).
\newblock \href {https://doi.org/10.2172/1596042} {\path{doi:10.2172/1596042}}.

\bibitem{green2012widefield}
J.~Green, P.~Schechter, C.~Baltay, R.~Bean, D.~Bennett, R.~Brown, C.~Conselice,
  M.~Donahue, X.~Fan, B.~S. Gaudi, C.~Hirata, J.~Kalirai, T.~Lauer, B.~Nichol,
  N.~Padmanabhan, S.~Perlmutter, B.~Rauscher, J.~Rhodes, T.~Roellig, D.~Stern,
  T.~Sumi, A.~Tanner, Y.~Wang, D.~Weinberg, E.~Wright, N.~Gehrels, R.~Sambruna,
  W.~Traub, J.~Anderson, K.~Cook, P.~Garnavich, L.~Hillenbrand, Z.~Ivezic,
  E.~Kerins, J.~Lunine, P.~McDonald, M.~Penny, M.~Phillips, G.~Rieke, A.~Riess,
  R.~van~der Marel, R.~K. Barry, E.~Cheng, D.~Content, R.~Cutri, R.~Goullioud,
  K.~Grady, G.~Helou, C.~Jackson, J.~Kruk, M.~Melton, C.~Peddie, N.~Rioux,
  M.~Seiffert, Wide-field infrared survey telescope (wfirst) final report
  (2012).
\newblock \href {http://arxiv.org/abs/1208.4012} {\path{arXiv:1208.4012}}.

\bibitem{NSheat:Mirror:Goldman:2022rth}
I.~Goldman, R.~N. Mohapatra, S.~Nussinov, Y.~Zhang, {Neutron-Mirror-Neutron
  Oscillation and Neutron Star Cooling}, Phys. Rev. Lett. 129~(6) (2022)
  061103.
\newblock \href {http://arxiv.org/abs/2208.03771} {\path{arXiv:2208.03771}},
  \href {https://doi.org/10.1103/PhysRevLett.129.061103}
  {\path{doi:10.1103/PhysRevLett.129.061103}}.

\bibitem{Davoudiasl:2023peu}
H.~Davoudiasl, {Stellar Signals of a Baryon-Number-Violating Long-Range Force}
  (4 2023).
\newblock \href {http://arxiv.org/abs/2304.06071} {\path{arXiv:2304.06071}}.

\bibitem{GWLimitsAxionCAJOHare}
\url{https://cajohare.github.io/AxionLimits/docs/fa.html}.

\bibitem{Croon:2017zcu}
D.~Croon, A.~E. Nelson, C.~Sun, D.~G.~E. Walker, Z.-Z. Xianyu, {Hidden-Sector
  Spectroscopy with Gravitational Waves from Binary Neutron Stars}, Astrophys.
  J. Lett. 858~(1) (2018) L2.
\newblock \href {http://arxiv.org/abs/1711.02096} {\path{arXiv:1711.02096}},
  \href {https://doi.org/10.3847/2041-8213/aabe76}
  {\path{doi:10.3847/2041-8213/aabe76}}.

\bibitem{Alexander:2018qzg}
S.~Alexander, E.~McDonough, R.~Sims, N.~Yunes, {Hidden-Sector Modifications to
  Gravitational Waves From Binary Inspirals}, Class. Quant. Grav. 35~(23)
  (2018) 235012.
\newblock \href {http://arxiv.org/abs/1808.05286} {\path{arXiv:1808.05286}},
  \href {https://doi.org/10.1088/1361-6382/aaeb5c}
  {\path{doi:10.1088/1361-6382/aaeb5c}}.

\bibitem{Ellis2018}
J.~Ellis, A.~Hektor, G.~Hutsi, K.~Kannike, L.~Marzola, M.~Raidal, V.~Vaskonen,
  \href{http://dx.doi.org/10.1016/j.physletb.2018.04.048}{Search for dark
  matter effects on gravitational signals from neutron star mergers}, Physics
  Letters B 781 (2018) 607?610.
\newblock \href {https://doi.org/10.1016/j.physletb.2018.04.048}
  {\path{doi:10.1016/j.physletb.2018.04.048}}.
\newline\urlprefix\url{http://dx.doi.org/10.1016/j.physletb.2018.04.048}

\bibitem{Emma:2022xjs}
M.~Emma, F.~Schianchi, F.~Pannarale, V.~Sagun, T.~Dietrich, {Numerical
  Simulations of Dark Matter Admixed Neutron Star Binaries}, Particles 5~(3)
  (2022) 273--286.
\newblock \href {http://arxiv.org/abs/2206.10887} {\path{arXiv:2206.10887}},
  \href {https://doi.org/10.3390/particles5030024}
  {\path{doi:10.3390/particles5030024}}.

\bibitem{Ruter:2023uzc}
H.~R. R\"uter, V.~Sagun, W.~Tichy, T.~Dietrich, {Quasi-equilibrium
  configurations of binary systems of dark matter admixed neutron stars} (1
  2023).
\newblock \href {http://arxiv.org/abs/2301.03568} {\path{arXiv:2301.03568}}.

\bibitem{Bauswein:2020kor}
A.~Bauswein, G.~Guo, J.-H. Lien, Y.-H. Lin, M.-R. Wu, {Compact dark objects in
  neutron star mergers}, Phys. Rev. D 107~(8) (2023) 083002.
\newblock \href {http://arxiv.org/abs/2012.11908} {\path{arXiv:2012.11908}},
  \href {https://doi.org/10.1103/PhysRevD.107.083002}
  {\path{doi:10.1103/PhysRevD.107.083002}}.

\bibitem{Fabbrichesi:2019ema}
M.~Fabbrichesi, A.~Urbano, {Charged neutron stars and observational tests of a
  dark force weaker than gravity}, JCAP 06 (2020) 007.
\newblock \href {http://arxiv.org/abs/1902.07914} {\path{arXiv:1902.07914}},
  \href {https://doi.org/10.1088/1475-7516/2020/06/007}
  {\path{doi:10.1088/1475-7516/2020/06/007}}.

\bibitem{Dror:2019uea}
J.~A. Dror, R.~Laha, T.~Opferkuch, {Probing muonic forces with neutron star
  binaries}, Phys. Rev. D 102~(2) (2020) 023005.
\newblock \href {http://arxiv.org/abs/1909.12845} {\path{arXiv:1909.12845}},
  \href {https://doi.org/10.1103/PhysRevD.102.023005}
  {\path{doi:10.1103/PhysRevD.102.023005}}.

\bibitem{Hook:2017psm}
A.~Hook, J.~Huang, {Probing axions with neutron star inspirals and other
  stellar processes}, JHEP 06 (2018) 036.
\newblock \href {http://arxiv.org/abs/1708.08464} {\path{arXiv:1708.08464}},
  \href {https://doi.org/10.1007/JHEP06(2018)036}
  {\path{doi:10.1007/JHEP06(2018)036}}.

\bibitem{Choi:2018axi}
H.~G. Choi, S.~Jung, {New probe of dark matter-induced fifth force with neutron
  star inspirals}, Phys. Rev. D 99~(1) (2019) 015013.
\newblock \href {http://arxiv.org/abs/1810.01421} {\path{arXiv:1810.01421}},
  \href {https://doi.org/10.1103/PhysRevD.99.015013}
  {\path{doi:10.1103/PhysRevD.99.015013}}.

\bibitem{Lambiase:2023hpq}
G.~Lambiase, T.~K. Poddar, {Pulsar kicks in ultralight dark matter background
  induced by neutrino oscillation} (7 2023).
\newblock \href {http://arxiv.org/abs/2307.05229} {\path{arXiv:2307.05229}}.

\bibitem{Diamond:2021ekg}
M.~D. Diamond, G.~Marques-Tavares, {Gamma-Ray Flashes from Dark Photons in
  Neutron Star Mergers}, Phys. Rev. Lett. 128~(21) (2022) 211101.
\newblock \href {http://arxiv.org/abs/2106.03879} {\path{arXiv:2106.03879}},
  \href {https://doi.org/10.1103/PhysRevLett.128.211101}
  {\path{doi:10.1103/PhysRevLett.128.211101}}.

\bibitem{Diamond:2023cto}
M.~Diamond, D.~F.~G. Fiorillo, G.~Marques-Tavares, I.~Tamborra, E.~Vitagliano,
  {Multimessenger Constraints on Radiatively Decaying Axions from GW170817} (5
  2023).
\newblock \href {http://arxiv.org/abs/2305.10327} {\path{arXiv:2305.10327}}.

\bibitem{PTAGW1}
G.~Hobbs, S.~Dai, {Gravitational wave research using pulsar timing arrays},
  Natl. Sci. Rev. 4~(5) (2017) 707--717.
\newblock \href {http://arxiv.org/abs/1707.01615} {\path{arXiv:1707.01615}},
  \href {https://doi.org/10.1093/nsr/nwx126} {\path{doi:10.1093/nsr/nwx126}}.

\bibitem{PTAGW2}
P.~K. {Dahal}, {Review of pulsar timing array for gravitational wave research},
  Journal of Astrophysics and Astronomy 41~(1) (2020) 8.
\newblock \href {http://arxiv.org/abs/2002.01954} {\path{arXiv:2002.01954}},
  \href {https://doi.org/10.1007/s12036-020-9625-y}
  {\path{doi:10.1007/s12036-020-9625-y}}.

\bibitem{Seto:2007kj}
N.~Seto, A.~Cooray, {Searching for primordial black hole dark matter with
  pulsar timing arrays}, Astrophys. J. Lett. 659 (2007) L33--L36.
\newblock \href {http://arxiv.org/abs/astro-ph/0702586}
  {\path{arXiv:astro-ph/0702586}}, \href {https://doi.org/10.1086/516570}
  {\path{doi:10.1086/516570}}.

\bibitem{Siegel:2007fz}
E.~R. Siegel, M.~P. Hertzberg, J.~N. Fry, {Probing Dark Matter Substructure
  with Pulsar Timing}, Mon. Not. Roy. Astron. Soc. 382 (2007) 879.
\newblock \href {http://arxiv.org/abs/astro-ph/0702546}
  {\path{arXiv:astro-ph/0702546}}, \href
  {https://doi.org/10.1111/j.1365-2966.2007.12435.x}
  {\path{doi:10.1111/j.1365-2966.2007.12435.x}}.

\bibitem{Baghram:2011is}
S.~Baghram, N.~Afshordi, K.~M. Zurek, {Prospects for Detecting Dark Matter Halo
  Substructure with Pulsar Timing}, Phys. Rev. D 84 (2011) 043511.
\newblock \href {http://arxiv.org/abs/1101.5487} {\path{arXiv:1101.5487}},
  \href {https://doi.org/10.1103/PhysRevD.84.043511}
  {\path{doi:10.1103/PhysRevD.84.043511}}.

\bibitem{Seto2012}
K.~{Kashiyama}, N.~{Seto}, {Enhanced exploration for primordial black holes
  using pulsar timing arrays}, MNRAS 426~(2) (2012) 1369--1373.
\newblock \href {http://arxiv.org/abs/1208.4101} {\path{arXiv:1208.4101}},
  \href {https://doi.org/10.1111/j.1365-2966.2012.21935.x}
  {\path{doi:10.1111/j.1365-2966.2012.21935.x}}.

\bibitem{Kashiyama:2018gsh}
K.~Kashiyama, M.~Oguri, {Detectability of Small-Scale Dark Matter Clumps with
  Pulsar Timing Arrays} (1 2018).
\newblock \href {http://arxiv.org/abs/1801.07847} {\path{arXiv:1801.07847}}.

\bibitem{Clark:2015sha}
H.~A. Clark, G.~F. Lewis, P.~Scott, {Investigating dark matter substructure
  with pulsar timing \textendash{} I. Constraints on ultracompact minihaloes},
  Mon. Not. Roy. Astron. Soc. 456~(2) (2016) 1394--1401, [Erratum:
  Mon.Not.Roy.Astron.Soc. 464, 2468 (2017)].
\newblock \href {http://arxiv.org/abs/1509.02938} {\path{arXiv:1509.02938}},
  \href {https://doi.org/10.1093/mnras/stv2743}
  {\path{doi:10.1093/mnras/stv2743}}.

\bibitem{Clark:2015tha}
H.~A. Clark, G.~F. Lewis, P.~Scott, {Investigating dark matter substructure
  with pulsar timing \textendash{} II. Improved limits on small-scale
  cosmology}, Mon. Not. Roy. Astron. Soc. 456~(2) (2016) 1402--1409, [Erratum:
  Mon.Not.Roy.Astron.Soc. 464, 955--956 (2017)].
\newblock \href {http://arxiv.org/abs/1509.02941} {\path{arXiv:1509.02941}},
  \href {https://doi.org/10.1093/mnras/stv2529}
  {\path{doi:10.1093/mnras/stv2529}}.

\bibitem{Schutz:2016khr}
K.~Schutz, A.~Liu, {Pulsar timing can constrain primordial black holes in the
  LIGO mass window}, Phys. Rev. D 95~(2) (2017) 023002.
\newblock \href {http://arxiv.org/abs/1610.04234} {\path{arXiv:1610.04234}},
  \href {https://doi.org/10.1103/PhysRevD.95.023002}
  {\path{doi:10.1103/PhysRevD.95.023002}}.

\bibitem{PTA:Dror:2019twh}
J.~A. Dror, H.~Ramani, T.~Trickle, K.~M. Zurek, {Pulsar Timing Probes of
  Primordial Black Holes and Subhalos}, Phys. Rev. D 100~(2) (2019) 023003.
\newblock \href {http://arxiv.org/abs/1901.04490} {\path{arXiv:1901.04490}},
  \href {https://doi.org/10.1103/PhysRevD.100.023003}
  {\path{doi:10.1103/PhysRevD.100.023003}}.

\bibitem{Ramani:2020hdo}
H.~Ramani, T.~Trickle, K.~M. Zurek, {Observability of Dark Matter Substructure
  with Pulsar Timing Correlations}, JCAP 12 (2020) 033.
\newblock \href {http://arxiv.org/abs/2005.03030} {\path{arXiv:2005.03030}},
  \href {https://doi.org/10.1088/1475-7516/2020/12/033}
  {\path{doi:10.1088/1475-7516/2020/12/033}}.

\bibitem{Lee:2021zqw}
V.~S.~H. Lee, S.~R. Taylor, T.~Trickle, K.~M. Zurek, {Bayesian Forecasts for
  Dark Matter Substructure Searches with Mock Pulsar Timing Data}, JCAP 08
  (2021) 025.
\newblock \href {http://arxiv.org/abs/2104.05717} {\path{arXiv:2104.05717}},
  \href {https://doi.org/10.1088/1475-7516/2021/08/025}
  {\path{doi:10.1088/1475-7516/2021/08/025}}.

\bibitem{Lee:2020wfn}
V.~S.~H. Lee, A.~Mitridate, T.~Trickle, K.~M. Zurek, {Probing Small-Scale Power
  Spectra with Pulsar Timing Arrays}, JHEP 06 (2021) 028.
\newblock \href {http://arxiv.org/abs/2012.09857} {\path{arXiv:2012.09857}},
  \href {https://doi.org/10.1007/JHEP06(2021)028}
  {\path{doi:10.1007/JHEP06(2021)028}}.

\bibitem{NANOGrav:2017wvv}
Z.~Arzoumanian, et~al., {The NANOGrav 11-year Data Set: High-precision timing
  of 45 Millisecond Pulsars}, Astrophys. J. Suppl. 235~(2) (2018) 37.
\newblock \href {http://arxiv.org/abs/1801.01837} {\path{arXiv:1801.01837}},
  \href {https://doi.org/10.3847/1538-4365/aab5b0}
  {\path{doi:10.3847/1538-4365/aab5b0}}.

\bibitem{NANOGRAV:2018hou}
Z.~Arzoumanian, et~al., {The NANOGrav 11-year Data Set: Pulsar-timing
  Constraints On The Stochastic Gravitational-wave Background}, Astrophys. J.
  859~(1) (2018) 47.
\newblock \href {http://arxiv.org/abs/1801.02617} {\path{arXiv:1801.02617}},
  \href {https://doi.org/10.3847/1538-4357/aabd3b}
  {\path{doi:10.3847/1538-4357/aabd3b}}.

\bibitem{Khmelnitsky:2013lxt}
A.~Khmelnitsky, V.~Rubakov, {Pulsar timing signal from ultralight scalar dark
  matter}, JCAP 02 (2014) 019.
\newblock \href {http://arxiv.org/abs/1309.5888} {\path{arXiv:1309.5888}},
  \href {https://doi.org/10.1088/1475-7516/2014/02/019}
  {\path{doi:10.1088/1475-7516/2014/02/019}}.

\bibitem{ULDMvPTA:Omiya:2023bio}
H.~Omiya, K.~Nomura, J.~Soda, {Hellings-Downs curve deformed by ultralight
  vector dark matter}, Phys. Rev. D 108~(10) (2023) 104006.
\newblock \href {http://arxiv.org/abs/2307.12624} {\path{arXiv:2307.12624}},
  \href {https://doi.org/10.1103/PhysRevD.108.104006}
  {\path{doi:10.1103/PhysRevD.108.104006}}.

\bibitem{ULDMvPTA:Armaleo:2020yml}
J.~M. Armaleo, D.~L\'opez~Nacir, F.~R. Urban, {Pulsar timing array constraints
  on spin-2 ULDM}, JCAP 09 (2020) 031.
\newblock \href {http://arxiv.org/abs/2005.03731} {\path{arXiv:2005.03731}},
  \href {https://doi.org/10.1088/1475-7516/2020/09/031}
  {\path{doi:10.1088/1475-7516/2020/09/031}}.

\bibitem{ULDMvPTA:Wu:2023dnp}
Y.-M. Wu, Z.-C. Chen, Q.-G. Huang, {Pulsar timing residual induced by
  ultralight tensor dark matter}, JCAP 09 (2023) 021.
\newblock \href {http://arxiv.org/abs/2305.08091} {\path{arXiv:2305.08091}},
  \href {https://doi.org/10.1088/1475-7516/2023/09/021}
  {\path{doi:10.1088/1475-7516/2023/09/021}}.

\bibitem{ULDMvNANOGrav:Porayko:2014rfa}
N.~K. Porayko, K.~A. Postnov, {Constraints on ultralight scalar dark matter
  from pulsar timing}, Phys. Rev. D 90~(6) (2014) 062008.
\newblock \href {http://arxiv.org/abs/1408.4670} {\path{arXiv:1408.4670}},
  \href {https://doi.org/10.1103/PhysRevD.90.062008}
  {\path{doi:10.1103/PhysRevD.90.062008}}.

\bibitem{ULDMvPPTA:Porayko:2018sfa}
N.~K. Porayko, et~al., {Parkes Pulsar Timing Array constraints on ultralight
  scalar-field dark matter}, Phys. Rev. D 98~(10) (2018) 102002.
\newblock \href {http://arxiv.org/abs/1810.03227} {\path{arXiv:1810.03227}},
  \href {https://doi.org/10.1103/PhysRevD.98.102002}
  {\path{doi:10.1103/PhysRevD.98.102002}}.

\bibitem{ULDMvEPTA:Smarra:2023ljf}
C.~Smarra, et~al., {The second data release from the European Pulsar Timing
  Array: VI. Challenging the ultralight dark matter paradigm} (6 2023).
\newblock \href {http://arxiv.org/abs/2306.16228} {\path{arXiv:2306.16228}}.

\bibitem{Stadnik:2014cea}
Y.~V. Stadnik, V.~V. Flambaum, {Searches for topological defect dark matter via
  nongravitational signatures}, Phys. Rev. Lett. 113~(15) (2014) 151301.
\newblock \href {http://arxiv.org/abs/1405.5337} {\path{arXiv:1405.5337}},
  \href {https://doi.org/10.1103/PhysRevLett.113.151301}
  {\path{doi:10.1103/PhysRevLett.113.151301}}.

\bibitem{EPTA:Desvignes:2016yex}
G.~Desvignes, et~al., {High-precision timing of 42 millisecond pulsars with the
  European Pulsar Timing Array}, Mon. Not. Roy. Astron. Soc. 458~(3) (2016)
  3341--3380.
\newblock \href {http://arxiv.org/abs/1602.08511} {\path{arXiv:1602.08511}},
  \href {https://doi.org/10.1093/mnras/stw483}
  {\path{doi:10.1093/mnras/stw483}}.

\bibitem{PPTA}
R.~N. {Manchester}, G.~{Hobbs}, M.~{Bailes}, W.~A. {Coles}, W.~{van Straten},
  M.~J. {Keith}, R.~M. {Shannon}, N.~D.~R. {Bhat}, A.~{Brown}, S.~G.
  {Burke-Spolaor}, D.~J. {Champion}, A.~{Chaudhary}, R.~T. {Edwards},
  G.~{Hampson}, A.~W. {Hotan}, A.~{Jameson}, F.~A. {Jenet}, M.~J. {Kesteven},
  J.~{Khoo}, J.~{Kocz}, K.~{Maciesiak}, S.~{Oslowski}, V.~{Ravi}, J.~R.
  {Reynolds}, J.~M. {Sarkissian}, J.~P.~W. {Verbiest}, Z.~L. {Wen}, W.~E.
  {Wilson}, D.~{Yardley}, W.~M. {Yan}, X.~P. {You}, {The Parkes Pulsar Timing
  Array Project}, PASA 30 (2013) e017.
\newblock \href {http://arxiv.org/abs/1210.6130} {\path{arXiv:1210.6130}},
  \href {https://doi.org/10.1017/pasa.2012.017}
  {\path{doi:10.1017/pasa.2012.017}}.

\bibitem{NANOGrav:Brazier:2019mmu}
A.~Brazier, et~al., {The NANOGrav Program for Gravitational Waves and
  Fundamental Physics} (8 2019).
\newblock \href {http://arxiv.org/abs/1908.05356} {\path{arXiv:1908.05356}}.

\bibitem{InPTA:Tarafdar:2022toa}
P.~Tarafdar, et~al., {The Indian Pulsar Timing Array: First data release},
  Publ. Astron. Soc. Austral. 39 (2022) e053.
\newblock \href {http://arxiv.org/abs/2206.09289} {\path{arXiv:2206.09289}},
  \href {https://doi.org/10.1017/pasa.2022.46}
  {\path{doi:10.1017/pasa.2022.46}}.

\bibitem{IPTA:Perera:2019sca}
B.~B.~P. Perera, et~al., {The International Pulsar Timing Array: Second data
  release}, Mon. Not. Roy. Astron. Soc. 490~(4) (2019) 4666--4687.
\newblock \href {http://arxiv.org/abs/1909.04534} {\path{arXiv:1909.04534}},
  \href {https://doi.org/10.1093/mnras/stz2857}
  {\path{doi:10.1093/mnras/stz2857}}.

\bibitem{MeerTime:Bailes:2018azh}
M.~Bailes, et~al., {MeerTime - the MeerKAT Key Science Program on Pulsar
  Timing}, PoS MeerKAT2016 (2018) 011.
\newblock \href {http://arxiv.org/abs/1803.07424} {\path{arXiv:1803.07424}},
  \href {https://doi.org/10.22323/1.277.0011} {\path{doi:10.22323/1.277.0011}}.

\bibitem{FASTPTA:Hobbs:2014tqa}
G.~Hobbs, S.~Dai, R.~N. Manchester, R.~M. Shannon, M.~Kerr, K.~J. Lee, R.~Xu,
  {The Role of FAST in Pulsar Timing Arrays}, Res. Astron. Astrophys. 19~(2)
  (2019) 020.
\newblock \href {http://arxiv.org/abs/1407.0435} {\path{arXiv:1407.0435}},
  \href {https://doi.org/10.1088/1674-4527/19/2/20}
  {\path{doi:10.1088/1674-4527/19/2/20}}.

\bibitem{SKAPTA1}
R.~{Smits}, M.~{Kramer}, B.~{Stappers}, D.~R. {Lorimer}, J.~{Cordes},
  A.~{Faulkner}, {Pulsar searches and timing with the square kilometre array},
  AAP 493~(3) (2009) 1161--1170.
\newblock \href {http://arxiv.org/abs/0811.0211} {\path{arXiv:0811.0211}},
  \href {https://doi.org/10.1051/0004-6361:200810383}
  {\path{doi:10.1051/0004-6361:200810383}}.

\bibitem{SKAPTA2:Rosado:2015epa}
P.~A. Rosado, A.~Sesana, J.~Gair, {Expected properties of the first
  gravitational wave signal detected with pulsar timing arrays}, Mon. Not. Roy.
  Astron. Soc. 451~(3) (2015) 2417--2433.
\newblock \href {http://arxiv.org/abs/1503.04803} {\path{arXiv:1503.04803}},
  \href {https://doi.org/10.1093/mnras/stv1098}
  {\path{doi:10.1093/mnras/stv1098}}.

\bibitem{EPTAInPTA:2023fyk}
J.~Antoniadis, et~al., {The second data release from the European Pulsar Timing
  Array - III. Search for gravitational wave signals}, Astron. Astrophys. 678
  (2023) A50.
\newblock \href {http://arxiv.org/abs/2306.16214} {\path{arXiv:2306.16214}},
  \href {https://doi.org/10.1051/0004-6361/202346844}
  {\path{doi:10.1051/0004-6361/202346844}}.

\bibitem{PPTA:2023gzh}
D.~J. Reardon, et~al., {Search for an Isotropic Gravitational-wave Background
  with the Parkes Pulsar Timing Array}, Astrophys. J. Lett. 951~(1) (2023) L6.
\newblock \href {http://arxiv.org/abs/2306.16215} {\path{arXiv:2306.16215}},
  \href {https://doi.org/10.3847/2041-8213/acdd02}
  {\path{doi:10.3847/2041-8213/acdd02}}.

\bibitem{CPTA:Xu:2023wog}
H.~Xu, et~al., {Searching for the Nano-Hertz Stochastic Gravitational Wave
  Background with the Chinese Pulsar Timing Array Data Release I}, Res. Astron.
  Astrophys. 23~(7) (2023) 075024.
\newblock \href {http://arxiv.org/abs/2306.16216} {\path{arXiv:2306.16216}},
  \href {https://doi.org/10.1088/1674-4527/acdfa5}
  {\path{doi:10.1088/1674-4527/acdfa5}}.

\bibitem{MeerKAT:Miles:2024seg}
M.~T. Miles, et~al., {The MeerKAT Pulsar Timing Array: The first search for
  gravitational waves with the MeerKAT radio telescope}, Mon. Not. Roy. Astron.
  Soc. 536~(2) (2025) 1489--1500.
\newblock \href {http://arxiv.org/abs/2412.01153} {\path{arXiv:2412.01153}},
  \href {https://doi.org/10.1093/mnras/stae2571}
  {\path{doi:10.1093/mnras/stae2571}}.

\bibitem{Depta:2023qst}
P.~F. Depta, K.~Schmidt-Hoberg, C.~Tasillo, {Do pulsar timing arrays observe
  merging primordial black holes?} (6 2023).
\newblock \href {http://arxiv.org/abs/2306.17836} {\path{arXiv:2306.17836}}.

\bibitem{Inomata:2023zup}
K.~Inomata, K.~Kohri, T.~Terada, {The Detected Stochastic Gravitational Waves
  and Sub-Solar Primordial Black Holes} (6 2023).
\newblock \href {http://arxiv.org/abs/2306.17834} {\path{arXiv:2306.17834}}.

\bibitem{Gouttenoire:2023nzr}
Y.~Gouttenoire, S.~Trifinopoulos, G.~Valogiannis, M.~Vanvlasselaer,
  {Scrutinizing the Primordial Black Holes Interpretation of PTA Gravitational
  Waves and JWST Early Galaxies} (7 2023).
\newblock \href {http://arxiv.org/abs/2307.01457} {\path{arXiv:2307.01457}}.

\bibitem{Ghoshal:2023fhh}
A.~Ghoshal, A.~Strumia, {Probing the Dark Matter density with gravitational
  waves from super-massive binary black holes} (6 2023).
\newblock \href {http://arxiv.org/abs/2306.17158} {\path{arXiv:2306.17158}}.

\bibitem{Shen:2023pan}
Z.-Q. Shen, G.-W. Yuan, Y.-Y. Wang, Y.-Z. Wang, {Dark Matter Spike surrounding
  Supermassive Black Holes Binary and the nanohertz Stochastic Gravitational
  Wave Background} (6 2023).
\newblock \href {http://arxiv.org/abs/2306.17143} {\path{arXiv:2306.17143}}.

\bibitem{Borah:2023sbc}
D.~Borah, S.~Jyoti~Das, R.~Samanta, {Inflationary origin of gravitational waves
  with \textbackslash{}textit{Miracle-less WIMP} dark matter in the light of
  recent PTA results} (7 2023).
\newblock \href {http://arxiv.org/abs/2307.00537} {\path{arXiv:2307.00537}}.

\bibitem{Broadhurst:2023tus}
T.~Broadhurst, C.~Chen, T.~Liu, K.-F. Zheng, {Binary Supermassive Black Holes
  Orbiting Dark Matter Solitons: From the Dual AGN in UGC4211 to NanoHertz
  Gravitational Waves} (6 2023).
\newblock \href {http://arxiv.org/abs/2306.17821} {\path{arXiv:2306.17821}}.

\bibitem{Xiao:2023dbb}
Y.~Xiao, J.~M. Yang, Y.~Zhang, {Implications of Nano-Hertz Gravitational Waves
  on Electroweak Phase Transition in the Singlet Dark Matter Model} (7 2023).
\newblock \href {http://arxiv.org/abs/2307.01072} {\path{arXiv:2307.01072}}.

\bibitem{Anchordoqui:2023tln}
L.~A. Anchordoqui, I.~Antoniadis, D.~Lust, {Fuzzy Dark Matter, the Dark
  Dimension, and the Pulsar Timing Array Signal} (7 2023).
\newblock \href {http://arxiv.org/abs/2307.01100} {\path{arXiv:2307.01100}}.

\bibitem{Kitajima:2023vre}
N.~Kitajima, K.~Nakayama, {Nanohertz gravitational waves from cosmic strings
  and dark photon dark matter} (6 2023).
\newblock \href {http://arxiv.org/abs/2306.17390} {\path{arXiv:2306.17390}}.

\bibitem{Pani:2015qhr}
P.~Pani, {Binary pulsars as dark-matter probes}, Phys. Rev. D 92~(12) (2015)
  123530.
\newblock \href {http://arxiv.org/abs/1512.01236} {\path{arXiv:1512.01236}},
  \href {https://doi.org/10.1103/PhysRevD.92.123530}
  {\path{doi:10.1103/PhysRevD.92.123530}}.

\bibitem{KumarPoddar:2019jxe}
T.~Kumar~Poddar, S.~Mohanty, S.~Jana, {Constraints on ultralight axions from
  compact binary systems}, Phys. Rev. D 101~(8) (2020) 083007.
\newblock \href {http://arxiv.org/abs/1906.00666} {\path{arXiv:1906.00666}},
  \href {https://doi.org/10.1103/PhysRevD.101.083007}
  {\path{doi:10.1103/PhysRevD.101.083007}}.

\bibitem{Blas:2016ddr}
D.~Blas, D.~L. Nacir, S.~Sibiryakov, {Ultralight Dark Matter Resonates with
  Binary Pulsars}, Phys. Rev. Lett. 118~(26) (2017) 261102.
\newblock \href {http://arxiv.org/abs/1612.06789} {\path{arXiv:1612.06789}},
  \href {https://doi.org/10.1103/PhysRevLett.118.261102}
  {\path{doi:10.1103/PhysRevLett.118.261102}}.

\bibitem{Blas:2019hxz}
D.~Blas, D.~L\'opez~Nacir, S.~Sibiryakov, {Secular effects of ultralight dark
  matter on binary pulsars}, Phys. Rev. D 101~(6) (2020) 063016.
\newblock \href {http://arxiv.org/abs/1910.08544} {\path{arXiv:1910.08544}},
  \href {https://doi.org/10.1103/PhysRevD.101.063016}
  {\path{doi:10.1103/PhysRevD.101.063016}}.

\bibitem{DiLuzio:2021pxd}
L.~Di~Luzio, B.~Gavela, P.~Quilez, A.~Ringwald, {An even lighter QCD axion},
  JHEP 05 (2021) 184.
\newblock \href {http://arxiv.org/abs/2102.00012} {\path{arXiv:2102.00012}},
  \href {https://doi.org/10.1007/JHEP05(2021)184}
  {\path{doi:10.1007/JHEP05(2021)184}}.

\bibitem{Berezhiani:2020zck}
Z.~Berezhiani, R.~Biondi, M.~Mannarelli, F.~Tonelli, {Neutron-mirror neutron
  mixing and neutron stars}, Eur. Phys. J. C 81~(11) (2021) 1036.
\newblock \href {http://arxiv.org/abs/2012.15233} {\path{arXiv:2012.15233}},
  \href {https://doi.org/10.1140/epjc/s10052-021-09806-1}
  {\path{doi:10.1140/epjc/s10052-021-09806-1}}.

\bibitem{Berryman:2023rmh}
J.~M. Berryman, S.~Gardner, M.~Zakeri, {How Macroscopic Limits on Neutron Star
  Baryon Loss Yield Microscopic Limits on Non-Standard-Model Baryon Decay} (5
  2023).
\newblock \href {http://arxiv.org/abs/2305.13377} {\path{arXiv:2305.13377}}.

\bibitem{Gardner:2023wyl}
S.~Gardner, M.~Zakeri, {Probing Dark Sectors with Neutron Stars}, Universe
  10~(2) (2024) 67.
\newblock \href {http://arxiv.org/abs/2311.13649} {\path{arXiv:2311.13649}},
  \href {https://doi.org/10.3390/universe10020067}
  {\path{doi:10.3390/universe10020067}}.

\bibitem{Kouvaris:2014rja}
C.~Kouvaris, M.~A. Perez-Garcia, {Can Dark Matter explain the Braking Index of
  Neutron Stars?}, Phys. Rev. D 89~(10) (2014) 103539.
\newblock \href {http://arxiv.org/abs/1401.3644} {\path{arXiv:1401.3644}},
  \href {https://doi.org/10.1103/PhysRevD.89.103539}
  {\path{doi:10.1103/PhysRevD.89.103539}}.

\bibitem{Huang:2015gta}
X.~Huang, X.-P. Zheng, W.-H. Wang, S.-Z. Li, {Constraints on millicharged
  particles by neutron stars}, Phys. Rev. D 91~(12) (2015) 123513.
\newblock \href {http://arxiv.org/abs/1509.07620} {\path{arXiv:1509.07620}},
  \href {https://doi.org/10.1103/PhysRevD.91.123513}
  {\path{doi:10.1103/PhysRevD.91.123513}}.

\bibitem{Foster:2020pgt}
J.~W. Foster, Y.~Kahn, O.~Macias, Z.~Sun, R.~P. Eatough, V.~I. Kondratiev,
  W.~M. Peters, C.~Weniger, B.~R. Safdi, {Green Bank and Effelsberg Radio
  Telescope Searches for Axion Dark Matter Conversion in Neutron Star
  Magnetospheres}, Phys. Rev. Lett. 125~(17) (2020) 171301.
\newblock \href {http://arxiv.org/abs/2004.00011} {\path{arXiv:2004.00011}},
  \href {https://doi.org/10.1103/PhysRevLett.125.171301}
  {\path{doi:10.1103/PhysRevLett.125.171301}}.

\bibitem{Battye:2021yue}
R.~A. Battye, J.~Darling, J.~I. McDonald, S.~Srinivasan, {Towards robust
  constraints on axion dark matter using PSR J1745-2900}, Phys. Rev. D 105~(2)
  (2022) L021305.
\newblock \href {http://arxiv.org/abs/2107.01225} {\path{arXiv:2107.01225}},
  \href {https://doi.org/10.1103/PhysRevD.105.L021305}
  {\path{doi:10.1103/PhysRevD.105.L021305}}.

\bibitem{Tinyakov:2021lnt}
P.~Tinyakov, M.~Pshirkov, S.~Popov, {Astroparticle Physics with Compact
  Objects}, Universe 7~(11) (2021) 401.
\newblock \href {http://arxiv.org/abs/2110.12298} {\path{arXiv:2110.12298}},
  \href {https://doi.org/10.3390/universe7110401}
  {\path{doi:10.3390/universe7110401}}.

\bibitem{Lai:2006af}
D.~Lai, J.~Heyl, {Probing Axions with Radiation from Magnetic Stars}, Phys.
  Rev. D 74 (2006) 123003.
\newblock \href {http://arxiv.org/abs/astro-ph/0609775}
  {\path{arXiv:astro-ph/0609775}}, \href
  {https://doi.org/10.1103/PhysRevD.74.123003}
  {\path{doi:10.1103/PhysRevD.74.123003}}.

\bibitem{Pshirkov:2007st}
M.~S. Pshirkov, S.~B. Popov, {Conversion of Dark matter axions to photons in
  magnetospheres of neutron stars}, J. Exp. Theor. Phys. 108 (2009) 384--388.
\newblock \href {http://arxiv.org/abs/0711.1264} {\path{arXiv:0711.1264}},
  \href {https://doi.org/10.1134/S1063776109030030}
  {\path{doi:10.1134/S1063776109030030}}.

\bibitem{Hook:2018iia}
A.~Hook, Y.~Kahn, B.~R. Safdi, Z.~Sun, {Radio Signals from Axion Dark Matter
  Conversion in Neutron Star Magnetospheres}, Phys. Rev. Lett. 121~(24) (2018)
  241102.
\newblock \href {http://arxiv.org/abs/1804.03145} {\path{arXiv:1804.03145}},
  \href {https://doi.org/10.1103/PhysRevLett.121.241102}
  {\path{doi:10.1103/PhysRevLett.121.241102}}.

\bibitem{Huang:2018lxq}
F.~P. Huang, K.~Kadota, T.~Sekiguchi, H.~Tashiro, {Radio telescope search for
  the resonant conversion of cold dark matter axions from the magnetized
  astrophysical sources}, Phys. Rev. D 97~(12) (2018) 123001.
\newblock \href {http://arxiv.org/abs/1803.08230} {\path{arXiv:1803.08230}},
  \href {https://doi.org/10.1103/PhysRevD.97.123001}
  {\path{doi:10.1103/PhysRevD.97.123001}}.

\bibitem{Safdi:2018oeu}
B.~R. Safdi, Z.~Sun, A.~Y. Chen, {Detecting Axion Dark Matter with Radio Lines
  from Neutron Star Populations}, Phys. Rev. D 99~(12) (2019) 123021.
\newblock \href {http://arxiv.org/abs/1811.01020} {\path{arXiv:1811.01020}},
  \href {https://doi.org/10.1103/PhysRevD.99.123021}
  {\path{doi:10.1103/PhysRevD.99.123021}}.

\bibitem{Battye:2019aco}
R.~A. Battye, B.~Garbrecht, J.~I. McDonald, F.~Pace, S.~Srinivasan, {Dark
  matter axion detection in the radio/mm-waveband}, Phys. Rev. D 102~(2) (2020)
  023504.
\newblock \href {http://arxiv.org/abs/1910.11907} {\path{arXiv:1910.11907}},
  \href {https://doi.org/10.1103/PhysRevD.102.023504}
  {\path{doi:10.1103/PhysRevD.102.023504}}.

\bibitem{Foster:2022fxn}
J.~W. Foster, S.~J. Witte, M.~Lawson, T.~Linden, V.~Gajjar, C.~Weniger, B.~R.
  Safdi, {Extraterrestrial Axion Search with the Breakthrough Listen Galactic
  Center Survey} (2 2022).
\newblock \href {http://arxiv.org/abs/2202.08274} {\path{arXiv:2202.08274}}.

\bibitem{Darling:2020plz}
J.~Darling, {Search for Axionic Dark Matter Using the Magnetar PSR J1745-2900},
  Phys. Rev. Lett. 125~(12) (2020) 121103.
\newblock \href {http://arxiv.org/abs/2008.01877} {\path{arXiv:2008.01877}},
  \href {https://doi.org/10.1103/PhysRevLett.125.121103}
  {\path{doi:10.1103/PhysRevLett.125.121103}}.

\bibitem{Darling:2020uyo}
J.~Darling, {New Limits on Axionic Dark Matter from the Magnetar PSR
  J1745-2900}, Astrophys. J. Lett. 900~(2) (2020) L28.
\newblock \href {http://arxiv.org/abs/2008.11188} {\path{arXiv:2008.11188}},
  \href {https://doi.org/10.3847/2041-8213/abb23f}
  {\path{doi:10.3847/2041-8213/abb23f}}.

\bibitem{ExtremeBaryakhtar:2022hbu}
M.~Baryakhtar, et~al., {Dark Matter In Extreme Astrophysical Environments}, in:
  {2022 Snowmass Summer Study}, 2022.
\newblock \href {http://arxiv.org/abs/2203.07984} {\path{arXiv:2203.07984}}.

\bibitem{Hardy:2022ufh}
E.~Hardy, N.~Song, {Listening for Dark Photon Radio from the Galactic Centre}
  (12 2022).
\newblock \href {http://arxiv.org/abs/2212.09756} {\path{arXiv:2212.09756}}.

\bibitem{Battye:2023oac}
R.~A. Battye, M.~J. Keith, J.~I. McDonald, S.~Srinivasan, B.~W. Stappers,
  P.~Weltevrede, {Searching for Time-Dependent Axion Dark Matter Signals in
  Pulsars} (3 2023).
\newblock \href {http://arxiv.org/abs/2303.11792} {\path{arXiv:2303.11792}}.

\bibitem{Millar:2021gzs}
A.~J. Millar, S.~Baum, M.~Lawson, M.~C.~D. Marsh, {Axion-photon conversion in
  strongly magnetised plasmas}, JCAP 11 (2021) 013.
\newblock \href {http://arxiv.org/abs/2107.07399} {\path{arXiv:2107.07399}},
  \href {https://doi.org/10.1088/1475-7516/2021/11/013}
  {\path{doi:10.1088/1475-7516/2021/11/013}}.

\bibitem{Edwards:2019tzf}
T.~D.~P. Edwards, M.~Chianese, B.~J. Kavanagh, S.~M. Nissanke, C.~Weniger,
  {Unique Multimessenger Signal of QCD Axion Dark Matter}, Phys. Rev. Lett.
  124~(16) (2020) 161101.
\newblock \href {http://arxiv.org/abs/1905.04686} {\path{arXiv:1905.04686}},
  \href {https://doi.org/10.1103/PhysRevLett.124.161101}
  {\path{doi:10.1103/PhysRevLett.124.161101}}.

\bibitem{Leroy:2019ghm}
M.~Leroy, M.~Chianese, T.~D.~P. Edwards, C.~Weniger, {Radio Signal of
  Axion-Photon Conversion in Neutron Stars: A Ray Tracing Analysis}, Phys. Rev.
  D 101~(12) (2020) 123003.
\newblock \href {http://arxiv.org/abs/1912.08815} {\path{arXiv:1912.08815}},
  \href {https://doi.org/10.1103/PhysRevD.101.123003}
  {\path{doi:10.1103/PhysRevD.101.123003}}.

\bibitem{Witte:2021arp}
S.~J. Witte, D.~Noordhuis, T.~D.~P. Edwards, C.~Weniger, {Axion-photon
  conversion in neutron star magnetospheres: The role of the plasma in the
  Goldreich-Julian model}, Phys. Rev. D 104~(10) (2021) 103030.
\newblock \href {http://arxiv.org/abs/2104.07670} {\path{arXiv:2104.07670}},
  \href {https://doi.org/10.1103/PhysRevD.104.103030}
  {\path{doi:10.1103/PhysRevD.104.103030}}.

\bibitem{Battye:2021xvt}
R.~A. Battye, B.~Garbrecht, J.~I. McDonald, S.~Srinivasan, {Radio line
  properties of axion dark matter conversion in neutron stars}, JHEP 09 (2021)
  105.
\newblock \href {http://arxiv.org/abs/2104.08290} {\path{arXiv:2104.08290}},
  \href {https://doi.org/10.1007/JHEP09(2021)105}
  {\path{doi:10.1007/JHEP09(2021)105}}.

\bibitem{Kavanagh:2020gcy}
B.~J. Kavanagh, T.~D.~P. Edwards, L.~Visinelli, C.~Weniger, {Stellar Disruption
  of Axion Miniclusters in the Milky Way} (11 2020).
\newblock \href {http://arxiv.org/abs/2011.05377} {\path{arXiv:2011.05377}}.

\bibitem{Buckley:2020fmh}
J.~H. Buckley, P.~S.~B. Dev, F.~Ferrer, F.~P. Huang, {Fast radio bursts from
  axion stars moving through pulsar magnetospheres}, Phys. Rev. D 103~(4)
  (2021) 043015.
\newblock \href {http://arxiv.org/abs/2004.06486} {\path{arXiv:2004.06486}},
  \href {https://doi.org/10.1103/PhysRevD.103.043015}
  {\path{doi:10.1103/PhysRevD.103.043015}}.

\bibitem{Nurmi:2021xds}
S.~Nurmi, E.~D. Schiappacasse, T.~T. Yanagida, {Radio signatures from
  encounters between neutron stars and QCD-axion minihalos around primordial
  black~holes}, JCAP 09 (2021) 004.
\newblock \href {http://arxiv.org/abs/2102.05680} {\path{arXiv:2102.05680}},
  \href {https://doi.org/10.1088/1475-7516/2021/09/004}
  {\path{doi:10.1088/1475-7516/2021/09/004}}.

\bibitem{Bai:2021nrs}
Y.~Bai, X.~Du, Y.~Hamada, {Diluted axion star collisions with neutron stars},
  JCAP 01~(01) (2022) 041.
\newblock \href {http://arxiv.org/abs/2109.01222} {\path{arXiv:2109.01222}},
  \href {https://doi.org/10.1088/1475-7516/2022/01/041}
  {\path{doi:10.1088/1475-7516/2022/01/041}}.

\bibitem{Witte:2022cjj}
S.~J. Witte, S.~Baum, M.~Lawson, M.~C.~D. Marsh, A.~J. Millar, G.~Salinas,
  {Transient Radio Lines from Axion Miniclusters and Axion Stars} (12 2022).
\newblock \href {http://arxiv.org/abs/2212.08079} {\path{arXiv:2212.08079}}.

\bibitem{Anzuini:2022bqd}
F.~Anzuini, J.~A. Pons, A.~G\'omez-Ba\~n\'on, P.~D. Lasky, F.~Bianchini,
  A.~Melatos, {Magnetic dynamo caused by axions in neutron stars} (11 2022).
\newblock \href {http://arxiv.org/abs/2211.10863} {\path{arXiv:2211.10863}}.

\bibitem{Niikura:2019kqi}
H.~Niikura, M.~Takada, S.~Yokoyama, T.~Sumi, S.~Masaki, {Constraints on
  Earth-mass primordial black holes from OGLE 5-year microlensing events},
  Phys. Rev. D 99~(8) (2019) 083503.
\newblock \href {http://arxiv.org/abs/1901.07120} {\path{arXiv:1901.07120}},
  \href {https://doi.org/10.1103/PhysRevD.99.083503}
  {\path{doi:10.1103/PhysRevD.99.083503}}.

\bibitem{topologicaldefect:Murayama:2009nj}
H.~Murayama, J.~Shu, {Topological Dark Matter}, Phys. Lett. B 686 (2010)
  162--165.
\newblock \href {http://arxiv.org/abs/0905.1720} {\path{arXiv:0905.1720}},
  \href {https://doi.org/10.1016/j.physletb.2010.02.037}
  {\path{doi:10.1016/j.physletb.2010.02.037}}.

\bibitem{topologicaldefect:DereviankoPospelov:2013oaa}
A.~Derevianko, M.~Pospelov, {Hunting for topological dark matter with atomic
  clocks}, Nature Phys. 10 (2014) 933.
\newblock \href {http://arxiv.org/abs/1311.1244} {\path{arXiv:1311.1244}},
  \href {https://doi.org/10.1038/nphys3137} {\path{doi:10.1038/nphys3137}}.

\bibitem{hyperburstPage:2022ikz}
D.~Page, J.~Homan, M.~Nava-Callejas, Y.~Cavecchi, M.~V. Beznogov, N.~Degenaar,
  R.~Wijnands, A.~S. Parikh, {A
  \textquotedblleft{}Hyperburst\textquotedblright{} in the MAXI
  J0556\textendash{}332 Neutron Star: Evidence for a New Type of Thermonuclear
  Explosion}, Astrophys. J. 933~(2) (2022) 216.
\newblock \href {http://arxiv.org/abs/2202.03962} {\path{arXiv:2202.03962}},
  \href {https://doi.org/10.3847/1538-4357/ac72a8}
  {\path{doi:10.3847/1538-4357/ac72a8}}.

\bibitem{superChWD:SNLS:2006ics}
D.~A. Howell, et~al., {The type Ia supernova SNLS-03D3bb from a
  super-Chandrasekhar-mass white dwarf star}, Nature 443 (2006) 308.
\newblock \href {http://arxiv.org/abs/astro-ph/0609616}
  {\path{arXiv:astro-ph/0609616}}, \href {https://doi.org/10.1038/nature05103}
  {\path{doi:10.1038/nature05103}}.

\bibitem{superChWD:Hachisu2012}
I.~{Hachisu}, M.~{Kato}, H.~{Saio}, K.~{Nomoto}, {A Single Degenerate
  Progenitor Model for Type Ia Supernovae Highly Exceeding the Chandrasekhar
  Mass Limit}, APJ 744~(1) (2012) 69.
\newblock \href {http://arxiv.org/abs/1106.3510} {\path{arXiv:1106.3510}},
  \href {https://doi.org/10.1088/0004-637X/744/1/69}
  {\path{doi:10.1088/0004-637X/744/1/69}}.

\bibitem{superChWD:Das:2013gd}
U.~Das, B.~Mukhopadhyay, {New mass limit for white dwarfs: super-Chandrasekhar
  type Ia supernova as a new standard candle}, Phys. Rev. Lett. 110~(7) (2013)
  071102.
\newblock \href {http://arxiv.org/abs/1301.5965} {\path{arXiv:1301.5965}},
  \href {https://doi.org/10.1103/PhysRevLett.110.071102}
  {\path{doi:10.1103/PhysRevLett.110.071102}}.

\bibitem{superCh:Zuraiq:2024ypo}
Z.~Zuraiq, A.~Kumar, A.~J. Hackett, S.~Bhattarai, C.~A. Tout, B.~Mukhopadhyay,
  {Simulating super-Chandrasekhar white dwarfs}, 2024.
\newblock \href {http://arxiv.org/abs/2411.18692} {\path{arXiv:2411.18692}}.

\bibitem{superChWD:instab:Coelho:2013bba}
J.~G. Coelho, R.~M. Marinho, M.~Malheiro, R.~Negreiros, J.~A. Rueda,
  R.~Ruffini, D.~L. C\'aceres, {Dynamical instability of white dwarfs and
  breaking of spherical symmetry under the presence of extreme magnetic
  fields}, Astrophys. J. 794~(1) (2014) 86.
\newblock \href {http://arxiv.org/abs/1306.4658} {\path{arXiv:1306.4658}},
  \href {https://doi.org/10.1088/0004-637X/794/1/86}
  {\path{doi:10.1088/0004-637X/794/1/86}}.

\bibitem{superChWD:instab:Chamel:2013tfa}
N.~Chamel, A.~F. Fantina, P.~J. Davis, {Stability of super-Chandrasekhar
  magnetic white dwarfs}, Phys. Rev. D 88~(8) (2013) 081301.
\newblock \href {http://arxiv.org/abs/1306.3444} {\path{arXiv:1306.3444}},
  \href {https://doi.org/10.1103/PhysRevD.88.081301}
  {\path{doi:10.1103/PhysRevD.88.081301}}.

\end{thebibliography}
